\documentclass[12pt,preprintnumbers,pra,superscriptaddress,aps,nofootinbib,tightenlines,floatfix]{revtex4-1}
\usepackage[normalem]{ulem}

\usepackage{amsmath,amssymb}
\usepackage{graphicx}
\usepackage[utf8]{inputenc}
\usepackage{csquotes}
\usepackage{hyperref}
\hypersetup{colorlinks=true,citecolor=TealBlue,linkcolor=BrickRed,urlcolor=TealBlue}
\usepackage{url}

\usepackage[dvipsnames,usenames]{xcolor}

\newcommand{\Tr}{\ensuremath{\text{Tr}}}

\usepackage{enumitem}
\setlist{
  listparindent=\parindent,
  parsep=0pt,
}

\date{07 July 2021}

\begin{document}

\title{
Standard Model Physics
and the Digital Quantum Revolution: Thoughts about the Interface}


\author{Natalie Klco}
\email{natklco@caltech.edu}
\affiliation{Institute for Quantum Information and Matter and Walter Burke Institute for Theoretical Physics, California Institute of Technology, Pasadena CA 91125, USA.}
\author{Alessandro Roggero}
\email{roggero@uw.edu}
\affiliation{InQubator for Quantum Simulation (IQuS), Department of Physics, University of Washington, Seattle, WA 98195, USA.}
\affiliation{Address after July 2021: Dipartimento di Fisica, University of Trento, via Sommarive 14, I–38123, Povo, Trento, Italy.}
\author{Martin J.~Savage}
\email{mjs5@uw.edu}
\affiliation{InQubator for Quantum Simulation (IQuS), Department of Physics, University of Washington, Seattle, WA 98195, USA.}

\preprint{IQuS@UW-21-007}

\begin{abstract}
Advances in  isolating, controlling and entangling quantum systems are transforming what was once a curious feature of quantum mechanics into a vehicle for disruptive scientific and technological progress.
Pursuing the vision articulated by Feynman, a concerted effort across many areas of research and development is introducing prototypical digital quantum devices into the computing ecosystem available to domain scientists.
Through interactions with these early quantum devices, the abstract vision of exploring classically-intractable quantum systems
is evolving toward becoming a tangible reality.
Beyond catalyzing these technological advances, entanglement is enabling parallel progress as
a diagnostic for quantum correlations and
as an organizational tool, both
guiding improved understanding of quantum many-body systems and quantum field theories defining and emerging from the Standard Model.
From the perspective of three domain science theorists,
this article compiles \emph{thoughts about the interface} on entanglement, complexity, and quantum simulation
in an effort to contextualize recent NISQ-era progress with the scientific objectives of nuclear and high-energy physics.
 \end{abstract}

\maketitle
\tableofcontents
\markboth{}{}


\section{Introduction}

Quantum computation, simulation, and communication utilize the coherence and inherent non-locality of entanglement in quantum mechanics to store, transmit, encrypt and process quantum information with the fundamental particles and forces of nature described by the Standard Model (SM).
Advances in understanding, creating and manipulating entanglement and coherence in the laboratory,
through broad-based collaborative research programs at universities, national laboratories and technology companies,
have recently led to first demonstrations of a quantum advantage~\footnote{
A quantum advantage has been achieved
for a particular calculation when
the measured time to solution (with a given precision)
on a quantum device is less than that achievable with the most capable classical computers.} using superconducting quantum devices~\cite{Google48651,Zhong1460,wu2021strong}.
This remarkable milestone in computing,
heralds a paradigm-shifting change in how we manipulate and engage with information.
Its accomplishment was a
{\it tour de force}
in the design and fabrication of materials,
in quantum algorithms, software and circuitry,
in engineering systems controls and device fabrication,
and in the integration and isolation of classical and quantum mechanical systems.
It ushers in the second impact era of quantum mechanics---after revolutionizing our understanding of the subatomic world, now the controlled and coherent manipulation of entangled quantum states offers to revolutionize broad scientific applications from computing, simulation and theoretical developments to sensing and experiment.

First highlighted in the famous EPR paper in 1935~\cite{PhysRev.47.777},
Bell's 1964 work~\cite{PhysicsPhysiqueFizika.1.195} provided  inequalities  to quantitatively probe entanglement, a distinctive attribute of quantum mechanics.
The first experimental confirmation of violations of classical predictions in correlations between observables, in a way consistent with quantum mechanics and the nonlocality of entanglement,
was performed in 1972 by Clauser and Freedman~\cite{PhysRevLett.28.938} using atomic transitions.
Since that time, remarkable experiments probing entanglement, and more generally quantum mechanics, have been performed e.g., Refs.~\cite{PhysRevLett.49.91,PhysRevLett.49.1804}.
This growing literature has closed loop holes in the interpretation of the early experiments and continues to
inspire remarkable theoretical progress.
The meticulous design and formidable experimental prowess applied to these works have firmly established that the subatomic rules governing our universe are at least as complex as quantum mechanics, and  have been instrumental in reaching the present state of quantum information and quantum computing.

In the early 1980's,
Feynman
and others~\cite{5392446,5391327,Benioff1980,Manin1980,Feynman1982,Fredkin1982,Feynman1986,doi:10.1063/1.881299} recognized that simulations of
all but the simplest features of
complex quantum systems
lie beyond the capabilities of classical computation, and ultimately require
simulations using quantum systems themselves i.e., quantum computers, sparking extensive activity in the area of quantum information~\cite{williamsNASAconference}~\footnote{See Ref.~\cite{Preskill:2021apy} for a recent discussion of the origins of quantum computing.}.
Precisely controlled systems of cold-atoms, trapped-ions, annealers, optical, SRF and superconducting qubit devices, and related programming languages, are becoming available to scientists to begin exploring analog, digital and hybrid simulations of simple quantum many-body (QMB) systems and quantum field theories (QFTs).
These devices fall into the category of Noisy Intermediate Scale Quantum (NISQ) devices~\cite{Preskill2018quantumcomputingin}.
They are of modest size and of \enquote{OK} fidelity.
Their quantum registers are not well isolated from the environment, operations applied to the registers are imperfect, and they operate more like quantum experiments with associated systematic errors to be quantified than, say, a laptop.
There are limited (or no) protocols or redundancies included in the hardware or software
to correct for errors introduced into the wavefunction due to imperfect isolation and operation, such as phase-flips or spin-flips, i.e. there is no intrinsic error-correction and the devices are not fault tolerant e.g., Ref.~\cite{gottesman2009introduction}.
While quantum annealing devices~\footnote{
Quantum annealers, such as the D-Wave systems~\cite{Dwave}, utilize quantum tunneling to adiabatically relax to the ground state, or nearby states, of systems mapped onto
a spin model~\cite{2000quant.ph..1106F} e.g., Refs.~\cite{King_2018,Ajagekar_2020}.}
are now operating with thousands of qubits~\cite{Dwave},
systems that implement a universal quantum gate set e.g., Ref.~\cite{doi:10.1080/23746149.2018.1457981,PRXQuantum.2.017001},
such as trapped ion, superconducting qubit, Rydberg atom systems and more,
are currently in the tens of qubits range, and are expected to scale to hundreds in the next few years.
Rather than relying upon software error correction with large physical-to-logical qubit ratios, there are also major research efforts focused on  non-Abelian anyons to provide topological robustness~\cite{Kitaev_2003}, including by Microsoft~\cite{MicrosoftTopo} and the Quantum Science Center at Oak Ridge National Laboratory~\cite{QSC}.
While it is understood that high-precision calculations at scale of direct comparison to experiment are not expected during the NISQ era,
which is expected to extend over the next several years, there are good motivations to pursue a quantum advantage for observables that scale poorly with classical computation, such as real-time evolution, the structure of finite density systems and key aspects of entangled systems~\cite{NSACQISNP}.
Activities are starting toward better understanding the complexity (in the \emph{formal computer science} meaning of this word) of important problems in the SM, to determine those that may be efficiently addressed with quantum simulation.

The potential of quantum computing for scientific applications is becoming widely appreciated.
For an ideal quantum computer with an $n$-qubit quantum register, the dimensionality of its Hilbert-space is $d=2^n$.
For $n=299$, $d\sim 10^{90}$, a number larger than the number of atoms in our universe, $\sim 10^{86}$.
The exponential growth of the number of states with the number of qubits parallels, for obvious reasons, the growth of the number of states required to describe $A$-body systems.
States in a Hilbert space are accessed through unitary operations among the qubits,
and while exponential time is required to access all states~\cite{PhysRevLett.106.170501}, physically relevant states for a local Hamiltonian are generally accessible through time evolution in polynomial time from other such states.
Though a quantum advantage remains to be obtained for
a scientific application, there is a threshold in $A$ beyond which quantum simulations become the only known avenue with the potential to achieve the requisite precision.
This threshold depends upon the system, the process to be computed, the desired precision, and the efficiency of classical algorithms and implementations on high-performance computing (HPC) systems. As such, one expects to see different quantum advantage thresholds for e.g., lattice gauge theory (LGT), nuclear structure and reactions, and coherent neutrino evolution.

Nuclear and high-energy physics research is focused on the nature of a diverse array of fundamental and emergent QFTs and QMB systems, defined by unique interactions and properties.
The SM~\cite{Glashow:1961tr,Higgs:1964pj,Weinberg:1967tq,Salam:1968rm,Politzer:1973fx,Gross:1973id} and beyond,
constructed around global and local symmetries and associated particle content,
underlies the properties and dynamics of these complex systems.
Analytic and computational techniques in QMB systems and QFTs developed during the $20^{\rm th}$-Century,
proved central to establishing the SM,
in providing predictions for processes beyond experimental reach,
and in defining and using low-energy effective field theories (EFTs) and phenomenological models to describe an array of phenomena in strongly interacting, correlated systems of mesons, nucleons and hyperons.
Classes of computations projected to require beyond exascale classical resources  have been identified~\cite{Habib:2016sce,osti_1369223},
with common attributes  including finite density, real-time non-equilibrium dynamics, inelasticities, the need for large Hilbert spaces, quantum coherence and other intrinsically quantum features.
This circa 2016 recognition is daunting if only a classical HPC path forward were possible,
but exciting in light of the coinciding events of early quantum devices becoming more widely accessible and the first QFT simulation performed on a quantum device~\cite{Martinez:2016yna}.
Soon after superconducting quantum devices became available to scientists via the cloud~\cite{IBMQ,Karalekas_2020}, the first calculation of the deuteron binding energy~\cite{Dumitrescu:2018njn} was accomplished by a multi-disciplinary team at Oak Ridge National Laboratory, and
similar computations of light nuclei with EFT interactions soon followed~\cite{Lu:2018pjk,Shehab:2019gfn,PhysRevD.101.074038,yeter2020practical,PhysRevA.103.042405}.

Our article is framed from the perspective of quantum simulation of SM quantities, though is relevant for  quantum systems more broadly through communications protocols, sensing, error correction, and generic large-scale quantum computation.
Beginning with the acknowledgement that nonlocality of information is fueling the quantum revolution, Section~\ref{sec:entanglement} illustrates the polycephalic form of entanglement at the interface---including weaving an organizational fabric between QMBs and QFTs and the quantum devices that will simulate them, and illuminating hierarchies in low-energy EFTs and their phases.
Predictive capabilities for the SM in the first era of quantum mechanics advanced, in part, through the construction of effective interactions and EFTs, providing \emph{classically tractable} leading order calculations that can be systematically improved with perturbation theory e.g., perturbative quantum chromodynamics (QCD).
While most EFTs have been built around local interactions and vanishing entanglement, the prospect of capable quantum computers motivates applying this mindset toward devising \emph{quantumly tractable} leading order calculations that naturally incorporate entanglement for quantum simulation at-scale.
Where possible, the practical utility of such organizations is to extend the scientific reach of bounded-error quantum simulation, mitigating na\"ive asymptotic complexity classifications, as discussed in Section~\ref{sec:complexityclasses}.
Progress over the last few decades has established an extensive {\it toolbox} of quantum protocols
that can be used, adapted and improved for
quantum simulations of SM quantities.
In Sections~\ref{sec:QTech} and~\ref{sec:QCNPapps}, we present an overview of relevant algorithms and techniques, progress that has been made in their application to SM simulations, and three SM areas that are being pursued with available quantum devices.


\section{Entanglement
in Quantum Few-Body, Many-Body, and Field Theories
}
\label{sec:entanglement}

Beyond purist motivations of understanding the non-local correlations within our fundamental descriptions of nature, studying the structure of entanglement within the SM is expected to provide guidance both in the formulation of theoretical frameworks and in the design of associated quantum  or classical simulations.
Because entanglement is a basis-dependent property, sensitive to the way in which a physical system is mapped onto quantum degrees of freedom, a detailed understanding of the entanglement properties expressed in subatomic interactions with natural bases will inform every level of quantum simulation, from initial decisions of the quantum description to the specifications for architectural co-design.

\subsection{Identifying and Characterizing Entanglement}
A necessary and sufficient condition for identifying quantum entanglement can be expressed in terms of the separability of the density matrix describing the union of two or more Hilbert spaces $\mathcal{H} = \cup_i \mathcal{H}_i$.
For the specific case of  $\mathcal{H} = \mathcal{H}_A \cup \mathcal{H}_B$, the two Hilbert spaces may be e.g., spatial regions of a field, separate momentum scales, valence and sea quantum number contributions, or a spatial region of a nucleon and its complement with a size set by the probing momentum transfer.
If the density matrix can be written as a classical mixture of separable density matrices, $\rho_{A\cup B} = \sum_i p_i \ \rho_{A,i} \otimes \rho_{B,i}$ with $p_i \geq 0$, then the system is separable and not entangled.
If the density matrix cannot be written in this way, the system is entangled.
The existence of entangled states, flouting attempts to describe their correlations classically, is culpable for the exponential growth of Hilbert spaces (and thus classical computational resources) required to describe QMB systems.
To date, every theoretical attempt to evade this exponential state space required by non-classical correlations has been ruled out by experiment, the most famous example being that of nature's consistent response, through the violation of Bell's inequalities, that a local hidden variable description is inconsistent with observation.

While experiment indicates that a complete evasion of the exponential growth of Hilbert space is not likely to provide a successful description of nature, it is possible to systematically delay the onset of this burdensome scaling.
The language of one concrete and powerful example~\cite{Banuls:2018jag} contains at its heart the pure-state Schmidt decomposition,
\begin{equation}
|\Psi\rangle = \sum_{k = 1}^{\chi_m} \lambda_k |\Psi_{A,k}\rangle \otimes |\Psi_{B,k}\rangle \ \ \ ,
\label{eq:schmidtdecomp}
\end{equation}
the expansion of a pure quantum state in terms of the separate eigenvectors of its $A$- and $B$-space reduced density matrices.
For pure states, the two density matrix spectra, $\vec{\lambda}$, will have equal non-zero eigenvalues of number limited by the dimensionality of the smaller Hilbert space, $\chi_m$.
Only for $\chi_m$ = 1, in which local pure states remain, is the state separable across the chosen bipartition.
Entanglement-motivated approximations to the state in Eq.~\eqref{eq:schmidtdecomp} can be achieved by introducing a truncation $\chi \leq \chi_m$, with a viable and affable entanglement measure identified as $\chi$'s logarithm.
Despite the opacity of entanglement structure when an $n$-qubit state is written in the computational basis,
$|\Psi\rangle = \sum \psi_{\vec{b}}\  |b_1\rangle \otimes \cdots \otimes |b_{n}\rangle $ with $ b_i \in \{0,1\} $,
an entanglement-driven hierarchical structure of the wavefunction can be exposed through a series of sequential Schmidt decompositions ~\cite{2003PhRvL..91n7902V,Vidal:2003lvx},
\begin{equation}
  \psi_{\vec{b}} =  \Gamma_{\alpha_1}^{(b_1)} \lambda_{\alpha_1} \Gamma_{\alpha_1 \alpha_2}^{(b_2)} \lambda_{\alpha_2} \cdots \lambda_{\alpha_{n-1}} \Gamma_{\alpha_{n-1}}^{(b_{n})} \ \ \ .
\end{equation}
Here, the $\Gamma$'s are tensors of maximal dimension $\chi$ characterizing the single-qubit Hilbert space and the $\lambda$'s are locally-contracted vectors with a dimensionality of $\chi$ at the bipartition.
These $\lambda$-vectors, serving as entanglement conduits between local qubit Hilbert spaces, may be truncated, leading to a family of systematically improvable approximations to $|\psi\rangle$ with reduced entanglement.
Rather than the $2^{n}$ complex numbers capturing the exact wavefunction, the truncated description requires a memory that scales $\sim \chi^2 n$, linearly in the number of qubits but exponentially in the entanglement measure mentioned above.
Efficiently representing low-entanglement quantum states by truncating their local Schmidt decompositions underpins the robust technology of tensor networks~\cite{Banuls:2018jag}, continuing to provide leading capabilities in low spatial dimensions and for short time evolutions where local bipartite entanglement is naturally constrained.

Though the clear separability criterion above seems at first simple and innocuous, determining whether a particular density matrix can be transformed into a separable mixture is, in general, a difficult and open area of research.
In particular, there is no known necessary and sufficient criterion for separability that is also efficiently computable as the dimensionality and/or number of Hilbert spaces becomes large.
Unfortunately, \emph{large} in the previous sentence happens to be around three qubits, a pair of three-state {\it qutrits}, or the presence of a four-state {\it qudit}.
It may be no surprise that describing the complex non-local correlations that quantum systems are capable of expressing is not easily achieved through a single observable or criterion.
Rather, analogous to the Bayesian approach to data analysis through classical probabilities as a \enquote{dialogue with the data}~\cite{doi:10.1142/9789812774187_0010}, the exploration and characterization of entanglement in quantum mechanical systems often requires a collection of entanglement criteria, each one perhaps being necessary \emph{or} sufficient, whose combination of fragmented information may be synthesized into more complete logical conclusions.

For example, a quantum generalization of the Shannon entropy leads to the entanglement entropy, $\mathcal{S} = - Tr[\rho_A \log \rho_A]$, used to explore the entanglement of bipartite states that are globally pure.
The entanglement entropy is calculated in terms of the reduced density matrix, $\rho_A = \Tr_B \rho_{A\cup B}$, and is symmetric with respect to the system $A, B$ that is chosen to be traced over.
Because entanglement entropy heralds the presence of entanglement by alerting us to mixedness in reduced density matrices (pure states exhibit $\mathcal{S} = 0$), the interpretation of this entropy becomes ineffectual when the global state is already mixed.
For the language of mixed states, one may turn to conditional entropy and mutual information that combine otherwise-ambiguous marginal distributions into more informative relative entropies.
The important caveats continue as such mutual information criteria are sensitive to classical correlations in addition to quantum correlations; it is possible to find a non-zero mutual information for a classically mixed unentangled state,  and thus non-zero mutual information is necessary though insufficient for detecting entanglement.
To address this, one may look yet further to the family of positive transformations, those that retain positive eigenvalues of the density matrix.  By acting locally (subsystem) with such a transformation, any emergence of a negative eigenvalue must herald inseparability.  A common positive transformation to consider, due to its relatively amiable computability, is the transpose~\cite{HORODECKI19961,PhysRevLett.84.2726,PhysRevA.65.032314,PhysRevLett.95.090503,HORODECKI1997333,PhysRevLett.80.5239}.  Physically, the \emph{partial transpose} implements a local momentum inversion that will produce a valid, positive density matrix if the two regions distinguished by the transpose are unentangled.  Though the presence of negative eigenvalues, non-zero \emph{negativity}, is unperturbed by classical correlations,  negativity is sufficient for detecting inseparability, but it is not necessary~\cite{HORODECKI1997333,PhysRevLett.80.5239,PhysRevLett.85.2657,2001PhRvL..86.3658W}.  Aptly named, the multitude of entanglement or separability \emph{witnesses}~\cite{TERHAL2000319}, will individually fail to identify the properties they are designed to observe in a subset of Hilbert space; their strength, however, often lies in collaboration among multiple witnesses and deductive reasoning.  Following Murphy's law, and in the spirit of EFTs, it is usually a safe default to assume: if states of particular entanglement properties can exist (e.g., positive under partial transposition but inseparable), they do.

In addition to entanglement witnesses that conclusively herald but inconclusively exclude entanglement or separability, is a category of measures regarding entanglement more fundamentally as a resource for quantum information protocols.
Examples of such measures are the entanglement of formation~\cite{Wooters1998,Hill1997} or the distillable entanglement~\cite{Bennett_1996a,Bennett_1996b}, quantifying the asymptotic ratio of maximally entangled resource states necessary to produce an ensemble of the state or the ratio capable of being distilled back from a given ensemble through local operations, respectively.
While operational measures provide welcomed clear physical interpretations, they are commonly computationally prohibitive.
It is thus common to combine heralding measures with operational measures, e.g.,  the logarithmic negativity is an upper bound to the distillable entanglement~\cite{PhysRevA.65.032314,PhysRevLett.90.027901}.
Consistent with the above creation of entanglement measures in a collaborative array connected to e.g., operational communications protocols, it is likely that developing the dialogue for exploring the entanglement of QMB and QFT systems will entail the creation of an extended array of entanglement measures, inspired by the natural symmetries and entanglement fluctuating interactions of the subatomic world.

\subsection{The Role of Entanglement}

Though the landscape for quantifying the structure of entanglement in QMB and QFT systems can seem daunting and precarious at times, this connection has already begun and will continue to provide novel perspectives of entanglement...
\subsubsection{...as an organizational principle:}

    The successes of tensor network methods in the simulation of low dimensional quantum lattice models~\cite{Banuls:2018jag} arise from the use of an entanglement-centric computational framework.  By forming a basis of variational states truncated by the entanglement entropy crossing any bipartition of the space, the exponential computational demands associated with the inclusion of entangled states can be systematically mitigated and included only as necessary e.g., at long times in dynamical scattering simulations~\cite{Milsted:2020jmf}.  These successes motivate the formation of other structural hierarchies guided by entanglement.

  The success of nuclear shell-models is due, in part, to the wavefunction of many nuclei being close to a tensor product of a valence model-space wavefunction with that of an inert closed-shell core. Inspired by the advances arising upon incorporation of non-local correlations into classical computational methods, modern nuclear structure calculations, using phenomenological and QCD-based chiral nuclear forces combined with sophisticated nuclear many-body numerical algorithms and codes, are now investigating the role of entanglement as an organizational principle.  Around 2015, the role of two-orbital mutual information for organizing bases of nucleon orbitals in medium mass nuclei~\cite{Legeza:2015fja} was examined.  Extensive work by Gorton and Johnson in 2018~\cite{gorton18a} examined the entanglement entropy in light, medium and heavy nuclei using the {\tt BIGSTICK} code~\cite{johnson2018bigstick}.  Inspired by this work and the multidimensional structure of entanglement, a detailed exploration of Helium nuclei using configuration-interaction calculations for a range of single particle bases of various complexities, studied the single-orbital entanglement entropy, mutual information and two-orbital negativity~\cite{Robin:2020aeh}. Starting from matrix elements in a harmonic oscillator basis derived from chiral interactions, and self-consistently evolved  to include two-body correlations in identifying a single particle basis, the localized entanglement within a nucleus was observed to become manifest in such a natural basis e.g., decoupling the two-neutron halo of $^6$He from its $^4$He core at the level of fundamental quantum correlations~\cite{Robin:2020aeh}.
  These works show that entanglement observables are capable of providing  insight into otherwise-obscure structures that may be leveraged for a  computational advantage in exploiting future hybrid quantum-classical architectures. Subspaces with large entanglement are amenable to quantum computation, and integrated with other such subspaces using classical computation (as the quantum correlations between subspaces is small, by construction).

    Complementary to the role of entanglement as an organizational principle is the role of entanglement fluctuations in constraining the forms of dynamical interactions or in illuminating a hierarchy among operators.   For example, in high energy scattering processes at tree-level in QED, a ubiquity of maximally entangled helicity states at particular final scattering angles was observed~\cite{Cervera-Lierta:2017tdt}.  Promoting this observation to an organizational principle led to the intriguing realization that demanding maximally entangled final states to be achievable in QED and weak interactions produces, in the space of theoretical deformations to these fundamental vertices, a potential landscape that appears to be extremized by the SM.  The capability of an entanglement extremization principle to constrain the structure of interactions, similar to more familiar rotational symmetries and gauge invariance, inspires further investigation into the role of quantum correlations in the theoretical description of fundamental interactions.

    An exploration of low-energy elastic, s-wave scattering of 2- and 3-flavor nuclei through the strong interaction has demonstrated another connection between emergent symmetries and entanglement structure~\cite{Beane:2018oxh}.    Rather than the capability to produce maximally entangled final states, this work quantified the operational entanglement power~\cite{Zanardi:2000zz,mahdavi2011cross} of the $S$-matrix, finding that the presence of spin-flavor symmetries coincide with parameter regimes of diminished entanglement \emph{fluctuations}.  When symmetry in this system is heightened e.g., through Wigner's SU(4) symmetry, the singlet/triplet phase shifts vanishing/residing at unitarity, or at a kinematical $\pi/2$ phase shift difference where spin spaces are exchanged (SWAP operator~\cite{Beane:2018oxh,Low:2021ufv}), the entanglement power of the $S$ matrix becomes suppressed.  Promoting this observation to an organizational principle led to an entanglement motivation for the emergent SU(16) symmetry calculated in Lattice QCD at heavy pion mass~\cite{Wagman:2017tmp} by
    establishing a conjecture~\cite{Beane:2018oxh} that the dynamical suppression of entanglement fluctuations may drive emergent symmetries.
    This observation has the potential to
    inform the hierarchy of local operators used to design EFTs of nuclei and hypernuclei, extending beyond those from large-$N_c$ alone.

    While central to quantum communication,
    studies of entanglement and separability in Gaussian systems also describe the nonlocality in harmonic chains~\cite{Srednicki_1993,Audenaert:2002xfl,2004PhRvA..70e2329B,cerf2007quantum,PhysRevA.80.012325,Coser_2017,DiGiulio:2019cxv} or in the free lattice field theories
    directly relevant to high-energy and nuclear physics.
    More generally, entanglement studies in field theories have spanned the areas of conformal field theories (CFTs)~\cite{Calabrese:2004eu,Calabrese:2009qy,Calabrese:2009ez,Calabrese:2012ew,Calabrese:2012nk,Ruggiero:2018hyl} and near CFTs such as pQCD~\cite{Ho:2015rga,Baker:2017wtt,Tu:2019ouv,Kharzeev:2021yyf},
    AdS/CFT and gravity-dualities~\cite{Maldacena:1997re,Maldacena:2001kr,Ryu:2006bv,Ryu:2006ef},
    lattice simulations~\cite{Klco:2020rga, Klco:2021biu},
    the SM~\cite{Cervera-Lierta:2017tdt}
    and low-energy EFTs~\cite{Beane:2018oxh,Beane:2019loz,Beane:2020wjl}.
    Aiming to understand the microscopic mechanism that incorporates entanglement into EFT construction, through perhaps a small expansion parameter or generation of a mass scale, the exponential decay of distillable entanglement between disjoint regions in the massless non-interacting scalar field vacuum can be studied~\cite{PhysRevA.80.012325, Klco:2020rga, Klco:2021biu}.  In doing so,  a dimension-independent UV-IR connection indicating that the entanglement or inseparability at long distances is governed by the high momentum modes of the field is found~\cite{2004PhRvA..70e2329B,Klco:2021biu}.  Extending a result of quantum information to the realm of field theories, the observation that Gaussian states provide minimal-entanglement approximations~\cite{2006PhRvL..96h0502W} opens the possibility that such features will persist in the presence of interactions, though further exploration is required. Being a framework built upon an expansion in local operators with inherent decoupling upon the appearance of a separation of scales, EFTs may require novel methods in order to incorporate the fundamental features of non-localities.
\subsubsection{...as an order parameter for symmetry breaking:}

    A beneficial interaction has developed in recent years between quantum information, condensed matter, and SM physics. One influential direction of progress has been in the utilization of entanglement structures to identify phase transitions, provide sensitivity to probe symmetries or non-perturbative phenomena such confinement, and classify states of matter, particularly those exhibiting non-trivial topology~\cite{Klebanov:2007ws,Wen_2019}.  Because separability reflects correlations between Hilbert spaces on all length scales, entanglement measures offer valuable non-local probes of quantum mechanical structure complementary to the local observables commonly leveraged in EFTs.

    As an example in the nucleon, the scale-dependent breaking of chiral symmetry, as the valence sector interacts with the parton sea, can be tracked by observing entanglement entropy~\cite{Beane:2019loz}.  Specifically, by analyzing contributions to the  nucleon state vector on a null-plane, where the internal helicity degrees of freedom can be separated from kinematical contributions, quantum correlations arising in the presence of collective excitations perturbing the separable large-$N_c$ descriptions are an indicator of chiral symmetry breaking.  As high-energy sea partons are incorporated into the nucleon description, the entanglement between the valence sector and the sea increases.  This entanglement saturates near its maximum value, blurring the distinction of the valence spin that dominates only for a low-energy parton sea and suggesting a fundamental understanding of the small valence contributions to the nucleon spin.
    In the chiral multiplet basis, it was shown explicitly that chiral symmetry is broken only when the nucleon state is entangled, and therefore entanglement is an order parameter for chiral symmetry breaking.

    In the SM beyond the nucleon, early  quantum  simulations of  neutrino flavor dynamics are beginning to highlight the role played by dynamical phase transitions in collective neutrino oscillations~\cite{roggero2021entanglement,roggero2021dynamical}, coinciding with a modification in the structure of entanglement. Notably, these first explorations were made possible by exploiting the capabilities of tensor networks to efficiently compress many-body states with low levels of bipartite entanglement~\cite{2003PhRvL..91n7902V}.
    These systems will be discussed further in Section~\ref{sec:QCNPapps}.

\subsubsection{...as insight into the structure of hadrons:}

    From heavy quark effective theory (HQET)~\cite{Isgur:1989vq}, to the large-$N_c$ expansion, to parton distribution functions (PDFs), coherent quantum effects are often successfully incorporated by expanding around a basis that is effectively classical.
    The successes of nuclear structure and reaction calculations,
    with basis states of (color-singlet) nucleon degrees of freedom and a hierarchy of multi-nucleon interactions,
    have been a testament to the nucleon's role as a fundamental building block in the finely-tuned transition from quarks and gluons to the predominantly classical world of macroscopic physics,  i.e., nuclei have more structure than would be expected of a quark-gluon droplet.
    This perspective supports the sensibility of beginning with factorizable or tensor-product structures at leading order and systematically incorporating entanglement effects at sub-leading orders e.g., suppressed by inverse powers of a heavy quark mass, number of colors, or parton momenta.   It is the experimental accessibility and control of sub-leading quantum coherent effects that is likely to become emblematic of 21$^{st}$ century explorations.

    In the last decade, increased attention has been placed on the factorization treatment of PDFs in non-Abelian gauge theories~\cite{Collins:1989gx,Collins:2007nk,Rogers:2010dm,Cherednikov:2010tr,Buffing:2013dxa,Schafer:2014xpa,Buffing:2015aha,Adare:2016bug,Zhou:2017mpw}.  This investigation has been guided by observed correlations and asymmetries that elude description when  transverse momentum PDFs are assumed independent (factorizable) for each final state hadron.  In a phenomenon that has become known as \emph{color entanglement}, two-gluon process provide the opportunity for a coherent exchange of color degrees of freedom.  This coherent exchange produces quantum mechanical correlations between hadronic Hilbert spaces, and thus an inseparability in the PDFs that characterize them.

    From another perspective,  entanglement can be considered inherent to hadronic structure produced either dynamically or through the enforcement of symmetries~\cite{Kovner:2015hga,Kharzeev:2017qzs,Kovner:2018rbf,Tu:2019ouv,Peschanski:2019yah,Castorina:2020cro,Ramos:2020kaj,Duan:2020jkz,Gotsman:2020bjc,H1:2020zpd,Baty:2021ugw,Kharzeev:2021yyf}.   For example, spatial entanglement between regions of a proton delineated by the volume probed in deep inelastic scattering~\cite{Kharzeev:2017qzs} can be considered.  In small Bjorken $x$ regimes expected to be accessible with the Electron Ion Collider (EIC), entanglement entropy is predicted to become maximal and naturally governed by a simple logarithm of the gluon distribution.  Further evidence for partonic entanglement arising from the color-singlet structure of hadrons is expected to be accessible in LHC proton-proton collisions~\cite{Tu:2019ouv}.   A deeper understanding of the role of quantum correlations in hadronic structure is expected as quantum coherent effects
    become more accessible to experiment.

\subsubsection{...as a source of local thermalization:}

    Connected to its natural role of delocalizing quantum information, the generation of entanglement has been linked to the local emergence of statistical mechanics in unitarily evolving closed quantum systems (see e.g., Refs.~\cite{PhysRevA.43.2046,Srednicki_1994,Srednicki_1999,Eisert_2015,D_Alessio_2016,Anza2017,Anza_2018}). Recent experimental demonstration in a Bose-Einstein condensate of Rubidium atoms in a 2D optical lattice has supported this connection as local observables became statistically governed while the global quantum state remained pure~\cite{2016Sci...353..794K}.  More generally, the role of coherent evolutions has been linked to the achievement of statistical distributions more rapidly than would be expected through semi-classical approaches alone~\cite{Bell:2003mg,2003EL.....62..615G,2003PhRvA..67e2109G,2003SPIE.5111....1G,PhysRevA.70.022308}.  In the sense that non-local correlations provide a mechanism of quantum mechanical mixing in addition to scattering events, the timescale of collisions need not limit the time scale of locally emergent thermalization.  A natural application of this phenomenon is in complex, non-equilibrium systems, such as heavy-ion collisions, where thermalization timescales appear to be much shorter than na\"ively expected by cross section analysis.  Early studies in this direction suggest that entanglement may be providing mechanisms of thermalization necessary to describe experimental observations of deep inelastic scattering and heavy-ion collisions~\cite{Muller:2011ra,Akkelin:2013jsa,Ho:2015rga,Kharzeev:2017qzs,Berges:2017hne,Muller:2017vnp,Baker:2017wtt,Berges:2018cny,Feal:2018ptp,Iskander:2020rkb}, probing mixed-state subsystems of globally entangled states.

\subsubsection{...as an indicator of geometry:}

    Connections between geometry and entanglement are plentiful:
    from area laws in low-energy states of 1-dim gapped Hamiltonians~\cite{Hastings_2007,Eisert_2010} and the role of geodesics guiding complexity in the design of unitary quantum circuits~\cite{10.5555/2016985.2016986}, to classifications of correlations through entanglement polytopes~\cite{PhysRevLett.99.250405,Walter1205} and potentially the form of spacetime itself~\cite{SwingleSpacetimeEntanglement}.
    Holographic perspectives on the over-completeness of degrees of freedom in field theory descriptions~\cite{tHooft:1993dmi,Susskind:1994vu} have led to concrete examples of bulk theories of quantum gravity in hyperbolically curved space being exactly dual to a boundary CFT in one dimension lower, known broadly as the AdS/CFT correspondence~\cite{Maldacena:1997re}.
    Extending the Bekenstein-Hawking formula~\cite{Bekenstein:1973ur,Hawking:1974sw}, connecting the entropy of a black hole to the surface area of its event horizon, to the framework of holographic dualities has produced an intriguing conjecture, the Ryu-Takayanagi formula~\cite{Maldacena:2001kr,Ryu:2006bv,Ryu:2006ef}, connecting the entanglement entropy of a region in CFT to the surface area of its associated geodesic in the dual gravitational bulk.
    Beyond opportunities for tractable explorations in quantum gravity, such connections illuminate the dispersion of information into higher-dimensional embeddings---a common strategy in subatomic physics, though not usually described in this language.

    One notable recent exploration connecting  holographic dualities to geometry and entanglement for nuclei~\cite{Beane:2020wjl} has done so through the $S$-matrix formulation of scattering, with explicit examples in the $s$-wave nucleon-nucleon system.
    Rather than the EFT action, this approach focuses upon trajectories of the $S$-matrix on an abstract surface defined by unitarity.
    The entanglement power~\cite{Zanardi:2000zz,mahdavi2011cross}, providing a quantification of operator entanglement, serves as a measure of distance to non-entangling subspaces and as a source of non-trivial curvature to $S$-matrix trajectories.
    Regarding the role of entanglement and symmetry complementary to the hierarchy of local operators allows this formulation to address non-local interactions, corresponding to deviations into the bulk from inelastic effects, with a novel perspective reminiscent of holographic dualities.

\subsubsection{...as guidance to quantum simulation design:}
\label{subsubsec:entguideqsim}

    While the impact of the entanglement-guided style of thinking is expected to apply to quantum simulation design for QMB systems broadly, most concrete examples have been demonstrated in the simulation of quantum fields.
    In the theoretical planning for the quantum simulation of the scalar field, it is straightforward to realize that the entanglement structure of the  vacuum state of the free field is trivial if represented in a basis of momentum modes---a collection of tensor product oscillators and the starting point for perturbative calculations in the language of Feynman diagrams.  Considering further the entanglement structure that will dynamically grow upon the incorporation of local self-interactions indicates the production of entanglement between any momentum-conserving collections of mode oscillators.  For any representation of momentum oscillators mapped onto physical quantum degrees of freedom for computational implementation, the self interactions necessarily produce momentum-space entanglement~\cite{Balasubramanian:2011wt,DBLP:journals/qic/JordanLP14} that requires interactions in the Hamiltonian that are beyond nearest neighbor~\footnote{Note that while genuine $n$-point Green's functions cannot be reduced into 2-point contributions
    e.g., 3- and 4-nucleon forces cannot be represented by 2-nucleon forces, any unitary operator can be decomposed into nearest-neighbor two-qubit entangling operations.}.
   Motivated by a desire to mitigate the entanglement structure demands on quantum hardware, a variety of bases for digitizing the scalar field upon a spatial lattice or with single particle states have also been explored~\cite{Jordan1130,DBLP:journals/qic/JordanLP14,somma2016quantum,PhysRevLett.121.110504,PhysRevA.98.042312,PhysRevA.99.052335,Barata:2020jtq}, focusing on the necessary entangling operators for the preparation of low-energy states and the time evolution of the field.   It is expected that developing intuition in the entanglement structure of dynamical fields will guide the design of quantum simulations of more complex fields, including those with non-Abelian gauge symmetries  (see Section~\ref{subsec:mappingnonabelian}).

    Classical tensor network methods~\cite{Banuls:2018jag}  or strategies for reorganizing unitary quantum circuits~\cite{Klco:2019xro} provide viable paths for incorporating simple area-law or symmetry-embedded entanglement at lower computational cost, incorporating additional entanglement perturbatively for the approximate initialization of entangled quantum states.  Distinct from the conduit-style treatment of entanglement in classical tensor networks, truncating in entanglement traversing all possible bipartitions of the space, one can imagine leveraging correlation structures to truncate instead in the size or non-locality of entangling operators utilized in quantum simulation.  For example, guided by the knowledge of exponentially localized spatial correlations in the ground states of massive fields, preparation circuits can be constructed  in which the relevance of long-distance quantum gates are exponentially suppressed~\cite{Klco:2019yrb,Klco:2020aud}. This approach is analogous to that used in the Approximate Quantum Fourier Transform (AQFT)~\cite{PhysRevA.54.139}.  When mapping such fields to quantum architectures through a position-space distribution of qubits, the locality of correlations in the simulated field becomes manifest in the hardware implementation---interactions performed experimentally also enjoy exponential localization.  The above examples provide  inspiration for the synergy expected between fundamental studies of the structure of entanglement in QFT or QMB systems and the design of their efficient simulation using quantum devices.

\subsubsection{...as a harbinger of computational complexity:}

Absent in the above discussions is the role of entanglement in distinguishing computational complexity or in allowing for a parametric reduction in the temporal and/or spatial resources demanded by a calculation.
While  quantum speed-ups have a dependence upon the presence of sufficiently scaling entanglement that follows practically by definition~(see e.g., Refs.~\cite{braunstein2000speed,Jozsa_2003,2003PhRvL..91n7902V,PhysRevLett.100.030504} for more precise elaboration), the desired statement that entanglement is the source of quantum advantage is perhaps misleadingly broad---some categories of highly entangled states can be efficiently represented and simulated by classical computational architectures, occasionally challenging the succinct identification of quantum advantages.

The canonical counterexample to the desired simplified understanding equating entanglement and computational complexity is the stabilizer formalism~\cite{Gottesman:1997zz,Gottesman:1998hu}, $S_i|\psi_S\rangle = |\psi_S\rangle$, describing quantum states $|\psi_S\rangle$ that are stabilized in the $+1$ eigensector of a commuting set of stabilizers, $\vec{S}$.
The $2^{n}$ elements of $\vec{S}$ can be parameterized in the  basis of tensor product (extended) Pauli operators, $P = \pm \sigma_1\otimes \sigma_2 \otimes \cdots \otimes \sigma_{n}$ with $\sigma_i \in \{\mathbb{I}, X, Y, Z\}$, and uniquely identify the vector $|\psi_S\rangle$.  For example, the three-qubit ($n = 3$) state $|GHZ\rangle = \frac{|000\rangle + |111\rangle}{\sqrt{2}}$ can be described by a set of 8 stabilizers, $\vec{S} = \{ \mathbb{I}_3, XXX,  Z\mathbb{I}Z, ZZ\mathbb{I}, \mathbb{I}ZZ, -XYY, -YXY, -YYX \}$.
At first, this appears to be an exacerbation of memory requirements.  However, leveraging the group structure of the stabilizers, $S$ can be captured by its $n$ generators e.g., $G_S = \{XXX, ZZ\mathbb{I}, Z\mathbb{I}Z \}$.
Constructing a classical binary basis assigning two classical bits (with values 0,1,2,3) to each qubit Hilbert space to distinguish the four possible Pauli operators, along with an additional bit for the sign of each generator leads to the classical storage of $G_S$ in $(2n+1)n$ classical bits.    Applying a quantum circuit, $U$, to a stabilizer state can be achieved by simply transforming each of the generators in $G_S$ as $g_i \rightarrow g_i' =  Ug_iU^\dagger$.
The crucial feature of the stabilizer formalism that enables its classical efficiency throughout computation is that each  transformed generator $g_i'$ remains a single tensor product of Pauli operators, $UPU^\dagger = P'$, \emph{if} the circuit is in the Clifford subgroup $U \in \{ H, S, {\rm CNOT}\}$ (which also generates the Pauli operators)~\footnote{$\{ H, S, {\rm CNOT}\}$ are the Hadamard, S-phase, and controlled-NOT gates, as discussed in Ref.~\cite{9781107002173}.}.
Thus, any circuit {\it acting on a stabilizer state} that is comprised of $\{ H, S, {\rm CNOT}, X, Y, Z\}$ gates is efficiently classically simulatable~\cite{Gottesman:1997zz,Gottesman:1998hu}.
This includes the creation of highly entangled states e.g., GHZ states and their $n$-qubit extensions, as well as non-trivial quantum protocols e.g., quantum teleportation.
Of course, to produce universal quantum computation one requires additionally the quarter phase $T$-gate~\cite{Kitaev_1997,2005quant.ph..5030D}, understood in this context to proliferate tensor product Pauli operators, $TPT^\dagger = \sum P'$, and thus forces one to abandon the classically efficient representation.
It is for this reason that $T$-gate count can be a meaningful quantity characterizing simulation complexity, providing a source of departure from a classically efficient computational strategy.
Furthermore, though the state space of the stabilizer formalism is highly restricted
(e.g., any non-zero amplitude must have the same value and all measurements must be either deterministic or exhibit a random 50\% probability)
and will not capture the elaborate wavefunctions of
SM systems, the formalism is an important and accessible language for the exploration of quantum error correction codes~\cite{Gottesman:1997zz}.
The high degrees of structure and symmetry presented by stabilizer states e.g., measurement constraints or restrictions to a finite octahedron of points on the Bloch sphere, further support the above discussion that symmetry is important when pursuing a connection between entanglement and computational complexity.
As a constructive example of the Gottesman-Knill theorem (see Ref.~\cite{Aaronson_2004}), this formalism demonstrates that there exists a subset of quantum computation that is encapsulated within the complexity of classical computation and yet includes the production of states highly entangled in the computational basis.
From a complementary perspective, as mentioned above, the organizational structure of tensor networks suggests a necessary condition that entanglement grows more rapidly than logarithmically with the system size in order to support exponential computational advantages~\cite{2003PhRvL..91n7902V,PhysRevLett.100.030504}.
Thus, to truly understand the role of experimentally controlled entanglement in expanding the accessibility of non-equilibrium properties and dynamics in QMB and QFT systems, it will be necessary, as demonstrated in the growing literature on entanglement measures, to push well beyond binary classifications of separability to understand how to work with, rather than against, the complex and subtle structure of entanglement.

One of the challenges we face, as theorists \emph{at the interface}, comes from the fact that much of our QFT tool-set developed in the $20^{\rm th}$-Century is based around the locality of interactions
and the enormous success of the operator product expansion in understanding high-momentum processes used to establish the SM.
Our experience with systematically improvable calculation frameworks
for non-local quantities is limited, one example of which is the inclusion of QED in finite-volume lattice QCD+QED calculations, see e.g., Refs.~\cite{10.1143/PTP.120.413,Borsanyi_2015,FODOR2016245,Hansen_2018,Davoudi:2018qpl}.
This points to a necessity of learning new ways to think about our calculations, including developing intuition about the inclusion and organization of non-localities and entanglement.
Einstein is quoted to have said \enquote{To raise new questions, new possibilities, to regard old problems from a new angle, requires creative imagination and marks real advance in science.}
From this new angle of quantum information, the necessary creative imagination can be sourced naturally from the  reality of nature at the microscopic scale.
Celebrating the versatile impact that such non-perturbative inspiration can provide, it is likely that the new perspectives developed in the process of exploring quantum simulation will provide valuable insights throughout existing techniques and algorithms, both quantum and classical e.g.,~Ref.~\cite{Osborne_2002}.


\section{The Vision of Quantum Simulation}
\label{sec:Qsim}

As the rules of quantum mechanics remain unchanged over vast energy scales,  the precise control of atomic-scale quantum devices is expected to provide naturally matched complexity for expressing quantum systems from QCD to chemistry.
This notion of intersimulatability is fundamentally connected with the concept of universality of a computational device: a universal computer is capable of  efficiently simulating the behavior of a different computer belonging to the same class, or one contained therein.
As such, important aspects of QFT and QMB systems are anticipated to become accessible when incorporating quantum degrees of freedom directly into computational architectures.

\subsection{\enquote{Gedanken} Scaling of Quantum Resources}
\label{sec:sim}
\noindent

Feynman introduced the concept of a  universal quantum simulator as a computational device constructed with quantum mechanical degrees of freedom (such as spins) that would be capable of simulating the behavior of compatible different quantum systems~\cite{Feynman1982,Feynman1986}. The question that remained open was how many classes of universal quantum simulators are there or, in other words, what controls the intersimulatability of different quantum systems?
To make this question more  concrete, a physical quantum system can be approximately encoded in a finite number of discrete degrees of freedom to be representable with a finite number of computing elements.
The efficiency of a simulation is then determined by the scaling of required resources with increasing system size.
In particular,
{\it efficient} computational strategies are those with resources increasing
at most as a polynomial in the size of the system.
The physical intuition followed by Feynman was that a quantum simulator constructed with locally-interacting quantum degrees of freedom should be able to efficiently simulate the behavior of a large class of other quantum systems also described by locally-interacting quantum degrees of freedom. This intuition was finally proved correct by Lloyd~\cite{Lloyd1073},
who was able to show that, indeed, mutual locality allows mutual simulation. The key insight was to configure the quantum computer to approximately simulate the dynamics over short time-intervals while controlling the total error accumulated during the entire evolution.
The simulation strategy can accommodate fermion statistics, and requires a  compute time that is appealingly proportional to the total time
of the internal simulation.
Prior to this milestone result,
Shor's algorithm for factoring~\cite{365700} was the
main candidate shown to provide a future quantum advantage,
na{\"i}vely quite distant from Feynman's initial vision.

Efficient simulation strategies on universal quantum computers have been formulated for a variety of SM systems from quantum chemistry~\cite{PhysRevA.90.022305,Reiher7555} to fermion-lattice models~\cite{PhysRevA.92.062318,PhysRevLett.120.110501,Kivlichan2020improvedfault},
including pionless EFT~\cite{Kaplan:1998tg,Kaplan:1998we,vanKolck:1998bw} on a lattice~\cite{PhysRevD.101.074038}, and some relativistic QFTs~\cite{PhysRevA.73.022328,Jordan1130,DBLP:journals/qic/JordanLP14,Shaw2020quantumalgorithms,PhysRevD.102.094501}.
Once again, physical intuition suggests that to simulate  a quantum system with $n$ degrees of freedom interacting locally for a time $T$, one would need a quantum device with $\mathcal{O}(n)$ degrees of freedom and perform $\mathcal{O}(T)$ time steps, for a total of $\mathcal{O}(nT)$ operations.
For a system on a lattice, this is equivalent to the expectation that the time needed for simulation would be independent of the volume of the system.
First suggested in Ref.~\cite{DBLP:journals/qic/JordanLP14},
this idea was ultimately shown to be correct with a further proof that the strategy has optimal asymptotic scaling with the number of operations~\cite{doi:10.1137/18M1231511}.
The main caveat is that, in order for the simulation time to be independent of the volume, the degrees of freedom in the quantum simulator should be organized on a lattice with the same dimensionality as the simulated system, e.g.,
a 2-dim array of spins interacting locally will not in general be sufficient to simulate a 3-dim system with local interactions in time $\mathcal{O}(T)$ that is independent of $n$.
The introduction of fermions usually leads to geometrically non-local interactions and, in general,
simulation run times for quantum algorithms are expected to scale as $\mathcal{O}(poly(n)T)$, still an exponential speed-up with respect to classical simulation strategies.

\subsection{Quantum Algorithm Development: \enquote{Gedanken} to Reality}
The first proposal for a universal quantum simulation algorithm in Ref.~\cite{Lloyd1073} used the Lie-Trotter~\cite{10.2307/2033649} approximation for the time evolution operator, which requires a number of operations scaling as $\mathcal{O}(T^2/\epsilon)$
(per measurement)
in order to ensure a maximum error $\epsilon$ in a simulation of time $T$.
Higher order Suzuki-type integrators~\cite{doi:10.1063/1.529425}  enjoy a better scaling $\mathcal{O}(5^{k}T^{1+\frac{1}{k}}/\epsilon^{\frac{1}{k}})$ with $k\in\mathbb{N}$ the approximation order.
Due to the exponentially increasing prefactor, these approximations are typically used  only with $k=1,2$ (e.g., Ref.~\cite{Childs9456} for a more detailed discussion). More recently, alternative approaches to simulating real-time evolution with a reduced dependence on $\epsilon$
have been proposed, enabled by increased qubit overhead.
Important examples are techniques based on the linear combination of unitaries (LCU)~\cite{Childs2012} and the Taylor expansion~\cite{PhysRevLett.114.090502,7354428}, which achieve an exponential improvement in the error scaling as $\mathcal{O}(T\frac{\log^2(1/\epsilon)}{\log\log(1/\epsilon)})$, and also techniques based on Qubitization and Quantum Signal Processing (QSP)~\cite{PhysRevLett.118.010501,Low2019hamiltonian}, which are able to achieve the optimal scaling $\mathcal{O}(c_T T + c_\epsilon  \log(1/\epsilon))$. Better than $\mathcal{O}(T)$ scaling is forbidden by the no-fast-forward theorem {\it unless} some of the structural properties  of the Hamiltonian are used in the design of the simulation scheme~\cite{Berry2007,Atia2017}, i.e. codesign.
It is important to point out that we do not expect to be able to implement generic unitary transformations~\cite{Knill:1995kz} and that being able to construct a qubit or qudit  Hamiltonian describing a physical system is a necessary but not sufficient condition for the existence of an efficient simulation strategy (see, e.g., Ref.~\cite{Childs2010}).

Besides second quantization schemes, an alternative mapping can be obtained using first quantization language where the particle's statistics are included by explicit symmetrization~\cite{PhysRevLett.79.2586,Aspuru-Guzik1704}.
The main advantage of working in this basis presents itself in situations where the number of particles $\eta$ in the system is conserved and discretization over a large number of fermionic modes $N$ is required---in second quantization the required number of qubits is at least $n=N$ (using e.g., the JW or BK mappings described in Sec.~\ref{subsec:fermions} below),
while in first quantization this can be reduced exponentially to $n=\mathcal{O}(\eta\log(N))$ instead~\cite{Babbush2019} when
$N\gg\eta\gg 1$.
This reduction in Hilbert space complexity is typically accompanied by an increase in time complexity and, for some  important situations, the two descriptions have comparable simulation performance (see, e.g., Ref.~\cite{su2020nearly}).
Analogous symmetry or conserved quantity projections are also under development in simulations of gauge field theories (see Section~\ref{subsec:mappingnonabelian}), and are found to increase classical preprocessing requirements.

More recently, growing interest in reducing the number of quantum gates required for realistic simulations has led to an increasing appreciation of the potential benefits of including stochastic components to  simulation algorithms.
Notable ideas in this direction are, for example, randomizing the order of exponentials in the Trotter expansion of the evolution operator, resulting in effective higher-order integrators~\cite{Childs2019fasterquantum}, and utilizing the fully stochastic compilation of Trotter evolution with the quantum stochastic drift protocol (QDRIFT)~\cite{PhysRevLett.123.070503} and its extensions~\cite{chen2020quantum,faehrmann2021randomizing,Berry2020timedependent}.
Further developments toward realistic simulations have leveraged EFT renormalization group (RG) ideas to prioritize the low energy subspace in dynamical simulations~\cite{2020arXiv200602660S}.

Due to the finite gate fidelities of real quantum devices, an inverse relationship exists between the number of qubits in a time-evolved system and the number of Trotter steps feasible within the \enquote{gate fidelity coherence time}.
As discussed by  Martonosi and  Roetteler~\cite{martonosi2019steps}, different algorithms for simulations with different target objectives will perform optimally on different architectures.
In particular, the  scaling of the number of gates with increasing system size for a given algorithm will determine performance on near-term devices with given error rates.
The important connection between gate errors and number of qubits for pursuing computation has led to the exploration of more holistic metrics than qubit number or error rate alone e.g., the quantum volume~\cite{Moll_2018}.
This metric characterizes the performance of \enquote{square} circuits e.g., if a 5 qubit depth 5 circuit provides reliable results (so defined from a random quantum circuit), but a 6-qubit, depth 6 circuit does not, then the quantum volume is $2^5 = 32$.

\begin{figure}[!ht]
\centering
\includegraphics[width=0.98\columnwidth]{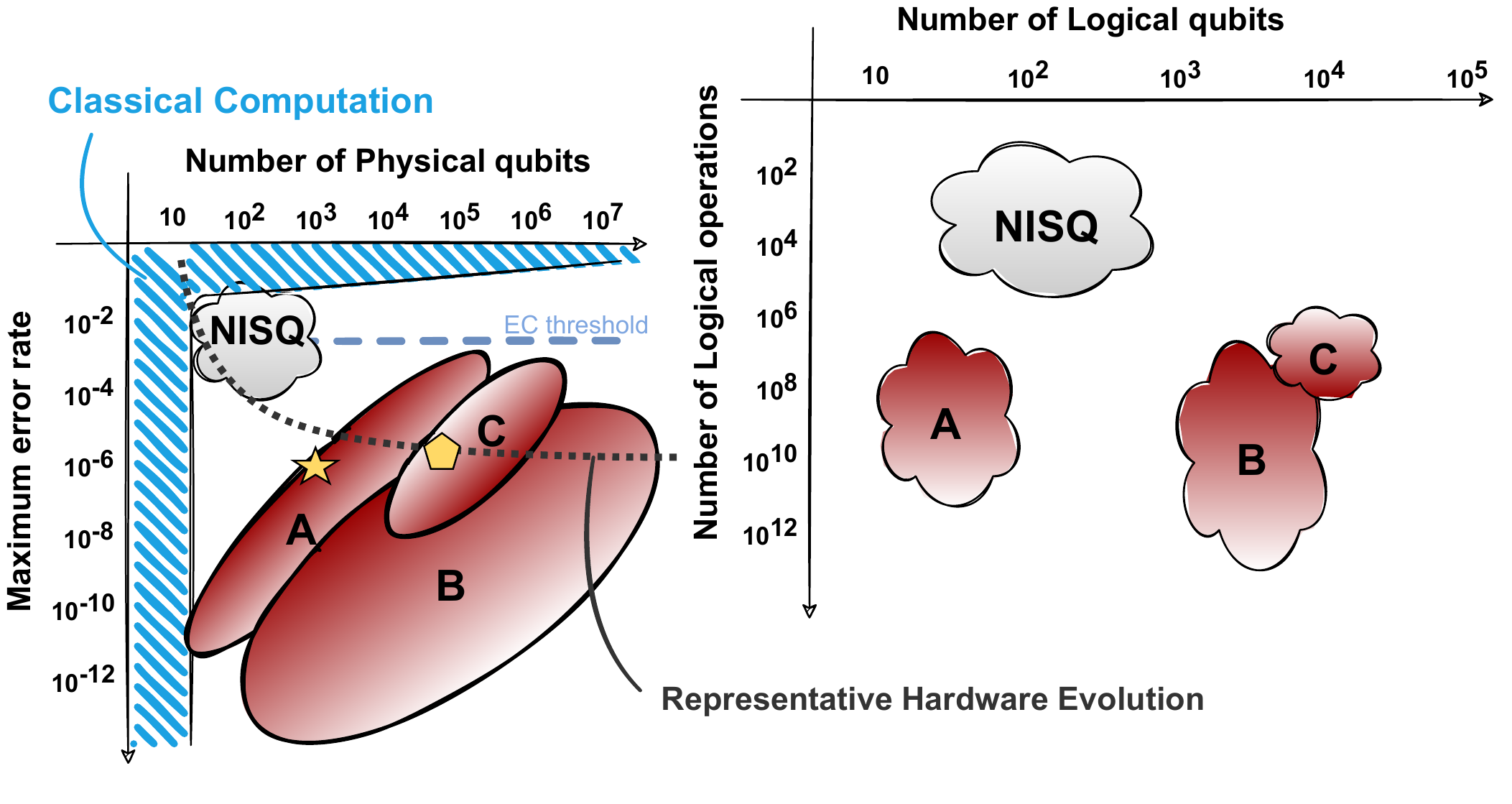}
\caption{
Schematic representation of simulation configurations defined in terms of their logical  (right panel) and  physical (left panel)  resource requirements on the device used for implementation.
See the text for further discussions of this diagram and example problems belonging to each region.
For this diagram,  NISQ devices are considered without error correction.
This figure is inspired by those of Refs.~\cite{martonosi2019steps,googlebristlecone}.
}
\label{fig:fig1}
\end{figure}

In light of the possibility of using error correction schemes to encode quantum information in a redundant way allowing for a direct reduction of the effective error rate afflicting a given simulation, it's important to realize that quantum algorithms requiring seemingly very different logical quantum resources could be implemented successfully on the same general purpose quantum machine. In order to illustrate this point, we show in the right hand panel of Figure~\ref{fig:fig1} three indicative configurations of computational problems (denoted with labels $A$,$B$ and $C$) requiring different logical resources. The gray region indicates instead the regime in logical resources we expect to be available in the near term NISQ era: a few hundred to a thousand logical qubits and a few thousand logical operations. By means of encoding, the problem classes defined at the logical level can be mapped to wide regions in terms of physical resources. Provided the error rate in the physical device is smaller than the threshold for error correction (indicated with a dashed blue line on the left panel of Figure~\ref{fig:fig1}), it is possible to trade a larger number of physical qubits for lower error rates. As a concrete example, the configurations of problems denoted as $A$ in Figure~\ref{fig:fig1} could be implemented successfully on both a small quantum device with $\mathcal{O}(10^2)$ physical qubits and very small error rates $\mathcal{O}(10^{-12})$ or equivalently on a large quantum device with $\mathcal{O}(10^5)$ physical qubits and much larger error rates $\mathcal{O}(10^{-4})$. Due to the formidable engineering challenges in reducing the limiting noise level to arbitrarily low levels, a commonly adopted strategy in the pursuit of universal quantum computing devices e.g., as discussed by technology companies~\cite{googlebristlecone},  is to first achieve error rates a few order of magnitude below the error correction threshold, and then develop larger qubit arrays that will enable scalable fault tolerance. A representative path for this hardware development is shown as the dotted black line in Figure~\ref{fig:fig1}.

While estimates of classical computing resources are well established in many areas, such activities for quantum computing are just beginning.
Research teams have undertaken extended periods of resource estimation and coordinated algorithm development
tied to present-day simulation capabilities
e.g., Refs.~\cite{PhysRevA.90.022305,Reiher7555,Babbush2018LowDepthQS,lee2020efficient},
to reduce naive quantum resource requirements from, in some cases,  billions of device-years to a handful of device-days.
One of the common standards for performance measures and resource estimation is the Hubbard model and other spin systems with \enquote{simple} interactions, e.g., Ref.~\cite{Childs9456}. At this point is important to highlight the fact that problem instances depicted in both panels of Figure~\ref{fig:fig1} are necessarily narrowly defined problems with clearly stated observables and target precision goals, examples of problems belonging to the 3 configurations depicted are:
\begin{itemize}
    \item configuration A: computing the ground-state fermion condensate in the Lattice Schwinger model with $N=16-64$ lattice sites and a electric field cutoff $\Lambda=2-8$ using an optimized scheme without amplitude estimation~\cite{Shaw2020quantumalgorithms}
    \item configuration B: estimating the frequency dependent, inclusive response function with resolutions $\Delta\omega=10-100$ MeV, at one value of the momentum transfer for a system of $A=\mathcal{O}(50)$ nucleons described using Lattice EFT~\cite{LEE2009117} with a minimal basis defined on a $10^3$ lattice~\cite{PhysRevD.101.074038}
    \item configuration C: minimal instance of a problem in financial derivative pricing showing a quantum advantage~\cite{Chakrabarti_2021}. The estimate is that a gate depth of $54\times 10^6$ on 8K logical qubits will be required.
\end{itemize}
With rapidly increasing focus on developing quantum algorithms to address scientific applications, and on dedicated and general purpose hardware, benchmarks and estimates of quantum resources requirements are expected to evolve rapidly.
In some instances, the estimated resources will increase as our knowledge improves and na{\"i}vety recedes, but  reductions in resource requirements for quantum advantages in scientific applications are anticipated as methodologies become more sophisticated and devices advance.

The flexibility enabled by error correction mappings inspires the viability of engineering universal quantum devices able to tackle a variety of problems with widely different computational requirements at the logical level. For our three problem configurations, for example, a quantum device characterized by the yellow pentagon in the left panel on Figure~\ref{fig:fig1} would be able to execute instances from all three configurations by using various levels of error correction. On the other hand, we expect near term applications in the NISQ era to use little or no error correction and the gray region of logical resources in the right panel of Figure~\ref{fig:fig1} will be mapped, almost unaltered, onto the physical resources. One complementary route to early success in obtaining a quantum advantage for a problem of scientific interest is the adoption of ideas of codesign to develop special purpose architectures tailored to specific problems~\footnote{
In the development of high-performance classical computing
in the US, the {\tt QCDSP} and {\tt QCDOC} systems~\cite{Christ2011} were specifically codesigned  for lattice QCD, with on-chip memory and low-latency communication fabric.  These systems constituted the precursor to IBM's successful {\tt Blue-Gene} series of HPC systems.
There were parallel developments
in Europe, with the {\tt APE} machines~\cite{Marinari1986}.
}.
Codesign of special purpose machines has the potential of drastically impacting the mapping between logical computational resources and physical ones displayed in Figure~\ref{fig:fig1}. For instance, it could be possible to design a quantum device with the physical properties denoted by the yellow star in Figure~\ref{fig:fig1}, which would be able to perform simulations of problems belonging to configuration $A$ more efficiently but unable to tackle problems belonging to other configurations.

The arrival of heterogeneous classical architectures led to an extended period of time during which scientific applications were ported from CPU-only to CPU-GPU hybrids, and now, most of the performance resides on the GPUs and not CPUs.
We anticipate the same sort of transition in the introduction of QPUs into heterogeneous computing~\cite{PRXQuantum.2.020343}.
Even assuming that there is a polynomial speed up to parts of the calculations, there is benefit in identifying  parts of  calculations that would be accelerated by a QPU.
In particular, computations of highly entangled subsystems could be performed on QPUs, while the classically efficient elements of the computational workflow, and re-assembly of the results from the QPUs,  could be performed with classical computers.
Analogous to the widespread adoption of GPU frameworks in scientific computing aided by the standardization of the hardware interface, standardization at both the low-level API and abstracted hardware-specific compilers will provide portability and high-level approaches to quantum algorithmic design. A variety of different framework and standards are currently being explored e.g., Cirq~\cite{Cirq}, Qiskit~\cite{gadi_aleksandrowicz_2019_2562111}, Qiskit Pulse~\cite{Alexander_2020}, OpenQASM~\cite{cross2017open},  xacc~\cite{McCaskey_2020}, Q\#~\cite{Qsharp}.

Systematic uncertainty quantification generally requires an ensemble of simulations with strategically chosen input parameters such that extrapolation to asymptotic configurations beyond
the capability of a single simulation configuration
(a {\it hero run})
can be achieved.
This ensemble maps out a surface in parameter space
that enables extrapolations and interpolations to reliably estimate observables of interest and their uncertainties e.g., Ref.~\cite{PhysRevX.7.021050,PhysRevLett.119.180509,Kandala_2019},
injecting another layer of analysis to resource estimation.
As is the case for classical calculations and experimental design, designing an optimal manifold of simulations to achieve the desired precision in observables requires planning and simulations of the simulations~\footnote{
While hero runs are not the pure focus of simulation, they may be the relevant mode of operation for other quantum applications.}.

Part of designing the simulation workflow involves identifying and optimally interleaving such calibrations with the physics simulations.
While it is obvious that the nature of the calibration is hardware dependent, they will also depend upon the structure of the quantum circuits being executed.
Simultaneous adoption of a suite of error mitigation protocols~\cite{PhysRevX.7.021050,PhysRevLett.119.180509,PhysRevX.8.031027,Kandala_2019,Dumitrescu:2018njn} has been used to assess the reliability of the extraction of physical quantities and their sensitivity to the specific noise properties of the device being utilized, e.g. Ref.~\cite{PhysRevD.101.074038}.
As NISQ-devices currently resemble experiments more than classical computers,
device calibrations that are temporally correlated (integrated)
with physics calculations
are valuable to {\it in vivo} monitor and
subsequently identify physics data accumulated during periods in which the device is performing within specified parameters.
An example of such \enquote{cuts} is provided in Ref.~\cite{Klco:2019xro}, where calibration circuits that measured the {\it in-medium} fidelity of a Hadamard gate were included to provide  contextualized performance information in initializing symmetric wavefunctions.

Another important aspect of workflow depends upon the (user defined) partitioning of
the simulation into classical and quantum components, dependent on the performance of the QPU relative to the CPU+GPU, and the efficiency of communication between them.
The current state of the NISQ ecosystem mandates that scientific applications must include a diverse range of integrated controls and
diagnostics of the quantum hardware,
both accessible and inaccessible to the users.
In addition to making available the first cloud-accessible quantum computers, IBM~Q~Experience~\cite{IBMQ} provides regular calibration data for each of their superconducting quantum systems, built-in functionality for some error mitigation protocols, and access to pulse-shaping for gate optimization through their {\tt Qiskit}~\cite{gadi_aleksandrowicz_2019_2562111} programming environment---demonstrating a successful path (and potential template) for co-design collaborations with universities and national laboratories toward scientific applications.

\section{Complexity Classes in Context: Asymptopia Versus Reality}
\label{sec:complexityclasses}

\subsection{Asymptopia: The Utopia of Infinite Resources}
\label{ssec:Asymptopia}
No discussion of quantum resources would be complete without considering complexity classes in the formal sense of computer science, see e.g., Refs.~\cite{watrous2008quantum,10.1145/167088.167097,doi:10.1142/9789810248185_0004,complexityhighlevel}.
The scaling of computational resources required to determine a given observable with increasing system size for {\it asymptotically} large systems and resources defines the complexity class within which the observable resides.
For instance, the {\bf P} class (PTIME) contains all problems that require computational time
scaling  as a polynomial of systems size (on a deterministic Turing machine), and is loosely defined by the set of problems that can be solved efficiently.
In some cases, the assignment of problems to complexity classes depends on the performance of known algorithms, though not all classifications provide constructive guidance.
Because classical complexity classes, defined with respect to Turing machines (deterministic or non-deterministic), do not capture the scaling of problems addressed with quantum devices, additional \emph{quantum} complexity classes have been introduced.
For example, {\bf BQP} is the complexity class containing problems with solutions of bounded-error requiring  polynomial-scaling
time using quantum resources.  Since a quantum circuit can simulate a classical one, both {\bf P}, and it's probabilistic generalization {\bf BPP}, are contained in {\bf BQP}.
Jordan, Krovi, Lee and Preskill~\cite{Jordan_2018} have shown that scattering within
interacting scalar QFT with external classical sources lies within {\bf BQP-complete}, indicating that all problems in {\bf BQP} can be mapped to scattering in scalar QFT with polynomial-scaling time to solution,
and further that the scattering problem itself is in {\bf BQP} and thus efficiently simulatable quantum mechanically.

While it would have been convenient if all scientifically interesting problems were in {\bf BPP} or {\bf BQP}, the unfortunate fact~\footnote{This fact may be rather fortunate if this complexity is essential to support life.} is that they are likely not.
Many important problems lie in {\bf NP} (nondeterministic polynomial time), requiring beyond polynomial resources to solve (assuming {\bf P }$\neq$~{\bf NP}) but only {\bf P} resources to verify solutions.
The quantum generalization of {\bf NP} with polynomially intractable problems is denoted by {\bf QMA}, and problems belonging to this class are not expected to be solved efficiently even with a quantum computer (unless {\bf BQP}$=${\bf QMA}). Similar to the {\bf NP}-completeness of finding the ground state energy of a classical Ising spin glass due to an exponentially vanishing gap to the first excited state (see e.g., Ref.~\cite{Barahona_1982,10.1145/335305.335316,Troyer_2005}), the problem of determining the ground state energy of a $k$-body Hamiltonian is {\bf QMA-complete} for $k\geq2$~\cite{10.1007/978-3-540-30538-5_31}. A $k$-body Hamiltonian (often referred to as a $k$-local Hamiltonian in the computer science literature, though no sense of spatial locality is intended~\footnote{
The most interesting aspects of $k$-local$_{\rm cs}$  (as defined by computer scientists) systems to physicists are likely the $k$-nonlocal$_{\rm phys}$ attributes induced by entanglement.})
is defined as an operator that admits a decomposition into a polynomial number of terms acting individually at most on $k$-qubits. Interestingly, the $2$-body Hamiltonian problem was shown to be {\bf QMA-complete}~\cite{10.1007/978-3-540-30538-5_31} also when restricted to spatially local interactions on a 2D grid. These results show that it will not be possible to solve efficiently for ground-state energies of even simple Hamiltonians in general, similar to the {\bf NP-hardness} of a general solution to the sign problem in Quantum Monte Carlo methods~\cite{Troyer_2005}.

It is important to point out that membership to a particular complexity class is a statement about the worst-case instance of a particular problem, and does not necessarily reflect the average- or best-case. A great example of this dichotomy between average case complexity and worst-case complexity is the Minesweeper game: in this game, players are given a 2D lattice with $L^2$ sites containing $M$ mines.
When a site without a mine is uncovered, the site is assigned a numerical value corresponding to the number of neighboring mines.
The goal is to uncover all sites not containing mines using inference from an initial set of uncovered sites.
This problem has been shown to be NP-complete~\cite{Kaye2000} (or more precisely co-NP-complete~\cite{Scott2011}) and therefore extremely hard to win in the worst case scenario.
For those who have played Minesweeper, this result may clash with practical experience, in which the game can often be solved very quickly. An important structural property of the Minesweeper game is the mine density $\rho=M/L^2$, and it was recently argued that a complexity phase transition occurs once the mine density exceeds a critical value $\rho_c$~\cite{dempsey_et_al:LIPIcs:2020:12773}. At small mine density, the game is typically  easy and inference can be carried out locally on the lattice. As the critical mine density is approached,  successful inference of the mine's location requires consideration of lattice patches approaching the full volume of the lattice, causing solutions to become expensive to find. This \enquote{frustration effect} shows up in certain regimes of a class of {\bf NP-complete} problems, for which a prototypical example is the phase transition in random k-SAT problems~\cite{1999Natur.400..133M}.
This discussion draws obvious analogies with chiral symmetry breaking, confinement, and the QCD phase diagram, where transitions are accompanied by the delocalization of natural degrees of freedom capturing the microscopic properties of quarks and gluons.

\subsection{Bounded Error: Uncertainties and Approximate Simulations}

To effectively connect quantum simulations of SM observables to experiment, a complete quantification of uncertainties is required.
Lattice QCD calculations have established well-defined uncertainty quantification methodologies~(for a review, see Ref.~\cite{Beane:2014oea}), in addition to demonstrating the benefits of independent teams of researchers and code developments,
and comparable access to independent computing environments.
Lattice QCD is a low-energy EFT of QCD that can faithfully reproduce low-energy QCD observables.
Such simulations involve discretizing volumes of spacetime, introducing a lattice spacing, and bounding a spatial extent in each of the four spacetime directions. Therefore,  observables will necessarily deviate from their QCD value.
In addition, as simulations only involve dimensionless quantities, an overall length scale is required to be determined,
along with input values for the quark masses.
This is typically achieved by making comparisons with a small number of precisely known experimental quantities, such as hadron masses.
Such tunings are imperfect and iterative, and introduce further deviations from QCD values.
EFTs,
such as $p$-regime chiral perturbation theory ($\chi$PT),
play an essential role in post-computational processing of simulation results to make reliable predictions of QCD with fully quantified uncertainties (see, e.g., Ref.~\cite{Golterman} for mesons and Ref.~\cite{Savage:2006et} for nuclei and multi-nucleon scattering systems).
For small enough lattice spacings, compared with the QCD scale and external kinematics,
the Symanzik action~\cite{Symanzik:1983dc,Symanzik:1983gh} includes the structure of the gauge-invariant operators that may be introduced into the simulation (dependent upon the nature of the field discretization) to compensate for the finite lattice spacing in each direction,  or alternately dictate the nature of the power-series in the lattice-spacing(s).
While the quark-mass dependence of a number of low-energy observables is known from $\chi$PT, or HQET, calculations are typically performed in the neighbourhood of the physical point (or the mass points of interest) and interpolations are performed to mitigate tuning errors.
On top of systematic errors, there are statistical errors and the range of behaviors of observables with sampling.
The extraction of energies and observables requires sufficient
statistics (sampling of gauge configurations) to recover correlation functions with sufficient precision to be able to fit relevant quantities with controlled estimates of mean values, statistical uncertainties and systematic errors.

For quantum simulations, there are further systematic errors associated with the quantum hardware and the qubit representation of the simulated system.
Mapping bosons onto a finite Hilbert space introduces  digitization errors and field truncation errors, to be quantified and/or mitigated.
There are errors associated with state-preparation (similar to source structure in Euclidean space, which may inform quantum simulations), and measurement, and both are at early stages of being understood and quantified.
Other sources of error that must be quantified are device noise and field pixelation (operator smearing) on quantum correlations
(i.e., the structure of entanglement between separated regions of a latticized field, e.g. Ref.~\cite{Klco:2020rga,Klco:2021biu}).
The systematic and statistical uncertainties associated with a
quantum simulation (determined by the device, algorithms and theoretical framework) can be written as,
\begin{equation}
\epsilon^{(t)} \sim
\epsilon_{\rm th.} +
\epsilon_a +
\epsilon_L +
\epsilon_\theta +
\epsilon_M +
\epsilon_{\rm Meas.} +
\epsilon_{\rm ops.} +
\epsilon_{\rm digital} +
\epsilon_\Lambda +
\epsilon_{\rm Trotter} +
\epsilon_{\rm Noise}
\  ,
\end{equation}
where
$
\{\epsilon_{\rm th.},
\epsilon_a,
\epsilon_L,
\epsilon_\theta,
\epsilon_M,
\epsilon_{\rm Meas.},
\epsilon_{\rm ops.} \}
$,
associated with the theory, lattice spacing, volumes, parameter tuning, scale setting, measurements and operator structures, are familiar (but some will be handled  differently)  from classical simulations.
The
$\epsilon_{\rm digital},
\epsilon_\Lambda,
\epsilon_{\rm Trotter},
\epsilon_{\rm Noise}
$
associated with field digitization and mappings, truncations in field space, Trotterized time evolution, and device noise,
are further sources of uncertainties that need to be quantified.

The statistical behavior of an observable, and how it is constructed from ensemble measurements, is another important feature of algorithmic design.
With direct access to the wavefunction of a system not practical (tomography scaling exponentially with system size), optimizing observable estimators is essential for efficient use of quantum simulation (as it is  classically).
Cancellations between contributions to expectation values can create sign problems at the measurement stage of a calculation, exacerbating
the number of circuit executions required to achieve a given precision e.g., Ref.~\cite{PhysRevA.101.022328}.
Because the statistical structure of the measurements is mapping (basis) dependent, it is possible that  circuit design and algorithms with less favorable asymptotic scaling  may provide compensating enhanced statistical convergence.

\subsection{Reality: Hacking Bounded-Error Quantum Simulations}

The complexity classes of SM observables will determine the observables that can be computed with arbitrary precision using combinations of classical and quantum resources.  However, if they lie outside {\bf P} or {\bf BQP}, it does not mean that significant progress, and potentially all of the required progress,
cannot be made in addressing important problems,  gaining significant physical insights and even proposing new experiments.
Explicitly, some observables formally in {\bf P} may scale with a sufficiently high-order polynomial that they are unreachable with the requisite precision for foreseeable computers, while some in {\bf NP} or {\bf QMA} may scale benignly for small or modest system sizes, enabling computations with ``sufficient'' precision to
have near-term or long-term impact.
The impact of any given simulation is limited by its precision, not only of the computation, but also of the theory input into the simulation.
Scientific simulations have progressed organically with simulation providing results of  precision  $\epsilon^{(t)}$ (where the superscript $t$ denotes total uncertainty at the time of the simulation).  Once $\epsilon^{(t)}$ is specified, there is freedom to define approximate (perturbatively-close) theoretical inputs and tolerances of the computation, to furnish  a result that lies within $\epsilon^{(t)}$ of the true result.
So, for bounded-error simulations,
with a target tolerance $\epsilon^{(t)}$,
as is the case with quantum devices, the  input theory (model or EFT) can deviate from the target theory, provided that the induced errors are $\le \epsilon^{(t)}$.
This builds upon the recognition that only a subset of possible quantum observables are required to be simulated e.g., low-energy scattering amplitudes and not amplitudes up to the Planck scale, which leads to the employment of EFTs and phenomenological models~\footnote{
An implicit distinction in motivations for developing simulations of arbitrary Hamiltonians and domain science applications.
}.

Leveraging lower-complexity leading order interactions has enabled classical simulations to make significant progress in addressing problems formally beyond the reach of classical computing.
The SM has many examples of QMB and QFT systems that are amenable to such perturbative expansions centered on lower-complexity leading orders:
\begin{itemize}
    \item
Density Functional Theory (DFT) is a highly successful framework for nuclear structure and reactions (see e.g., Ref.~\cite{doi:10.1080/23746149.2020.1740061}). The foundational result upon which DFT is based is the Hoenenberg-Kohn theorem~\cite{PhysRev.136.B864}, which states that the problem of determining the ground state energy of QMB or QFT systems can be replaced by an equivalent optimization problem over one-body density matrices $\rho$ using an appropriate functional $F[\rho]$ as a cost function. Central to the accuracy of the method is the availability of accurate approximations to the functional $F$ since, once this is available, the optimization can be usually carried out efficiently in {\bf P}.
At first glance, this seems at odds with the observation that estimating ground state energies of even simple $2$-body Hamiltonians is {\bf QMA-complete}.
Indeed, it is possible to show that, even though a DFT calculation is usually tractable once the functional $F$ is known, finding an accurate functional is  a {\bf QMA-complete} problem~\cite{Schuch_2009}.
However, with the guidance of physical understanding and experiment, approximate functionals of sufficient quality to reproduce observables like binding energies within a given accuracy can often be found---nature hacking complexity.
The {\bf QMA} classification in this case is thus a statement about the intrinsic difficulty of designing functionals capable of arbitrary precision in general.
    \item
Low-energy nuclear forces approximately manifest
Wigner's  emergent SU(4) spin-flavor symmetry~\cite{PhysRev.51.106,PhysRev.51.947} (which also emerges from large-N$_c$ considerations~\cite{Kaplan:1995yg}, vanishing entanglement power considerations~\cite{Beane:2018oxh}, and from lattice QCD calculations~\cite{Wagman:2017tmp}).
Quantum Monte Carlo simulations of multi-nucleon system ground states, e.g., using the pionless EFT,
generally exhibit a sign problem. Because the sign problem vanishes in the limit of SU(4) symmetry,  successful hybrid schemes exist involving \enquote{pre-conditioning} systems with the SU(4) symmetric interactions followed by implementation of the full interaction,  mitigate the impact of the sign problem~\cite{Lee:2016sxr}. This process identifies a leading order part of the {\bf QMA} calculation that is in {\bf BPP}, and perturbatively estimates the full result to fixed precision.
    \item
Nuclear structure has been tremendously impacted by Green's Function Monte Carlo (GFMC) calculations of light nuclei (see Ref.~\cite{RevModPhys.87.1067} for a review). Despite requiring exponentially increasing resources with baryon number,
for an extended period of time,
GFMC was able to provide accurate results for system sizes not reachable with polynomially-scaling algorithms, such as Coupled Cluster.
This balance of practical computational power has almost entirely reversed in recent years due to the adoption of  low-resolution interactions (derived from chiral-EFT~\cite{Weinberg:1990rz,Weinberg:1991um,Ordonez:1995rz,Kaplan:1996xu,Kaplan:1998tg,Kaplan:1998we,Epelbaum:1998hg,Machleidt_2011})
and RG techniques~\cite{Kaplan:1998tg,Kaplan:1998we,Bogner:2003wn} (see, e.g.,  Ref.~\cite{Hergert_2020} for a recent review).
\item
Current lattice QCD calculations indicate that:
\begin{itemize}
    \item Computing the mass of the pion to a given precision can be accomplished efficiently at scale, consistent with residing in {\bf BPP}
    \item The exponentially degrading signal-to-noise at late times in nucleon and multi-baryon correlation functions~\cite{Parisi:1983ae,Lepage:1989hd,Beane:2008dv} (a sign problem) currently suggests~\cite{Troyer_2005} that the
Quantum Monte Carlo computation of the binding energy of nuclei lies outside of {\bf P} and {\bf BPP}.
\end{itemize}
Through its connection to the
Hubbard model, finding the ground state of the pionless EFT is among the hardest problems in {\bf QMA}~\cite{Schuch_2009,ogorman2021electronic},
and therefore, it is likely that lattice QCD calculations of nuclei
lie in {\bf QMA} or beyond in {\bf PSPACE}.
However,  calculations of energies and properties of smaller nuclei, which have been demonstrated~\cite{Beane:2012vq}, permit the bounded-error extraction of counterterms in nuclear  EFTs~\cite{Barnea:2013uqa,Kirscher:2015yda,Contessi:2017rww,Bansal:2017pwn}. These results may be then used to compute the energies, properties, and low-energy dynamics of somewhat larger nuclei~\cite{Savage:2006et} (for reviews on this subject, e.g., Ref.~\cite{Beane:2008dv,Beane:2010em,Savage:2016egr,Drischler:2019xuo,Davoudi:2020ngi,Aoki:2020bew}).
\item
Heavy-quark systems (exhibiting heavy-quark symmetry for $m_Q\rightarrow\infty$) are comprised of one or more heavy quarks, $Q$, and light degrees of freedom.   Leading order wavefunctions of $Q\overline{l}$-mesons are a tensor product, $|Q\overline{l}\rangle \sim |Q\rangle\otimes |\overline{l}\rangle$.
The higher-order $1/m_Q$ contributions to observables
can subsequently be included in perturbation theory from this wavefunction.
Powerful theorems, such as the Adellmolo-Gatto theorem~\cite{Ademollo:1964sr} and Luke's theorem~\cite{Luke:1990eg},
ensure that leading order results in the symmetry breaking parameters are known without model dependence, and that symmetry breaking in certain matrix elements is suppressed by two powers of the breaking parameter.
Further, any model dependence or parameters that need to be computed elsewhere enter at higher orders in the perturbative expansion(s),
and  associated uncertainties are thus parametrically smaller than leading order.
\item The parton model characterizes hadronic structure and processes at high momentum transfer.
The scale invariance of QCD in this regime, due to asymptotic freedom at short distances, allows many QCD processes to factorize at leading order in pQCD, providing a foundation for the systematic inclusion of higher order corrections.
In such situations, classical computing has been proven effective in computing higher-order matrix elements with the required precision to advance experimental programs.
\end{itemize}
For quantum simulations, this {\it hacking} of bounded-error complexity classes through perturbative changes to the theory
is only now beginning.
For example,
precision first principles calculations of fragmentation lie beyond the reach of classical computing.
It is also expected to lie outside of {\bf BQP},
due to final state complexities,
but perturbatively addressable with simulations
that lie within {\bf BQP}.
The use of Soft-Collinear EFT (SCET)~\cite{Bauer:2000ew,Bauer:2000yr,Bauer:2001ct} may enable simulations within desired  error bounds (though this remains to be demonstrated)~\cite{Bauer:2021gup}.
For such hacks to be successful, problems should be broken up into a large part that resides within {\bf BQP} and can be performed efficiently on a quantum device,
and a perturbative part (containing the terms that place the problem outside of {\bf BQP}).

\subsection{Bottom Line: Enhancing Leading Order Complexity}

Remarkable progress has been made toward understanding nature through the SM even for problems that reside in complexity classes formally beyond the capability of classical or quantum computers.
This progress has been enabled, to a large degree, by
\emph{points of enhanced symmetry  exhibiting reduced complexity}.
In such instances, nature has been \enquote{kind} to us by providing Hamiltonians that can be separated into \enquote{large} and \enquote{small} parts for the observables of interest.
This separation has been enabled by the identification of approximate symmetries of the systems, with the \enquote{large} part(s) invariant under the symmetries, and the \enquote{small} part(s) breaking the symmetries.
Generally, and in hindsight, these symmetries are identified by eliminating entanglement and identifying perturbatively close tensor-product states.
However, examples of expanding about leading-order entangled states are known~\cite{Kaplan:1998tg,Kaplan:1998we,vanKolck:1998bw}, resulting from a non-trivial fixed-point in RG flow.
For classical computing, the challenge has been to find symmetries that place leading order approximations into {\bf P} or {\bf BPP} with a controlled perturbative expansion that sufficiently converges.
For quantum computing, the leading order symmetries need not eliminate entanglement.
Instead, incorporation of quantum devices will allow the advantage of addressing important {\bf QMA-complete} SM applications from the standpoint of leading order approximations residing in {\bf BQP}.
A nice example of this strategy can be found in low-energy nuclear structure where nucleon-nucleon correlations can be self-consistently included into a more effective single nucleon basis~\cite{Robin:2020aeh}, perhaps distilling the leading-order consequences of entanglement from a {\bf QMA}-complete problem into one that is in {\bf BQP}.
Broadly, identification of the leading order entangled systems of enhanced symmetry presents an exciting challenge.

At this early stage of development of quantum simulation, asymptotic scaling does not conclusively determine the reach of present day or near-term calculations, including  beyond the NISQ-era into the period where error-correction becomes routine.
While a better understanding of the formal asymptotic scaling of classical and quantum resources required to address challenges facing the SM is important,  it will not be the sole consideration in identifying goals for  near-term quantum simulation, as important progress
may be achievable, possibly even with demonstrations of quantum advantages, for problems that lie outside {\bf BQP}.

\section{Quantum Techniques and the Standard Model}
\label{sec:QTech}

\subsection{Mapping Quantum Fields to Quantum Degrees of Freedom}

The mapping one chooses to represent a quantum system of interest with quantum computational degrees of freedom will dramatically impact the quantum resources required.  In particular, the representation will affect the cost of all steps in a quantum simulation from the entanglement structures created in initial state preparation, to the interactions needed to perform time evolutions, to the distribution of information desired to be probed in the final measurement procedure.
It is expected that the basis-dependent property of entanglement may guide the development of intuition in mapping fields onto quantum degrees of freedom (see Section~\ref{subsubsec:entguideqsim}). Furthermore, exploring the repercussions of various representations informs theoretical flexibility to codesign simulations.

A basic question to consider in quantum simulation design
(that has begun to be addressed in the application to fields) is \emph{Where is the Line}, referring to the flexible partitioning of computational responsibility between the quantum device and the large-scale classical resources to which it will be symbiotically married.  One category of explorations shifting this line has been in state preparation e.g.,  through the use of quantum devices to optimize the design of interpolating operators for classical LGTs~\cite{Avkhadiev:2019niu}, through the use of classical snapshots of a volume set by the correlation length scale to inform state preparation on quantum volumes capturing many correlation length scales~\cite{Klco:2019yrb,Klco:2020aud}, or through purely quantum techniques inspired by the structure of classical tensor network design~\cite{Moosavian:2019rxg}.  A second category of explorations shifting \emph{the line} is in the basic representation of the field.
The influential papers of Jordan, Lee, and Preskill (JLP) illustrated an efficient approach for expressing real-time dynamics of scattering events quantum mechanically~\cite{Jordan1130,DBLP:journals/qic/JordanLP14}.  With polynomial resources in particle number, energy, precision, and volume, a microscopic description of interacting scalar field theory represented by local qubit degrees of freedom was shown to produce final state probability distributions that reflect, to systematically controlled precision, the exact probability distributions that would be experimentally observed.
Further inspiring the use of quantum devices for this purpose, recent quantification of computational complexity in tensor network simulation of inelastic false-vacuum bubble scattering illustrated a growth of entanglement at high energies and multiple scatterings~\cite{Milsted:2020jmf}, clearly connecting quantum effects to the voracious consumption of classical computational resources in scattering processes.   As algorithms for the quantum implementation of microscopic time evolution operators continue to progress, a complementary perspective from EFT descriptions is under development~\cite{Provasoli:2019tzz,Bauer:2019qxa,Bauer:2021gup},  utilizing quantum devices to address non-perturbative physics isolated in EFT descriptions of gauge field dynamics e.g., of parton showers, high multiplicity final states, or soft functions of SCET.
As a reliable methodology for systematically isolating energy regimes,  the demonstrated synergy between EFTs and microscopic quantum simulations of fields is expected to reduce quantum resource requirements by multiple orders of magnitude and offer a clear route for integration of quantum devices into existing computational frameworks of experimental impact.
Conveniently, the dynamical techniques developing for the microscopic simulation of gauge theories are closely related (often identical) to those leveraged for the calculation of real-time non-perturbative matrix elements within EFT descriptions.  As such, explorations considering different quantum-classical partition lines may be explored in parallel, to mutual benefit, without concern of obsolescence dependent on the trajectory of quantum hardware development.

\subsubsection{Fermions (quarks, neutrinos, electrons, nucleons)}
\label{subsec:fermions}

A standard approach for the mapping of fermion degrees of freedom onto qubits is the Jordan-Wigner (JW) transformation~\cite{Jordan1928}, which associates one qubit with each fermionic mode in the system. Fermi statistics are enforced by mapping the creation/annihilation mode operators of the fermions into products of Pauli matrices acting on multiple spins~\cite{PhysRevA.64.022319,PhysRevA.65.042323},
\begin{equation}
c^\dagger_i = \prod_{k=0}^{i-1} Z_k \sigma^+_i\quad c_i = \prod_{k=0}^{i-1} Z_k\sigma^-_i\ \ ,
\end{equation}
where the labelling underscores a global ordering of the fermionic modes. In this notation, $Z_j$ is the Pauli-$Z$ matrix acting on the qubit representing the $j^{\rm th}$ fermionic mode, while $\sigma^{\pm}_j=(X_j\pm iY_j)/2$ act as bosonic creation/annihilation operators.

Due to the underlying 1-dim structure in the JW transformation, even for locally interacting systems in more than 1-dim, the resulting qubit Hamiltonian will typically contain long range interactions acting non-trivially on $\mathcal{O}(n)$ qubits, which prevents the use of algorithms that are optimal only for local theories e.g., Ref.~\cite{doi:10.1137/18M1231511}.
This problem of induced non-locality caused by the mapping of fermions has received a considerable attention.
An  alternative approach is the Bravyi-Kitaev (BK) mapping~\cite{BRAVYI2002210,doi:10.1063/1.4768229} which, similar to JW, requires the same number of qubits as fermion modes, but reduces the induced non-locality exponentially to $\mathcal{O}(\log(n))$.
There are also various mappings employing additional auxiliary-qubits with the goal of reducing this non-locality by exploiting the connectivity of the underlying Hamiltonian~\cite{Verstraete_2005,PhysRevLett.95.176407,PhysRevA.95.032332,Steudtner_2018,derby2020compact}.
Notably, some fermion mappings might present different advantages than the locality of the resulting spin interactions e.g., the ternary tree construction from Ref.~\cite{Jiang2020optimalfermionto} allows for an efficient extraction of $k$-qubit fermionic density matrices, while the BK-superfast transformation~\cite{BRAVYI2002210,doi:10.1063/1.5019371}  has useful error-correcting properties~\cite{PhysRevApplied.12.064041} but requires additional auxiliary qubits.
Finally, it is important to point out that the presence of long strings of Pauli operators in JW mappings do not necessarily increase the computational cost in some situations e.g., fermionic swap networks~\cite{PhysRevLett.120.110501}.

\subsubsection{Scalar Fields (The BQP-Complete ``Gold Standard'')}

Encoding continuous quantum fields, such as scalar and gauge fields, into quantum simulations presents different challenges and opportunities than the inclusion of fermions.
Wavefunctions can be encoded in the amplitudes of states in a Hilbert space i.e., the states provide a binary representation of the digitized argument of the wavefunction while their amplitudes provide the corresponding value of the wavefunction.
To be concrete, let us consider scalar fields as their mapping onto qubits is the best understood scenario in the context of QFT.

The objective of any such mapping is that observables of the continuum field theory are recovered in the limit of vanishing discretization of spacetime and of the field.
The pioneering work of JLP~\cite{Jordan1130,DBLP:journals/qic/JordanLP14}
established a mapping that mirrors that of classical lattice scalar field theory, but with the continuous field values at each spatial lattice site replaced by a quantum register of $n_Q$, describing $2^{n_Q}$ values of the field.  For a lattice of spatial extent $L$ in each direction, the total number of qubits for such a simulation is $n = n_Q L^3$.
In the JLP basis, states in the computational Hilbert space correspond to eigenstates of the field operator
$\hat\phi |\phi\rangle = \phi |\phi\rangle$ at each lattice site.
After rescaling the Hamiltonian, a (symmetric~\cite{PhysRevA.99.052335}) mapping can be defined in terms of the maximum value of the field,
$\phi(k) = \phi_{\rm max} \left( -1 + 2k/(2^{n_Q}-1)\right)$,
where $k=0,1,... 2^{n_Q}-1$.
A simulation of high-fidelity requires that, at some level of precision, the support of the field wavefunction at each site, $\psi(\phi({\bf x}))$, is contained within  site registers, requiring an iterative tuning of the actual mapping based upon low-resource simulations for the interacting QFT, or direct evaluation for the non-interacting theory.
At this point, a connection can be made with the
Nyquist-Shannon sampling theorem.
With uniform sampling of $\phi({\bf x})$ in both field and field conjugate momentum spaces over the regions of support, the digitized representation of $\phi$ and static/dynamical observables can be computed with precision improving exponentially with increasing number of states at each site~\cite{Jordan1130,DBLP:journals/qic/JordanLP14,somma2016quantum,PhysRevLett.121.110504,PhysRevA.98.042312,PhysRevA.99.052335}.
This translates into a qubit requirement scaling as $n_Q \sim \log\log \left(1/\epsilon\right)$, where $\epsilon$ is the upper limit of tolerable systematic error.
The systematic errors remaining from the spatial discretization that are
encountered in classical
 lattice field theory simulations are also present, requiring
$n \sim \log\left(1/\epsilon\right)$.
One of the enjoyable features of this position-space mapping is that on-site and nearest-neighbor interactions are required for simulation, and only a small number of \enquote{layers} of unitary operations are required for a complete step in Trotterized time-evolution~\cite{Jordan1130,DBLP:journals/qic/JordanLP14}.
Instead of eigenstates of $\hat\phi$ to define the field at each site, states digitizing Hermite polynomials~\cite{somma2016quantum,PhysRevLett.121.110504,PhysRevA.98.042312} or directly associated with eigenstates of a harmonic oscillator could be used~\cite{PhysRevA.99.052335}.  The oscillators could be tuned to optimize support of $\psi(\phi({\bf x}))$ with low-resource simulations, and the dimensionality of the space could be used to define the
lowest $2^{n_Q}$ levels of the truncated oscillator.

The direct use of spin systems for mapping fields onto quantum registers leverages the fact that the minimal requirement for a legitimate mapping is recovering the correct continuum physics in the limit of vanishing discretization and digitization.
In particular, the approach to the continuum limit is generally mapping dependent.
For scalar field theory (with on-site wavefunctions that have bounded support in field and conjugate-momentum space, up to exponentially small corrections), the Nyquist-Shannon sampling theorem provides double-exponential convergence, with respect to an increasing number of qubits per site, of field digitization.
This supplements the single-exponential convergence of simulating the field with conventional lattice systematic uncertainties.
An alternate spin-system motivated mapping with one or two qubits per site makes implicit use of a space-spin RG-blocking to recover the space and field continuum limits.
By mixing space-space and field-space, the convergence of this latter mapping is expected to be single-exponential.
As discussed previously, this anticipated RG-flow is at the heart of QLMs~\cite{Chandrasekharan:1996ih,Brower:1997ha}.
Identifying efficient mappings for NISQ and beyond quantum simulations is an important line of investigation, and is already starting to provide some interesting results, e.g., Ref.~\cite{Zache:2021lrh}.
Classical simulations of a 1-dim
Heisenberg comb with two qubits at each lattice site suggest that
the continuum limit of the asymptotically free  nonlinear O(3) sigma model could emerge,
and that NISQ devices may have utility in further demonstrations~\cite{Singh:2019uwd,Bhattacharya:2020gpm}.

\subsubsection{Abelian and Non-Abelian Gauge Fields: Toward Lattice QCD}
\label{subsec:mappingnonabelian}
A main target of quantum simulation
is QCD, and in particular the observables that cannot be accessed with sufficient precision using classical computing~\cite{NSACQISNP}.
The simulation of gauge theories requires the incorporation of abstract spaces and local symmetries at lattice sites and the links between them (a thorough review of quantum simulation of LGTs can be found in Ref.~\cite{Banuls:2019bmf}).  It is well understood how to address the internal symmetries and different forms of inter-site couplings for classical simulations, and also for quantum simulation using lattice Hamiltonians, but conclusive determinations of optimal field representations and algorithms for implementation on quantum devices continue to be developed (for a recent review, see e.g., Ref.~\cite{Zohar:2021nyc}), in parallel with ongoing developments in the devices themselves.
The wisdom gained and algorithms explored
in the process of simulating scalar field theory
are expected to be valuable in developing simulations of, e.g., SU(N) Yang-Mills gauge theories, QCD and O(N) models.

In the case of gauge-field theories,
the action-angle variables are one convenient
path for mapping the Hamiltonian to a discrete Hilbert space.
While one can digitize directly the angle variables in a way analogous to the scalar field in a curved higher-dimensional space, for instance, using discrete subgroups of SU(N)~\cite{PhysRevD.22.2465,PhysRevD.24.3319,GROSSE198177,Alexandru:2019nsa}, making a somewhat direct connection to path integral Monte Carlo methods in Euclidean space, the action variables alternatively correspond to irreducible representations of the gauge group and provide a convenient means to define the Hamiltonian and states associated with the link operator, i.e. eigenstates of the electric operator.
The Kogut-Susskind Hamiltonian~\cite{PhysRevD.11.395,RevModPhys.51.659}, formulated in terms of the Casimir and plaquette operators acting on a lattice of links defined in the electric basis,
is such a construction for SU(N) gauge field theories, and has been the focus of much recent work.
First examined in the context of SU(3) quantum simulation by Byrnes and Yamamoto~\cite{PhysRevA.73.022328},  tensors defining the irreps of each local
SU(N) can be mapped to the states in local Hilbert spaces i.e., tensor indices mapped to quantum registers localized on each link.
For a given lattice spacing (coupling), the dimensionality of the local Hilbert space can be increased to achieve the desired precision in calculating observables of interest, with decreasing lattice spacing demanding an increasing number of irreps of SU(3) on each link in the electric basis.
After exploiting the local gauge symmetry to integrate over the abstract space at each site~\cite{Banuls_2020,PhysRevD.101.074512,PhysRevD.103.094501}, Yang-Mills theory can be written entirely in terms of wavefunctions of link variables defined by a basis of integers, $(p,q)$, that define the number of upper and lower indices of each irrep of SU(3). Truncation of each index register at some maximum value, $\Lambda_{p,q}$, causes the SU(3) lattice to become defined by connected pairs of $(\Lambda_{p,q}+1)$-dimensional Hilbert spaces with nearest neighbor connectivity~\cite{PhysRevD.103.094501}.   Such requirements point naturally to considering qudits, as opposed to qubits, assigned to each link.
Most architectures currently being pursued for qubit design aim to isolate two-level quantum systems from higher-dimensional spaces, allowing natural access to qudit structures in e.g.,  trapped-ions~\cite{osti_20636540,Low_2020}, photonics~\cite{lu2020quantum}, ultra-cold atoms and molecules~\cite{Sawant_2020}, and SRF cavities~\cite{Holland:2019zju,SRFFNAL}.
Whether providing intermediate theoretical organization or asymptotic scaling advantages~\cite{10.1145/3307650.3322253}, it appears that codesign of such systems for application to QCD, using a mapping of the Kogut-Susskind Hamiltonian with integrated local gauge spaces, may be an interesting path forward.

Beyond this originally considered multiplet basis of electric irreps, a cornucopia of bases and strategies for quantum simulation with local non-Abelian gauge symmetry continues to be developed---an abundance crucial for the hardware-software codesign necessary to efficiently interface with rapidly developing quantum technologies and crucial for assessing the effects of systematics when calculations extend beyond the reach of classical confirmation.
One alternate formulation of non-Abelian LGT amenable for quantum simulation (though not originally designed for this application) is quantum link models (QLMs)~\cite{HORN1981149,ORLAND1990647,Chandrasekharan_1997,Brower:1997ha,Brower:2003vy,Wiese:2006kp,Uwe:CERN_2020}, with a number of concrete proposals for implementation on quantum  devices e.g., Refs.~\cite{PhysRevLett.115.240502,Bernien2017,Shen:2020coq,Surace_2020}.
The idea of QLMs is that the intrinsic building blocks of the lattice are few-spin systems with appropriate symmetry and commutation relations.
While the link Hilbert spaces are never extended as a function of the coupling, making them attractive for near term quantum simulation~\cite{Stannigel:2013zka,Mezzacapo:2015bra,Singh:2019uwd, singh2019qubit,Surace:2019dtp, Bhattacharya:2020gpm,Shen:2020coq,VanDamme:2020rur},
effective degrees of freedom of a continuous field at each spatial site in a corresponding Kogut-Susskind lattice may be interpreted as distributed across a coupling-dependent volume of the lattice.
Though their approach to the continuum and the associated systematic errors remain to be quantified within simulations, RG-flow of QLMs is expected to recover the predictions of QCD if they reside in the same universality class and with appropriate simulation tunings to the critical point.
Progress has been made more broadly using spin systems as sandboxes for exploring algorithms and techniques to facilitate quantum simulations of field theories.
While ground state preparation is of considerable interest for spin systems, they also admit scattering of wavepackets or kinks, and allow for detailed studies of the evolution of entanglement and energy density e.g.,~\cite{Tan:2019kya,Verdel:2019chj,Milsted:2020jmf,Gustafson:2021imb}. Spin systems also provide a laboratory for building \enquote{particle detectors} within simulations.
Recent work using trapped ion and superconducting qubits systems
has shown that measuring time delays in wavepacket scattering between interacting and non-interacting theories can be used to extract elastic phase shifts even with NISQ devices, a step toward calculating S-matrix elements from real-time dynamics~\cite{Gustafson:2021imb}.
Interestingly, the entanglement entropy is found to increase through multiple elastic collisions; adding a cubic interaction enabled the production of ballistic particles during kink collisions, providing parallels with fragmentation in hadronic collisions~\cite{Milsted:2020jmf}.
Studies of confinement between magnetic domain walls have been carried out in spin models, where it has been shown that confinement suppresses information propagation~\cite{Tan:2019kya}, with obvious impacts beyond the SM.

Another approach for reducing the near-term qubit requirements for representing gauge fields in 1-dim, and the approach used in the first quantum simulation of the Schwinger model~\cite{Martinez:2016yna}, is to integrate out the gauge field, trading non-local interactions for only fermionic degrees of freedom.
This simplification expresses the lattice only through its fermionic degrees of freedom, alleviating the necessity of digitizing the continuous manifold of the local gauge field.
While the purely-fermionic formulation is only available in 1-dim (where the gauge field is non-dynamical)~\cite{Zohar:2021nyc}, it has been demonstrated that a formulation in terms of purely bosonic degrees of freedom is both possible and likely to be experimentally fruitful~\cite{Zohar:2018cwb,Zohar:2019ygc,aidelsburger2021cold}.
There are also harmonic-oscillator-inspired formulations of gauge theories that may be amenable to quantum simulation including orbifold lattices~\cite{Buser:2020cvn} and Schwinger bosons~\cite{Schwinger:1965,Mathur:2010wc}.
The latter of these formulations has evolved into the loop-string-hadron~(LSH) formulation~\cite{Raychowdhury:2018osk,Raychowdhury_2020} utilizing fundamentally gauge invariant degrees of freedom to analytically trade non-Abelian Gauss law constraints for Abelian ones, a feature that is likely to be advantageous in experimental connections e.g., cold atom quantum simulators  have demonstrated reliable expression of local Abelian symmetries~\cite{Gonzalez-Cuadra:2017lvz,Mil:2019pbt,Ott:2020ycj,Dasgupta:2020itb}.
Also underway are explorations utilizing the light front formalism~\cite{Kreshchuk:2020dla,Echevarria:2020wct,Kreshchuk:2020aiq,Kreshchuk:2020kcz}, or using measurement-based techniques and quantum sensors to extract correlation functions and the generating functional~\cite{Zache:2019xkx,Bermudez:2017yrq,Martin-Vazquez:2021mxc}.
Furthermore, there are hybrid proposals that naturally map the gauge symmetry onto a fundamental conservation of angular momentum in the architecture of ultracold atoms~\cite{Zohar:2013zla,Zohar:2012xf}, or that map the bosonic modes of the gauge field to local phonon modes in the architecture of trapped ions~\cite{Davoudi:2021ney}.
Beyond the basic realization that quantum devices will always be embedded in large classical ecosystems for device operation, error correction, and optimal use of quantum resources, such works illuminate the potential of hybrid and heterogeneous architectures as potentially advantageous for simulations of LGTs.

In addition to the quantum resources required for performance of time evolution on idealized quantum hardware, the choice of basis or field representation can dramatically affect the ratio of gauge-invariant to gauge-variant Hilbert space, modifying the sensitivity that a calculation will have to inevitable errors and noise experienced by the quantum device.
For example, gauge field integration available in (1+1)-dim removes all gauge-variant space from the computational Hilbert space of the hardware~\cite{Martinez:2016yna}, while the local group integrations of the multiplet basis or transformations to gauge-invariant degrees of freedom with Abelian constraints in the LSH formulation both dramatically reduce, but do not completely remove, the gauge violating Hilbert space.
Beyond simple dimensionality, mappings can vary in their available techniques for error mitigation.
In particular, paralleling analyses of quantum error correction codes, the array of gauge field quantum representations will exhibit varying \enquote{code distances}, or error separations between gauge-invariant states through the gauge-variant volume, determining the ability to reliably correct gauge-violating errors if they are detected~\cite{Stryker:2018efp}.
Of particular interest in this direction has been the dynamical generation or protection of gauge invariance from coherent sources of error~\cite{Goldman:2014xja,Halimeh:2019svu,Kasper:2020owz,Lamm:2020jwv,Halimeh:2020ecg,VanDamme:2020rur,Halimeh:2020djb,VanDamme:2021teo} through the introduction of low-body potential terms, creating effective Hamiltonians with the desired gauge invariance emerging robustly.
Optimizing theoretical and software strategies for the simulation of gauge theories in the high-dimensional landscape of state preparation, time evolution, and measurement resources, robustness to quantum noise, and isolation of the gauge invariant space will be an active process in perpetual communication with progress in experiment and device design.

\subsection{Quantum Fields for Quantum Information}
\label{ssec:fields4QIS}
As a purist motivation for focusing on the quantum simulation of fields, it is expected that the quantum mechanical complexity at play in atomic scale and larger systems will be commensurate with that
in subatomic and smaller systems.
Thus, precision control of the former may enable efficient calculation of emergent properties of the latter, in particular with resources that no longer scale exponentially with volume.
A concrete example of this commensurate scaling can be seen in the identification of vacuum-to-vacuum matrix elements in a self-interacting scalar field with classical external sources to be a {\bf BQP-complete} problem~\cite{Jordan_2018}.
In this direction, the relevance of fields for quantum simulation is not limited to relativistic fields describing the dynamics of the building blocks of the SM, but also applies to non-relativistic systems, such as EFT descriptions of low-energy physics.
From a less-purist perspective, the variety of qubit arrays that are being developed in laboratories around the world e.g., 2-dim arrays of implanted atoms in instrumented silicon substrates, ultracold atoms in optical lattices, ion lattices, superconducting circuit ladders and arrays of Josephson junctions,
are the closest physical systems to latticized quantum fields that have been created to date.
As such,  the {\bf BQP-completeness} of the scalar field~\cite{Jordan_2018} potentially indicates a natural language for quantum computation---any problem (physics or beyond) that can be efficiently calculated on a quantum device can be efficiently mapped to a scattering problem in scalar field theory, which itself is efficient to implement on a quantum device~\cite{Jordan1130,DBLP:journals/qic/JordanLP14}.
One step further lies the connection with  quantum error detection/correction, where the quantum information used to compute is embedded in topological excitations of a gauge field or a bulk holographic dual~\cite{Almheiri:2014lwa,Pastawski_2015}, both providing a redundancy to enable robustness to local environmental perturbations.
So, whether one is interested in fundamental simulations of nature or large-scale quantum computations more broadly, the QFT language provides a natural framework.

Typically considered to be in the realm of condensed matter, quantum spin liquids represent an important connection between research areas, bringing together QFTs, spin models and the distribution and protection of quantum information (for a review, see Ref.~\cite{Savary_2016}).
While there are a number of candidate systems to provide logical qubits and quantum memories, as reviewed in Ref.~\cite{Campbell_2017},
Kitaev's Toric Code~\cite{Kitaev_2003} (and subsequent surface codes exploring alternate boundary conditions~\cite{kitaev1997quantum,Bravyi:1998sy,Dennis_2002,PhysRevA.86.032324}) represents perhaps the first and clearest connection between these areas of research.
In its simplest form, a 2-dim system of spin-${\frac{1}{2}}$ spins defining the
links of a square lattice is subject to a Hamiltonian with vertex operators (stars), plaquette operators, and periodic boundary conditions.
The ground state of the system is topologically ordered and consequently highly entangled, supporting two logical qubits defined by the action of Wilson-loop operators in the two directions.
These topological states are robust against errors (local perturbations) below some threshold, which can be corrected by measuring $n-4$ independent stabilizers (commuting star and plaquette operators) and using decoders to determine an effective set of operations to restore the system.
Reducing the number of independent stabilizers allows an increased dimension of the code space at the expense of error robustness.
The quasi-particle picture is a convenient way to understand the behavior of these systems and errors.   Localized errors create quasi-particles in the system and error correction corresponds to annihilating quasi particles, returning the systems back into the code space (ideally without inadvertently producing a logical error).
Excitations above these ground states are (non-local) pairs of electric charges and pairs of magnetic vortices. The anyonic nature of these excitations and their braiding points to the  high-degree of underlying entanglement of the ground state(s), and as such,  local operations are unable to resolve which state of the (logical) ground-state space the  system is in.
An auxillary spin can be included at each lattice vertex to transform this system into a $Z_2$ LGT, of similar form as the Kogut-Susskind Hamiltonian for SU(N) LGTs (for a review, see Ref.~\cite{kitaev2009topological}), and generalizing in detail to SU(N) gauge theories~\cite{Mathur:2016cko}.
The Toric Code is also established in 3-dim and 4-dim using hypercubic (and other)  Euclidean space lattice geometries.
A major challenge to the execution of circuits on logical qubits is the required action of gates across the physical qubits defining the logical qubits.  This development is mature for the Toric Code, see e.g., Refs.~\cite{Dennis_2002,Jochym_O_Connor_2021}, and also for other encodings, such as color codes~\cite{PhysRevLett.97.180501}.

Interestingly from an SM physics perspective,
higher-dimension SU(N) LGTs, due to the phenomena of confinement manifest in non-Abelian gauge theories,
have utility in defining logical qubits due to their innate mechanisms to localize qubit errors within \enquote{hadrons}~\cite{PhysRevX.5.031043,Brown_2016} (a similar effect can be induced by Anderson localization~\cite{Wootton2010LocalizationAQ}).  Unlike the Toric code, where electric charges and magnetic vortices are unconfined and error correction requires multiple rounds of stabilizer measurements,
confinement keeps SU(N) Gauss-law-violating vertex charges induced by qubit errors from propagating \enquote{too far} from each other, restricting them to approximately within the confinement volume.
A possible exception to this is
in instances where multiple color-singlet \enquote{hadrons} can be formed from the SU(N) charges and wrap \enquote{around the world}, an effect that is suppressed at small lattice spacings and is assessed to be topologically benign due to confinement.
As a result of this connection, the techniques and technology that continue to be developed for classical LGTs  are likely to be of direct impact upon logical qubits and error-correcting protocols, e.g., Ref.~\cite{Andrist_2011,Brown2016}.

Another lattice system that has been developed to define logical qubits is the 2-dim honeycomb lattice with spin-$\frac{1}{ 2}$ spins at each vertex and with link-direction-dependent couplings between spins~\cite{kitaev2006anyons}.
This system can be written in terms of  Majorana fermions with both gapped and gapless phases.  Imposing a magnetic field on the gapless phase induces a gap and non-trivial topological structures  with chiral edge states carrying Chern numbers.
Edge states, and their deep connection to topological structure are commonly used to simulate chiral fermions in LGT calculations---domain wall fermions employ a 5th dimension to isolate a fermion of definite chirality at the domain wall on the 4-dim boundary~\cite{Kaplan_1992,Kaplan_youtube_Harvard2020}.

\subsection{Preparing Wavefunctions: Ground States and Finite-Density}

While efficient non-dynamical approaches to state preparation can be determined for systems with sufficient structure, the ability to simulate the real-time dynamics of a quantum system is commonly a useful resource in studying static properties, e.g., the ground-state energy density. There are three main classes of approaches to attain this goal: adiabatic state preparation, the use of energy filters, and variational approaches. Using adiabatic evolution, it is possible to drive a quantum state from the ground-state of an initial Hamiltonian $H_i$ to the ground-state of a final Hamiltonian $H_f$ of interest~\cite{2000quant.ph..1106F,RevModPhys.90.015002}. Given a procedure to initialize a quantum-computer in the ground-state of $H_i$, an appropriate adiabatic path can be designed to produce a final state after time evolution with a large overlap onto the target ground state. The scale for the evolution time $T$ is set by the smallest energy gap $\Delta E$ between the instantaneous ground-state and first excited state throughout the adiabatic path.
Several strategies have been proposed to optimize the efficiency of this scheme, ranging from explicit coupling to an external
\enquote{bath} of degrees of freedom~\cite{2017arXiv170908250K,LEE2020135536,PhysRevResearch.2.023214}, gap amplification techniques~\cite{doi:10.1137/120871997},
and specially designed adiabatic paths that leverage previous knowledge of the excitation spectrum to avoid closing gaps during the adiabatic evolution (e.g., Refs.~\cite{PhysRevA.92.062318,du2021ab}).
This method has been applied to lattice field theory simulations~\cite{Jordan1130,DBLP:journals/qic/JordanLP14} and applications to the nuclear many-body problem have recently started to be developed~\cite{du2021ab}.

Quantum Phase Estimation (QPE)~\cite{1995quant.ph.11026K,PhysRevLett.79.2586,PhysRevLett.83.5162} or it's more recent variants~\cite{10.5555/2600508.2600515,PhysRevLett.117.010503}, is a candidate technique for computing ground-state properties with a quantum computer at scale. This versatile approach can be used to measure the eigenvalues of the Hamiltonian operator $H$ by exploiting the ability to simulate real-time evolution under the operator $U(t)=\exp(-itH)$ to implement an energy filter of resolution $\delta H=\mathcal{O}(1/T)$, with $T$ the longest time interval required for the algorithm. For sufficiently long propagation times $T$ the probability to measure the ground-state energy when performing QPE on an initial trial state $\vert\Psi_T\rangle$ is proportional to it's overlap $|\langle\Phi_0|\Psi_T\rangle|^2$ with the ground-state~\cite{ovrum2007quantum}. It is therefore crucial to be able to initialize the quantum device in a high-overlap trial state $\vert\Psi_T\rangle$, while keeping the complexity of this initialization low
for ease in simulating the real-time dynamics of QPE within the limited coherence time of the devices. These trial states are typically obtained using the variational approaches described below, but both adiabatic state preparation or a preliminary coarse energy filter could be employed. Similar to the calculation of dynamical response functions, polynomial expansion methods using Qubitization/QSP can be employed  to  define alternative unitary operators to use in place of $U(t)$~\cite{PhysRevLett.121.010501} or directly to implement more efficient energy filters for ground-state projection~\cite{Lin2020nearoptimalground}. Recent proposals for ground-state preparation algorithms like the Rodeo algorithm~\cite{choi2021rodeo} or simulated imaginary-time propagation~\cite{turro2021imaginary} also belong to the class of energy filters, though their resource scalings are currently less understood.

Variational approaches address the ground-state preparation problem through an (in general non-convex) optimization. This is achieved by constructing quantum states using a fixed quantum circuit with free parameters and  using an appropriate variational principle to find a good approximation of the ground-state by optimization. Typically the expectation value of the energy is used as a cost function in the optimization e.g., the Variational Quantum Eigensolver (VQE)~\cite{Peruzzo_2014,McClean_2016,AspuruGuzikSim2019}, which directly optimizes over the parameters of a general unitary operator.    Alternatively,  the Quantum Approximate Optimization Algorithm (QAOA)~\cite{farhi2014quantum} mimicks the adiabatic state preparation described above, but optimizes a parametrization of the adiabatic path with the goal of reducing the dependence on energy gaps in the spectrum.
Other cost functions have also been considered e.g., the action generating imaginary time evolution~\cite{McArdle_2019} or few-body correlations~\cite{Motta_2019}. The two main advantages of variational approaches are the relatively small number of quantum operations required (at least for simple parametrizations) and the inherent robustness to miscalibration of the quantum device and to noise.  An intuitive way to understand this last feature is to realize that the various noise sources will effectively shift the location of the global minimum for a given parametrization, but this effect may be partially compensated by the optimization procedure~\cite{PhysRevX.6.031007}. Variational algorithms are heuristic in nature.
On one hand, it would be classically hard to simulate the measurement statistics given a sufficiently complex circuit structure.
On the other hand however, finding the optimal parameters using non-convex optimization is in general {\bf NP-hard} and we do not expect to be able to reach the globally optimal set of parameters for all instances of the problem~\cite{bittel2021training}. The {\bf NP-hardness} of these variational algorithms is a natural occurrence not necessarily specific to quantum algorithms e.g., the corresponding {\bf NP-hardness} of finding the optimal mean-field solution using the Hartree-Fock method~\cite{Schuch_2009}, and  it does not necessarily imply a limitation in the applicability of variational algorithms in practice (see Sec.~\ref{sec:complexityclasses}), {\bf NP-hardness} is a statement about the behavior in worst-case instances and is not necessarily indicative of the average complexity, especially if additional structure is present in the problem.

Given their versatility, limited use of quantum resources and resilience to device noise, Variational Quantum Algorithms (VQA) are a flourishing area of research (see Ref.~\cite{cerezo2020variational} for a recent review). VQA will likely be important as a stepping stone towards preparing high-accuracy trial states to be used as initial conditions for more sophisticated quantum algorithms.  This is similar to the use of Variational Monte Carlo as a precursor to more accurate, and expensive, GFMC calculations of nuclei (see Ref.~\cite{RevModPhys.87.1067} for a review). In addition to such compounding applications, another possible use of VQA is as components for Quantum Machine Learning~\cite{Biamonte2017} tasks (possibly coupled to classical neural networks e.g., Ref.~\cite{filipek2021quantum}), such as characterizing the output states generated by a quantum algorithm or even directly as a way to perform inference on classical data (for a recent review of applications to SM problems, see Ref.~\cite{Guan_2021}).
Presently, it is not clear if an exponential advantage could be obtained for the latter task (see, e.g., Refs.~\cite{2015NatPh..11..291A,Wiebe2020}).

A dominant component of the cost in variational techniques comes from the number of repetitions required to estimate the cost function (and/or it's gradient) used for VQE, and considerable effort has been devoted in the recent past to design approaches to reduce this cost.
These include optimization of commuting sets of operators that can be measured simultaneously and techniques using information from short-time dynamical evolution~\cite{PhysRevA.101.022328,PhysRevA.103.042405,doi:10.1063/1.5141458,O_Brien_2019,AspuruGuzikSim2019,PRXQuantum.2.020317,Funcke:2020vkw,Huggins_2021,guzman2021predicting}.
One last class of techniques developed recently for efficient estimation of general expectation values uses random sampling before measurement to drastically reduce the number of measurements required for convergence~\cite{Brydges260,PhysRevA.99.052323}.
For example, by informing a 'classical shadow'~\cite{10.1145/3188745.3188802,Huang_2020} using the results of random measurements,
the expectation value of $M$ distinct observables can be estimated using only $\mathcal{O}(log(M))$ measurements on a quantum device.

\subsection{Probing Quantum Systems in Real Time: Scattering and Inelasticities}
\label{ssec:real-time}

Extracting reliable dynamical information from QFTs and QMB systems is an extremely challenging task due to sign problems that generically plague real-time path integrals.
This difficulty is particularly
impactful in the SM since the vast majority of experimental information is obtained through scattering processes.
Progress continues to be made in developing more efficient implementations of time evolution for QFTs and QMB systems, as outlined in Section~\ref{sec:Qsim},
but
there is a further level of sophistication that is required in designing time-dependent calculations performed on a device to optimally address the physics problems at hand.

A common approach to evaluate scattering cross sections in the linear response regime
using classical computing
is to first compute Euclidean-time correlation functions, readily available from imaginary-time path integrals, and then attempt a numerical inversion to real-time. This process is ill-posed, in the sense that small errors in the Euclidean correlators can lead to large errors in the result of the inversion, and various heuristic approaches have been proposed to extract dynamical properties with minimal bias (see e.g., Refs.~\cite{PhysRevLett.103.210403,PhysRevLett.111.182003,PhysRevLett.117.082501}). A similar approach popular in few-body nuclear physics is to attempt a numerical inversion of integral transforms that are better behaved than the Laplace transform e.g., the Lorentz~\cite{Barnea_2010} or the Sumudu integral transforms~\cite{PhysRevB.88.094302}, with the goal of simplifying the inversion procedure. As we will see further below, similar ideas will play an important role in proposed quantum algorithms for real-time dynamics.

Using the efficient simulation strategies presented in the previous section, quantum computers are expected to be able to simulate efficiently the real-time dynamics of systems with local interactions.
As perhaps the most straightforward application of these ideas, direct simulations of Hamiltonian dynamics starting from a reference state have recently appeared e.g., the probability of pair production and chiral dynamics in the Schwinger model~\cite{Martinez:2016yna,PhysRevA.98.032331,Kharzeev:2020kgc}, the three-body contact density in a simple model for the triton~\cite{PhysRevD.101.074038}, the time-dependent Coulomb excitation of the deuteron colliding with an heavy-ion~\cite{du2020quantum}, the spin dynamics of a pair of nucleons interacting through a tensor interaction~\cite{Holland:2019zju} and the entanglement and flavor evolution in collective neutrino oscillations~\cite{yeteraydeniz2021collective,hall2021simulation}.

Particularly important for comparison with experiments is the calculation of scattering cross sections in the linear response regime, where the interaction between system and probe is well approximated by a single vertex. Apart from kinematical factors, the full energy dependence of the cross-section is contained in the the response function $R(\omega)=FT\left[\langle O(0)O(t)\rangle\right]$. In this expression $FT$ denotes the Fourier transform and $\langle O(0)O(t)\rangle$ is the real-time two-point function of the vertex operator $O$ evaluated on the ground state of the target. A direct calculation of this two point function is possible in both digital quantum devices~\cite{PhysRevA.65.042323} and more general quantum simulators~\cite{Baez_2020}. In practice, the Fourier transform is performed using the two-point function signal up to a maximum time $T$, leading to a resolution in the frequency domain $\Delta\omega=\mathcal{O}(1/T)$. Since the dominant cost of the digital algorithm is the implementation of the real-time evolution operator for a total time $T$, the final cost of these types of approaches scale as $\mathcal{O}(1/\Delta\omega)$. This agrees well with the no-fast-forward theorem mentioned above, and doesn't exclude the possibility of an important speed up if structural information of the system's Hamiltonian is used in the design of the algorithm. It is important to remember that the Minkowski-space correlation functions extracted from any finite volume quantum computation need to be properly extrapolated to the continuum. Contrary to their Euclidean time counterparts, e.g. Refs.~\cite{LUSCHER1991531,Luscher1986,Luscher1986b}, these real-time correlators have been shown to pose additional challenges in this extrapolation for certain kinematical regimes, and practical strategies for mitigating the problem have been proposed~\cite{PhysRevD.103.014506}.

A more direct approach to scattering matrix elements, inspired by the integral transform techniques used in few-body nuclear theory, can be formulated to work directly in the frequency domain avoiding the approximation of the Fourier transform in both the UV and IR~\cite{PhysRevC.100.034610}. The main idea is to apply an approximate projector into a frequency band, in the same spirit as the Lorentz Integral Transform method of Ref.~\cite{Barnea_2010}, followed by a direct measurement of the population in that band to reveal the response function strength. Besides a better control of the frequency-space errors, this type of approach  opens the way for characterization of semi-exclusive information about the final state of the reaction since, after the application of the energy filter, the register of qubits representing the target is mapped to the set of final states. This is an important step towards a more complete characterization of semi-exclusive cross sections e.g., those required to extract neutrino parameters from long baseline experiments like DUNE, and first estimates of the quantum resources required for this goal have appeared~\cite{PhysRevD.101.074038}, suggesting further improvements are needed for near term applications on real devices. The scheme can be generalized by considering different frequency projectors constructed either using the real-time unitary operator (see e.g., Refs.~\cite{Somma_2019,PRXQuantum.2.020321}) or more general expansions in orthogonal polynomials~\cite{PhysRevA.102.022409,PhysRevA.102.022408}, which allow for superior scaling with the target precision.

\section{Standard Model Applications: Select Implementations on Digital Quantum Hardware}
\label{sec:QCNPapps}

For the purposes of the following discussions, we consider energy scales higher than those relevant to chemistry and lower than (roughly) the TeV scale.  As a result,  we will not present the significant number of impressive results that have been obtained for molecular ground state energies and reactions,
nor results related to quantum gravity, holography, baryogenesis,  or other physics beyond the SM.
Rather than duplicating the comprehensive set of reviews that contextualize the now-extensive literature,
we present three broadly defined areas of SM research that are beginning to be addressed with NISQ simulations, and which are likely to form the backbone of SM quantum simulation research.

\subsection{Lattice Gauge Field Theory Dynamics}

\begin{figure}
\centering
\includegraphics[width=0.98\columnwidth]{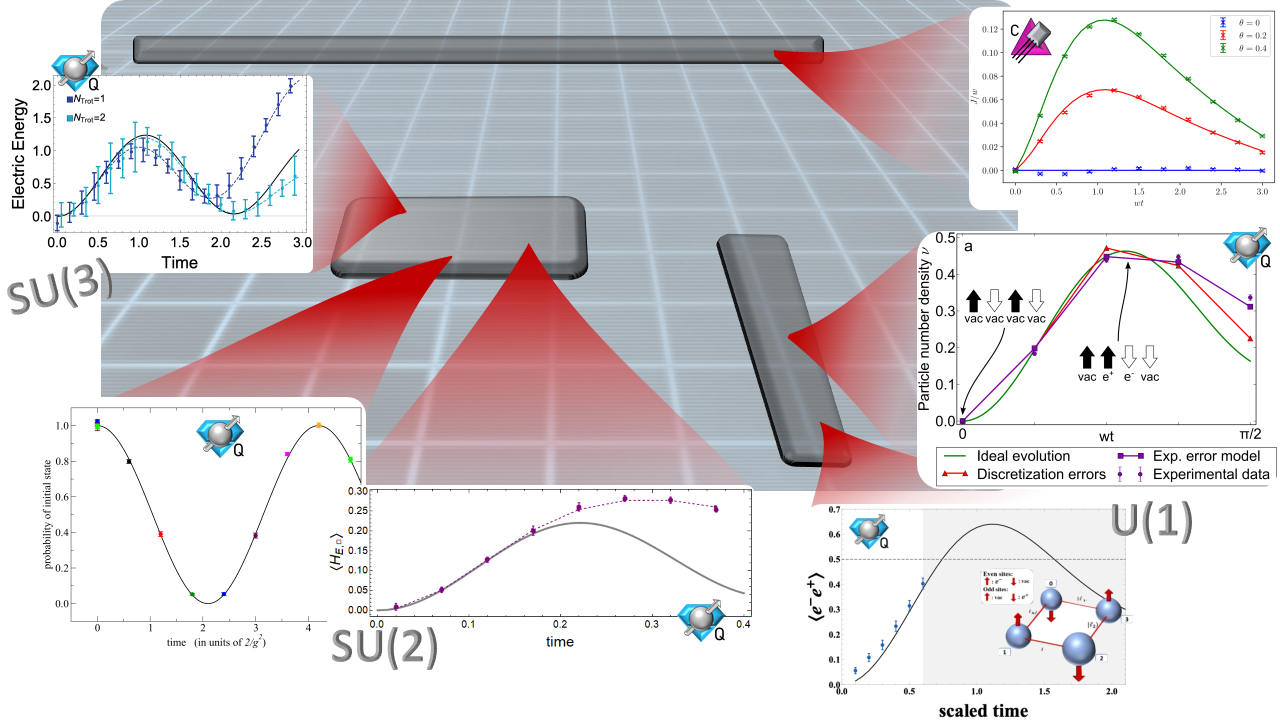}
\caption{
The real-time evolution of  1+1 dim QED and Yang-Mills gauge theories on small lattices.
The right-side panels show results obtained for the Schwinger model: (top) the vector current after a $\theta$-quench obtained using IBM's classical simulators (2020)~\cite{Kharzeev:2020kgc}, and
 pair-production on a four-site (staggered) lattice obtained using (middle) Innsbruck's trapped ion systems (2016)~\cite{Martinez:2016yna}, and
(bottom) IBM's quantum devices (2018)~\cite{PhysRevA.98.032331}.
The lower-left panels show, for two-plaquette systems in SU(2), (right) the time dependence of the local electric energy obtained using IBM's quantum devices (2019)~\cite{PhysRevD.101.074512}, and
(left) the strong-coupling vacuum persistence probability
obtained using D-wave's quantum annealing systems (2021)~\cite{Rahman:2021yse}.
The upper-left panel shows
the electric energy of two plaquettes of SU(3) obtained using IBM's quantum devices (2021)~\cite{PhysRevD.103.094501}.
{\scriptsize [
(right-top): Reprinted figure with permission from Dmitri E. Kharzeev, Physical Review Research, 2, 023342, 2020. Copyright (2020) by the American Physical Society.
(right-middle): Reprinted with permission from Christine Muschik.
(lower-left-left): Reprinted with permission from Randy Lewis.
]}
}
\label{fig:RealTimeSchwinger}
\end{figure}

The close connection between spin models that can be used to define logical qubits and LGTs (see Section~\ref{ssec:fields4QIS}) might suggest that we dedicate much of this section to discussing results related to the Toric code, quantum spin liquids~\cite{Savary_2016} and related systems.
Instead, we focus on hardware implementations explicitly targeting the SM.
The first digital quantum simulation of real-time dynamics in a LGT~\cite{Martinez:2016yna} not only demonstrated the progress that had been achieved in the control and manipulation of quantum hardware, but provided tangible inspiration that the vision of digital quantum simulation for microscopic descriptions of nature was becoming a reality.
Complementary to larger U(1) simulations on analog quantum devices
and classical simulations, e.g., Refs.~\cite{Banuls:2013jaa,Buyens:2016hhu,Banuls:2016lkq,Funcke:2019zna,Yang_2020}, this digital experiment~\cite{Martinez:2016yna}  was performed on four trapped calcium ions representing the fermionic degrees of freedom at two spatial sites of latticized 1+1 dim quantum electrodynamics (QED), also known as the lattice Schwinger model.
First proposed in Ref.~\cite{Hauke:2013jga},
this work showed the dynamical generation of $e^+e^-$ pairs emerging from the trivial vacuum under time evolution, as shown in Figure~\ref{fig:RealTimeSchwinger}~\footnote{
 In each figure panel of this section, icons from Ref.~\cite{Klco:2019xro} are assigned to indicate the type of compute device that generated the results, i.e. a classical computer without a noise model, a classical computer using a noise model, or a quantum device.  These icons and their descriptions can be found at \href{https://iqus.uw.edu/resources/icons/}{https://iqus.uw.edu/resources/icons/}.
}.
Encouragingly, even with the $\sim 50$ quantum gates necessary for each step in the time evolution, control and coherence was demonstrated  up to four Trotter steps with a survival probability in the physical, zero-charge subspace (six of the $2^4$ fermionic configurations) of greater than 70\%, allowing for error-mitigating post-selection while maintaining sufficient statistics.
As a theory in one spatial dimension, the Schwinger model is  a system where a number of dynamical attributes can be addressed with precision using modern analytical and numerical tools, such as tensor networks, e.g.,  Refs.~\cite{Banuls:2013jaa,Buyens:2016hhu,Buyens:2016ecr,Pichler:2015yqa,Butt:2019uul,Banuls_2020}.
However, continuing to perform quantum simulations of this and similar theories aids the benchmarking of quantum devices, and the development of algorithmic techniques.

While 1+1 dim QED is not of direct interest to fundamental physics, it shares attributes with QCD, such as charge screening, a fermion condensate and a spectrum of composite particles (\enquote{hadrons}), including two-body and three-body bound states (\enquote{nuclei}).
Of particular note, it has non-trivial topological structure through a $\theta$-term, proportional to the electric field at non-zero coupling, and also a chiral chemical potential~\cite{Kharzeev:2020kgc}.
Early studies of the impact of time-dependent $\theta(t)$ and $\theta$-quenches have been undertaken using classical simulations of quantum devices~\cite{Kharzeev:2020kgc,PhysRevLett.122.050403} (see also related work with quantum link models~\cite{PhysRevLett.122.250401}), shown in  Figure~\ref{fig:RealTimeSchwinger}.
However, in contrast to 3-dim gauge theories, the U(1) gauge field of the Schwinger model, like any gauge field in 1+1 dim, is not dynamical; constrained by Gauss's law at each lattice site, it is dependent only on the boundary conditions, which conveniently allows an entirely fermionic representation.
Since its first implementation, an array of simulations and proposals for simulations (e.g., Refs.~\cite{Zohar:2016iic,Muschik:2016tws,Gonzalez-Cuadra:2017lvz,PhysRevA.98.032331,Banuls:2019bmf,Luo:2019vmi,aidelsburger2021cold}) of U(1) gauge theories using different types of quantum devices have been established, see Figure~\ref{fig:RealTimeSchwinger}.
Though the long-range interactions arising from the non-dynamical gauge field are amenable to trapped ion implementation~\cite{Hauke:2013jga,Martinez:2016yna,Davoudi:2019bhy},
the utilization of superconductors inspired the exploration of an alternate approach of classical pre-processing for global and local symmetry projections~\cite{PhysRevA.98.032331}.
This early example served to emphasize the extent to which the design of a digital quantum simulation of LGTs must consider the physics of the underlying quantum device for near-term progress.

First considerations have been made in determining the Schwinger Model's asymptotic scaling~\cite{Shaw2020quantumalgorithms}, and further detailed studies continue to provide important information, e.g., Refs.~\cite{Kaplan:2018vnj,Stryker:2018efp,Magnifico:2019kyj,Davoudi:2019bhy,Stryker:2021asy,Halimeh:2020ecg,Halimeh:2020djb,VanDamme:2020rur,Davoudi:2021ney},
including  detecting and correcting violations of Gauss's law and mitigating the effects of evolution into gauge-variant parts of the hardware Hilbert space.
While these and future studies reveal important aspects that will be encountered going forward, the fact that the gauge field is not dynamical limits the implications for simulations of the SM.
Formalisms for simulating QED in higher dimensions have been put in place for a number of systems, e.g.,  Refs.~\cite{Zohar:2011cw,Zohar:2012ay,Tagliacozzo:2012vg,Zohar:2012ts,Wiese:2013uua,Marcos:2014lda,Kuno:2014npa,Bazavov:2015kka,Kasper:2015cca,Brennen:2015pgn,Kuno:2016xbf,Zohar:2016iic,Kasper:2016mzj,Gonzalez-Cuadra:2017lvz,Ott:2020ycj,Paulson:2020zjd,Kan:2021nyu,aidelsburger2021cold}, but remain to be executed on a quantum device.

The last several years has seen a number of exciting developments in establishing frameworks for simulations of non-Abelian gauge theories, starting with the pioneering work of Byrnes and Yamamoto~\cite{PhysRevA.73.022328}.
That work builds upon the Kogut-Susskind Hamiltonian~\cite{PhysRevD.11.395} for SU(N) Yang-Mills and first suggestions for implementation by Zohar, Cirac and Reznik~\cite{Zohar:2012xf},
which in turn built upon an extensive literature on Hamiltonian formulations of Yang-Mills gauge theories and QCD, e.g., Refs.~\cite{Wilson:1994fk,Heinzl:1995jn,PhysRevD.31.2020,LIGTERINK2000983c,LIGTERINK2000215}.
While there are a number of formulations of non-Abelian gauge theories for hardware implementation (see Section~\ref{subsec:mappingnonabelian}), the real-time dynamics (on small lattices) of
only one construction has so far been simulated on quantum devices, using IBM's quantum systems~\cite{PhysRevD.101.074512,PhysRevD.103.094501} and a D-wave annealing device~\cite{Rahman:2021yse}, as shown in Figure~\ref{fig:RealTimeSchwinger}.
The local gauge invariance at each lattice site has been exploited to implement the non-Abelian constraints explicitly by integrating over the gauge group~\cite{Banuls:2017ena,PhysRevD.101.074512,PhysRevD.103.094501}.
This reduces the dimensionality of the hardware Hilbert space to
capture only the irrep of each link, and introduces controlled-plaquette operators.
Simulations that have been performed on quantum devices have employed global bases for one and two plaquettes.
With the knowledge gained from these small systems, performing simulations with more than two plaquettes in higher dimensions is a priority (for a recent review see Ref.~\cite{paulson2020simulating}).

\subsection{Structure and Reactions of Nuclei}

\begin{figure}[!ht]
\centering
\includegraphics[width=0.98\columnwidth]{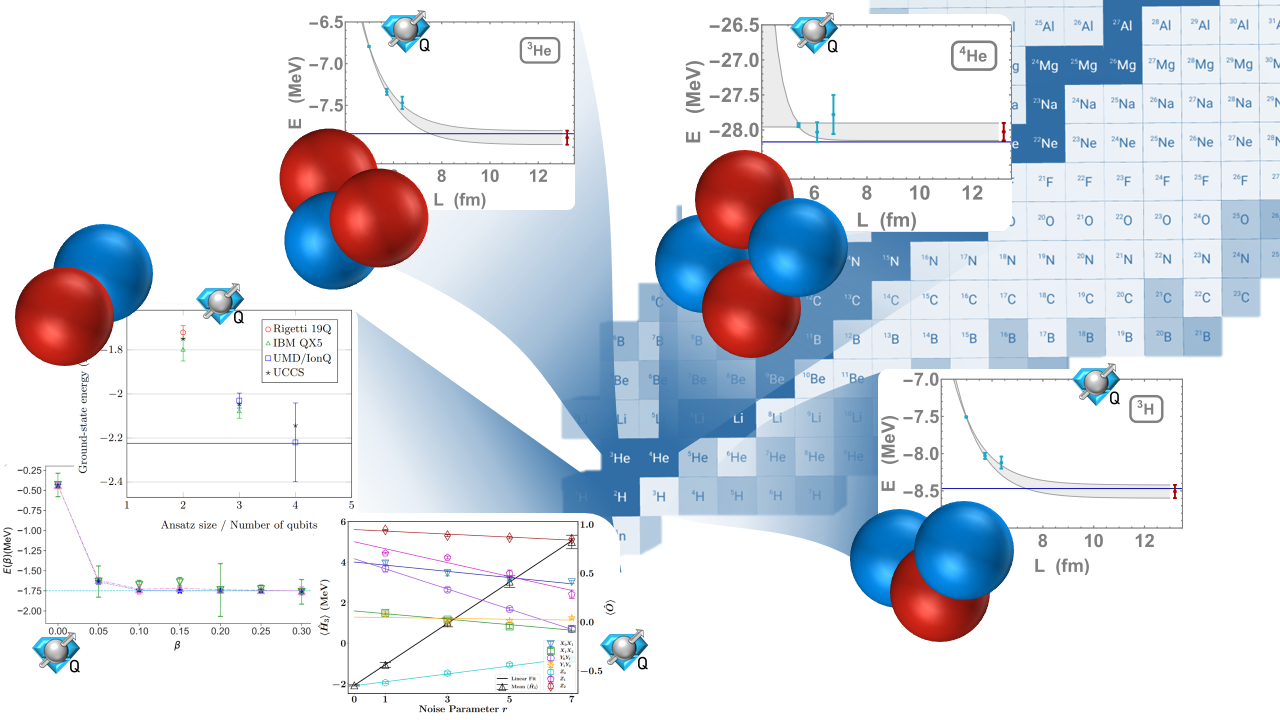}
\caption{Ground state energies of light nuclei
from quantum simulations.
Lower-left panels show results for the deuteron addressed with (right) VQE on superconducting qubits~\cite{Dumitrescu:2018njn} (top) VQE on trapped ions~\cite{Shehab:2019gfn} and (left) QITE simulation on IBM's quantum devices~\cite{yeter2020practical}.
Three remaining panels show VQE results for the triton, $^3$He,  and $^4$He using a quantum frequency processor~\cite{Lu:2018pjk}.
{\scriptsize [
(lower-left-right): Reprinted figure  with permission from Eugene Dumitrescu, Physical Review Letters, 120, 210501, 2018. Copyright (2018) by the American Physical Society.
(lower-left-left): Reprinted figure with permission from Raphael Pooser, \cite{creativecommons4}.
(lower-left-top): Reprinted figure  with permission from Raphael Pooser, Physical Review A, 100, 062319, 2019. Copyright (2019) by the American Physical Society.
]}
}
\label{fig:NuclStruct}
\end{figure}

Nuclear structure and reactions reside in an important low-energy sector of the SM, with
precision targets expected to become achievable with future quantum simulation capabilities.
As illustrated in Figure~\ref{fig:NuclStruct}, the first ground-state calculation of a nuclear system was a VQE simulation of the deuteron using a harmonic oscillator basis~\cite{Dumitrescu:2018njn} with the superconducting quantum devices of IBM and Rigetti. Extensions of the approach soon followed with implementations of the same model on trapped-ion quantum computers~\cite{Shehab:2019gfn}, explorations of different mode-to-qubit encodings using the Gray code~\cite{PhysRevA.103.042405}, and calculations of the energies of larger systems with $A\leq4$~\cite{Lu:2018pjk}. These latter calculations, in large Hilbert spaces containing up to 68 states, were made possible by the use of an all-optical quantum device called a Quantum Frequency Processor (QFP), which allows for an entire calculation to be formulated using a single $d$-dim qudit whose state is manipulated using standard quantum-optics elements. The depth achievable with these devices is unfortunately limited by photon loss and, more generally, the qudit encoding employed requires  resources scaling exponentially with the number of orbitals. However, specialized photonic systems like the QFP may have a role in the global ecosystem of quantum technologies for SM physics.
The use of variational approaches based on imaginary-time evolution like QITE was also studied recently in Ref.~\cite{yeter2020practical}. This class of techniques, at least in principle, could be helpful not just as a way of preparing approximations to ground-states (by working at low temperatures) but also open the possibility of studying finite temperature properties in systems like neutron star matter.

Particularly interesting is the possibility to study real-time properties of nuclear systems in thermal equilibrium e.g., transport coefficients in strongly correlated nuclear matter.
In general, these calculations will require knowledge of Minkowski-space correlation functions, which are not easily available with classical methods.
It would have been  \enquote{nice} if
Euclidean time correlations functions could be readily obtained  from Monte Carlo simulations and analytically continued to real time (see e.g., Ref.~\cite{Meyer_2011}). As discussed  in Sec.~\ref{ssec:real-time}, this procedure is in general ill-posed for finite precision correlators (e.g., those extracted from MC sampling), but nevertheless in some situations it is possible to gain considerable insight into transport properties by analyzing features of the Euclidean correlation functions (see e.g., Ref.~\cite{Roggero_2016}). However, a large-scale, high-fidelity quantum computer is expected to be able to evaluate real-time correlators accurately. Depending on the particular problem at hand, it could also be advantageous to work directly in frequency space using well-behaved integral kernels e.g., the Fejer~\cite{PhysRevC.100.034610} or Gaussian~\cite{PhysRevA.102.022409}. Similar in spirit are approaches to extract Green's functions from a quantum simulation~\cite{PhysRevD.102.094505}, which uses the expectation values $F(t)=\langle\Psi|e^{-iHt}|\Psi\rangle$ evaluated over a range of times to approximate  $g(\omega,\eta)=\langle\Psi|\frac{1}{\omega-H+i\eta}|\Psi\rangle$, which in the limit $\eta\to0$ gives the exact Green's function. This Green's function algorithm was investigated through a simple model of the $\phi\to2\chi$ decay of a (fictitious) heavy scalar $\phi$ into two lighter scalar particles $\chi$
using one of IBM's superconducting quantum devices~\cite{PhysRevD.102.094505}, as shown in  Figure~\ref{fig:NuclStruct}.
Variational approaches with reduced computational cost have also been proposed recently~\cite{PhysRevResearch.2.033281}, possibly opening the way to exploration of more realistic simulation on near term devices.
\begin{figure}[!ht]
\centering
\includegraphics[width=0.98\columnwidth]{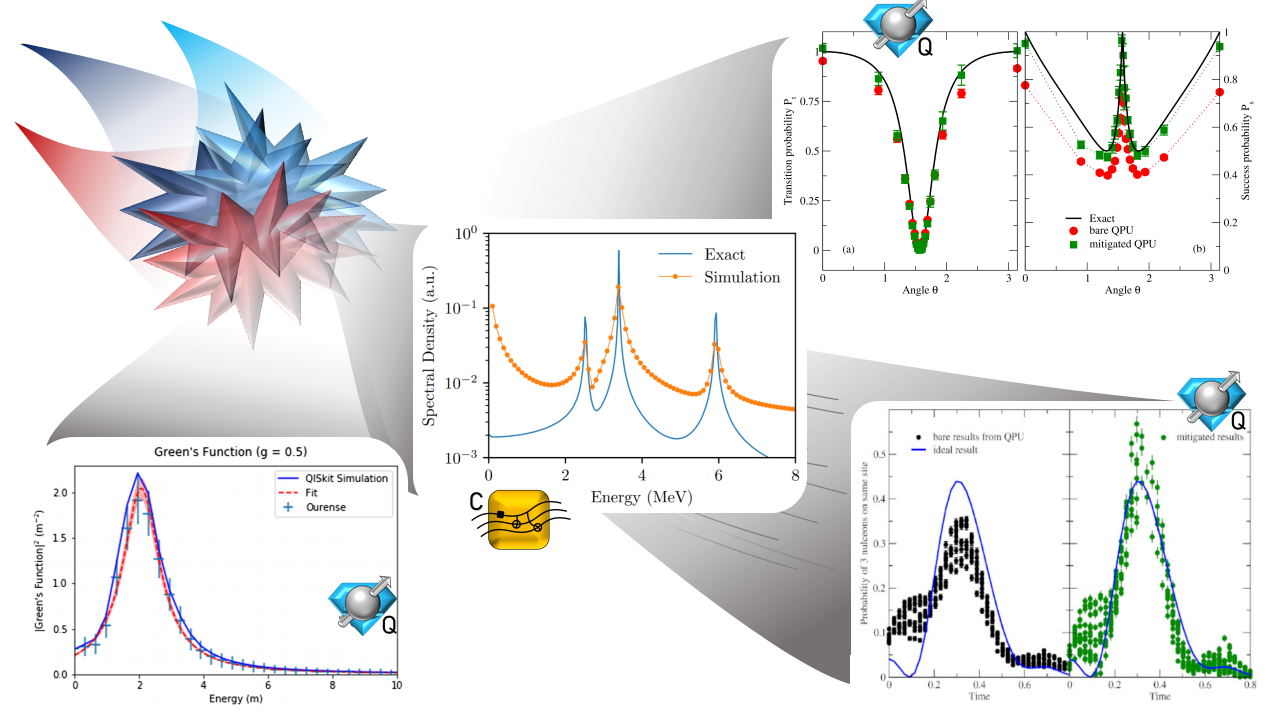}
\caption{Progress in addressing reaction dynamics using quantum devices: (lower left) the magnitude of the extracted Green's function for heavy scalar decay~\cite{PhysRevD.102.094505}, (upper-right)  the $np\leftrightarrow d\gamma$ transition probability and the success probability of the algorithm~\cite{PhysRevC.102.064624} , (lower-right) the
bare and mitigated probabilities of observing three site-localized nucleons as a function of time in a simplified model of scattering with linear response~\cite{PhysRevD.101.074038}, and (center) the spectral density of the neutron-neutron spin-space persistence probability extracted from time evolution~\cite{Holland:2019zju}.
 {\scriptsize  [
(lower-left): Reprinted figure with permission from Anthony Ciavarella, Physical Review D, 102, 094505, 2020. Copyright (2020) by the American Physical Society.
(center): Reprinted figure with permission from Sofia Quaglioni, Physical Review A, 101, 062307, 2020. Copyright (2020) by the American Physical Society.
]}
}
\label{fig:NuclReact}
\end{figure}

Another advantage of frequency space methods is their ability to provide access to information about the possible final states of a nuclear reaction, thus opening the way to a better characterization of semi-exclusive scattering cross sections. This capability will be important in situations where reliable theoretical predictions for the properties of final states is required as input to achieve the physics goals of an experiment. An example of this is long-baseline neutrino oscillation detectors, e.g., DUNE, which are designed to constrain neutrino properties like masses, mixing-angles and the CP-violating phase in the lepton sector. Even though these parameters are  captured by the inclusive scattering of neutrinos off nuclei in the detector (followed possibly by a particle shower), the strong energy dependence of the inclusive cross section, together with the wide range of neutrino energies in the beam, requires an explicit reconstruction of the incoming neutrino energy on an event-by-event basis (using information from the observed final states).
For Argon based detectors like DUNE, it is possible to reconstruct the trajectories of charged particles, and in principle reconstruct the kinematic parameters of the reaction. In practice, however, additional theoretical input is required in order to properly take into account the possible presence of neutral particles in the final state. The dominant quantum resource required for the quantum simulation of these processes is connected with the implementation of the energy filters used to approximately project into intervals of the excitation energy.
A first estimation of the quantum resources required to perform a simulation with possible impact on the physics goals of DUNE was carried out recently~\cite{PhysRevD.101.074038}, using leading-order pionless EFT on a lattice as a minimal choice for realistic nuclear interactions. These results suggest that, at least for the Fejer kernel, a minimally realistic simulation of scattering from $^{40}$Ar will require $\mathcal{O}(10^8)$ one- and two-qubit operations.
Using recent improvements in both integral kernels~\cite{PhysRevA.102.022409} and more efficient representations of the real-time evolution operator for similar models~\cite{su2020nearly}, this estimate could be reduced below $\mathcal{O}(10^{6})$,  leaving realistic studies of nuclear scattering out of reach for NISQ devices.

Thinking beyond the universal gate set towards custom-designed operators, SRF-cavity quantum devices, under development at Lawrence Livermore National Laboratory (LLNL), Fermi National Accelerator Laboratory (FNAL) and elsewhere, are  superconducting quantum devices capable of operating  high dimensional Hilbert spaces per cavity.
Time-ordered radio-frequency signals are introduced into the cavities to accomplish unitary transformations between the many transmon-coupled states supported in high-Q cavities, describing the time evolution of the QMB system of interest.
Classical simulations performed to support the LLNL cavity devices show that entangled spin systems, relevant to low-energy nuclear physics, can be efficiently evolved in time with high fidelity~\cite{Holland:2019zju}.
Figure~\ref{fig:NuclReact} shows the classically-simulated frequency-space evolution of the exact neutron-neutron spin persistence probability during time evolution.

Beyond the preparation of a good approximation $|\Psi_0\rangle$ to the ground state of the target nucleus, techniques working directly in frequency space~\cite{PhysRevC.100.034610} require one more non-trivial step: applying the vertex operator $\hat{O}$ describing the interaction between probe and target while maintaining a normalized state $|\Phi_O\rangle \propto \hat{O}|\Psi_0\rangle$.
The main complication in this step is that the vertex operator $\hat{O}$ is Hermitian but not necessarily unitary.
A detailed comparison of two  approaches to carry out this step has been made~\cite{PhysRevC.102.064624} using a schematic representation of the $np\leftrightarrow d\gamma$ reactions.
The first method~\cite{PhysRevC.100.034610} proposed using real time evolution under the action of the vertex operator $\hat{O}$ controlled by the state of an additional ancilla qubit,
while the second method~\cite{Childs2012} is based on the general linear combination of unitaries (LCU) scheme, developed predominantly for fault tolerant quantum devices (as in general it requires  control logic and large ancilla registers).
Results obtained in Ref.~\cite{PhysRevC.102.064624} and shown in Figure~\ref{fig:NuclReact}, indicate that the computational cost of LCU can be substantially reduced by a careful choice of operator basis and by optimizing the implementation with respect to the limited connectivity available in the device. The net result is that the LCU-based state preparation is expected to outperform the more memory efficient time-dependent algorithm in both the fidelity and success probability in most cases.

\subsection{Collective Neutrino Oscillations}
Neutrinos initially produced in a flavor eigenstate can experience flavor oscillations, with frequencies determined by its energy and the mass-squared gaps between mass eigenstates.
The flavor evolution of a neutrino \enquote{cloud} can be modified substantially through weak interaction with a background e.g., the MSW effect generated by interactions with a background of leptons~\cite{Wolfenstein:1977ue,Mikheev:1986gs}.
In astrophysical settings with large neutrino densities e.g., core-collapse supernovae and the early universe, neutral-current neutrino-neutrino interactions can become important and modify substantially the transport of flavor in these environments.
Neutrino-neutrino interactions can lead to collective neutrino oscillations, and these modes can have important effects on the dynamics of extreme astrophysical environments and the early universe~\cite{1987ApJ...322..795F,Savage:1990by}.
Considering only the leading order neutrino-neutrino interactions in  forward scattering, and neglecting momentum-changing interactions suppressed as $G_F^2$, the dynamics  act only on the flavor degrees of freedom, which can be represented by a collection of $SU(N_f)$ spins.
In this framework, the neutrino momenta are fixed and only the spin (flavor) polarization evolves in time. In the simplified approximation with  $N_f=2$ flavors, the model Hamiltonian is equivalent to an all-to-all Heisenberg model with an interaction strength proportional to the neutrino density~\cite{Pehlivan2011}.

In general, solving for the exact real-time dynamics of these models is out of reach to direct methods based on exact diagonalization.   In highly symmetric homogeneous situations, the neutrino Hamiltonian becomes integrable and the Bethe ansatz can be used~\cite{Pehlivan2011,PhysRevD.99.123013}. For more general conditions,  mean field approximations are usually employed (see e.g., Refs.~\cite{Duan:2010bg,Chakraborty2016b} for recent reviews). Due to the widespread use of mean-field simulation techniques, understanding the accuracy of this approximation in capturing the dominant effects in collective oscillations is therefore essential~\cite{Bell:2003mg,Friedland2003,Cervia:2019,Rrapaj2020}.
Due to the complexity of the calculations, however, the systems sizes that can be explored with these direct approaches is limited to $\mathcal{O}(10)$, making it difficult to understand extrapolations to the thermodynamic limit.
\begin{figure}[!ht]
\centering
\includegraphics[width=0.98\columnwidth]{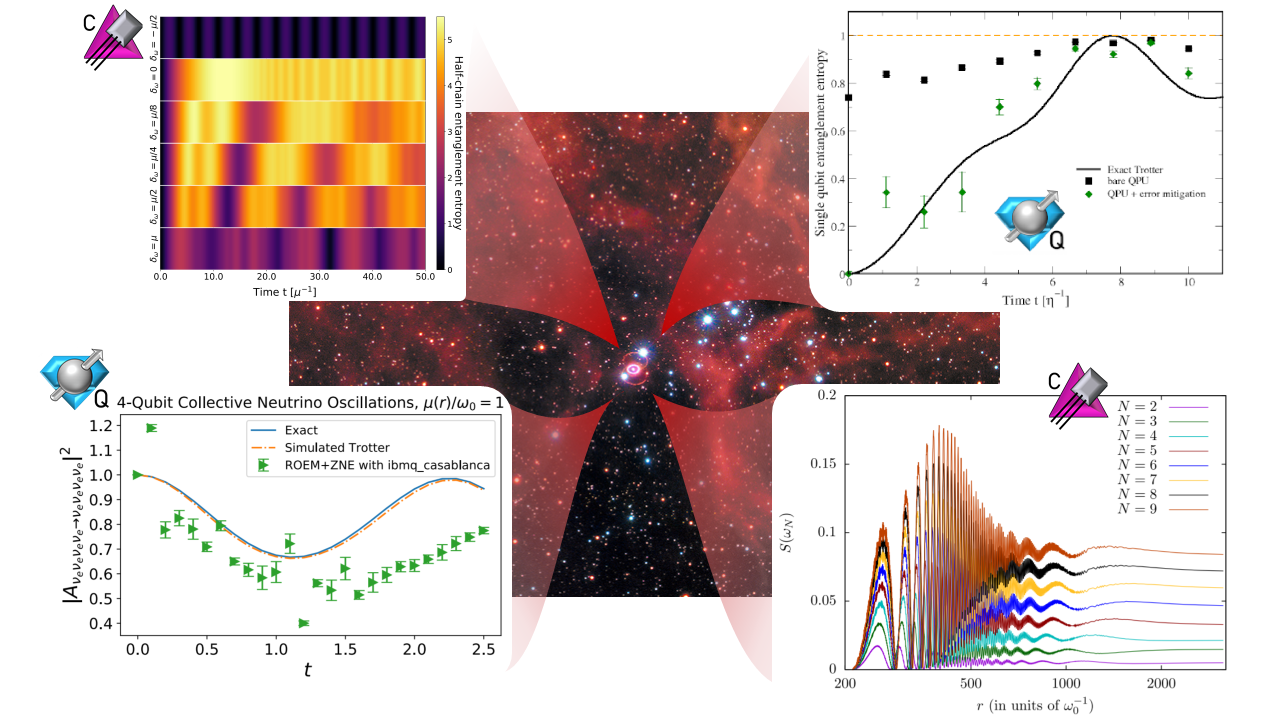}
\caption{Highlights of recent works on collective neutrino oscillations: (upper-left) maximum entanglement entropy in a system of $N_\nu=96$ neutrinos for various energy asymmetries computed using MPS~\cite{roggero2021dynamical},
(upper-right) single-qubit entanglement entropy extracted from a real-time simulation of $N_\nu=4$ neutrinos using IBM's quantum devices~\cite{hall2021simulation},
(lower-right) evolution of entanglement entropy for the most energetic neutrino in systems of size $N_\nu=2-9$~\cite{Cervia:2019},
(lower-left) flavor survival probability extracted from simulations of $N_\nu=4$ neutrinos~\cite{yeteraydeniz2021collective}.
 {\scriptsize  [
 Background image from NASA shows SN 1987A (in the center): Link: \href{https://www.nasa.gov/feature/goddard/2017/the-dawn-of-a-new-era-for-supernova-1987a}{NASA} Credits: NASA, ESA, R. Kirshner (Harvard-Smithsonian Center for Astrophysics and Gordon and Betty Moore Foundation), and M. Mutchler and R. Avila (STScI).
(lower-left): Reprinted figure with permission Raphael Pooser.
(lower-right): Reprinted figure with permission from Calvin Johnson , Physical Review D, 100, 083001, 2019. Copyright (2019) by the American Physical Society.
]}
}
\label{fig:Neutrinos}
\end{figure}
One venue where quantum technologies can help in breaking this computational barrier is in simulations of the flavor dynamics.  First explorations of this possibility with superconducting qubit devices on small systems with up to $N_\nu=4$ neutrino amplitudes have been performed~\cite{hall2021simulation,yeteraydeniz2021collective}. One of the main bottlenecks for efficient implementations of these models is the need to implement the all-to-all pairwise interactions on devices with limited connectivity.
A general procedure to side-step this problem has been proposed~\cite{hall2021simulation} by means of a swap-network construction that can be implemented efficiently even on devices with linear nearest-neighbor connectivity.
The detectable difference from the mean field approximation
may result from non-trivial evolution of entanglement  and different techniques to reliably extract information about different entanglement measures from a digital quantum simulation have been presented in Ref.~\cite{hall2021simulation}.
A complementary approach is to exploit the long-range nature of interactions in trapped-ion systems~\cite{RevModPhys.93.025001} to carry out analog simulations of flavor evolution, by generalizing earlier work on dynamical phase transitions in Ising models~\cite{Zhang2017}.

First explorations of employing an MPS ansatz for the many-body neutrino wave functions have been carried out~\cite{roggero2021entanglement,roggero2021dynamical}. These calculations show that, at least for conditions with a high degree of symmetry, the bipartite entanglement in a neutrino cloud starting as a product state grows  slowly with system size as $\approx\log(N_\nu)$. This allows for controllable simulations of more than a hundred neutrino amplitudes and the identification of a link between the appearance of instabilities in the flavor evolution and the presence of a dynamical phase transition in the underlying spin model~\cite{roggero2021dynamical}.
This connection has the potential of providing a general framework to understand the conditions leading to collective oscillation phenomena without resorting to a mean-field approximation.
A particularly exciting prospect of using MPS-based techniques for studies on collective neutrino oscillations is the ability to detect the presence of configurations capable of generating substantial levels of entanglement by checking the convergence of the calculations.
The identification of these special conditions, if they exist, will allow a focusing of quantum resources on the simulation of these instances, which will be out of the reach of tensor-network-based methods.

\section{Closing}
\label{sec:outlook}

This article has highlighted a few key aspects of integrating Standard Model research and advances in quantum information,
inspired by rapid improvements in the control of entanglement and coherence in the laboratory leading to the first quantum devices available for computation.
The heightened appreciation of non-local correlations and distributed quantum information,
in particular the multifaceted role of entanglement throughout the Standard Model,
is anticipated to broadly inspire new paradigms of calculation design, accelerate the development of scalable quantum technologies, and propel aspects of nuclear and high-energy research through the 21st Century.
While capturing only an incomplete slice of research {\it at the interface}, it is hoped that the reflections within this article will stimulate further discussions throughout the diverse and growing quantum community.

\section{Acknowledgements}
We would like to thank our friends and collaborators for providing a stimulating and thriving {\it quantum village}
 from which this article emerged, and ask for their forgiveness in failing to do justice to their insights and accomplishments.
The work of Natalie Klco is supported in part by the Walter Burke Institute for Theoretical Physics, and by the U.S. Department of Energy Office of Science, Office of Advanced Scientific Computing Research, (DE-SC0020290), and Office of High Energy Physics DEACO2-07CH11359.
The work of Alessandro Roggero and Martin Savage were supported in part by the
U.S. Department of Energy, Office of Science, Office
of Nuclear Physics, InQubator for Quantum Simulation
(IQuS) under Award Number DOE (NP) Award DESC0020970.

\bibliography{biblio}

\begin{thebibliography}{521}%
\makeatletter
\providecommand \@ifxundefined [1]{%
 \@ifx{#1\undefined}
}%
\providecommand \@ifnum [1]{%
 \ifnum #1\expandafter \@firstoftwo
 \else \expandafter \@secondoftwo
 \fi
}%
\providecommand \@ifx [1]{%
 \ifx #1\expandafter \@firstoftwo
 \else \expandafter \@secondoftwo
 \fi
}%
\providecommand \natexlab [1]{#1}%
\providecommand \enquote  [1]{``#1''}%
\providecommand \bibnamefont  [1]{#1}%
\providecommand \bibfnamefont [1]{#1}%
\providecommand \citenamefont [1]{#1}%
\providecommand \href@noop [0]{\@secondoftwo}%
\providecommand \href [0]{\begingroup \@sanitize@url \@href}%
\providecommand \@href[1]{\@@startlink{#1}\@@href}%
\providecommand \@@href[1]{\endgroup#1\@@endlink}%
\providecommand \@sanitize@url [0]{\catcode `\\12\catcode `\$12\catcode
  `\&12\catcode `\#12\catcode `\^12\catcode `\_12\catcode `\%12\relax}%
\providecommand \@@startlink[1]{}%
\providecommand \@@endlink[0]{}%
\providecommand \url  [0]{\begingroup\@sanitize@url \@url }%
\providecommand \@url [1]{\endgroup\@href {#1}{\urlprefix }}%
\providecommand \urlprefix  [0]{URL }%
\providecommand \Eprint [0]{\href }%
\providecommand \doibase [0]{http://dx.doi.org/}%
\providecommand \selectlanguage [0]{\@gobble}%
\providecommand \bibinfo  [0]{\@secondoftwo}%
\providecommand \bibfield  [0]{\@secondoftwo}%
\providecommand \translation [1]{[#1]}%
\providecommand \BibitemOpen [0]{}%
\providecommand \bibitemStop [0]{}%
\providecommand \bibitemNoStop [0]{.\EOS\space}%
\providecommand \EOS [0]{\spacefactor3000\relax}%
\providecommand \BibitemShut  [1]{\csname bibitem#1\endcsname}%
\let\auto@bib@innerbib\@empty
\bibitem [{\citenamefont {Arute}\ \emph {et~al.}(2019)\citenamefont {Arute},
  \citenamefont {Arya}, \citenamefont {Babbush}, \citenamefont {Bacon},
  \citenamefont {Bardin}, \citenamefont {Barends}, \citenamefont {Biswas},
  \citenamefont {Boixo}, \citenamefont {Brandao}, \citenamefont {Buell},
  \citenamefont {Burkett}, \citenamefont {Chen}, \citenamefont {Chen},
  \citenamefont {Chiaro}, \citenamefont {Collins}, \citenamefont {Courtney},
  \citenamefont {Dunsworth}, \citenamefont {Farhi}, \citenamefont {Foxen},
  \citenamefont {Fowler}, \citenamefont {Gidney}, \citenamefont {Giustina},
  \citenamefont {Graff}, \citenamefont {Guerin}, \citenamefont {Habegger},
  \citenamefont {Harrigan}, \citenamefont {Hartmann}, \citenamefont {Ho},
  \citenamefont {Hoffmann}, \citenamefont {Huang}, \citenamefont {Humble},
  \citenamefont {Isakov}, \citenamefont {Jeffrey}, \citenamefont {Jiang},
  \citenamefont {Kafri}, \citenamefont {Kechedzhi}, \citenamefont {Kelly},
  \citenamefont {Klimov}, \citenamefont {Knysh}, \citenamefont {Korotkov},
  \citenamefont {Kostritsa}, \citenamefont {Landhuis}, \citenamefont
  {Lindmark}, \citenamefont {Lucero}, \citenamefont {Lyakh}, \citenamefont
  {Mandrà}, \citenamefont {McClean}, \citenamefont {McEwen}, \citenamefont
  {Megrant}, \citenamefont {Mi}, \citenamefont {Michielsen}, \citenamefont
  {Mohseni}, \citenamefont {Mutus}, \citenamefont {Naaman}, \citenamefont
  {Neeley}, \citenamefont {Neill}, \citenamefont {Niu}, \citenamefont {Ostby},
  \citenamefont {Petukhov}, \citenamefont {Platt}, \citenamefont {Quintana},
  \citenamefont {Rieffel}, \citenamefont {Roushan}, \citenamefont {Rubin},
  \citenamefont {Sank}, \citenamefont {Satzinger}, \citenamefont {Smelyanskiy},
  \citenamefont {Sung}, \citenamefont {Trevithick}, \citenamefont
  {Vainsencher}, \citenamefont {Villalonga}, \citenamefont {White},
  \citenamefont {Yao}, \citenamefont {Yeh}, \citenamefont {Zalcman},
  \citenamefont {Neven},\ and\ \citenamefont {Martinis}}]{Google48651}%
  \BibitemOpen
  \bibfield  {author} {\bibinfo {author} {\bibfnamefont {F.}~\bibnamefont
  {Arute}}, \bibinfo {author} {\bibfnamefont {K.}~\bibnamefont {Arya}},
  \bibinfo {author} {\bibfnamefont {R.}~\bibnamefont {Babbush}}, \bibinfo
  {author} {\bibfnamefont {D.}~\bibnamefont {Bacon}}, \bibinfo {author}
  {\bibfnamefont {J.}~\bibnamefont {Bardin}}, \bibinfo {author} {\bibfnamefont
  {R.}~\bibnamefont {Barends}}, \bibinfo {author} {\bibfnamefont
  {R.}~\bibnamefont {Biswas}}, \bibinfo {author} {\bibfnamefont
  {S.}~\bibnamefont {Boixo}}, \bibinfo {author} {\bibfnamefont
  {F.}~\bibnamefont {Brandao}}, \bibinfo {author} {\bibfnamefont
  {D.}~\bibnamefont {Buell}}, \bibinfo {author} {\bibfnamefont
  {B.}~\bibnamefont {Burkett}}, \bibinfo {author} {\bibfnamefont
  {Y.}~\bibnamefont {Chen}}, \bibinfo {author} {\bibfnamefont {J.}~\bibnamefont
  {Chen}}, \bibinfo {author} {\bibfnamefont {B.}~\bibnamefont {Chiaro}},
  \bibinfo {author} {\bibfnamefont {R.}~\bibnamefont {Collins}}, \bibinfo
  {author} {\bibfnamefont {W.}~\bibnamefont {Courtney}}, \bibinfo {author}
  {\bibfnamefont {A.}~\bibnamefont {Dunsworth}}, \bibinfo {author}
  {\bibfnamefont {E.}~\bibnamefont {Farhi}}, \bibinfo {author} {\bibfnamefont
  {B.}~\bibnamefont {Foxen}}, \bibinfo {author} {\bibfnamefont
  {A.}~\bibnamefont {Fowler}}, \bibinfo {author} {\bibfnamefont {C.~M.}\
  \bibnamefont {Gidney}}, \bibinfo {author} {\bibfnamefont {M.}~\bibnamefont
  {Giustina}}, \bibinfo {author} {\bibfnamefont {R.}~\bibnamefont {Graff}},
  \bibinfo {author} {\bibfnamefont {K.}~\bibnamefont {Guerin}}, \bibinfo
  {author} {\bibfnamefont {S.}~\bibnamefont {Habegger}}, \bibinfo {author}
  {\bibfnamefont {M.}~\bibnamefont {Harrigan}}, \bibinfo {author}
  {\bibfnamefont {M.}~\bibnamefont {Hartmann}}, \bibinfo {author}
  {\bibfnamefont {A.}~\bibnamefont {Ho}}, \bibinfo {author} {\bibfnamefont
  {M.~R.}\ \bibnamefont {Hoffmann}}, \bibinfo {author} {\bibfnamefont
  {T.}~\bibnamefont {Huang}}, \bibinfo {author} {\bibfnamefont
  {T.}~\bibnamefont {Humble}}, \bibinfo {author} {\bibfnamefont
  {S.}~\bibnamefont {Isakov}}, \bibinfo {author} {\bibfnamefont
  {E.}~\bibnamefont {Jeffrey}}, \bibinfo {author} {\bibfnamefont
  {Z.}~\bibnamefont {Jiang}}, \bibinfo {author} {\bibfnamefont
  {D.}~\bibnamefont {Kafri}}, \bibinfo {author} {\bibfnamefont
  {K.}~\bibnamefont {Kechedzhi}}, \bibinfo {author} {\bibfnamefont
  {J.}~\bibnamefont {Kelly}}, \bibinfo {author} {\bibfnamefont
  {P.}~\bibnamefont {Klimov}}, \bibinfo {author} {\bibfnamefont
  {S.}~\bibnamefont {Knysh}}, \bibinfo {author} {\bibfnamefont
  {A.}~\bibnamefont {Korotkov}}, \bibinfo {author} {\bibfnamefont
  {F.}~\bibnamefont {Kostritsa}}, \bibinfo {author} {\bibfnamefont
  {D.}~\bibnamefont {Landhuis}}, \bibinfo {author} {\bibfnamefont
  {M.}~\bibnamefont {Lindmark}}, \bibinfo {author} {\bibfnamefont
  {E.}~\bibnamefont {Lucero}}, \bibinfo {author} {\bibfnamefont
  {D.}~\bibnamefont {Lyakh}}, \bibinfo {author} {\bibfnamefont
  {S.}~\bibnamefont {Mandrà}}, \bibinfo {author} {\bibfnamefont {J.~R.}\
  \bibnamefont {McClean}}, \bibinfo {author} {\bibfnamefont {M.}~\bibnamefont
  {McEwen}}, \bibinfo {author} {\bibfnamefont {A.}~\bibnamefont {Megrant}},
  \bibinfo {author} {\bibfnamefont {X.}~\bibnamefont {Mi}}, \bibinfo {author}
  {\bibfnamefont {K.}~\bibnamefont {Michielsen}}, \bibinfo {author}
  {\bibfnamefont {M.}~\bibnamefont {Mohseni}}, \bibinfo {author} {\bibfnamefont
  {J.}~\bibnamefont {Mutus}}, \bibinfo {author} {\bibfnamefont
  {O.}~\bibnamefont {Naaman}}, \bibinfo {author} {\bibfnamefont
  {M.}~\bibnamefont {Neeley}}, \bibinfo {author} {\bibfnamefont
  {C.}~\bibnamefont {Neill}}, \bibinfo {author} {\bibfnamefont {M.~Y.}\
  \bibnamefont {Niu}}, \bibinfo {author} {\bibfnamefont {E.}~\bibnamefont
  {Ostby}}, \bibinfo {author} {\bibfnamefont {A.}~\bibnamefont {Petukhov}},
  \bibinfo {author} {\bibfnamefont {J.}~\bibnamefont {Platt}}, \bibinfo
  {author} {\bibfnamefont {C.}~\bibnamefont {Quintana}}, \bibinfo {author}
  {\bibfnamefont {E.~G.}\ \bibnamefont {Rieffel}}, \bibinfo {author}
  {\bibfnamefont {P.}~\bibnamefont {Roushan}}, \bibinfo {author} {\bibfnamefont
  {N.}~\bibnamefont {Rubin}}, \bibinfo {author} {\bibfnamefont
  {D.}~\bibnamefont {Sank}}, \bibinfo {author} {\bibfnamefont {K.~J.}\
  \bibnamefont {Satzinger}}, \bibinfo {author} {\bibfnamefont {V.}~\bibnamefont
  {Smelyanskiy}}, \bibinfo {author} {\bibfnamefont {K.~J.}\ \bibnamefont
  {Sung}}, \bibinfo {author} {\bibfnamefont {M.}~\bibnamefont {Trevithick}},
  \bibinfo {author} {\bibfnamefont {A.}~\bibnamefont {Vainsencher}}, \bibinfo
  {author} {\bibfnamefont {B.}~\bibnamefont {Villalonga}}, \bibinfo {author}
  {\bibfnamefont {T.}~\bibnamefont {White}}, \bibinfo {author} {\bibfnamefont
  {Z.~J.}\ \bibnamefont {Yao}}, \bibinfo {author} {\bibfnamefont
  {P.}~\bibnamefont {Yeh}}, \bibinfo {author} {\bibfnamefont {A.}~\bibnamefont
  {Zalcman}}, \bibinfo {author} {\bibfnamefont {H.}~\bibnamefont {Neven}}, \
  and\ \bibinfo {author} {\bibfnamefont {J.}~\bibnamefont {Martinis}},\ }\href
  {https://www.nature.com/articles/s41586-019-1666-5} {\bibfield  {journal}
  {\bibinfo  {journal} {Nature}\ }\textbf {\bibinfo {volume} {574}},\ \bibinfo
  {pages} {505–510} (\bibinfo {year} {2019})}\BibitemShut {NoStop}%
\bibitem [{\citenamefont {Zhong}\ \emph {et~al.}(2020)\citenamefont {Zhong},
  \citenamefont {Wang}, \citenamefont {Deng}, \citenamefont {Chen},
  \citenamefont {Peng}, \citenamefont {Luo}, \citenamefont {Qin}, \citenamefont
  {Wu}, \citenamefont {Ding}, \citenamefont {Hu}, \citenamefont {Hu},
  \citenamefont {Yang}, \citenamefont {Zhang}, \citenamefont {Li},
  \citenamefont {Li}, \citenamefont {Jiang}, \citenamefont {Gan}, \citenamefont
  {Yang}, \citenamefont {You}, \citenamefont {Wang}, \citenamefont {Li},
  \citenamefont {Liu}, \citenamefont {Lu},\ and\ \citenamefont
  {Pan}}]{Zhong1460}%
  \BibitemOpen
  \bibfield  {author} {\bibinfo {author} {\bibfnamefont {H.-S.}\ \bibnamefont
  {Zhong}}, \bibinfo {author} {\bibfnamefont {H.}~\bibnamefont {Wang}},
  \bibinfo {author} {\bibfnamefont {Y.-H.}\ \bibnamefont {Deng}}, \bibinfo
  {author} {\bibfnamefont {M.-C.}\ \bibnamefont {Chen}}, \bibinfo {author}
  {\bibfnamefont {L.-C.}\ \bibnamefont {Peng}}, \bibinfo {author}
  {\bibfnamefont {Y.-H.}\ \bibnamefont {Luo}}, \bibinfo {author} {\bibfnamefont
  {J.}~\bibnamefont {Qin}}, \bibinfo {author} {\bibfnamefont {D.}~\bibnamefont
  {Wu}}, \bibinfo {author} {\bibfnamefont {X.}~\bibnamefont {Ding}}, \bibinfo
  {author} {\bibfnamefont {Y.}~\bibnamefont {Hu}}, \bibinfo {author}
  {\bibfnamefont {P.}~\bibnamefont {Hu}}, \bibinfo {author} {\bibfnamefont
  {X.-Y.}\ \bibnamefont {Yang}}, \bibinfo {author} {\bibfnamefont {W.-J.}\
  \bibnamefont {Zhang}}, \bibinfo {author} {\bibfnamefont {H.}~\bibnamefont
  {Li}}, \bibinfo {author} {\bibfnamefont {Y.}~\bibnamefont {Li}}, \bibinfo
  {author} {\bibfnamefont {X.}~\bibnamefont {Jiang}}, \bibinfo {author}
  {\bibfnamefont {L.}~\bibnamefont {Gan}}, \bibinfo {author} {\bibfnamefont
  {G.}~\bibnamefont {Yang}}, \bibinfo {author} {\bibfnamefont {L.}~\bibnamefont
  {You}}, \bibinfo {author} {\bibfnamefont {Z.}~\bibnamefont {Wang}}, \bibinfo
  {author} {\bibfnamefont {L.}~\bibnamefont {Li}}, \bibinfo {author}
  {\bibfnamefont {N.-L.}\ \bibnamefont {Liu}}, \bibinfo {author} {\bibfnamefont
  {C.-Y.}\ \bibnamefont {Lu}}, \ and\ \bibinfo {author} {\bibfnamefont {J.-W.}\
  \bibnamefont {Pan}},\ }\href {\doibase 10.1126/science.abe8770} {\bibfield
  {journal} {\bibinfo  {journal} {Science}\ }\textbf {\bibinfo {volume}
  {370}},\ \bibinfo {pages} {1460} (\bibinfo {year} {2020})}\BibitemShut
  {NoStop}%
\bibitem [{\citenamefont {Wu}\ \emph {et~al.}(2021)\citenamefont {Wu},
  \citenamefont {Bao}, \citenamefont {Cao}, \citenamefont {Chen}, \citenamefont
  {Chen}, \citenamefont {Chen}, \citenamefont {Chung}, \citenamefont {Deng},
  \citenamefont {Du}, \citenamefont {Fan}, \citenamefont {Gong}, \citenamefont
  {Guo}, \citenamefont {Guo}, \citenamefont {Guo}, \citenamefont {Han},
  \citenamefont {Hong}, \citenamefont {Huang}, \citenamefont {Huo},
  \citenamefont {Li}, \citenamefont {Li}, \citenamefont {Li}, \citenamefont
  {Li}, \citenamefont {Liang}, \citenamefont {Lin}, \citenamefont {Lin},
  \citenamefont {Qian}, \citenamefont {Qiao}, \citenamefont {Rong},
  \citenamefont {Su}, \citenamefont {Sun}, \citenamefont {Wang}, \citenamefont
  {Wang}, \citenamefont {Wu}, \citenamefont {Xu}, \citenamefont {Yan},
  \citenamefont {Yang}, \citenamefont {Yang}, \citenamefont {Ye}, \citenamefont
  {Yin}, \citenamefont {Ying}, \citenamefont {Yu}, \citenamefont {Zha},
  \citenamefont {Zhang}, \citenamefont {Zhang}, \citenamefont {Zhang},
  \citenamefont {Zhang}, \citenamefont {Zhao}, \citenamefont {Zhao},
  \citenamefont {Zhou}, \citenamefont {Zhu}, \citenamefont {Lu}, \citenamefont
  {Peng}, \citenamefont {Zhu},\ and\ \citenamefont {Pan}}]{wu2021strong}%
  \BibitemOpen
  \bibfield  {author} {\bibinfo {author} {\bibfnamefont {Y.}~\bibnamefont
  {Wu}}, \bibinfo {author} {\bibfnamefont {W.-S.}\ \bibnamefont {Bao}},
  \bibinfo {author} {\bibfnamefont {S.}~\bibnamefont {Cao}}, \bibinfo {author}
  {\bibfnamefont {F.}~\bibnamefont {Chen}}, \bibinfo {author} {\bibfnamefont
  {M.-C.}\ \bibnamefont {Chen}}, \bibinfo {author} {\bibfnamefont
  {X.}~\bibnamefont {Chen}}, \bibinfo {author} {\bibfnamefont {T.-H.}\
  \bibnamefont {Chung}}, \bibinfo {author} {\bibfnamefont {H.}~\bibnamefont
  {Deng}}, \bibinfo {author} {\bibfnamefont {Y.}~\bibnamefont {Du}}, \bibinfo
  {author} {\bibfnamefont {D.}~\bibnamefont {Fan}}, \bibinfo {author}
  {\bibfnamefont {M.}~\bibnamefont {Gong}}, \bibinfo {author} {\bibfnamefont
  {C.}~\bibnamefont {Guo}}, \bibinfo {author} {\bibfnamefont {C.}~\bibnamefont
  {Guo}}, \bibinfo {author} {\bibfnamefont {S.}~\bibnamefont {Guo}}, \bibinfo
  {author} {\bibfnamefont {L.}~\bibnamefont {Han}}, \bibinfo {author}
  {\bibfnamefont {L.}~\bibnamefont {Hong}}, \bibinfo {author} {\bibfnamefont
  {H.-L.}\ \bibnamefont {Huang}}, \bibinfo {author} {\bibfnamefont {Y.-H.}\
  \bibnamefont {Huo}}, \bibinfo {author} {\bibfnamefont {L.}~\bibnamefont
  {Li}}, \bibinfo {author} {\bibfnamefont {N.}~\bibnamefont {Li}}, \bibinfo
  {author} {\bibfnamefont {S.}~\bibnamefont {Li}}, \bibinfo {author}
  {\bibfnamefont {Y.}~\bibnamefont {Li}}, \bibinfo {author} {\bibfnamefont
  {F.}~\bibnamefont {Liang}}, \bibinfo {author} {\bibfnamefont
  {C.}~\bibnamefont {Lin}}, \bibinfo {author} {\bibfnamefont {J.}~\bibnamefont
  {Lin}}, \bibinfo {author} {\bibfnamefont {H.}~\bibnamefont {Qian}}, \bibinfo
  {author} {\bibfnamefont {D.}~\bibnamefont {Qiao}}, \bibinfo {author}
  {\bibfnamefont {H.}~\bibnamefont {Rong}}, \bibinfo {author} {\bibfnamefont
  {H.}~\bibnamefont {Su}}, \bibinfo {author} {\bibfnamefont {L.}~\bibnamefont
  {Sun}}, \bibinfo {author} {\bibfnamefont {L.}~\bibnamefont {Wang}}, \bibinfo
  {author} {\bibfnamefont {S.}~\bibnamefont {Wang}}, \bibinfo {author}
  {\bibfnamefont {D.}~\bibnamefont {Wu}}, \bibinfo {author} {\bibfnamefont
  {Y.}~\bibnamefont {Xu}}, \bibinfo {author} {\bibfnamefont {K.}~\bibnamefont
  {Yan}}, \bibinfo {author} {\bibfnamefont {W.}~\bibnamefont {Yang}}, \bibinfo
  {author} {\bibfnamefont {Y.}~\bibnamefont {Yang}}, \bibinfo {author}
  {\bibfnamefont {Y.}~\bibnamefont {Ye}}, \bibinfo {author} {\bibfnamefont
  {J.}~\bibnamefont {Yin}}, \bibinfo {author} {\bibfnamefont {C.}~\bibnamefont
  {Ying}}, \bibinfo {author} {\bibfnamefont {J.}~\bibnamefont {Yu}}, \bibinfo
  {author} {\bibfnamefont {C.}~\bibnamefont {Zha}}, \bibinfo {author}
  {\bibfnamefont {C.}~\bibnamefont {Zhang}}, \bibinfo {author} {\bibfnamefont
  {H.}~\bibnamefont {Zhang}}, \bibinfo {author} {\bibfnamefont
  {K.}~\bibnamefont {Zhang}}, \bibinfo {author} {\bibfnamefont
  {Y.}~\bibnamefont {Zhang}}, \bibinfo {author} {\bibfnamefont
  {H.}~\bibnamefont {Zhao}}, \bibinfo {author} {\bibfnamefont {Y.}~\bibnamefont
  {Zhao}}, \bibinfo {author} {\bibfnamefont {L.}~\bibnamefont {Zhou}}, \bibinfo
  {author} {\bibfnamefont {Q.}~\bibnamefont {Zhu}}, \bibinfo {author}
  {\bibfnamefont {C.-Y.}\ \bibnamefont {Lu}}, \bibinfo {author} {\bibfnamefont
  {C.-Z.}\ \bibnamefont {Peng}}, \bibinfo {author} {\bibfnamefont
  {X.}~\bibnamefont {Zhu}}, \ and\ \bibinfo {author} {\bibfnamefont {J.-W.}\
  \bibnamefont {Pan}},\ }\href@noop {} {\enquote {\bibinfo {title} {Strong
  quantum computational advantage using a superconducting quantum processor},}\
  } (\bibinfo {year} {2021}),\ \Eprint {http://arxiv.org/abs/2106.14734}
  {arXiv:2106.14734 [quant-ph]} \BibitemShut {NoStop}%
\bibitem [{\citenamefont {Einstein}\ \emph {et~al.}(1935)\citenamefont
  {Einstein}, \citenamefont {Podolsky},\ and\ \citenamefont
  {Rosen}}]{PhysRev.47.777}%
  \BibitemOpen
  \bibfield  {author} {\bibinfo {author} {\bibfnamefont {A.}~\bibnamefont
  {Einstein}}, \bibinfo {author} {\bibfnamefont {B.}~\bibnamefont {Podolsky}},
  \ and\ \bibinfo {author} {\bibfnamefont {N.}~\bibnamefont {Rosen}},\ }\href
  {\doibase 10.1103/PhysRev.47.777} {\bibfield  {journal} {\bibinfo  {journal}
  {Phys. Rev.}\ }\textbf {\bibinfo {volume} {47}},\ \bibinfo {pages} {777}
  (\bibinfo {year} {1935})}\BibitemShut {NoStop}%
\bibitem [{\citenamefont {Bell}(1964)}]{PhysicsPhysiqueFizika.1.195}%
  \BibitemOpen
  \bibfield  {author} {\bibinfo {author} {\bibfnamefont {J.~S.}\ \bibnamefont
  {Bell}},\ }\href {\doibase 10.1103/PhysicsPhysiqueFizika.1.195} {\bibfield
  {journal} {\bibinfo  {journal} {Physics Physique Fizika}\ }\textbf {\bibinfo
  {volume} {1}},\ \bibinfo {pages} {195} (\bibinfo {year} {1964})}\BibitemShut
  {NoStop}%
\bibitem [{\citenamefont {Freedman}\ and\ \citenamefont
  {Clauser}(1972)}]{PhysRevLett.28.938}%
  \BibitemOpen
  \bibfield  {author} {\bibinfo {author} {\bibfnamefont {S.~J.}\ \bibnamefont
  {Freedman}}\ and\ \bibinfo {author} {\bibfnamefont {J.~F.}\ \bibnamefont
  {Clauser}},\ }\href {\doibase 10.1103/PhysRevLett.28.938} {\bibfield
  {journal} {\bibinfo  {journal} {Phys. Rev. Lett.}\ }\textbf {\bibinfo
  {volume} {28}},\ \bibinfo {pages} {938} (\bibinfo {year} {1972})}\BibitemShut
  {NoStop}%
\bibitem [{\citenamefont {Aspect}\ \emph
  {et~al.}(1982{\natexlab{a}})\citenamefont {Aspect}, \citenamefont
  {Grangier},\ and\ \citenamefont {Roger}}]{PhysRevLett.49.91}%
  \BibitemOpen
  \bibfield  {author} {\bibinfo {author} {\bibfnamefont {A.}~\bibnamefont
  {Aspect}}, \bibinfo {author} {\bibfnamefont {P.}~\bibnamefont {Grangier}}, \
  and\ \bibinfo {author} {\bibfnamefont {G.}~\bibnamefont {Roger}},\ }\href
  {\doibase 10.1103/PhysRevLett.49.91} {\bibfield  {journal} {\bibinfo
  {journal} {Phys. Rev. Lett.}\ }\textbf {\bibinfo {volume} {49}},\ \bibinfo
  {pages} {91} (\bibinfo {year} {1982}{\natexlab{a}})}\BibitemShut {NoStop}%
\bibitem [{\citenamefont {Aspect}\ \emph
  {et~al.}(1982{\natexlab{b}})\citenamefont {Aspect}, \citenamefont
  {Dalibard},\ and\ \citenamefont {Roger}}]{PhysRevLett.49.1804}%
  \BibitemOpen
  \bibfield  {author} {\bibinfo {author} {\bibfnamefont {A.}~\bibnamefont
  {Aspect}}, \bibinfo {author} {\bibfnamefont {J.}~\bibnamefont {Dalibard}}, \
  and\ \bibinfo {author} {\bibfnamefont {G.}~\bibnamefont {Roger}},\ }\href
  {\doibase 10.1103/PhysRevLett.49.1804} {\bibfield  {journal} {\bibinfo
  {journal} {Phys. Rev. Lett.}\ }\textbf {\bibinfo {volume} {49}},\ \bibinfo
  {pages} {1804} (\bibinfo {year} {1982}{\natexlab{b}})}\BibitemShut {NoStop}%
\bibitem [{\citenamefont {Landauer}(1961)}]{5392446}%
  \BibitemOpen
  \bibfield  {author} {\bibinfo {author} {\bibfnamefont {R.}~\bibnamefont
  {Landauer}},\ }\href {\doibase 10.1147/rd.53.0183} {\bibfield  {journal}
  {\bibinfo  {journal} {IBM Journal of Research and Development}\ }\textbf
  {\bibinfo {volume} {5}},\ \bibinfo {pages} {183} (\bibinfo {year}
  {1961})}\BibitemShut {NoStop}%
\bibitem [{\citenamefont {Bennett}(1973)}]{5391327}%
  \BibitemOpen
  \bibfield  {author} {\bibinfo {author} {\bibfnamefont {C.~H.}\ \bibnamefont
  {Bennett}},\ }\href {\doibase 10.1147/rd.176.0525} {\bibfield  {journal}
  {\bibinfo  {journal} {IBM Journal of Research and Development}\ }\textbf
  {\bibinfo {volume} {17}},\ \bibinfo {pages} {525} (\bibinfo {year}
  {1973})}\BibitemShut {NoStop}%
\bibitem [{\citenamefont {Benioff}(1980)}]{Benioff1980}%
  \BibitemOpen
  \bibfield  {author} {\bibinfo {author} {\bibfnamefont {P.}~\bibnamefont
  {Benioff}},\ }\href {\doibase 10.1007/BF01011339} {\bibfield  {journal}
  {\bibinfo  {journal} {Journal of Statistical Physics}\ }\textbf {\bibinfo
  {volume} {22}},\ \bibinfo {pages} {563} (\bibinfo {year} {1980})}\BibitemShut
  {NoStop}%
\bibitem [{\citenamefont {Manin}(1980)}]{Manin1980}%
  \BibitemOpen
  \bibfield  {author} {\bibinfo {author} {\bibfnamefont {Y.}~\bibnamefont
  {Manin}},\ }\href@noop {} {\bibfield  {journal} {\bibinfo  {journal}
  {Sovetskoye Radio, Moscow"}\ }\textbf {\bibinfo {volume} {128}} (\bibinfo
  {year} {1980})}\BibitemShut {NoStop}%
\bibitem [{\citenamefont {Feynman}(1982)}]{Feynman1982}%
  \BibitemOpen
  \bibfield  {author} {\bibinfo {author} {\bibfnamefont {R.~P.}\ \bibnamefont
  {Feynman}},\ }\href {\doibase 10.1007/BF02650179} {\bibfield  {journal}
  {\bibinfo  {journal} {International Journal of Theoretical Physics}\ }\textbf
  {\bibinfo {volume} {21}},\ \bibinfo {pages} {467} (\bibinfo {year}
  {1982})}\BibitemShut {NoStop}%
\bibitem [{\citenamefont {Fredkin}\ and\ \citenamefont
  {Toffoli}(1982)}]{Fredkin1982}%
  \BibitemOpen
  \bibfield  {author} {\bibinfo {author} {\bibfnamefont {E.}~\bibnamefont
  {Fredkin}}\ and\ \bibinfo {author} {\bibfnamefont {T.}~\bibnamefont
  {Toffoli}},\ }\href {\doibase 10.1007/bf01857727} {\bibfield  {journal}
  {\bibinfo  {journal} {International Journal of Theoretical Physics}\ }\textbf
  {\bibinfo {volume} {21}},\ \bibinfo {pages} {219} (\bibinfo {year}
  {1982})}\BibitemShut {NoStop}%
\bibitem [{\citenamefont {Feynman}(1986)}]{Feynman1986}%
  \BibitemOpen
  \bibfield  {author} {\bibinfo {author} {\bibfnamefont {R.~P.}\ \bibnamefont
  {Feynman}},\ }\href {\doibase 10.1007/BF01886518} {\bibfield  {journal}
  {\bibinfo  {journal} {Foundations of Physics}\ }\textbf {\bibinfo {volume}
  {16}},\ \bibinfo {pages} {507} (\bibinfo {year} {1986})}\BibitemShut
  {NoStop}%
\bibitem [{\citenamefont {Landauer}(1991)}]{doi:10.1063/1.881299}%
  \BibitemOpen
  \bibfield  {author} {\bibinfo {author} {\bibfnamefont {R.}~\bibnamefont
  {Landauer}},\ }\href {\doibase 10.1063/1.881299} {\bibfield  {journal}
  {\bibinfo  {journal} {Physics Today}\ }\textbf {\bibinfo {volume} {44}},\
  \bibinfo {pages} {23} (\bibinfo {year} {1991})},\ \Eprint
  {http://arxiv.org/abs/https://doi.org/10.1063/1.88129}
  {https://doi.org/10.1063/1.88129} \BibitemShut {NoStop}%
\bibitem [{\citenamefont {Williams}(1998)}]{williamsNASAconference}%
  \BibitemOpen
  \bibfield  {author} {\bibinfo {author} {\bibfnamefont {C.~P.}\ \bibnamefont
  {Williams}},\ }\href {https://doi.org/10.1007/3-540-49208-9} {\enquote
  {\bibinfo {title} {{Quantum Computing and Quantum Communications: First NASA
  International Conference, QCQC’98 Palm Springs, California, USA February
  17–20, Selected Papers}},}\ } (\bibinfo {year} {1998})\BibitemShut
  {NoStop}%
\bibitem [{\citenamefont {Preskill}(2021)}]{Preskill:2021apy}%
  \BibitemOpen
  \bibfield  {author} {\bibinfo {author} {\bibfnamefont {J.}~\bibnamefont
  {Preskill}}\ }(\bibinfo {year} {2021})\ \Eprint
  {http://arxiv.org/abs/2106.10522} {arXiv:2106.10522 [quant-ph]} \BibitemShut
  {NoStop}%
\bibitem [{\citenamefont {Preskill}(2018)}]{Preskill2018quantumcomputingin}%
  \BibitemOpen
  \bibfield  {author} {\bibinfo {author} {\bibfnamefont {J.}~\bibnamefont
  {Preskill}},\ }\href {\doibase 10.22331/q-2018-08-06-79} {\bibfield
  {journal} {\bibinfo  {journal} {{Quantum}}\ }\textbf {\bibinfo {volume}
  {2}},\ \bibinfo {pages} {79} (\bibinfo {year} {2018})}\BibitemShut {NoStop}%
\bibitem [{\citenamefont {Gottesman}(2009)}]{gottesman2009introduction}%
  \BibitemOpen
  \bibfield  {author} {\bibinfo {author} {\bibfnamefont {D.}~\bibnamefont
  {Gottesman}},\ }\href@noop {} {\enquote {\bibinfo {title} {An introduction to
  quantum error correction and fault-tolerant quantum computation},}\ }
  (\bibinfo {year} {2009}),\ \Eprint {http://arxiv.org/abs/0904.2557}
  {arXiv:0904.2557 [quant-ph]} \BibitemShut {NoStop}%
\bibitem [{Dwa()}]{Dwave}%
  \BibitemOpen
  \href@noop {} {\enquote {\bibinfo {title} {{D-Wave Systems}},}\ }\bibinfo
  {howpublished} {\url{https://www.dwavesys.com/}}\BibitemShut {NoStop}%
\bibitem [{\citenamefont {{Farhi}}\ \emph {et~al.}(2000)\citenamefont
  {{Farhi}}, \citenamefont {{Goldstone}}, \citenamefont {{Gutmann}},\ and\
  \citenamefont {{Sipser}}}]{2000quant.ph..1106F}%
  \BibitemOpen
  \bibfield  {author} {\bibinfo {author} {\bibfnamefont {E.}~\bibnamefont
  {{Farhi}}}, \bibinfo {author} {\bibfnamefont {J.}~\bibnamefont
  {{Goldstone}}}, \bibinfo {author} {\bibfnamefont {S.}~\bibnamefont
  {{Gutmann}}}, \ and\ \bibinfo {author} {\bibfnamefont {M.}~\bibnamefont
  {{Sipser}}},\ }\href@noop {} {\bibfield  {journal} {\bibinfo  {journal}
  {arXiv e-prints}\ ,\ \bibinfo {eid} {quant-ph/0001106}} (\bibinfo {year}
  {2000})},\ \Eprint {http://arxiv.org/abs/quant-ph/0001106}
  {arXiv:quant-ph/0001106 [quant-ph]} \BibitemShut {NoStop}%
\bibitem [{\citenamefont {King}\ \emph {et~al.}(2018)\citenamefont {King},
  \citenamefont {Carrasquilla}, \citenamefont {Raymond}, \citenamefont
  {Ozfidan}, \citenamefont {Andriyash}, \citenamefont {Berkley}, \citenamefont
  {Reis}, \citenamefont {Lanting}, \citenamefont {Harris}, \citenamefont
  {Altomare},\ and\ \citenamefont {et~al.}}]{King_2018}%
  \BibitemOpen
  \bibfield  {author} {\bibinfo {author} {\bibfnamefont {A.~D.}\ \bibnamefont
  {King}}, \bibinfo {author} {\bibfnamefont {J.}~\bibnamefont {Carrasquilla}},
  \bibinfo {author} {\bibfnamefont {J.}~\bibnamefont {Raymond}}, \bibinfo
  {author} {\bibfnamefont {I.}~\bibnamefont {Ozfidan}}, \bibinfo {author}
  {\bibfnamefont {E.}~\bibnamefont {Andriyash}}, \bibinfo {author}
  {\bibfnamefont {A.}~\bibnamefont {Berkley}}, \bibinfo {author} {\bibfnamefont
  {M.}~\bibnamefont {Reis}}, \bibinfo {author} {\bibfnamefont {T.}~\bibnamefont
  {Lanting}}, \bibinfo {author} {\bibfnamefont {R.}~\bibnamefont {Harris}},
  \bibinfo {author} {\bibfnamefont {F.}~\bibnamefont {Altomare}}, \ and\
  \bibinfo {author} {\bibnamefont {et~al.}},\ }\href {\doibase
  10.1038/s41586-018-0410-x} {\bibfield  {journal} {\bibinfo  {journal}
  {Nature}\ }\textbf {\bibinfo {volume} {560}},\ \bibinfo {pages} {456–460}
  (\bibinfo {year} {2018})}\BibitemShut {NoStop}%
\bibitem [{\citenamefont {Ajagekar}\ \emph {et~al.}(2020)\citenamefont
  {Ajagekar}, \citenamefont {Humble},\ and\ \citenamefont
  {You}}]{Ajagekar_2020}%
  \BibitemOpen
  \bibfield  {author} {\bibinfo {author} {\bibfnamefont {A.}~\bibnamefont
  {Ajagekar}}, \bibinfo {author} {\bibfnamefont {T.}~\bibnamefont {Humble}}, \
  and\ \bibinfo {author} {\bibfnamefont {F.}~\bibnamefont {You}},\ }\href
  {\doibase 10.1016/j.compchemeng.2019.106630} {\bibfield  {journal} {\bibinfo
  {journal} {Computers and Chemical Engineering}\ }\textbf {\bibinfo {volume}
  {132}},\ \bibinfo {pages} {106630} (\bibinfo {year} {2020})}\BibitemShut
  {NoStop}%
\bibitem [{\citenamefont {Lamata}\ \emph {et~al.}(2018)\citenamefont {Lamata},
  \citenamefont {Parra-Rodriguez}, \citenamefont {Sanz},\ and\ \citenamefont
  {Solano}}]{doi:10.1080/23746149.2018.1457981}%
  \BibitemOpen
  \bibfield  {author} {\bibinfo {author} {\bibfnamefont {L.}~\bibnamefont
  {Lamata}}, \bibinfo {author} {\bibfnamefont {A.}~\bibnamefont
  {Parra-Rodriguez}}, \bibinfo {author} {\bibfnamefont {M.}~\bibnamefont
  {Sanz}}, \ and\ \bibinfo {author} {\bibfnamefont {E.}~\bibnamefont
  {Solano}},\ }\href {\doibase 10.1080/23746149.2018.1457981} {\bibfield
  {journal} {\bibinfo  {journal} {Advances in Physics: X}\ }\textbf {\bibinfo
  {volume} {3}},\ \bibinfo {pages} {1457981} (\bibinfo {year} {2018})},\
  \Eprint {http://arxiv.org/abs/https://doi.org/10.1080/23746149.2018.1457981}
  {https://doi.org/10.1080/23746149.2018.1457981} \BibitemShut {NoStop}%
\bibitem [{\citenamefont {Alexeev}\ \emph {et~al.}(2021)\citenamefont
  {Alexeev}, \citenamefont {Bacon}, \citenamefont {Brown}, \citenamefont
  {Calderbank}, \citenamefont {Carr}, \citenamefont {Chong}, \citenamefont
  {DeMarco}, \citenamefont {Englund}, \citenamefont {Farhi}, \citenamefont
  {Fefferman}, \citenamefont {Gorshkov}, \citenamefont {Houck}, \citenamefont
  {Kim}, \citenamefont {Kimmel}, \citenamefont {Lange}, \citenamefont {Lloyd},
  \citenamefont {Lukin}, \citenamefont {Maslov}, \citenamefont {Maunz},
  \citenamefont {Monroe}, \citenamefont {Preskill}, \citenamefont {Roetteler},
  \citenamefont {Savage},\ and\ \citenamefont
  {Thompson}}]{PRXQuantum.2.017001}%
  \BibitemOpen
  \bibfield  {author} {\bibinfo {author} {\bibfnamefont {Y.}~\bibnamefont
  {Alexeev}}, \bibinfo {author} {\bibfnamefont {D.}~\bibnamefont {Bacon}},
  \bibinfo {author} {\bibfnamefont {K.~R.}\ \bibnamefont {Brown}}, \bibinfo
  {author} {\bibfnamefont {R.}~\bibnamefont {Calderbank}}, \bibinfo {author}
  {\bibfnamefont {L.~D.}\ \bibnamefont {Carr}}, \bibinfo {author}
  {\bibfnamefont {F.~T.}\ \bibnamefont {Chong}}, \bibinfo {author}
  {\bibfnamefont {B.}~\bibnamefont {DeMarco}}, \bibinfo {author} {\bibfnamefont
  {D.}~\bibnamefont {Englund}}, \bibinfo {author} {\bibfnamefont
  {E.}~\bibnamefont {Farhi}}, \bibinfo {author} {\bibfnamefont
  {B.}~\bibnamefont {Fefferman}}, \bibinfo {author} {\bibfnamefont {A.~V.}\
  \bibnamefont {Gorshkov}}, \bibinfo {author} {\bibfnamefont {A.}~\bibnamefont
  {Houck}}, \bibinfo {author} {\bibfnamefont {J.}~\bibnamefont {Kim}}, \bibinfo
  {author} {\bibfnamefont {S.}~\bibnamefont {Kimmel}}, \bibinfo {author}
  {\bibfnamefont {M.}~\bibnamefont {Lange}}, \bibinfo {author} {\bibfnamefont
  {S.}~\bibnamefont {Lloyd}}, \bibinfo {author} {\bibfnamefont {M.~D.}\
  \bibnamefont {Lukin}}, \bibinfo {author} {\bibfnamefont {D.}~\bibnamefont
  {Maslov}}, \bibinfo {author} {\bibfnamefont {P.}~\bibnamefont {Maunz}},
  \bibinfo {author} {\bibfnamefont {C.}~\bibnamefont {Monroe}}, \bibinfo
  {author} {\bibfnamefont {J.}~\bibnamefont {Preskill}}, \bibinfo {author}
  {\bibfnamefont {M.}~\bibnamefont {Roetteler}}, \bibinfo {author}
  {\bibfnamefont {M.~J.}\ \bibnamefont {Savage}}, \ and\ \bibinfo {author}
  {\bibfnamefont {J.}~\bibnamefont {Thompson}},\ }\href {\doibase
  10.1103/PRXQuantum.2.017001} {\bibfield  {journal} {\bibinfo  {journal} {PRX
  Quantum}\ }\textbf {\bibinfo {volume} {2}},\ \bibinfo {pages} {017001}
  (\bibinfo {year} {2021})}\BibitemShut {NoStop}%
\bibitem [{\citenamefont {Kitaev}(2003)}]{Kitaev_2003}%
  \BibitemOpen
  \bibfield  {author} {\bibinfo {author} {\bibfnamefont {A.}~\bibnamefont
  {Kitaev}},\ }\href {\doibase 10.1016/s0003-4916(02)00018-0} {\bibfield
  {journal} {\bibinfo  {journal} {Annals of Physics}\ }\textbf {\bibinfo
  {volume} {303}},\ \bibinfo {pages} {2–30} (\bibinfo {year}
  {2003})}\BibitemShut {NoStop}%
\bibitem [{\citenamefont {{Microsoft}}()}]{MicrosoftTopo}%
  \BibitemOpen
  \bibfield  {author} {\bibinfo {author} {\bibnamefont {{Microsoft}}},\
  }\href@noop {} {\enquote {\bibinfo {title} {{Topological Quantum Computing
  Research}},}\ }\bibinfo {howpublished}
  {\url{https://www.microsoft.com/en-us/research/project/topological-quantum-computing/}},\
  \bibinfo {note} {accessed: 2021-07-03}\BibitemShut {NoStop}%
\bibitem [{\citenamefont {{Oak Ridge National Laboratory}}()}]{QSC}%
  \BibitemOpen
  \bibfield  {author} {\bibinfo {author} {\bibnamefont {{Oak Ridge National
  Laboratory}}},\ }\href@noop {} {\enquote {\bibinfo {title} {{Quantum Science
  Center}},}\ }\bibinfo {howpublished} {\url{https://qscience.org/}},\ \bibinfo
  {note} {accessed: 2021-05-31}\BibitemShut {NoStop}%
\bibitem [{\citenamefont {{Nuclear Physics and Quantum Information Science, A
  Report from the NSAC Quantum Information Science Subcommittee
  }}(2019)}]{NSACQISNP}%
  \BibitemOpen
  \bibfield  {author} {\bibinfo {author} {\bibnamefont {{Nuclear Physics and
  Quantum Information Science, A Report from the NSAC Quantum Information
  Science Subcommittee }}},\ }\href
  {https://science.osti.gov/-/media/np/pdf/Reports/NSAC_QIS_Report.pdf} {\
  (\bibinfo {year} {{2019,}})}\BibitemShut {NoStop}%
\bibitem [{\citenamefont {Poulin}\ \emph {et~al.}(2011)\citenamefont {Poulin},
  \citenamefont {Qarry}, \citenamefont {Somma},\ and\ \citenamefont
  {Verstraete}}]{PhysRevLett.106.170501}%
  \BibitemOpen
  \bibfield  {author} {\bibinfo {author} {\bibfnamefont {D.}~\bibnamefont
  {Poulin}}, \bibinfo {author} {\bibfnamefont {A.}~\bibnamefont {Qarry}},
  \bibinfo {author} {\bibfnamefont {R.}~\bibnamefont {Somma}}, \ and\ \bibinfo
  {author} {\bibfnamefont {F.}~\bibnamefont {Verstraete}},\ }\href {\doibase
  10.1103/PhysRevLett.106.170501} {\bibfield  {journal} {\bibinfo  {journal}
  {Phys. Rev. Lett.}\ }\textbf {\bibinfo {volume} {106}},\ \bibinfo {pages}
  {170501} (\bibinfo {year} {2011})}\BibitemShut {NoStop}%
\bibitem [{\citenamefont {Glashow}(1961)}]{Glashow:1961tr}%
  \BibitemOpen
  \bibfield  {author} {\bibinfo {author} {\bibfnamefont {S.}~\bibnamefont
  {Glashow}},\ }\href {\doibase 10.1016/0029-5582(61)90469-2} {\bibfield
  {journal} {\bibinfo  {journal} {Nucl. Phys.}\ }\textbf {\bibinfo {volume}
  {22}},\ \bibinfo {pages} {579} (\bibinfo {year} {1961})}\BibitemShut
  {NoStop}%
\bibitem [{\citenamefont {Higgs}(1964)}]{Higgs:1964pj}%
  \BibitemOpen
  \bibfield  {author} {\bibinfo {author} {\bibfnamefont {P.~W.}\ \bibnamefont
  {Higgs}},\ }\href {\doibase 10.1103/PhysRevLett.13.508} {\bibfield  {journal}
  {\bibinfo  {journal} {Phys. Rev. Lett.}\ }\textbf {\bibinfo {volume} {13}},\
  \bibinfo {pages} {508} (\bibinfo {year} {1964})}\BibitemShut {NoStop}%
\bibitem [{\citenamefont {Weinberg}(1967)}]{Weinberg:1967tq}%
  \BibitemOpen
  \bibfield  {author} {\bibinfo {author} {\bibfnamefont {S.}~\bibnamefont
  {Weinberg}},\ }\href {\doibase 10.1103/PhysRevLett.19.1264} {\bibfield
  {journal} {\bibinfo  {journal} {Phys. Rev. Lett.}\ }\textbf {\bibinfo
  {volume} {19}},\ \bibinfo {pages} {1264} (\bibinfo {year}
  {1967})}\BibitemShut {NoStop}%
\bibitem [{\citenamefont {Salam}(1968)}]{Salam:1968rm}%
  \BibitemOpen
  \bibfield  {author} {\bibinfo {author} {\bibfnamefont {A.}~\bibnamefont
  {Salam}},\ }\href {\doibase 10.1142/9789812795915\_0034} {\bibfield
  {journal} {\bibinfo  {journal} {Conf. Proc. C}\ }\textbf {\bibinfo {volume}
  {680519}},\ \bibinfo {pages} {367} (\bibinfo {year} {1968})}\BibitemShut
  {NoStop}%
\bibitem [{\citenamefont {Politzer}(1973)}]{Politzer:1973fx}%
  \BibitemOpen
  \bibfield  {author} {\bibinfo {author} {\bibfnamefont {H.}~\bibnamefont
  {Politzer}},\ }\href {\doibase 10.1103/PhysRevLett.30.1346} {\bibfield
  {journal} {\bibinfo  {journal} {Phys. Rev. Lett.}\ }\textbf {\bibinfo
  {volume} {30}},\ \bibinfo {pages} {1346} (\bibinfo {year}
  {1973})}\BibitemShut {NoStop}%
\bibitem [{\citenamefont {Gross}\ and\ \citenamefont
  {Wilczek}(1973)}]{Gross:1973id}%
  \BibitemOpen
  \bibfield  {author} {\bibinfo {author} {\bibfnamefont {D.~J.}\ \bibnamefont
  {Gross}}\ and\ \bibinfo {author} {\bibfnamefont {F.}~\bibnamefont
  {Wilczek}},\ }\href {\doibase 10.1103/PhysRevLett.30.1343} {\bibfield
  {journal} {\bibinfo  {journal} {Phys. Rev. Lett.}\ }\textbf {\bibinfo
  {volume} {30}},\ \bibinfo {pages} {1343} (\bibinfo {year}
  {1973})}\BibitemShut {NoStop}%
\bibitem [{\citenamefont {Habib}\ \emph {et~al.}(2016)\citenamefont {Habib}
  \emph {et~al.}}]{Habib:2016sce}%
  \BibitemOpen
  \bibfield  {author} {\bibinfo {author} {\bibfnamefont {S.}~\bibnamefont
  {Habib}} \emph {et~al.},\ }\href@noop {} {\  (\bibinfo {year} {2016})},\
  \Eprint {http://arxiv.org/abs/1603.09303} {arXiv:1603.09303
  [physics.comp-ph]} \BibitemShut {NoStop}%
\bibitem [{\citenamefont {Carlson}\ \emph {et~al.}(2017)\citenamefont
  {Carlson}, \citenamefont {Savage}, \citenamefont {Gerber}, \citenamefont
  {Antypas}, \citenamefont {Bard}, \citenamefont {Coffey}, \citenamefont
  {Dart}, \citenamefont {Dosanjh}, \citenamefont {Hack}, \citenamefont {Monga},
  \citenamefont {Papka}, \citenamefont {Riley}, \citenamefont {Rotman},
  \citenamefont {Straatsma}, \citenamefont {Wells}, \citenamefont {Avakian},
  \citenamefont {Ayyad}, \citenamefont {Bass}, \citenamefont {Bazin},
  \citenamefont {Boehnlein}, \citenamefont {Bollen}, \citenamefont {Broussard},
  \citenamefont {Calder}, \citenamefont {Couch}, \citenamefont {Couture},
  \citenamefont {Cromaz}, \citenamefont {Detmold}, \citenamefont {Detwiler},
  \citenamefont {Duan}, \citenamefont {Edwards}, \citenamefont {Engel},
  \citenamefont {Fryer}, \citenamefont {Fuller}, \citenamefont {Gandolfi},
  \citenamefont {Gavalian}, \citenamefont {Georgobiani}, \citenamefont {Gupta},
  \citenamefont {Gyurjyan}, \citenamefont {Hausmann}, \citenamefont {Heyes},
  \citenamefont {Hix}, \citenamefont {ito}, \citenamefont {Jansen},
  \citenamefont {Jones}, \citenamefont {Joo}, \citenamefont {Kaczmarek},
  \citenamefont {Kasen}, \citenamefont {Kostin}, \citenamefont {Kurth},
  \citenamefont {Lauret}, \citenamefont {Lawrence}, \citenamefont {Lin},
  \citenamefont {Lin}, \citenamefont {Mantica}, \citenamefont {Maris},
  \citenamefont {Messer}, \citenamefont {Mittig}, \citenamefont {Mosby},
  \citenamefont {Mukherjee}, \citenamefont {Nam}, \citenamefont {navratil},
  \citenamefont {Nazarewicz}, \citenamefont {Ng}, \citenamefont {O'Donnell},
  \citenamefont {Orginos}, \citenamefont {Pellemoine}, \citenamefont
  {Petreczky}, \citenamefont {Pieper}, \citenamefont {Pinkenburg},
  \citenamefont {Plaster}, \citenamefont {Porter}, \citenamefont {Portillo},
  \citenamefont {Pratt}, \citenamefont {Purschke}, \citenamefont {Qiang},
  \citenamefont {Quaglioni}, \citenamefont {Richards}, \citenamefont {Roblin},
  \citenamefont {Schenke}, \citenamefont {Schiavilla}, \citenamefont
  {Schlichting}, \citenamefont {Schunck}, \citenamefont {Steinbrecher},
  \citenamefont {Strickland}, \citenamefont {Syritsyn}, \citenamefont {Terzic},
  \citenamefont {Varner}, \citenamefont {Vary}, \citenamefont {Wild},
  \citenamefont {Winter}, \citenamefont {Zegers}, \citenamefont {Zhang},
  \citenamefont {Ziegler},\ and\ \citenamefont {Zingale}}]{osti_1369223}%
  \BibitemOpen
  \bibfield  {author} {\bibinfo {author} {\bibfnamefont {J.}~\bibnamefont
  {Carlson}}, \bibinfo {author} {\bibfnamefont {M.~J.}\ \bibnamefont {Savage}},
  \bibinfo {author} {\bibfnamefont {R.}~\bibnamefont {Gerber}}, \bibinfo
  {author} {\bibfnamefont {K.}~\bibnamefont {Antypas}}, \bibinfo {author}
  {\bibfnamefont {D.}~\bibnamefont {Bard}}, \bibinfo {author} {\bibfnamefont
  {R.}~\bibnamefont {Coffey}}, \bibinfo {author} {\bibfnamefont
  {E.}~\bibnamefont {Dart}}, \bibinfo {author} {\bibfnamefont {S.}~\bibnamefont
  {Dosanjh}}, \bibinfo {author} {\bibfnamefont {J.}~\bibnamefont {Hack}},
  \bibinfo {author} {\bibfnamefont {I.}~\bibnamefont {Monga}}, \bibinfo
  {author} {\bibfnamefont {M.~E.}\ \bibnamefont {Papka}}, \bibinfo {author}
  {\bibfnamefont {K.}~\bibnamefont {Riley}}, \bibinfo {author} {\bibfnamefont
  {L.}~\bibnamefont {Rotman}}, \bibinfo {author} {\bibfnamefont
  {T.}~\bibnamefont {Straatsma}}, \bibinfo {author} {\bibfnamefont
  {J.}~\bibnamefont {Wells}}, \bibinfo {author} {\bibfnamefont
  {H.}~\bibnamefont {Avakian}}, \bibinfo {author} {\bibfnamefont
  {Y.}~\bibnamefont {Ayyad}}, \bibinfo {author} {\bibfnamefont {S.~A.}\
  \bibnamefont {Bass}}, \bibinfo {author} {\bibfnamefont {D.}~\bibnamefont
  {Bazin}}, \bibinfo {author} {\bibfnamefont {A.}~\bibnamefont {Boehnlein}},
  \bibinfo {author} {\bibfnamefont {G.}~\bibnamefont {Bollen}}, \bibinfo
  {author} {\bibfnamefont {L.~J.}\ \bibnamefont {Broussard}}, \bibinfo {author}
  {\bibfnamefont {A.}~\bibnamefont {Calder}}, \bibinfo {author} {\bibfnamefont
  {S.}~\bibnamefont {Couch}}, \bibinfo {author} {\bibfnamefont
  {A.}~\bibnamefont {Couture}}, \bibinfo {author} {\bibfnamefont
  {M.}~\bibnamefont {Cromaz}}, \bibinfo {author} {\bibfnamefont
  {W.}~\bibnamefont {Detmold}}, \bibinfo {author} {\bibfnamefont
  {J.}~\bibnamefont {Detwiler}}, \bibinfo {author} {\bibfnamefont
  {H.}~\bibnamefont {Duan}}, \bibinfo {author} {\bibfnamefont {R.}~\bibnamefont
  {Edwards}}, \bibinfo {author} {\bibfnamefont {J.}~\bibnamefont {Engel}},
  \bibinfo {author} {\bibfnamefont {C.}~\bibnamefont {Fryer}}, \bibinfo
  {author} {\bibfnamefont {G.~M.}\ \bibnamefont {Fuller}}, \bibinfo {author}
  {\bibfnamefont {S.}~\bibnamefont {Gandolfi}}, \bibinfo {author}
  {\bibfnamefont {G.}~\bibnamefont {Gavalian}}, \bibinfo {author}
  {\bibfnamefont {D.}~\bibnamefont {Georgobiani}}, \bibinfo {author}
  {\bibfnamefont {R.}~\bibnamefont {Gupta}}, \bibinfo {author} {\bibfnamefont
  {V.}~\bibnamefont {Gyurjyan}}, \bibinfo {author} {\bibfnamefont
  {M.}~\bibnamefont {Hausmann}}, \bibinfo {author} {\bibfnamefont
  {G.}~\bibnamefont {Heyes}}, \bibinfo {author} {\bibfnamefont {W.~R.}\
  \bibnamefont {Hix}}, \bibinfo {author} {\bibfnamefont {M.}~\bibnamefont
  {ito}}, \bibinfo {author} {\bibfnamefont {G.}~\bibnamefont {Jansen}},
  \bibinfo {author} {\bibfnamefont {R.}~\bibnamefont {Jones}}, \bibinfo
  {author} {\bibfnamefont {B.}~\bibnamefont {Joo}}, \bibinfo {author}
  {\bibfnamefont {O.}~\bibnamefont {Kaczmarek}}, \bibinfo {author}
  {\bibfnamefont {D.}~\bibnamefont {Kasen}}, \bibinfo {author} {\bibfnamefont
  {M.}~\bibnamefont {Kostin}}, \bibinfo {author} {\bibfnamefont
  {T.}~\bibnamefont {Kurth}}, \bibinfo {author} {\bibfnamefont
  {J.}~\bibnamefont {Lauret}}, \bibinfo {author} {\bibfnamefont
  {D.}~\bibnamefont {Lawrence}}, \bibinfo {author} {\bibfnamefont {H.-W.}\
  \bibnamefont {Lin}}, \bibinfo {author} {\bibfnamefont {M.}~\bibnamefont
  {Lin}}, \bibinfo {author} {\bibfnamefont {P.}~\bibnamefont {Mantica}},
  \bibinfo {author} {\bibfnamefont {P.}~\bibnamefont {Maris}}, \bibinfo
  {author} {\bibfnamefont {B.}~\bibnamefont {Messer}}, \bibinfo {author}
  {\bibfnamefont {W.}~\bibnamefont {Mittig}}, \bibinfo {author} {\bibfnamefont
  {S.}~\bibnamefont {Mosby}}, \bibinfo {author} {\bibfnamefont
  {S.}~\bibnamefont {Mukherjee}}, \bibinfo {author} {\bibfnamefont {H.~A.}\
  \bibnamefont {Nam}}, \bibinfo {author} {\bibfnamefont {P.}~\bibnamefont
  {navratil}}, \bibinfo {author} {\bibfnamefont {W.}~\bibnamefont
  {Nazarewicz}}, \bibinfo {author} {\bibfnamefont {E.}~\bibnamefont {Ng}},
  \bibinfo {author} {\bibfnamefont {T.}~\bibnamefont {O'Donnell}}, \bibinfo
  {author} {\bibfnamefont {K.}~\bibnamefont {Orginos}}, \bibinfo {author}
  {\bibfnamefont {F.}~\bibnamefont {Pellemoine}}, \bibinfo {author}
  {\bibfnamefont {P.}~\bibnamefont {Petreczky}}, \bibinfo {author}
  {\bibfnamefont {S.~C.}\ \bibnamefont {Pieper}}, \bibinfo {author}
  {\bibfnamefont {C.~H.}\ \bibnamefont {Pinkenburg}}, \bibinfo {author}
  {\bibfnamefont {B.}~\bibnamefont {Plaster}}, \bibinfo {author} {\bibfnamefont
  {R.~J.}\ \bibnamefont {Porter}}, \bibinfo {author} {\bibfnamefont
  {M.}~\bibnamefont {Portillo}}, \bibinfo {author} {\bibfnamefont
  {S.}~\bibnamefont {Pratt}}, \bibinfo {author} {\bibfnamefont {M.~L.}\
  \bibnamefont {Purschke}}, \bibinfo {author} {\bibfnamefont {J.}~\bibnamefont
  {Qiang}}, \bibinfo {author} {\bibfnamefont {S.}~\bibnamefont {Quaglioni}},
  \bibinfo {author} {\bibfnamefont {D.}~\bibnamefont {Richards}}, \bibinfo
  {author} {\bibfnamefont {Y.}~\bibnamefont {Roblin}}, \bibinfo {author}
  {\bibfnamefont {B.}~\bibnamefont {Schenke}}, \bibinfo {author} {\bibfnamefont
  {R.}~\bibnamefont {Schiavilla}}, \bibinfo {author} {\bibfnamefont
  {S.}~\bibnamefont {Schlichting}}, \bibinfo {author} {\bibfnamefont
  {N.}~\bibnamefont {Schunck}}, \bibinfo {author} {\bibfnamefont
  {P.}~\bibnamefont {Steinbrecher}}, \bibinfo {author} {\bibfnamefont
  {M.}~\bibnamefont {Strickland}}, \bibinfo {author} {\bibfnamefont
  {S.}~\bibnamefont {Syritsyn}}, \bibinfo {author} {\bibfnamefont
  {B.}~\bibnamefont {Terzic}}, \bibinfo {author} {\bibfnamefont
  {R.}~\bibnamefont {Varner}}, \bibinfo {author} {\bibfnamefont
  {J.}~\bibnamefont {Vary}}, \bibinfo {author} {\bibfnamefont {S.}~\bibnamefont
  {Wild}}, \bibinfo {author} {\bibfnamefont {F.}~\bibnamefont {Winter}},
  \bibinfo {author} {\bibfnamefont {R.}~\bibnamefont {Zegers}}, \bibinfo
  {author} {\bibfnamefont {H.}~\bibnamefont {Zhang}}, \bibinfo {author}
  {\bibfnamefont {V.}~\bibnamefont {Ziegler}}, \ and\ \bibinfo {author}
  {\bibfnamefont {M.}~\bibnamefont {Zingale}},\ }\href {\doibase
  10.2172/1369223} {\  (\bibinfo {year} {2017}),\ 10.2172/1369223}\BibitemShut
  {NoStop}%
\bibitem [{\citenamefont {Martinez}\ \emph {et~al.}(2016)\citenamefont
  {Martinez} \emph {et~al.}}]{Martinez:2016yna}%
  \BibitemOpen
  \bibfield  {author} {\bibinfo {author} {\bibfnamefont {E.~A.}\ \bibnamefont
  {Martinez}} \emph {et~al.},\ }\href {\doibase 10.1038/nature18318} {\bibfield
   {journal} {\bibinfo  {journal} {Nature}\ }\textbf {\bibinfo {volume}
  {534}},\ \bibinfo {pages} {516} (\bibinfo {year} {2016})},\ \Eprint
  {http://arxiv.org/abs/1605.04570} {arXiv:1605.04570 [quant-ph]} \BibitemShut
  {NoStop}%
\bibitem [{IBM()}]{IBMQ}%
  \BibitemOpen
  \href@noop {} {\enquote {\bibinfo {title} {{IBM Quantum}},}\ }\bibinfo
  {howpublished} {\url{https://quantum-computing.ibm.com/}, 2021}\BibitemShut
  {NoStop}%
\bibitem [{\citenamefont {Karalekas}\ \emph {et~al.}(2020)\citenamefont
  {Karalekas}, \citenamefont {Tezak}, \citenamefont {Peterson}, \citenamefont
  {Ryan}, \citenamefont {da~Silva},\ and\ \citenamefont
  {Smith}}]{Karalekas_2020}%
  \BibitemOpen
  \bibfield  {author} {\bibinfo {author} {\bibfnamefont {P.~J.}\ \bibnamefont
  {Karalekas}}, \bibinfo {author} {\bibfnamefont {N.~A.}\ \bibnamefont
  {Tezak}}, \bibinfo {author} {\bibfnamefont {E.~C.}\ \bibnamefont {Peterson}},
  \bibinfo {author} {\bibfnamefont {C.~A.}\ \bibnamefont {Ryan}}, \bibinfo
  {author} {\bibfnamefont {M.~P.}\ \bibnamefont {da~Silva}}, \ and\ \bibinfo
  {author} {\bibfnamefont {R.~S.}\ \bibnamefont {Smith}},\ }\href {\doibase
  10.1088/2058-9565/ab7559} {\bibfield  {journal} {\bibinfo  {journal} {Quantum
  Science and Technology}\ }\textbf {\bibinfo {volume} {5}},\ \bibinfo {pages}
  {024003} (\bibinfo {year} {2020})}\BibitemShut {NoStop}%
\bibitem [{\citenamefont {Dumitrescu}\ \emph {et~al.}(2018)\citenamefont
  {Dumitrescu}, \citenamefont {McCaskey}, \citenamefont {Hagen}, \citenamefont
  {Jansen}, \citenamefont {Morris}, \citenamefont {Papenbrock}, \citenamefont
  {Pooser}, \citenamefont {Dean},\ and\ \citenamefont
  {Lougovski}}]{Dumitrescu:2018njn}%
  \BibitemOpen
  \bibfield  {author} {\bibinfo {author} {\bibfnamefont {E.~F.}\ \bibnamefont
  {Dumitrescu}}, \bibinfo {author} {\bibfnamefont {A.~J.}\ \bibnamefont
  {McCaskey}}, \bibinfo {author} {\bibfnamefont {G.}~\bibnamefont {Hagen}},
  \bibinfo {author} {\bibfnamefont {G.~R.}\ \bibnamefont {Jansen}}, \bibinfo
  {author} {\bibfnamefont {T.~D.}\ \bibnamefont {Morris}}, \bibinfo {author}
  {\bibfnamefont {T.}~\bibnamefont {Papenbrock}}, \bibinfo {author}
  {\bibfnamefont {R.~C.}\ \bibnamefont {Pooser}}, \bibinfo {author}
  {\bibfnamefont {D.~J.}\ \bibnamefont {Dean}}, \ and\ \bibinfo {author}
  {\bibfnamefont {P.}~\bibnamefont {Lougovski}},\ }\href {\doibase
  10.1103/PhysRevLett.120.210501} {\bibfield  {journal} {\bibinfo  {journal}
  {Phys. Rev. Lett.}\ }\textbf {\bibinfo {volume} {120}},\ \bibinfo {pages}
  {210501} (\bibinfo {year} {2018})},\ \Eprint
  {http://arxiv.org/abs/1801.03897} {arXiv:1801.03897 [quant-ph]} \BibitemShut
  {NoStop}%
\bibitem [{\citenamefont {Lu}\ \emph {et~al.}(2019)\citenamefont {Lu},
  \citenamefont {Klco}, \citenamefont {Lukens}, \citenamefont {Morris},
  \citenamefont {Bansal}, \citenamefont {Ekstr\"om}, \citenamefont {Hagen},
  \citenamefont {Papenbrock}, \citenamefont {Weiner}, \citenamefont {Savage},\
  and\ \citenamefont {Lougovski}}]{Lu:2018pjk}%
  \BibitemOpen
  \bibfield  {author} {\bibinfo {author} {\bibfnamefont {H.-H.}\ \bibnamefont
  {Lu}}, \bibinfo {author} {\bibfnamefont {N.}~\bibnamefont {Klco}}, \bibinfo
  {author} {\bibfnamefont {J.~M.}\ \bibnamefont {Lukens}}, \bibinfo {author}
  {\bibfnamefont {T.~D.}\ \bibnamefont {Morris}}, \bibinfo {author}
  {\bibfnamefont {A.}~\bibnamefont {Bansal}}, \bibinfo {author} {\bibfnamefont
  {A.}~\bibnamefont {Ekstr\"om}}, \bibinfo {author} {\bibfnamefont
  {G.}~\bibnamefont {Hagen}}, \bibinfo {author} {\bibfnamefont
  {T.}~\bibnamefont {Papenbrock}}, \bibinfo {author} {\bibfnamefont {A.~M.}\
  \bibnamefont {Weiner}}, \bibinfo {author} {\bibfnamefont {M.~J.}\
  \bibnamefont {Savage}}, \ and\ \bibinfo {author} {\bibfnamefont
  {P.}~\bibnamefont {Lougovski}},\ }\href {\doibase
  10.1103/PhysRevA.100.012320} {\bibfield  {journal} {\bibinfo  {journal}
  {Phys. Rev. A}\ }\textbf {\bibinfo {volume} {100}},\ \bibinfo {pages}
  {012320} (\bibinfo {year} {2019})}\BibitemShut {NoStop}%
\bibitem [{\citenamefont {Shehab}\ \emph {et~al.}(2019)\citenamefont {Shehab},
  \citenamefont {Landsman}, \citenamefont {Nam}, \citenamefont {Zhu},
  \citenamefont {Linke}, \citenamefont {Keesan}, \citenamefont {Pooser},\ and\
  \citenamefont {Monroe}}]{Shehab:2019gfn}%
  \BibitemOpen
  \bibfield  {author} {\bibinfo {author} {\bibfnamefont {O.}~\bibnamefont
  {Shehab}}, \bibinfo {author} {\bibfnamefont {K.~A.}\ \bibnamefont
  {Landsman}}, \bibinfo {author} {\bibfnamefont {Y.}~\bibnamefont {Nam}},
  \bibinfo {author} {\bibfnamefont {D.}~\bibnamefont {Zhu}}, \bibinfo {author}
  {\bibfnamefont {N.~M.}\ \bibnamefont {Linke}}, \bibinfo {author}
  {\bibfnamefont {M.~J.}\ \bibnamefont {Keesan}}, \bibinfo {author}
  {\bibfnamefont {R.~C.}\ \bibnamefont {Pooser}}, \ and\ \bibinfo {author}
  {\bibfnamefont {C.~R.}\ \bibnamefont {Monroe}},\ }\href {\doibase
  10.1103/PhysRevA.100.062319} {\bibfield  {journal} {\bibinfo  {journal}
  {Phys. Rev. A}\ }\textbf {\bibinfo {volume} {100}},\ \bibinfo {pages}
  {062319} (\bibinfo {year} {2019})},\ \Eprint
  {http://arxiv.org/abs/1904.04338} {arXiv:1904.04338 [quant-ph]} \BibitemShut
  {NoStop}%
\bibitem [{\citenamefont {Roggero}\ \emph
  {et~al.}(2020{\natexlab{a}})\citenamefont {Roggero}, \citenamefont {Li},
  \citenamefont {Carlson}, \citenamefont {Gupta},\ and\ \citenamefont
  {Perdue}}]{PhysRevD.101.074038}%
  \BibitemOpen
  \bibfield  {author} {\bibinfo {author} {\bibfnamefont {A.}~\bibnamefont
  {Roggero}}, \bibinfo {author} {\bibfnamefont {A.~C.~Y.}\ \bibnamefont {Li}},
  \bibinfo {author} {\bibfnamefont {J.}~\bibnamefont {Carlson}}, \bibinfo
  {author} {\bibfnamefont {R.}~\bibnamefont {Gupta}}, \ and\ \bibinfo {author}
  {\bibfnamefont {G.~N.}\ \bibnamefont {Perdue}},\ }\href {\doibase
  10.1103/PhysRevD.101.074038} {\bibfield  {journal} {\bibinfo  {journal}
  {Phys. Rev. D}\ }\textbf {\bibinfo {volume} {101}},\ \bibinfo {pages}
  {074038} (\bibinfo {year} {2020}{\natexlab{a}})}\BibitemShut {NoStop}%
\bibitem [{\citenamefont {Yeter-Aydeniz}\ \emph {et~al.}(2020)\citenamefont
  {Yeter-Aydeniz}, \citenamefont {Pooser},\ and\ \citenamefont
  {Siopsis}}]{yeter2020practical}%
  \BibitemOpen
  \bibfield  {author} {\bibinfo {author} {\bibfnamefont {K.}~\bibnamefont
  {Yeter-Aydeniz}}, \bibinfo {author} {\bibfnamefont {R.~C.}\ \bibnamefont
  {Pooser}}, \ and\ \bibinfo {author} {\bibfnamefont {G.}~\bibnamefont
  {Siopsis}},\ }\href@noop {} {\bibfield  {journal} {\bibinfo  {journal} {npj
  Quantum Information}\ }\textbf {\bibinfo {volume} {6}},\ \bibinfo {pages} {1}
  (\bibinfo {year} {2020})}\BibitemShut {NoStop}%
\bibitem [{\citenamefont {Di~Matteo}\ \emph {et~al.}(2021)\citenamefont
  {Di~Matteo}, \citenamefont {McCoy}, \citenamefont {Gysbers}, \citenamefont
  {Miyagi}, \citenamefont {Woloshyn},\ and\ \citenamefont
  {Navr\'atil}}]{PhysRevA.103.042405}%
  \BibitemOpen
  \bibfield  {author} {\bibinfo {author} {\bibfnamefont {O.}~\bibnamefont
  {Di~Matteo}}, \bibinfo {author} {\bibfnamefont {A.}~\bibnamefont {McCoy}},
  \bibinfo {author} {\bibfnamefont {P.}~\bibnamefont {Gysbers}}, \bibinfo
  {author} {\bibfnamefont {T.}~\bibnamefont {Miyagi}}, \bibinfo {author}
  {\bibfnamefont {R.~M.}\ \bibnamefont {Woloshyn}}, \ and\ \bibinfo {author}
  {\bibfnamefont {P.}~\bibnamefont {Navr\'atil}},\ }\href {\doibase
  10.1103/PhysRevA.103.042405} {\bibfield  {journal} {\bibinfo  {journal}
  {Phys. Rev. A}\ }\textbf {\bibinfo {volume} {103}},\ \bibinfo {pages}
  {042405} (\bibinfo {year} {2021})}\BibitemShut {NoStop}%
\bibitem [{\citenamefont {Ba\~nuls}\ \emph {et~al.}(2018)\citenamefont
  {Ba\~nuls}, \citenamefont {Cichy}, \citenamefont {Cirac}, \citenamefont
  {Jansen},\ and\ \citenamefont {K\"uhn}}]{Banuls:2018jag}%
  \BibitemOpen
  \bibfield  {author} {\bibinfo {author} {\bibfnamefont {M.~C.}\ \bibnamefont
  {Ba\~nuls}}, \bibinfo {author} {\bibfnamefont {K.}~\bibnamefont {Cichy}},
  \bibinfo {author} {\bibfnamefont {J.~I.}\ \bibnamefont {Cirac}}, \bibinfo
  {author} {\bibfnamefont {K.}~\bibnamefont {Jansen}}, \ and\ \bibinfo {author}
  {\bibfnamefont {S.}~\bibnamefont {K\"uhn}},\ }\href {\doibase
  10.22323/1.334.0022} {\bibfield  {journal} {\bibinfo  {journal} {PoS}\
  }\textbf {\bibinfo {volume} {LATTICE2018}},\ \bibinfo {pages} {022} (\bibinfo
  {year} {2018})},\ \Eprint {http://arxiv.org/abs/1810.12838} {arXiv:1810.12838
  [hep-lat]} \BibitemShut {NoStop}%
\bibitem [{\citenamefont {{Vidal}}(2003)}]{2003PhRvL..91n7902V}%
  \BibitemOpen
  \bibfield  {author} {\bibinfo {author} {\bibfnamefont {G.}~\bibnamefont
  {{Vidal}}},\ }\href {\doibase 10.1103/PhysRevLett.91.147902} {\bibfield
  {journal} {\bibinfo  {journal} {Phys. Rev. Lett.}\ }\textbf {\bibinfo
  {volume} {91}},\ \bibinfo {eid} {147902} (\bibinfo {year} {2003})},\ \Eprint
  {http://arxiv.org/abs/quant-ph/0301063} {arXiv:quant-ph/0301063 [quant-ph]}
  \BibitemShut {NoStop}%
\bibitem [{\citenamefont {Vidal}(2004)}]{Vidal:2003lvx}%
  \BibitemOpen
  \bibfield  {author} {\bibinfo {author} {\bibfnamefont {G.}~\bibnamefont
  {Vidal}},\ }\href {\doibase 10.1103/PhysRevLett.93.040502} {\bibfield
  {journal} {\bibinfo  {journal} {Phys. Rev. Lett.}\ }\textbf {\bibinfo
  {volume} {93}},\ \bibinfo {pages} {040502} (\bibinfo {year} {2004})},\
  \Eprint {http://arxiv.org/abs/quant-ph/0310089} {arXiv:quant-ph/0310089}
  \BibitemShut {NoStop}%
\bibitem [{\citenamefont {Sivia}()}]{doi:10.1142/9789812774187_0010}%
  \BibitemOpen
  \bibfield  {author} {\bibinfo {author} {\bibfnamefont {D.~S.}\ \bibnamefont
  {Sivia}},\ }\enquote {\bibinfo {title} {Data analysis: A dialogue with the
  data},}\ in\ \href {\doibase 10.1142/9789812774187_0010} {\emph {\bibinfo
  {booktitle} {Advanced Mathematical and Computational Tools in Metrology
  VII}}},\ pp.\ \bibinfo {pages} {108--118}\BibitemShut {NoStop}%
\bibitem [{\citenamefont {Horodecki}\ \emph {et~al.}(1996)\citenamefont
  {Horodecki}, \citenamefont {Horodecki},\ and\ \citenamefont
  {Horodecki}}]{HORODECKI19961}%
  \BibitemOpen
  \bibfield  {author} {\bibinfo {author} {\bibfnamefont {M.}~\bibnamefont
  {Horodecki}}, \bibinfo {author} {\bibfnamefont {P.}~\bibnamefont
  {Horodecki}}, \ and\ \bibinfo {author} {\bibfnamefont {R.}~\bibnamefont
  {Horodecki}},\ }\href {\doibase
  https://doi.org/10.1016/S0375-9601(96)00706-2} {\bibfield  {journal}
  {\bibinfo  {journal} {Physics Letters A}\ }\textbf {\bibinfo {volume}
  {223}},\ \bibinfo {pages} {1} (\bibinfo {year} {1996})}\BibitemShut {NoStop}%
\bibitem [{\citenamefont {Simon}(2000)}]{PhysRevLett.84.2726}%
  \BibitemOpen
  \bibfield  {author} {\bibinfo {author} {\bibfnamefont {R.}~\bibnamefont
  {Simon}},\ }\href {\doibase 10.1103/PhysRevLett.84.2726} {\bibfield
  {journal} {\bibinfo  {journal} {Phys. Rev. Lett.}\ }\textbf {\bibinfo
  {volume} {84}},\ \bibinfo {pages} {2726} (\bibinfo {year}
  {2000})}\BibitemShut {NoStop}%
\bibitem [{\citenamefont {Vidal}\ and\ \citenamefont
  {Werner}(2002)}]{PhysRevA.65.032314}%
  \BibitemOpen
  \bibfield  {author} {\bibinfo {author} {\bibfnamefont {G.}~\bibnamefont
  {Vidal}}\ and\ \bibinfo {author} {\bibfnamefont {R.~F.}\ \bibnamefont
  {Werner}},\ }\href {\doibase 10.1103/PhysRevA.65.032314} {\bibfield
  {journal} {\bibinfo  {journal} {Phys. Rev. A}\ }\textbf {\bibinfo {volume}
  {65}},\ \bibinfo {pages} {032314} (\bibinfo {year} {2002})}\BibitemShut
  {NoStop}%
\bibitem [{\citenamefont {Plenio}(2005)}]{PhysRevLett.95.090503}%
  \BibitemOpen
  \bibfield  {author} {\bibinfo {author} {\bibfnamefont {M.~B.}\ \bibnamefont
  {Plenio}},\ }\href {\doibase 10.1103/PhysRevLett.95.090503} {\bibfield
  {journal} {\bibinfo  {journal} {Phys. Rev. Lett.}\ }\textbf {\bibinfo
  {volume} {95}},\ \bibinfo {pages} {090503} (\bibinfo {year}
  {2005})}\BibitemShut {NoStop}%
\bibitem [{\citenamefont {Horodecki}(1997)}]{HORODECKI1997333}%
  \BibitemOpen
  \bibfield  {author} {\bibinfo {author} {\bibfnamefont {P.}~\bibnamefont
  {Horodecki}},\ }\href {\doibase
  https://doi.org/10.1016/S0375-9601(97)00416-7} {\bibfield  {journal}
  {\bibinfo  {journal} {Physics Letters A}\ }\textbf {\bibinfo {volume}
  {232}},\ \bibinfo {pages} {333} (\bibinfo {year} {1997})}\BibitemShut
  {NoStop}%
\bibitem [{\citenamefont {Horodecki}\ \emph {et~al.}(1998)\citenamefont
  {Horodecki}, \citenamefont {Horodecki},\ and\ \citenamefont
  {Horodecki}}]{PhysRevLett.80.5239}%
  \BibitemOpen
  \bibfield  {author} {\bibinfo {author} {\bibfnamefont {M.}~\bibnamefont
  {Horodecki}}, \bibinfo {author} {\bibfnamefont {P.}~\bibnamefont
  {Horodecki}}, \ and\ \bibinfo {author} {\bibfnamefont {R.}~\bibnamefont
  {Horodecki}},\ }\href {\doibase 10.1103/PhysRevLett.80.5239} {\bibfield
  {journal} {\bibinfo  {journal} {Phys. Rev. Lett.}\ }\textbf {\bibinfo
  {volume} {80}},\ \bibinfo {pages} {5239} (\bibinfo {year}
  {1998})}\BibitemShut {NoStop}%
\bibitem [{\citenamefont {Horodecki}\ and\ \citenamefont
  {Lewenstein}(2000)}]{PhysRevLett.85.2657}%
  \BibitemOpen
  \bibfield  {author} {\bibinfo {author} {\bibfnamefont {P.}~\bibnamefont
  {Horodecki}}\ and\ \bibinfo {author} {\bibfnamefont {M.}~\bibnamefont
  {Lewenstein}},\ }\href {\doibase 10.1103/PhysRevLett.85.2657} {\bibfield
  {journal} {\bibinfo  {journal} {Phys. Rev. Lett.}\ }\textbf {\bibinfo
  {volume} {85}},\ \bibinfo {pages} {2657} (\bibinfo {year}
  {2000})}\BibitemShut {NoStop}%
\bibitem [{\citenamefont {{Werner, R. F. and Wolf, M.
  M.}}(2001)}]{2001PhRvL..86.3658W}%
  \BibitemOpen
  \bibfield  {author} {\bibinfo {author} {\bibnamefont {{Werner, R. F. and
  Wolf, M. M.}}},\ }\href {\doibase 10.1103/PhysRevLett.86.3658} {\bibfield
  {journal} {\bibinfo  {journal} {Phys. Rev. Lett.}\ }\textbf {\bibinfo
  {volume} {86}},\ \bibinfo {pages} {3658} (\bibinfo {year} {2001})},\ \Eprint
  {http://arxiv.org/abs/quant-ph/0009118} {arXiv:quant-ph/0009118 [quant-ph]}
  \BibitemShut {NoStop}%
\bibitem [{\citenamefont {Terhal}(2000)}]{TERHAL2000319}%
  \BibitemOpen
  \bibfield  {author} {\bibinfo {author} {\bibfnamefont {B.~M.}\ \bibnamefont
  {Terhal}},\ }\href {\doibase https://doi.org/10.1016/S0375-9601(00)00401-1}
  {\bibfield  {journal} {\bibinfo  {journal} {Physics Letters A}\ }\textbf
  {\bibinfo {volume} {271}},\ \bibinfo {pages} {319} (\bibinfo {year}
  {2000})}\BibitemShut {NoStop}%
\bibitem [{\citenamefont {Wootters}(1998)}]{Wooters1998}%
  \BibitemOpen
  \bibfield  {author} {\bibinfo {author} {\bibfnamefont {W.~K.}\ \bibnamefont
  {Wootters}},\ }\href {\doibase 10.1103/PhysRevLett.80.2245} {\bibfield
  {journal} {\bibinfo  {journal} {Phys. Rev. Lett.}\ }\textbf {\bibinfo
  {volume} {80}},\ \bibinfo {pages} {2245} (\bibinfo {year}
  {1998})}\BibitemShut {NoStop}%
\bibitem [{\citenamefont {Hill}\ and\ \citenamefont
  {Wootters}(1997)}]{Hill1997}%
  \BibitemOpen
  \bibfield  {author} {\bibinfo {author} {\bibfnamefont {S.}~\bibnamefont
  {Hill}}\ and\ \bibinfo {author} {\bibfnamefont {W.~K.}\ \bibnamefont
  {Wootters}},\ }\href {\doibase 10.1103/PhysRevLett.78.5022} {\bibfield
  {journal} {\bibinfo  {journal} {Phys. Rev. Lett.}\ }\textbf {\bibinfo
  {volume} {78}},\ \bibinfo {pages} {5022} (\bibinfo {year}
  {1997})}\BibitemShut {NoStop}%
\bibitem [{\citenamefont {Bennett}\ \emph
  {et~al.}(1996{\natexlab{a}})\citenamefont {Bennett}, \citenamefont
  {Brassard}, \citenamefont {Popescu}, \citenamefont {Schumacher},
  \citenamefont {Smolin},\ and\ \citenamefont {Wootters}}]{Bennett_1996a}%
  \BibitemOpen
  \bibfield  {author} {\bibinfo {author} {\bibfnamefont {C.~H.}\ \bibnamefont
  {Bennett}}, \bibinfo {author} {\bibfnamefont {G.}~\bibnamefont {Brassard}},
  \bibinfo {author} {\bibfnamefont {S.}~\bibnamefont {Popescu}}, \bibinfo
  {author} {\bibfnamefont {B.}~\bibnamefont {Schumacher}}, \bibinfo {author}
  {\bibfnamefont {J.~A.}\ \bibnamefont {Smolin}}, \ and\ \bibinfo {author}
  {\bibfnamefont {W.~K.}\ \bibnamefont {Wootters}},\ }\href {\doibase
  10.1103/physrevlett.76.722} {\bibfield  {journal} {\bibinfo  {journal}
  {Physical Review Letters}\ }\textbf {\bibinfo {volume} {76}},\ \bibinfo
  {pages} {722–725} (\bibinfo {year} {1996}{\natexlab{a}})}\BibitemShut
  {NoStop}%
\bibitem [{\citenamefont {Bennett}\ \emph
  {et~al.}(1996{\natexlab{b}})\citenamefont {Bennett}, \citenamefont
  {Bernstein}, \citenamefont {Popescu},\ and\ \citenamefont
  {Schumacher}}]{Bennett_1996b}%
  \BibitemOpen
  \bibfield  {author} {\bibinfo {author} {\bibfnamefont {C.~H.}\ \bibnamefont
  {Bennett}}, \bibinfo {author} {\bibfnamefont {H.~J.}\ \bibnamefont
  {Bernstein}}, \bibinfo {author} {\bibfnamefont {S.}~\bibnamefont {Popescu}},
  \ and\ \bibinfo {author} {\bibfnamefont {B.}~\bibnamefont {Schumacher}},\
  }\href {\doibase 10.1103/physreva.53.2046} {\bibfield  {journal} {\bibinfo
  {journal} {Physical Review A}\ }\textbf {\bibinfo {volume} {53}},\ \bibinfo
  {pages} {2046–2052} (\bibinfo {year} {1996}{\natexlab{b}})}\BibitemShut
  {NoStop}%
\bibitem [{\citenamefont {Audenaert}\ \emph {et~al.}(2003)\citenamefont
  {Audenaert}, \citenamefont {Plenio},\ and\ \citenamefont
  {Eisert}}]{PhysRevLett.90.027901}%
  \BibitemOpen
  \bibfield  {author} {\bibinfo {author} {\bibfnamefont {K.}~\bibnamefont
  {Audenaert}}, \bibinfo {author} {\bibfnamefont {M.~B.}\ \bibnamefont
  {Plenio}}, \ and\ \bibinfo {author} {\bibfnamefont {J.}~\bibnamefont
  {Eisert}},\ }\href {\doibase 10.1103/PhysRevLett.90.027901} {\bibfield
  {journal} {\bibinfo  {journal} {Phys. Rev. Lett.}\ }\textbf {\bibinfo
  {volume} {90}},\ \bibinfo {pages} {027901} (\bibinfo {year}
  {2003})}\BibitemShut {NoStop}%
\bibitem [{\citenamefont {Milsted}\ \emph {et~al.}(2020)\citenamefont
  {Milsted}, \citenamefont {Liu}, \citenamefont {Preskill},\ and\ \citenamefont
  {Vidal}}]{Milsted:2020jmf}%
  \BibitemOpen
  \bibfield  {author} {\bibinfo {author} {\bibfnamefont {A.}~\bibnamefont
  {Milsted}}, \bibinfo {author} {\bibfnamefont {J.}~\bibnamefont {Liu}},
  \bibinfo {author} {\bibfnamefont {J.}~\bibnamefont {Preskill}}, \ and\
  \bibinfo {author} {\bibfnamefont {G.}~\bibnamefont {Vidal}},\ }\href@noop {}
  {\  (\bibinfo {year} {2020})},\ \Eprint {http://arxiv.org/abs/2012.07243}
  {arXiv:2012.07243 [quant-ph]} \BibitemShut {NoStop}%
\bibitem [{\citenamefont {Legeza}\ \emph {et~al.}(2015)\citenamefont {Legeza},
  \citenamefont {Veis}, \citenamefont {Poves},\ and\ \citenamefont
  {Dukelsky}}]{Legeza:2015fja}%
  \BibitemOpen
  \bibfield  {author} {\bibinfo {author} {\bibfnamefont {O.}~\bibnamefont
  {Legeza}}, \bibinfo {author} {\bibfnamefont {L.}~\bibnamefont {Veis}},
  \bibinfo {author} {\bibfnamefont {A.}~\bibnamefont {Poves}}, \ and\ \bibinfo
  {author} {\bibfnamefont {J.}~\bibnamefont {Dukelsky}},\ }\href {\doibase
  10.1103/PhysRevC.92.051303} {\bibfield  {journal} {\bibinfo  {journal} {Phys.
  Rev. C}\ }\textbf {\bibinfo {volume} {92}},\ \bibinfo {pages} {051303}
  (\bibinfo {year} {2015})},\ \Eprint {http://arxiv.org/abs/1507.00161}
  {arXiv:1507.00161 [nucl-th]} \BibitemShut {NoStop}%
\bibitem [{\citenamefont {Gorton}(2018)}]{gorton18a}%
  \BibitemOpen
  \bibfield  {author} {\bibinfo {author} {\bibfnamefont {O.~C.}\ \bibnamefont
  {Gorton}},\ }\href
  {https://digitallibrary.sdsu.edu/islandora/object/sdsu\%3A22155} {\enquote
  {\bibinfo {title} {Efficient modeling of nuclei through coupling of proton
  and neutron wavefunctions},}\ } (\bibinfo {year} {2018})\BibitemShut
  {NoStop}%
\bibitem [{\citenamefont {Johnson}\ \emph {et~al.}(2018)\citenamefont
  {Johnson}, \citenamefont {Ormand}, \citenamefont {McElvain},\ and\
  \citenamefont {Shan}}]{johnson2018bigstick}%
  \BibitemOpen
  \bibfield  {author} {\bibinfo {author} {\bibfnamefont {C.~W.}\ \bibnamefont
  {Johnson}}, \bibinfo {author} {\bibfnamefont {W.~E.}\ \bibnamefont {Ormand}},
  \bibinfo {author} {\bibfnamefont {K.~S.}\ \bibnamefont {McElvain}}, \ and\
  \bibinfo {author} {\bibfnamefont {H.}~\bibnamefont {Shan}},\ }\href
  {https://arxiv.org/abs/1801.08432} {\enquote {\bibinfo {title} {Bigstick: A
  flexible configuration-interaction shell-model code},}\ } (\bibinfo {year}
  {2018}),\ \Eprint {http://arxiv.org/abs/1801.08432} {arXiv:1801.08432
  [physics.comp-ph]} \BibitemShut {NoStop}%
\bibitem [{\citenamefont {Robin}\ \emph {et~al.}(2021)\citenamefont {Robin},
  \citenamefont {Savage},\ and\ \citenamefont {Pillet}}]{Robin:2020aeh}%
  \BibitemOpen
  \bibfield  {author} {\bibinfo {author} {\bibfnamefont {C.}~\bibnamefont
  {Robin}}, \bibinfo {author} {\bibfnamefont {M.~J.}\ \bibnamefont {Savage}}, \
  and\ \bibinfo {author} {\bibfnamefont {N.}~\bibnamefont {Pillet}},\ }\href
  {\doibase 10.1103/PhysRevC.103.034325} {\bibfield  {journal} {\bibinfo
  {journal} {Phys. Rev. C}\ }\textbf {\bibinfo {volume} {103}},\ \bibinfo
  {pages} {034325} (\bibinfo {year} {2021})},\ \Eprint
  {http://arxiv.org/abs/2007.09157} {arXiv:2007.09157 [nucl-th]} \BibitemShut
  {NoStop}%
\bibitem [{\citenamefont {Cervera-Lierta}\ \emph {et~al.}(2017)\citenamefont
  {Cervera-Lierta}, \citenamefont {Latorre}, \citenamefont {Rojo},\ and\
  \citenamefont {Rottoli}}]{Cervera-Lierta:2017tdt}%
  \BibitemOpen
  \bibfield  {author} {\bibinfo {author} {\bibfnamefont {A.}~\bibnamefont
  {Cervera-Lierta}}, \bibinfo {author} {\bibfnamefont {J.~I.}\ \bibnamefont
  {Latorre}}, \bibinfo {author} {\bibfnamefont {J.}~\bibnamefont {Rojo}}, \
  and\ \bibinfo {author} {\bibfnamefont {L.}~\bibnamefont {Rottoli}},\ }\href
  {\doibase 10.21468/SciPostPhys.3.5.036} {\bibfield  {journal} {\bibinfo
  {journal} {SciPost Phys.}\ }\textbf {\bibinfo {volume} {3}},\ \bibinfo
  {pages} {036} (\bibinfo {year} {2017})},\ \Eprint
  {http://arxiv.org/abs/1703.02989} {arXiv:1703.02989 [hep-th]} \BibitemShut
  {NoStop}%
\bibitem [{\citenamefont {Beane}\ \emph {et~al.}(2019)\citenamefont {Beane},
  \citenamefont {Kaplan}, \citenamefont {Klco},\ and\ \citenamefont
  {Savage}}]{Beane:2018oxh}%
  \BibitemOpen
  \bibfield  {author} {\bibinfo {author} {\bibfnamefont {S.~R.}\ \bibnamefont
  {Beane}}, \bibinfo {author} {\bibfnamefont {D.~B.}\ \bibnamefont {Kaplan}},
  \bibinfo {author} {\bibfnamefont {N.}~\bibnamefont {Klco}}, \ and\ \bibinfo
  {author} {\bibfnamefont {M.~J.}\ \bibnamefont {Savage}},\ }\href {\doibase
  10.1103/PhysRevLett.122.102001} {\bibfield  {journal} {\bibinfo  {journal}
  {Phys. Rev. Lett.}\ }\textbf {\bibinfo {volume} {122}},\ \bibinfo {pages}
  {102001} (\bibinfo {year} {2019})},\ \Eprint
  {http://arxiv.org/abs/1812.03138} {arXiv:1812.03138 [nucl-th]} \BibitemShut
  {NoStop}%
\bibitem [{\citenamefont {Zanardi}\ \emph {et~al.}(2000)\citenamefont
  {Zanardi}, \citenamefont {Zalka},\ and\ \citenamefont
  {Faoro}}]{Zanardi:2000zz}%
  \BibitemOpen
  \bibfield  {author} {\bibinfo {author} {\bibfnamefont {P.}~\bibnamefont
  {Zanardi}}, \bibinfo {author} {\bibfnamefont {C.}~\bibnamefont {Zalka}}, \
  and\ \bibinfo {author} {\bibfnamefont {L.}~\bibnamefont {Faoro}},\ }\href
  {\doibase 10.1103/PhysRevA.62.030301} {\bibfield  {journal} {\bibinfo
  {journal} {Phys. Rev. A}\ }\textbf {\bibinfo {volume} {62}},\ \bibinfo
  {pages} {030301} (\bibinfo {year} {2000})},\ \Eprint
  {http://arxiv.org/abs/quant-ph/0005031} {arXiv:quant-ph/0005031} \BibitemShut
  {NoStop}%
\bibitem [{\citenamefont {Ballard}\ and\ \citenamefont
  {Wu}(2011)}]{mahdavi2011cross}%
  \BibitemOpen
  \bibfield  {author} {\bibinfo {author} {\bibfnamefont {A.~D.}\ \bibnamefont
  {Ballard}}\ and\ \bibinfo {author} {\bibfnamefont {Y.-S.}\ \bibnamefont
  {Wu}},\ }\enquote {\bibinfo {title} {Cross disciplinary advances in quantum
  computing},}\ \ (\bibinfo  {publisher} {American Mathematical Society},\
  \bibinfo {year} {2011})\ Chap.\ \bibinfo {chapter} {Cartan Decomposition and
  Entangleing Power of Braiding Quantum Gates}\BibitemShut {NoStop}%
\bibitem [{\citenamefont {Low}\ and\ \citenamefont
  {Mehen}(2021)}]{Low:2021ufv}%
  \BibitemOpen
  \bibfield  {author} {\bibinfo {author} {\bibfnamefont {I.}~\bibnamefont
  {Low}}\ and\ \bibinfo {author} {\bibfnamefont {T.}~\bibnamefont {Mehen}},\
  }\href@noop {} {\  (\bibinfo {year} {2021})},\ \Eprint
  {http://arxiv.org/abs/2104.10835} {arXiv:2104.10835 [hep-th]} \BibitemShut
  {NoStop}%
\bibitem [{\citenamefont {Wagman}\ \emph {et~al.}(2017)\citenamefont {Wagman},
  \citenamefont {Winter}, \citenamefont {Chang}, \citenamefont {Davoudi},
  \citenamefont {Detmold}, \citenamefont {Orginos}, \citenamefont {Savage},\
  and\ \citenamefont {Shanahan}}]{Wagman:2017tmp}%
  \BibitemOpen
  \bibfield  {author} {\bibinfo {author} {\bibfnamefont {M.~L.}\ \bibnamefont
  {Wagman}}, \bibinfo {author} {\bibfnamefont {F.}~\bibnamefont {Winter}},
  \bibinfo {author} {\bibfnamefont {E.}~\bibnamefont {Chang}}, \bibinfo
  {author} {\bibfnamefont {Z.}~\bibnamefont {Davoudi}}, \bibinfo {author}
  {\bibfnamefont {W.}~\bibnamefont {Detmold}}, \bibinfo {author} {\bibfnamefont
  {K.}~\bibnamefont {Orginos}}, \bibinfo {author} {\bibfnamefont {M.~J.}\
  \bibnamefont {Savage}}, \ and\ \bibinfo {author} {\bibfnamefont {P.~E.}\
  \bibnamefont {Shanahan}},\ }\href {\doibase 10.1103/PhysRevD.96.114510}
  {\bibfield  {journal} {\bibinfo  {journal} {Phys. Rev. D}\ }\textbf {\bibinfo
  {volume} {96}},\ \bibinfo {pages} {114510} (\bibinfo {year} {2017})},\
  \Eprint {http://arxiv.org/abs/1706.06550} {arXiv:1706.06550 [hep-lat]}
  \BibitemShut {NoStop}%
\bibitem [{\citenamefont {Srednicki}(1993)}]{Srednicki_1993}%
  \BibitemOpen
  \bibfield  {author} {\bibinfo {author} {\bibfnamefont {M.}~\bibnamefont
  {Srednicki}},\ }\href {\doibase 10.1103/physrevlett.71.666} {\bibfield
  {journal} {\bibinfo  {journal} {Physical Review Letters}\ }\textbf {\bibinfo
  {volume} {71}},\ \bibinfo {pages} {666–669} (\bibinfo {year}
  {1993})}\BibitemShut {NoStop}%
\bibitem [{\citenamefont {Audenaert}\ \emph {et~al.}(2002)\citenamefont
  {Audenaert}, \citenamefont {Eisert}, \citenamefont {Plenio},\ and\
  \citenamefont {Werner}}]{Audenaert:2002xfl}%
  \BibitemOpen
  \bibfield  {author} {\bibinfo {author} {\bibfnamefont {K.}~\bibnamefont
  {Audenaert}}, \bibinfo {author} {\bibfnamefont {J.}~\bibnamefont {Eisert}},
  \bibinfo {author} {\bibfnamefont {M.}~\bibnamefont {Plenio}}, \ and\ \bibinfo
  {author} {\bibfnamefont {R.}~\bibnamefont {Werner}},\ }\href {\doibase
  10.1103/PhysRevA.66.042327} {\bibfield  {journal} {\bibinfo  {journal} {Phys.
  Rev. A}\ }\textbf {\bibinfo {volume} {66}},\ \bibinfo {pages} {042327}
  (\bibinfo {year} {2002})},\ \Eprint {http://arxiv.org/abs/quant-ph/0205025}
  {arXiv:quant-ph/0205025} \BibitemShut {NoStop}%
\bibitem [{\citenamefont {{Botero}}\ and\ \citenamefont
  {{Reznik}}(2004)}]{2004PhRvA..70e2329B}%
  \BibitemOpen
  \bibfield  {author} {\bibinfo {author} {\bibfnamefont {A.}~\bibnamefont
  {{Botero}}}\ and\ \bibinfo {author} {\bibfnamefont {B.}~\bibnamefont
  {{Reznik}}},\ }\href {\doibase 10.1103/PhysRevA.70.052329} {\bibfield
  {journal} {\bibinfo  {journal} {Phys. Rev. A}\ }\textbf {\bibinfo {volume}
  {70}},\ \bibinfo {eid} {052329} (\bibinfo {year} {2004})},\ \Eprint
  {http://arxiv.org/abs/quant-ph/0403233} {arXiv:quant-ph/0403233 [quant-ph]}
  \BibitemShut {NoStop}%
\bibitem [{\citenamefont {Audenaert}\ \emph {et~al.}(2007)\citenamefont
  {Audenaert}, \citenamefont {Eisert},\ and\ \citenamefont
  {Plenio}}]{cerf2007quantum}%
  \BibitemOpen
  \bibfield  {author} {\bibinfo {author} {\bibfnamefont {K.~M.~R.}\
  \bibnamefont {Audenaert}}, \bibinfo {author} {\bibfnamefont {J.}~\bibnamefont
  {Eisert}}, \ and\ \bibinfo {author} {\bibfnamefont {M.~B.}\ \bibnamefont
  {Plenio}},\ }in\ \href@noop {} {\emph {\bibinfo {booktitle} {Quantum
  information with continuous variables of atoms and light}}},\ \bibinfo
  {editor} {edited by\ \bibinfo {editor} {\bibfnamefont {N.~J.}\ \bibnamefont
  {Cerf}}, \bibinfo {editor} {\bibfnamefont {G.}~\bibnamefont {Leuchs}}, \ and\
  \bibinfo {editor} {\bibfnamefont {E.~S.}\ \bibnamefont {Polzik}}}\ (\bibinfo
  {publisher} {Imperial College Press},\ \bibinfo {address} {London},\ \bibinfo
  {year} {2007})\ pp.\ \bibinfo {pages} {43--62}\BibitemShut {NoStop}%
\bibitem [{\citenamefont {Marcovitch}\ \emph {et~al.}(2009)\citenamefont
  {Marcovitch}, \citenamefont {Retzker}, \citenamefont {Plenio},\ and\
  \citenamefont {Reznik}}]{PhysRevA.80.012325}%
  \BibitemOpen
  \bibfield  {author} {\bibinfo {author} {\bibfnamefont {S.}~\bibnamefont
  {Marcovitch}}, \bibinfo {author} {\bibfnamefont {A.}~\bibnamefont {Retzker}},
  \bibinfo {author} {\bibfnamefont {M.~B.}\ \bibnamefont {Plenio}}, \ and\
  \bibinfo {author} {\bibfnamefont {B.}~\bibnamefont {Reznik}},\ }\href
  {\doibase 10.1103/PhysRevA.80.012325} {\bibfield  {journal} {\bibinfo
  {journal} {Phys. Rev. A}\ }\textbf {\bibinfo {volume} {80}},\ \bibinfo
  {pages} {012325} (\bibinfo {year} {2009})}\BibitemShut {NoStop}%
\bibitem [{\citenamefont {Coser}\ \emph {et~al.}(2017)\citenamefont {Coser},
  \citenamefont {De~Nobili},\ and\ \citenamefont {Tonni}}]{Coser_2017}%
  \BibitemOpen
  \bibfield  {author} {\bibinfo {author} {\bibfnamefont {A.}~\bibnamefont
  {Coser}}, \bibinfo {author} {\bibfnamefont {C.}~\bibnamefont {De~Nobili}}, \
  and\ \bibinfo {author} {\bibfnamefont {E.}~\bibnamefont {Tonni}},\ }\href
  {\doibase 10.1088/1751-8121/aa7902} {\bibfield  {journal} {\bibinfo
  {journal} {Journal of Physics A: Mathematical and Theoretical}\ }\textbf
  {\bibinfo {volume} {50}},\ \bibinfo {pages} {314001} (\bibinfo {year}
  {2017})}\BibitemShut {NoStop}%
\bibitem [{\citenamefont {Di~Giulio}\ and\ \citenamefont
  {Tonni}(2020)}]{DiGiulio:2019cxv}%
  \BibitemOpen
  \bibfield  {author} {\bibinfo {author} {\bibfnamefont {G.}~\bibnamefont
  {Di~Giulio}}\ and\ \bibinfo {author} {\bibfnamefont {E.}~\bibnamefont
  {Tonni}},\ }\href {\doibase 10.1088/1742-5468/ab7129} {\bibfield  {journal}
  {\bibinfo  {journal} {J. Stat. Mech.}\ }\textbf {\bibinfo {volume} {2003}},\
  \bibinfo {pages} {033102} (\bibinfo {year} {2020})},\ \Eprint
  {http://arxiv.org/abs/1911.07188} {arXiv:1911.07188 [cond-mat.stat-mech]}
  \BibitemShut {NoStop}%
\bibitem [{\citenamefont {Calabrese}\ and\ \citenamefont
  {Cardy}(2004)}]{Calabrese:2004eu}%
  \BibitemOpen
  \bibfield  {author} {\bibinfo {author} {\bibfnamefont {P.}~\bibnamefont
  {Calabrese}}\ and\ \bibinfo {author} {\bibfnamefont {J.~L.}\ \bibnamefont
  {Cardy}},\ }\href {\doibase 10.1088/1742-5468/2004/06/P06002} {\bibfield
  {journal} {\bibinfo  {journal} {Journal of Statistical Mechanics: Theory and
  Experiment}\ }\textbf {\bibinfo {volume} {0406}},\ \bibinfo {pages} {P06002}
  (\bibinfo {year} {2004})},\ \Eprint {http://arxiv.org/abs/hep-th/0405152}
  {arXiv:hep-th/0405152} \BibitemShut {NoStop}%
\bibitem [{\citenamefont {Calabrese}\ and\ \citenamefont
  {Cardy}(2009)}]{Calabrese:2009qy}%
  \BibitemOpen
  \bibfield  {author} {\bibinfo {author} {\bibfnamefont {P.}~\bibnamefont
  {Calabrese}}\ and\ \bibinfo {author} {\bibfnamefont {J.}~\bibnamefont
  {Cardy}},\ }\href {\doibase 10.1088/1751-8113/42/50/504005} {\bibfield
  {journal} {\bibinfo  {journal} {J. Phys. A}\ }\textbf {\bibinfo {volume}
  {42}},\ \bibinfo {pages} {504005} (\bibinfo {year} {2009})},\ \Eprint
  {http://arxiv.org/abs/0905.4013} {arXiv:0905.4013 [cond-mat.stat-mech]}
  \BibitemShut {NoStop}%
\bibitem [{\citenamefont {Calabrese}\ \emph {et~al.}(2009)\citenamefont
  {Calabrese}, \citenamefont {Cardy},\ and\ \citenamefont
  {Tonni}}]{Calabrese:2009ez}%
  \BibitemOpen
  \bibfield  {author} {\bibinfo {author} {\bibfnamefont {P.}~\bibnamefont
  {Calabrese}}, \bibinfo {author} {\bibfnamefont {J.}~\bibnamefont {Cardy}}, \
  and\ \bibinfo {author} {\bibfnamefont {E.}~\bibnamefont {Tonni}},\ }\href
  {\doibase 10.1088/1742-5468/2009/11/P11001} {\bibfield  {journal} {\bibinfo
  {journal} {J. Stat. Mech.}\ }\textbf {\bibinfo {volume} {0911}},\ \bibinfo
  {pages} {P11001} (\bibinfo {year} {2009})},\ \Eprint
  {http://arxiv.org/abs/0905.2069} {arXiv:0905.2069 [hep-th]} \BibitemShut
  {NoStop}%
\bibitem [{\citenamefont {Calabrese}\ \emph {et~al.}(2012)\citenamefont
  {Calabrese}, \citenamefont {Cardy},\ and\ \citenamefont
  {Tonni}}]{Calabrese:2012ew}%
  \BibitemOpen
  \bibfield  {author} {\bibinfo {author} {\bibfnamefont {P.}~\bibnamefont
  {Calabrese}}, \bibinfo {author} {\bibfnamefont {J.}~\bibnamefont {Cardy}}, \
  and\ \bibinfo {author} {\bibfnamefont {E.}~\bibnamefont {Tonni}},\ }\href
  {\doibase 10.1103/PhysRevLett.109.130502} {\bibfield  {journal} {\bibinfo
  {journal} {Phys. Rev. Lett.}\ }\textbf {\bibinfo {volume} {109}},\ \bibinfo
  {pages} {130502} (\bibinfo {year} {2012})},\ \Eprint
  {http://arxiv.org/abs/1206.3092} {arXiv:1206.3092 [cond-mat.stat-mech]}
  \BibitemShut {NoStop}%
\bibitem [{\citenamefont {Calabrese}\ \emph {et~al.}(2013)\citenamefont
  {Calabrese}, \citenamefont {Cardy},\ and\ \citenamefont
  {Tonni}}]{Calabrese:2012nk}%
  \BibitemOpen
  \bibfield  {author} {\bibinfo {author} {\bibfnamefont {P.}~\bibnamefont
  {Calabrese}}, \bibinfo {author} {\bibfnamefont {J.}~\bibnamefont {Cardy}}, \
  and\ \bibinfo {author} {\bibfnamefont {E.}~\bibnamefont {Tonni}},\ }\href
  {\doibase 10.1088/1742-5468/2013/02/P02008} {\bibfield  {journal} {\bibinfo
  {journal} {J. Stat. Mech.}\ }\textbf {\bibinfo {volume} {1302}},\ \bibinfo
  {pages} {P02008} (\bibinfo {year} {2013})},\ \Eprint
  {http://arxiv.org/abs/1210.5359} {arXiv:1210.5359 [cond-mat.stat-mech]}
  \BibitemShut {NoStop}%
\bibitem [{\citenamefont {Ruggiero}\ \emph {et~al.}(2018)\citenamefont
  {Ruggiero}, \citenamefont {Tonni},\ and\ \citenamefont
  {Calabrese}}]{Ruggiero:2018hyl}%
  \BibitemOpen
  \bibfield  {author} {\bibinfo {author} {\bibfnamefont {P.}~\bibnamefont
  {Ruggiero}}, \bibinfo {author} {\bibfnamefont {E.}~\bibnamefont {Tonni}}, \
  and\ \bibinfo {author} {\bibfnamefont {P.}~\bibnamefont {Calabrese}},\ }\href
  {\doibase 10.1088/1742-5468/aae5a8} {\bibfield  {journal} {\bibinfo
  {journal} {J. Stat. Mech.}\ }\textbf {\bibinfo {volume} {1811}},\ \bibinfo
  {pages} {113101} (\bibinfo {year} {2018})},\ \Eprint
  {http://arxiv.org/abs/1805.05975} {arXiv:1805.05975 [cond-mat.stat-mech]}
  \BibitemShut {NoStop}%
\bibitem [{\citenamefont {Ho}\ and\ \citenamefont {Hsu}(2016)}]{Ho:2015rga}%
  \BibitemOpen
  \bibfield  {author} {\bibinfo {author} {\bibfnamefont {C.~M.}\ \bibnamefont
  {Ho}}\ and\ \bibinfo {author} {\bibfnamefont {S.~D.~H.}\ \bibnamefont
  {Hsu}},\ }\href {\doibase 10.1142/S0217732316501108} {\bibfield  {journal}
  {\bibinfo  {journal} {Mod. Phys. Lett. A}\ }\textbf {\bibinfo {volume}
  {31}},\ \bibinfo {pages} {1650110} (\bibinfo {year} {2016})},\ \Eprint
  {http://arxiv.org/abs/1506.03696} {arXiv:1506.03696 [hep-th]} \BibitemShut
  {NoStop}%
\bibitem [{\citenamefont {Baker}\ and\ \citenamefont
  {Kharzeev}(2018)}]{Baker:2017wtt}%
  \BibitemOpen
  \bibfield  {author} {\bibinfo {author} {\bibfnamefont {O.~K.}\ \bibnamefont
  {Baker}}\ and\ \bibinfo {author} {\bibfnamefont {D.~E.}\ \bibnamefont
  {Kharzeev}},\ }\href {\doibase 10.1103/PhysRevD.98.054007} {\bibfield
  {journal} {\bibinfo  {journal} {Phys. Rev. D}\ }\textbf {\bibinfo {volume}
  {98}},\ \bibinfo {pages} {054007} (\bibinfo {year} {2018})},\ \Eprint
  {http://arxiv.org/abs/1712.04558} {arXiv:1712.04558 [hep-ph]} \BibitemShut
  {NoStop}%
\bibitem [{\citenamefont {Tu}\ \emph {et~al.}(2020)\citenamefont {Tu},
  \citenamefont {Kharzeev},\ and\ \citenamefont {Ullrich}}]{Tu:2019ouv}%
  \BibitemOpen
  \bibfield  {author} {\bibinfo {author} {\bibfnamefont {Z.}~\bibnamefont
  {Tu}}, \bibinfo {author} {\bibfnamefont {D.~E.}\ \bibnamefont {Kharzeev}}, \
  and\ \bibinfo {author} {\bibfnamefont {T.}~\bibnamefont {Ullrich}},\ }\href
  {\doibase 10.1103/PhysRevLett.124.062001} {\bibfield  {journal} {\bibinfo
  {journal} {Phys. Rev. Lett.}\ }\textbf {\bibinfo {volume} {124}},\ \bibinfo
  {pages} {062001} (\bibinfo {year} {2020})},\ \Eprint
  {http://arxiv.org/abs/1904.11974} {arXiv:1904.11974 [hep-ph]} \BibitemShut
  {NoStop}%
\bibitem [{\citenamefont {Kharzeev}\ and\ \citenamefont
  {Levin}(2021)}]{Kharzeev:2021yyf}%
  \BibitemOpen
  \bibfield  {author} {\bibinfo {author} {\bibfnamefont {D.~E.}\ \bibnamefont
  {Kharzeev}}\ and\ \bibinfo {author} {\bibfnamefont {E.}~\bibnamefont
  {Levin}},\ }\href@noop {} {\  (\bibinfo {year} {2021})},\ \Eprint
  {http://arxiv.org/abs/2102.09773} {arXiv:2102.09773 [hep-ph]} \BibitemShut
  {NoStop}%
\bibitem [{\citenamefont {Maldacena}(1998)}]{Maldacena:1997re}%
  \BibitemOpen
  \bibfield  {author} {\bibinfo {author} {\bibfnamefont {J.~M.}\ \bibnamefont
  {Maldacena}},\ }\href {\doibase 10.1023/A:1026654312961} {\bibfield
  {journal} {\bibinfo  {journal} {Adv. Theor. Math. Phys.}\ }\textbf {\bibinfo
  {volume} {2}},\ \bibinfo {pages} {231} (\bibinfo {year} {1998})},\ \Eprint
  {http://arxiv.org/abs/hep-th/9711200} {arXiv:hep-th/9711200} \BibitemShut
  {NoStop}%
\bibitem [{\citenamefont {Maldacena}(2003)}]{Maldacena:2001kr}%
  \BibitemOpen
  \bibfield  {author} {\bibinfo {author} {\bibfnamefont {J.~M.}\ \bibnamefont
  {Maldacena}},\ }\href {\doibase 10.1088/1126-6708/2003/04/021} {\bibfield
  {journal} {\bibinfo  {journal} {JHEP}\ }\textbf {\bibinfo {volume} {04}},\
  \bibinfo {pages} {021} (\bibinfo {year} {2003})},\ \Eprint
  {http://arxiv.org/abs/hep-th/0106112} {arXiv:hep-th/0106112} \BibitemShut
  {NoStop}%
\bibitem [{\citenamefont {Ryu}\ and\ \citenamefont
  {Takayanagi}(2006{\natexlab{a}})}]{Ryu:2006bv}%
  \BibitemOpen
  \bibfield  {author} {\bibinfo {author} {\bibfnamefont {S.}~\bibnamefont
  {Ryu}}\ and\ \bibinfo {author} {\bibfnamefont {T.}~\bibnamefont
  {Takayanagi}},\ }\href {\doibase 10.1103/PhysRevLett.96.181602} {\bibfield
  {journal} {\bibinfo  {journal} {Phys. Rev. Lett.}\ }\textbf {\bibinfo
  {volume} {96}},\ \bibinfo {pages} {181602} (\bibinfo {year}
  {2006}{\natexlab{a}})},\ \Eprint {http://arxiv.org/abs/hep-th/0603001}
  {arXiv:hep-th/0603001} \BibitemShut {NoStop}%
\bibitem [{\citenamefont {Ryu}\ and\ \citenamefont
  {Takayanagi}(2006{\natexlab{b}})}]{Ryu:2006ef}%
  \BibitemOpen
  \bibfield  {author} {\bibinfo {author} {\bibfnamefont {S.}~\bibnamefont
  {Ryu}}\ and\ \bibinfo {author} {\bibfnamefont {T.}~\bibnamefont
  {Takayanagi}},\ }\href {\doibase 10.1088/1126-6708/2006/08/045} {\bibfield
  {journal} {\bibinfo  {journal} {JHEP}\ }\textbf {\bibinfo {volume} {08}},\
  \bibinfo {pages} {045} (\bibinfo {year} {2006}{\natexlab{b}})},\ \Eprint
  {http://arxiv.org/abs/hep-th/0605073} {arXiv:hep-th/0605073} \BibitemShut
  {NoStop}%
\bibitem [{\citenamefont {Klco}\ and\ \citenamefont
  {Savage}(2021{\natexlab{a}})}]{Klco:2020rga}%
  \BibitemOpen
  \bibfield  {author} {\bibinfo {author} {\bibfnamefont {N.}~\bibnamefont
  {Klco}}\ and\ \bibinfo {author} {\bibfnamefont {M.~J.}\ \bibnamefont
  {Savage}},\ }\href {\doibase 10.1103/PhysRevD.103.065007} {\bibfield
  {journal} {\bibinfo  {journal} {Phys. Rev. D}\ }\textbf {\bibinfo {volume}
  {103}},\ \bibinfo {pages} {065007} (\bibinfo {year} {2021}{\natexlab{a}})},\
  \Eprint {http://arxiv.org/abs/2008.03647} {arXiv:2008.03647 [quant-ph]}
  \BibitemShut {NoStop}%
\bibitem [{\citenamefont {Klco}\ and\ \citenamefont
  {Savage}(2021{\natexlab{b}})}]{Klco:2021biu}%
  \BibitemOpen
  \bibfield  {author} {\bibinfo {author} {\bibfnamefont {N.}~\bibnamefont
  {Klco}}\ and\ \bibinfo {author} {\bibfnamefont {M.~J.}\ \bibnamefont
  {Savage}},\ }\href@noop {} {\  (\bibinfo {year} {2021}{\natexlab{b}})},\
  \Eprint {http://arxiv.org/abs/2103.14999} {arXiv:2103.14999 [hep-th]}
  \BibitemShut {NoStop}%
\bibitem [{\citenamefont {Beane}\ and\ \citenamefont
  {Ehlers}(2019)}]{Beane:2019loz}%
  \BibitemOpen
  \bibfield  {author} {\bibinfo {author} {\bibfnamefont {S.~R.}\ \bibnamefont
  {Beane}}\ and\ \bibinfo {author} {\bibfnamefont {P.}~\bibnamefont {Ehlers}},\
  }\href {\doibase 10.1142/S0217732320500480} {\bibfield  {journal} {\bibinfo
  {journal} {Mod. Phys. Lett. A}\ }\textbf {\bibinfo {volume} {35}},\ \bibinfo
  {pages} {2050048} (\bibinfo {year} {2019})},\ \Eprint
  {http://arxiv.org/abs/1905.03295} {arXiv:1905.03295 [hep-ph]} \BibitemShut
  {NoStop}%
\bibitem [{\citenamefont {Beane}\ and\ \citenamefont
  {Farrell}(2020)}]{Beane:2020wjl}%
  \BibitemOpen
  \bibfield  {author} {\bibinfo {author} {\bibfnamefont {S.~R.}\ \bibnamefont
  {Beane}}\ and\ \bibinfo {author} {\bibfnamefont {R.~C.}\ \bibnamefont
  {Farrell}},\ }\href@noop {} {\  (\bibinfo {year} {2020})},\ \Eprint
  {http://arxiv.org/abs/2011.01278} {arXiv:2011.01278 [hep-th]} \BibitemShut
  {NoStop}%
\bibitem [{\citenamefont {{Wolf}}\ \emph {et~al.}(2006)\citenamefont {{Wolf}},
  \citenamefont {{Giedke}},\ and\ \citenamefont
  {{Cirac}}}]{2006PhRvL..96h0502W}%
  \BibitemOpen
  \bibfield  {author} {\bibinfo {author} {\bibfnamefont {M.~M.}\ \bibnamefont
  {{Wolf}}}, \bibinfo {author} {\bibfnamefont {G.}~\bibnamefont {{Giedke}}}, \
  and\ \bibinfo {author} {\bibfnamefont {J.~I.}\ \bibnamefont {{Cirac}}},\
  }\href {\doibase 10.1103/PhysRevLett.96.080502} {\bibfield  {journal}
  {\bibinfo  {journal} {Phys Rev. Lett.}\ }\textbf {\bibinfo {volume} {96}},\
  \bibinfo {eid} {080502} (\bibinfo {year} {2006})},\ \Eprint
  {http://arxiv.org/abs/quant-ph/0509154} {arXiv:quant-ph/0509154 [quant-ph]}
  \BibitemShut {NoStop}%
\bibitem [{\citenamefont {Klebanov}\ \emph {et~al.}(2008)\citenamefont
  {Klebanov}, \citenamefont {Kutasov},\ and\ \citenamefont
  {Murugan}}]{Klebanov:2007ws}%
  \BibitemOpen
  \bibfield  {author} {\bibinfo {author} {\bibfnamefont {I.~R.}\ \bibnamefont
  {Klebanov}}, \bibinfo {author} {\bibfnamefont {D.}~\bibnamefont {Kutasov}}, \
  and\ \bibinfo {author} {\bibfnamefont {A.}~\bibnamefont {Murugan}},\ }\href
  {\doibase 10.1016/j.nuclphysb.2007.12.017} {\bibfield  {journal} {\bibinfo
  {journal} {Nucl. Phys. B}\ }\textbf {\bibinfo {volume} {796}},\ \bibinfo
  {pages} {274} (\bibinfo {year} {2008})},\ \Eprint
  {http://arxiv.org/abs/0709.2140} {arXiv:0709.2140 [hep-th]} \BibitemShut
  {NoStop}%
\bibitem [{\citenamefont {Wen}(2019)}]{Wen_2019}%
  \BibitemOpen
  \bibfield  {author} {\bibinfo {author} {\bibfnamefont {X.-G.}\ \bibnamefont
  {Wen}},\ }\href {\doibase 10.1126/science.aal3099} {\bibfield  {journal}
  {\bibinfo  {journal} {Science}\ }\textbf {\bibinfo {volume} {363}},\ \bibinfo
  {pages} {eaal3099} (\bibinfo {year} {2019})}\BibitemShut {NoStop}%
\bibitem [{\citenamefont
  {Roggero}(2021{\natexlab{a}})}]{roggero2021entanglement}%
  \BibitemOpen
  \bibfield  {author} {\bibinfo {author} {\bibfnamefont {A.}~\bibnamefont
  {Roggero}},\ }\href@noop {} {\enquote {\bibinfo {title} {Entanglement and
  many-body effects in collective neutrino oscillations},}\ } (\bibinfo {year}
  {2021}{\natexlab{a}}),\ \Eprint {http://arxiv.org/abs/2102.10188}
  {arXiv:2102.10188 [hep-ph]} \BibitemShut {NoStop}%
\bibitem [{\citenamefont {Roggero}(2021{\natexlab{b}})}]{roggero2021dynamical}%
  \BibitemOpen
  \bibfield  {author} {\bibinfo {author} {\bibfnamefont {A.}~\bibnamefont
  {Roggero}},\ }\href@noop {} {\enquote {\bibinfo {title} {Dynamical phase
  transitions in models of collective neutrino oscillations},}\ } (\bibinfo
  {year} {2021}{\natexlab{b}}),\ \Eprint {http://arxiv.org/abs/2103.11497}
  {arXiv:2103.11497 [hep-ph]} \BibitemShut {NoStop}%
\bibitem [{\citenamefont {Isgur}\ and\ \citenamefont
  {Wise}(1989)}]{Isgur:1989vq}%
  \BibitemOpen
  \bibfield  {author} {\bibinfo {author} {\bibfnamefont {N.}~\bibnamefont
  {Isgur}}\ and\ \bibinfo {author} {\bibfnamefont {M.~B.}\ \bibnamefont
  {Wise}},\ }\href {\doibase 10.1016/0370-2693(89)90566-2} {\bibfield
  {journal} {\bibinfo  {journal} {Phys. Lett. B}\ }\textbf {\bibinfo {volume}
  {232}},\ \bibinfo {pages} {113} (\bibinfo {year} {1989})}\BibitemShut
  {NoStop}%
\bibitem [{\citenamefont {Collins}\ \emph {et~al.}(1989)\citenamefont
  {Collins}, \citenamefont {Soper},\ and\ \citenamefont
  {Sterman}}]{Collins:1989gx}%
  \BibitemOpen
  \bibfield  {author} {\bibinfo {author} {\bibfnamefont {J.~C.}\ \bibnamefont
  {Collins}}, \bibinfo {author} {\bibfnamefont {D.~E.}\ \bibnamefont {Soper}},
  \ and\ \bibinfo {author} {\bibfnamefont {G.~F.}\ \bibnamefont {Sterman}},\
  }\href {\doibase 10.1142/9789814503266_0001} {\bibfield  {journal} {\bibinfo
  {journal} {Adv. Ser. Direct. High Energy Phys.}\ }\textbf {\bibinfo {volume}
  {5}},\ \bibinfo {pages} {1} (\bibinfo {year} {1989})},\ \Eprint
  {http://arxiv.org/abs/hep-ph/0409313} {arXiv:hep-ph/0409313} \BibitemShut
  {NoStop}%
\bibitem [{\citenamefont {Collins}\ and\ \citenamefont
  {Qiu}(2007)}]{Collins:2007nk}%
  \BibitemOpen
  \bibfield  {author} {\bibinfo {author} {\bibfnamefont {J.}~\bibnamefont
  {Collins}}\ and\ \bibinfo {author} {\bibfnamefont {J.-W.}\ \bibnamefont
  {Qiu}},\ }\href {\doibase 10.1103/PhysRevD.75.114014} {\bibfield  {journal}
  {\bibinfo  {journal} {Phys. Rev. D}\ }\textbf {\bibinfo {volume} {75}},\
  \bibinfo {pages} {114014} (\bibinfo {year} {2007})},\ \Eprint
  {http://arxiv.org/abs/0705.2141} {arXiv:0705.2141 [hep-ph]} \BibitemShut
  {NoStop}%
\bibitem [{\citenamefont {Rogers}\ and\ \citenamefont
  {Mulders}(2010)}]{Rogers:2010dm}%
  \BibitemOpen
  \bibfield  {author} {\bibinfo {author} {\bibfnamefont {T.~C.}\ \bibnamefont
  {Rogers}}\ and\ \bibinfo {author} {\bibfnamefont {P.~J.}\ \bibnamefont
  {Mulders}},\ }\href {\doibase 10.1103/PhysRevD.81.094006} {\bibfield
  {journal} {\bibinfo  {journal} {Phys. Rev. D}\ }\textbf {\bibinfo {volume}
  {81}},\ \bibinfo {pages} {094006} (\bibinfo {year} {2010})},\ \Eprint
  {http://arxiv.org/abs/1001.2977} {arXiv:1001.2977 [hep-ph]} \BibitemShut
  {NoStop}%
\bibitem [{\citenamefont {Cherednikov}\ and\ \citenamefont
  {Stefanis}(2012)}]{Cherednikov:2010tr}%
  \BibitemOpen
  \bibfield  {author} {\bibinfo {author} {\bibfnamefont {I.~O.}\ \bibnamefont
  {Cherednikov}}\ and\ \bibinfo {author} {\bibfnamefont {N.~G.}\ \bibnamefont
  {Stefanis}},\ }\href {\doibase 10.1142/S0217751X1250008X} {\bibfield
  {journal} {\bibinfo  {journal} {Int. J. Mod. Phys. A}\ }\textbf {\bibinfo
  {volume} {27}},\ \bibinfo {pages} {1250008} (\bibinfo {year} {2012})},\
  \Eprint {http://arxiv.org/abs/1010.4463} {arXiv:1010.4463 [hep-ph]}
  \BibitemShut {NoStop}%
\bibitem [{\citenamefont {Buffing}\ and\ \citenamefont
  {Mulders}(2014)}]{Buffing:2013dxa}%
  \BibitemOpen
  \bibfield  {author} {\bibinfo {author} {\bibfnamefont {M.~G.~A.}\
  \bibnamefont {Buffing}}\ and\ \bibinfo {author} {\bibfnamefont {P.~J.}\
  \bibnamefont {Mulders}},\ }\href {\doibase 10.1103/PhysRevLett.112.092002}
  {\bibfield  {journal} {\bibinfo  {journal} {Phys. Rev. Lett.}\ }\textbf
  {\bibinfo {volume} {112}},\ \bibinfo {pages} {092002} (\bibinfo {year}
  {2014})},\ \Eprint {http://arxiv.org/abs/1309.4681} {arXiv:1309.4681
  [hep-ph]} \BibitemShut {NoStop}%
\bibitem [{\citenamefont {Sch\"afer}\ and\ \citenamefont
  {Zhou}(2014)}]{Schafer:2014xpa}%
  \BibitemOpen
  \bibfield  {author} {\bibinfo {author} {\bibfnamefont {A.}~\bibnamefont
  {Sch\"afer}}\ and\ \bibinfo {author} {\bibfnamefont {J.}~\bibnamefont
  {Zhou}},\ }\href {\doibase 10.1103/PhysRevD.90.094012} {\bibfield  {journal}
  {\bibinfo  {journal} {Phys. Rev. D}\ }\textbf {\bibinfo {volume} {90}},\
  \bibinfo {pages} {094012} (\bibinfo {year} {2014})},\ \Eprint
  {http://arxiv.org/abs/1406.3198} {arXiv:1406.3198 [hep-ph]} \BibitemShut
  {NoStop}%
\bibitem [{\citenamefont {Buffing}\ and\ \citenamefont
  {Mulders}(2015)}]{Buffing:2015aha}%
  \BibitemOpen
  \bibfield  {author} {\bibinfo {author} {\bibfnamefont {M.~G.~A.}\
  \bibnamefont {Buffing}}\ and\ \bibinfo {author} {\bibfnamefont {P.~J.}\
  \bibnamefont {Mulders}},\ }\href {\doibase 10.1142/S2010194515600228}
  {\bibfield  {journal} {\bibinfo  {journal} {Int. J. Mod. Phys. Conf. Ser.}\
  }\textbf {\bibinfo {volume} {37}},\ \bibinfo {pages} {1560022} (\bibinfo
  {year} {2015})}\BibitemShut {NoStop}%
\bibitem [{\citenamefont {Adare}\ \emph {et~al.}(2017)\citenamefont {Adare}
  \emph {et~al.}}]{Adare:2016bug}%
  \BibitemOpen
  \bibfield  {author} {\bibinfo {author} {\bibfnamefont {A.}~\bibnamefont
  {Adare}} \emph {et~al.} (\bibinfo {collaboration} {PHENIX}),\ }\href
  {\doibase 10.1103/PhysRevD.95.072002} {\bibfield  {journal} {\bibinfo
  {journal} {Phys. Rev. D}\ }\textbf {\bibinfo {volume} {95}},\ \bibinfo
  {pages} {072002} (\bibinfo {year} {2017})},\ \Eprint
  {http://arxiv.org/abs/1609.04769} {arXiv:1609.04769 [hep-ex]} \BibitemShut
  {NoStop}%
\bibitem [{\citenamefont {Zhou}(2017)}]{Zhou:2017mpw}%
  \BibitemOpen
  \bibfield  {author} {\bibinfo {author} {\bibfnamefont {J.}~\bibnamefont
  {Zhou}},\ }\href {\doibase 10.1103/PhysRevD.96.114001} {\bibfield  {journal}
  {\bibinfo  {journal} {Phys. Rev. D}\ }\textbf {\bibinfo {volume} {96}},\
  \bibinfo {pages} {114001} (\bibinfo {year} {2017})},\ \Eprint
  {http://arxiv.org/abs/1706.02842} {arXiv:1706.02842 [hep-ph]} \BibitemShut
  {NoStop}%
\bibitem [{\citenamefont {Kovner}\ and\ \citenamefont
  {Lublinsky}(2015)}]{Kovner:2015hga}%
  \BibitemOpen
  \bibfield  {author} {\bibinfo {author} {\bibfnamefont {A.}~\bibnamefont
  {Kovner}}\ and\ \bibinfo {author} {\bibfnamefont {M.}~\bibnamefont
  {Lublinsky}},\ }\href {\doibase 10.1103/PhysRevD.92.034016} {\bibfield
  {journal} {\bibinfo  {journal} {Phys. Rev. D}\ }\textbf {\bibinfo {volume}
  {92}},\ \bibinfo {pages} {034016} (\bibinfo {year} {2015})},\ \Eprint
  {http://arxiv.org/abs/1506.05394} {arXiv:1506.05394 [hep-ph]} \BibitemShut
  {NoStop}%
\bibitem [{\citenamefont {Kharzeev}\ and\ \citenamefont
  {Levin}(2017)}]{Kharzeev:2017qzs}%
  \BibitemOpen
  \bibfield  {author} {\bibinfo {author} {\bibfnamefont {D.~E.}\ \bibnamefont
  {Kharzeev}}\ and\ \bibinfo {author} {\bibfnamefont {E.~M.}\ \bibnamefont
  {Levin}},\ }\href {\doibase 10.1103/PhysRevD.95.114008} {\bibfield  {journal}
  {\bibinfo  {journal} {Phys. Rev. D}\ }\textbf {\bibinfo {volume} {95}},\
  \bibinfo {pages} {114008} (\bibinfo {year} {2017})},\ \Eprint
  {http://arxiv.org/abs/1702.03489} {arXiv:1702.03489 [hep-ph]} \BibitemShut
  {NoStop}%
\bibitem [{\citenamefont {Kovner}\ \emph {et~al.}(2019)\citenamefont {Kovner},
  \citenamefont {Lublinsky},\ and\ \citenamefont {Serino}}]{Kovner:2018rbf}%
  \BibitemOpen
  \bibfield  {author} {\bibinfo {author} {\bibfnamefont {A.}~\bibnamefont
  {Kovner}}, \bibinfo {author} {\bibfnamefont {M.}~\bibnamefont {Lublinsky}}, \
  and\ \bibinfo {author} {\bibfnamefont {M.}~\bibnamefont {Serino}},\ }\href
  {\doibase 10.1016/j.physletb.2018.10.043} {\bibfield  {journal} {\bibinfo
  {journal} {Phys. Lett. B}\ }\textbf {\bibinfo {volume} {792}},\ \bibinfo
  {pages} {4} (\bibinfo {year} {2019})},\ \Eprint
  {http://arxiv.org/abs/1806.01089} {arXiv:1806.01089 [hep-ph]} \BibitemShut
  {NoStop}%
\bibitem [{\citenamefont {Peschanski}\ and\ \citenamefont
  {Seki}(2019)}]{Peschanski:2019yah}%
  \BibitemOpen
  \bibfield  {author} {\bibinfo {author} {\bibfnamefont {R.}~\bibnamefont
  {Peschanski}}\ and\ \bibinfo {author} {\bibfnamefont {S.}~\bibnamefont
  {Seki}},\ }\href {\doibase 10.1103/PhysRevD.100.076012} {\bibfield  {journal}
  {\bibinfo  {journal} {Phys. Rev. D}\ }\textbf {\bibinfo {volume} {100}},\
  \bibinfo {pages} {076012} (\bibinfo {year} {2019})},\ \Eprint
  {http://arxiv.org/abs/1906.09696} {arXiv:1906.09696 [hep-th]} \BibitemShut
  {NoStop}%
\bibitem [{\citenamefont {Castorina}\ \emph {et~al.}(2021)\citenamefont
  {Castorina}, \citenamefont {Iorio}, \citenamefont {Lanteri},\ and\
  \citenamefont {Luke\v{s}}}]{Castorina:2020cro}%
  \BibitemOpen
  \bibfield  {author} {\bibinfo {author} {\bibfnamefont {P.}~\bibnamefont
  {Castorina}}, \bibinfo {author} {\bibfnamefont {A.}~\bibnamefont {Iorio}},
  \bibinfo {author} {\bibfnamefont {D.}~\bibnamefont {Lanteri}}, \ and\
  \bibinfo {author} {\bibfnamefont {P.}~\bibnamefont {Luke\v{s}}},\ }\href
  {\doibase 10.1142/S0218301321500105} {\bibfield  {journal} {\bibinfo
  {journal} {Int. J. Mod. Phys. E}\ }\textbf {\bibinfo {volume} {30}},\
  \bibinfo {pages} {2150010} (\bibinfo {year} {2021})},\ \Eprint
  {http://arxiv.org/abs/2003.00112} {arXiv:2003.00112 [hep-ph]} \BibitemShut
  {NoStop}%
\bibitem [{\citenamefont {Ramos}\ and\ \citenamefont
  {Machado}(2020)}]{Ramos:2020kaj}%
  \BibitemOpen
  \bibfield  {author} {\bibinfo {author} {\bibfnamefont {G.~S.}\ \bibnamefont
  {Ramos}}\ and\ \bibinfo {author} {\bibfnamefont {M.~V.~T.}\ \bibnamefont
  {Machado}},\ }\href {\doibase 10.1103/PhysRevD.101.074040} {\bibfield
  {journal} {\bibinfo  {journal} {Phys. Rev. D}\ }\textbf {\bibinfo {volume}
  {101}},\ \bibinfo {pages} {074040} (\bibinfo {year} {2020})},\ \Eprint
  {http://arxiv.org/abs/2003.05008} {arXiv:2003.05008 [hep-ph]} \BibitemShut
  {NoStop}%
\bibitem [{\citenamefont {Duan}\ \emph {et~al.}(2020)\citenamefont {Duan},
  \citenamefont {Akkaya}, \citenamefont {Kovner},\ and\ \citenamefont
  {Skokov}}]{Duan:2020jkz}%
  \BibitemOpen
  \bibfield  {author} {\bibinfo {author} {\bibfnamefont {H.}~\bibnamefont
  {Duan}}, \bibinfo {author} {\bibfnamefont {C.}~\bibnamefont {Akkaya}},
  \bibinfo {author} {\bibfnamefont {A.}~\bibnamefont {Kovner}}, \ and\ \bibinfo
  {author} {\bibfnamefont {V.~V.}\ \bibnamefont {Skokov}},\ }\href {\doibase
  10.1103/PhysRevD.101.036017} {\bibfield  {journal} {\bibinfo  {journal}
  {Phys. Rev. D}\ }\textbf {\bibinfo {volume} {101}},\ \bibinfo {pages}
  {036017} (\bibinfo {year} {2020})},\ \Eprint
  {http://arxiv.org/abs/2001.01726} {arXiv:2001.01726 [hep-ph]} \BibitemShut
  {NoStop}%
\bibitem [{\citenamefont {Gotsman}\ and\ \citenamefont
  {Levin}(2020)}]{Gotsman:2020bjc}%
  \BibitemOpen
  \bibfield  {author} {\bibinfo {author} {\bibfnamefont {E.}~\bibnamefont
  {Gotsman}}\ and\ \bibinfo {author} {\bibfnamefont {E.}~\bibnamefont
  {Levin}},\ }\href {\doibase 10.1103/PhysRevD.102.074008} {\bibfield
  {journal} {\bibinfo  {journal} {Phys. Rev. D}\ }\textbf {\bibinfo {volume}
  {102}},\ \bibinfo {pages} {074008} (\bibinfo {year} {2020})},\ \Eprint
  {http://arxiv.org/abs/2006.11793} {arXiv:2006.11793 [hep-ph]} \BibitemShut
  {NoStop}%
\bibitem [{\citenamefont {Andreev}\ \emph {et~al.}(2021)\citenamefont {Andreev}
  \emph {et~al.}}]{H1:2020zpd}%
  \BibitemOpen
  \bibfield  {author} {\bibinfo {author} {\bibfnamefont {V.}~\bibnamefont
  {Andreev}} \emph {et~al.} (\bibinfo {collaboration} {H1}),\ }\href {\doibase
  10.1140/epjc/s10052-021-08896-1} {\bibfield  {journal} {\bibinfo  {journal}
  {Eur. Phys. J. C}\ }\textbf {\bibinfo {volume} {81}},\ \bibinfo {pages} {212}
  (\bibinfo {year} {2021})},\ \Eprint {http://arxiv.org/abs/2011.01812}
  {arXiv:2011.01812 [hep-ex]} \BibitemShut {NoStop}%
\bibitem [{\citenamefont {Baty}\ \emph {et~al.}(2021)\citenamefont {Baty},
  \citenamefont {Gardner},\ and\ \citenamefont {Li}}]{Baty:2021ugw}%
  \BibitemOpen
  \bibfield  {author} {\bibinfo {author} {\bibfnamefont {A.}~\bibnamefont
  {Baty}}, \bibinfo {author} {\bibfnamefont {P.}~\bibnamefont {Gardner}}, \
  and\ \bibinfo {author} {\bibfnamefont {W.}~\bibnamefont {Li}},\ }\href@noop
  {} {\  (\bibinfo {year} {2021})},\ \Eprint {http://arxiv.org/abs/2104.11735}
  {arXiv:2104.11735 [hep-ph]} \BibitemShut {NoStop}%
\bibitem [{\citenamefont {Deutsch}(1991)}]{PhysRevA.43.2046}%
  \BibitemOpen
  \bibfield  {author} {\bibinfo {author} {\bibfnamefont {J.~M.}\ \bibnamefont
  {Deutsch}},\ }\href {\doibase 10.1103/PhysRevA.43.2046} {\bibfield  {journal}
  {\bibinfo  {journal} {Phys. Rev. A}\ }\textbf {\bibinfo {volume} {43}},\
  \bibinfo {pages} {2046} (\bibinfo {year} {1991})}\BibitemShut {NoStop}%
\bibitem [{\citenamefont {Srednicki}(1994)}]{Srednicki_1994}%
  \BibitemOpen
  \bibfield  {author} {\bibinfo {author} {\bibfnamefont {M.}~\bibnamefont
  {Srednicki}},\ }\href {\doibase 10.1103/physreve.50.888} {\bibfield
  {journal} {\bibinfo  {journal} {Physical Review E}\ }\textbf {\bibinfo
  {volume} {50}},\ \bibinfo {pages} {888–901} (\bibinfo {year}
  {1994})}\BibitemShut {NoStop}%
\bibitem [{\citenamefont {Srednicki}(1999)}]{Srednicki_1999}%
  \BibitemOpen
  \bibfield  {author} {\bibinfo {author} {\bibfnamefont {M.}~\bibnamefont
  {Srednicki}},\ }\href {\doibase 10.1088/0305-4470/32/7/007} {\bibfield
  {journal} {\bibinfo  {journal} {Journal of Physics A: Mathematical and
  General}\ }\textbf {\bibinfo {volume} {32}},\ \bibinfo {pages} {1163–1175}
  (\bibinfo {year} {1999})}\BibitemShut {NoStop}%
\bibitem [{\citenamefont {Eisert}\ \emph {et~al.}(2015)\citenamefont {Eisert},
  \citenamefont {Friesdorf},\ and\ \citenamefont {Gogolin}}]{Eisert_2015}%
  \BibitemOpen
  \bibfield  {author} {\bibinfo {author} {\bibfnamefont {J.}~\bibnamefont
  {Eisert}}, \bibinfo {author} {\bibfnamefont {M.}~\bibnamefont {Friesdorf}}, \
  and\ \bibinfo {author} {\bibfnamefont {C.}~\bibnamefont {Gogolin}},\ }\href
  {\doibase 10.1038/nphys3215} {\bibfield  {journal} {\bibinfo  {journal}
  {Nature Physics}\ }\textbf {\bibinfo {volume} {11}},\ \bibinfo {pages}
  {124–130} (\bibinfo {year} {2015})}\BibitemShut {NoStop}%
\bibitem [{\citenamefont {D’Alessio}\ \emph {et~al.}(2016)\citenamefont
  {D’Alessio}, \citenamefont {Kafri}, \citenamefont {Polkovnikov},\ and\
  \citenamefont {Rigol}}]{D_Alessio_2016}%
  \BibitemOpen
  \bibfield  {author} {\bibinfo {author} {\bibfnamefont {L.}~\bibnamefont
  {D’Alessio}}, \bibinfo {author} {\bibfnamefont {Y.}~\bibnamefont {Kafri}},
  \bibinfo {author} {\bibfnamefont {A.}~\bibnamefont {Polkovnikov}}, \ and\
  \bibinfo {author} {\bibfnamefont {M.}~\bibnamefont {Rigol}},\ }\href
  {\doibase 10.1080/00018732.2016.1198134} {\bibfield  {journal} {\bibinfo
  {journal} {Advances in Physics}\ }\textbf {\bibinfo {volume} {65}},\ \bibinfo
  {pages} {239–362} (\bibinfo {year} {2016})}\BibitemShut {NoStop}%
\bibitem [{\citenamefont {Anza}\ and\ \citenamefont {Vedral}(2017)}]{Anza2017}%
  \BibitemOpen
  \bibfield  {author} {\bibinfo {author} {\bibfnamefont {F.}~\bibnamefont
  {Anza}}\ and\ \bibinfo {author} {\bibfnamefont {V.}~\bibnamefont {Vedral}},\
  }\href {\doibase 10.1038/srep44066} {\bibfield  {journal} {\bibinfo
  {journal} {Scientific Reports}\ }\textbf {\bibinfo {volume} {7}},\ \bibinfo
  {pages} {44066} (\bibinfo {year} {2017})}\BibitemShut {NoStop}%
\bibitem [{\citenamefont {Anza}\ \emph {et~al.}(2018)\citenamefont {Anza},
  \citenamefont {Gogolin},\ and\ \citenamefont {Huber}}]{Anza_2018}%
  \BibitemOpen
  \bibfield  {author} {\bibinfo {author} {\bibfnamefont {F.}~\bibnamefont
  {Anza}}, \bibinfo {author} {\bibfnamefont {C.}~\bibnamefont {Gogolin}}, \
  and\ \bibinfo {author} {\bibfnamefont {M.}~\bibnamefont {Huber}},\ }\href
  {\doibase 10.1103/physrevlett.120.150603} {\bibfield  {journal} {\bibinfo
  {journal} {Physical Review Letters}\ }\textbf {\bibinfo {volume} {120}}
  (\bibinfo {year} {2018}),\ 10.1103/physrevlett.120.150603}\BibitemShut
  {NoStop}%
\bibitem [{\citenamefont {{Kaufman}}\ \emph {et~al.}(2016)\citenamefont
  {{Kaufman}}, \citenamefont {{Tai}}, \citenamefont {{Lukin}}, \citenamefont
  {{Rispoli}}, \citenamefont {{Schittko}}, \citenamefont {{Preiss}},\ and\
  \citenamefont {{Greiner}}}]{2016Sci...353..794K}%
  \BibitemOpen
  \bibfield  {author} {\bibinfo {author} {\bibfnamefont {A.~M.}\ \bibnamefont
  {{Kaufman}}}, \bibinfo {author} {\bibfnamefont {M.~E.}\ \bibnamefont
  {{Tai}}}, \bibinfo {author} {\bibfnamefont {A.}~\bibnamefont {{Lukin}}},
  \bibinfo {author} {\bibfnamefont {M.}~\bibnamefont {{Rispoli}}}, \bibinfo
  {author} {\bibfnamefont {R.}~\bibnamefont {{Schittko}}}, \bibinfo {author}
  {\bibfnamefont {P.~M.}\ \bibnamefont {{Preiss}}}, \ and\ \bibinfo {author}
  {\bibfnamefont {M.}~\bibnamefont {{Greiner}}},\ }\href {\doibase
  10.1126/science.aaf6725} {\bibfield  {journal} {\bibinfo  {journal}
  {Science}\ }\textbf {\bibinfo {volume} {353}},\ \bibinfo {pages} {794}
  (\bibinfo {year} {2016})},\ \Eprint {http://arxiv.org/abs/1603.04409}
  {arXiv:1603.04409 [quant-ph]} \BibitemShut {NoStop}%
\bibitem [{\citenamefont {Bell}\ \emph {et~al.}(2003)\citenamefont {Bell},
  \citenamefont {Rawlinson},\ and\ \citenamefont {Sawyer}}]{Bell:2003mg}%
  \BibitemOpen
  \bibfield  {author} {\bibinfo {author} {\bibfnamefont {N.~F.}\ \bibnamefont
  {Bell}}, \bibinfo {author} {\bibfnamefont {A.~A.}\ \bibnamefont {Rawlinson}},
  \ and\ \bibinfo {author} {\bibfnamefont {R.~F.}\ \bibnamefont {Sawyer}},\
  }\href {\doibase 10.1016/j.physletb.2003.08.035} {\bibfield  {journal}
  {\bibinfo  {journal} {Phys. Lett. B}\ }\textbf {\bibinfo {volume} {573}},\
  \bibinfo {pages} {86} (\bibinfo {year} {2003})},\ \Eprint
  {http://arxiv.org/abs/hep-ph/0304082} {arXiv:hep-ph/0304082} \BibitemShut
  {NoStop}%
\bibitem [{\citenamefont {{Giovannetti}}\ \emph
  {et~al.}(2003{\natexlab{a}})\citenamefont {{Giovannetti}}, \citenamefont
  {{Lloyd}},\ and\ \citenamefont {{Maccone}}}]{2003EL.....62..615G}%
  \BibitemOpen
  \bibfield  {author} {\bibinfo {author} {\bibfnamefont {V.}~\bibnamefont
  {{Giovannetti}}}, \bibinfo {author} {\bibfnamefont {S.}~\bibnamefont
  {{Lloyd}}}, \ and\ \bibinfo {author} {\bibfnamefont {L.}~\bibnamefont
  {{Maccone}}},\ }\href {\doibase 10.1209/epl/i2003-00418-8} {\bibfield
  {journal} {\bibinfo  {journal} {EPL (Europhysics Letters)}\ }\textbf
  {\bibinfo {volume} {62}},\ \bibinfo {pages} {615} (\bibinfo {year}
  {2003}{\natexlab{a}})},\ \Eprint {http://arxiv.org/abs/quant-ph/0206001}
  {arXiv:quant-ph/0206001 [quant-ph]} \BibitemShut {NoStop}%
\bibitem [{\citenamefont {{Giovannetti}}\ \emph
  {et~al.}(2003{\natexlab{b}})\citenamefont {{Giovannetti}}, \citenamefont
  {{Lloyd}},\ and\ \citenamefont {{Maccone}}}]{2003PhRvA..67e2109G}%
  \BibitemOpen
  \bibfield  {author} {\bibinfo {author} {\bibfnamefont {V.}~\bibnamefont
  {{Giovannetti}}}, \bibinfo {author} {\bibfnamefont {S.}~\bibnamefont
  {{Lloyd}}}, \ and\ \bibinfo {author} {\bibfnamefont {L.}~\bibnamefont
  {{Maccone}}},\ }\href {\doibase 10.1103/PhysRevA.67.052109} {\bibfield
  {journal} {\bibinfo  {journal} {Phys. Rev. A}\ }\textbf {\bibinfo {volume}
  {67}},\ \bibinfo {eid} {052109} (\bibinfo {year} {2003}{\natexlab{b}})},\
  \Eprint {http://arxiv.org/abs/quant-ph/0210197} {arXiv:quant-ph/0210197
  [quant-ph]} \BibitemShut {NoStop}%
\bibitem [{\citenamefont {{Giovannetti}}\ \emph
  {et~al.}(2003{\natexlab{c}})\citenamefont {{Giovannetti}}, \citenamefont
  {{Lloyd}},\ and\ \citenamefont {{Maccone}}}]{2003SPIE.5111....1G}%
  \BibitemOpen
  \bibfield  {author} {\bibinfo {author} {\bibfnamefont {V.}~\bibnamefont
  {{Giovannetti}}}, \bibinfo {author} {\bibfnamefont {S.}~\bibnamefont
  {{Lloyd}}}, \ and\ \bibinfo {author} {\bibfnamefont {L.}~\bibnamefont
  {{Maccone}}},\ }in\ \href {\doibase 10.1117/12.507486} {\emph {\bibinfo
  {booktitle} {Fluctuations and Noise in Photonics and Quantum Optics}}},\
  \bibinfo {series} {Society of Photo-Optical Instrumentation Engineers (SPIE)
  Conference Series}, Vol.\ \bibinfo {volume} {5111},\ \bibinfo {editor}
  {edited by\ \bibinfo {editor} {\bibfnamefont {D.}~\bibnamefont {{Abbott}}},
  \bibinfo {editor} {\bibfnamefont {J.~H.}\ \bibnamefont {{Shapiro}}}, \ and\
  \bibinfo {editor} {\bibfnamefont {Y.}~\bibnamefont {{Yamamoto}}}}\ (\bibinfo
  {year} {2003})\ pp.\ \bibinfo {pages} {1--6},\ \Eprint
  {http://arxiv.org/abs/quant-ph/0303085} {arXiv:quant-ph/0303085 [quant-ph]}
  \BibitemShut {NoStop}%
\bibitem [{\citenamefont {Sawyer}(2004)}]{PhysRevA.70.022308}%
  \BibitemOpen
  \bibfield  {author} {\bibinfo {author} {\bibfnamefont {R.~F.}\ \bibnamefont
  {Sawyer}},\ }\href {\doibase 10.1103/PhysRevA.70.022308} {\bibfield
  {journal} {\bibinfo  {journal} {Phys. Rev. A}\ }\textbf {\bibinfo {volume}
  {70}},\ \bibinfo {pages} {022308} (\bibinfo {year} {2004})}\BibitemShut
  {NoStop}%
\bibitem [{\citenamefont {Muller}\ and\ \citenamefont
  {Schafer}(2011)}]{Muller:2011ra}%
  \BibitemOpen
  \bibfield  {author} {\bibinfo {author} {\bibfnamefont {B.}~\bibnamefont
  {Muller}}\ and\ \bibinfo {author} {\bibfnamefont {A.}~\bibnamefont
  {Schafer}},\ }\href {\doibase 10.1142/S0218301311020459} {\bibfield
  {journal} {\bibinfo  {journal} {Int. J. Mod. Phys. E}\ }\textbf {\bibinfo
  {volume} {20}},\ \bibinfo {pages} {2235} (\bibinfo {year} {2011})},\ \Eprint
  {http://arxiv.org/abs/1110.2378} {arXiv:1110.2378 [hep-ph]} \BibitemShut
  {NoStop}%
\bibitem [{\citenamefont {Akkelin}\ and\ \citenamefont
  {Sinyukov}(2014)}]{Akkelin:2013jsa}%
  \BibitemOpen
  \bibfield  {author} {\bibinfo {author} {\bibfnamefont {S.~V.}\ \bibnamefont
  {Akkelin}}\ and\ \bibinfo {author} {\bibfnamefont {Y.~M.}\ \bibnamefont
  {Sinyukov}},\ }\href {\doibase 10.1103/PhysRevC.89.034910} {\bibfield
  {journal} {\bibinfo  {journal} {Phys. Rev. C}\ }\textbf {\bibinfo {volume}
  {89}},\ \bibinfo {pages} {034910} (\bibinfo {year} {2014})},\ \Eprint
  {http://arxiv.org/abs/1309.4388} {arXiv:1309.4388 [nucl-th]} \BibitemShut
  {NoStop}%
\bibitem [{\citenamefont {Berges}\ \emph {et~al.}(2018)\citenamefont {Berges},
  \citenamefont {Floerchinger},\ and\ \citenamefont
  {Venugopalan}}]{Berges:2017hne}%
  \BibitemOpen
  \bibfield  {author} {\bibinfo {author} {\bibfnamefont {J.}~\bibnamefont
  {Berges}}, \bibinfo {author} {\bibfnamefont {S.}~\bibnamefont
  {Floerchinger}}, \ and\ \bibinfo {author} {\bibfnamefont {R.}~\bibnamefont
  {Venugopalan}},\ }\href {\doibase 10.1007/JHEP04(2018)145} {\bibfield
  {journal} {\bibinfo  {journal} {JHEP}\ }\textbf {\bibinfo {volume} {04}},\
  \bibinfo {pages} {145} (\bibinfo {year} {2018})},\ \Eprint
  {http://arxiv.org/abs/1712.09362} {arXiv:1712.09362 [hep-th]} \BibitemShut
  {NoStop}%
\bibitem [{\citenamefont {M\"uller}\ and\ \citenamefont
  {Sch\"afer}(2017)}]{Muller:2017vnp}%
  \BibitemOpen
  \bibfield  {author} {\bibinfo {author} {\bibfnamefont {B.}~\bibnamefont
  {M\"uller}}\ and\ \bibinfo {author} {\bibfnamefont {A.}~\bibnamefont
  {Sch\"afer}},\ }\href@noop {} {\  (\bibinfo {year} {2017})},\ \Eprint
  {http://arxiv.org/abs/1712.03567} {arXiv:1712.03567 [nucl-th]} \BibitemShut
  {NoStop}%
\bibitem [{\citenamefont {Berges}\ \emph {et~al.}(2019)\citenamefont {Berges},
  \citenamefont {Floerchinger},\ and\ \citenamefont
  {Venugopalan}}]{Berges:2018cny}%
  \BibitemOpen
  \bibfield  {author} {\bibinfo {author} {\bibfnamefont {J.}~\bibnamefont
  {Berges}}, \bibinfo {author} {\bibfnamefont {S.}~\bibnamefont
  {Floerchinger}}, \ and\ \bibinfo {author} {\bibfnamefont {R.}~\bibnamefont
  {Venugopalan}},\ }\href {\doibase 10.1016/j.nuclphysa.2018.12.008} {\bibfield
   {journal} {\bibinfo  {journal} {Nucl. Phys. A}\ }\textbf {\bibinfo {volume}
  {982}},\ \bibinfo {pages} {819} (\bibinfo {year} {2019})},\ \Eprint
  {http://arxiv.org/abs/1812.08120} {arXiv:1812.08120 [hep-th]} \BibitemShut
  {NoStop}%
\bibitem [{\citenamefont {Feal}\ \emph {et~al.}(2019)\citenamefont {Feal},
  \citenamefont {Pajares},\ and\ \citenamefont {Vazquez}}]{Feal:2018ptp}%
  \BibitemOpen
  \bibfield  {author} {\bibinfo {author} {\bibfnamefont {X.}~\bibnamefont
  {Feal}}, \bibinfo {author} {\bibfnamefont {C.}~\bibnamefont {Pajares}}, \
  and\ \bibinfo {author} {\bibfnamefont {R.~A.}\ \bibnamefont {Vazquez}},\
  }\href {\doibase 10.1103/PhysRevC.99.015205} {\bibfield  {journal} {\bibinfo
  {journal} {Phys. Rev. C}\ }\textbf {\bibinfo {volume} {99}},\ \bibinfo
  {pages} {015205} (\bibinfo {year} {2019})},\ \Eprint
  {http://arxiv.org/abs/1805.12444} {arXiv:1805.12444 [hep-ph]} \BibitemShut
  {NoStop}%
\bibitem [{\citenamefont {Iskander}\ \emph {et~al.}(2020)\citenamefont
  {Iskander}, \citenamefont {Pan}, \citenamefont {Tyler}, \citenamefont
  {Weber},\ and\ \citenamefont {Baker}}]{Iskander:2020rkb}%
  \BibitemOpen
  \bibfield  {author} {\bibinfo {author} {\bibfnamefont {G.}~\bibnamefont
  {Iskander}}, \bibinfo {author} {\bibfnamefont {J.}~\bibnamefont {Pan}},
  \bibinfo {author} {\bibfnamefont {M.}~\bibnamefont {Tyler}}, \bibinfo
  {author} {\bibfnamefont {C.}~\bibnamefont {Weber}}, \ and\ \bibinfo {author}
  {\bibfnamefont {O.~K.}\ \bibnamefont {Baker}},\ }\href {\doibase
  10.1016/j.physletb.2020.135948} {\bibfield  {journal} {\bibinfo  {journal}
  {Phys. Lett. B}\ }\textbf {\bibinfo {volume} {811}},\ \bibinfo {pages}
  {135948} (\bibinfo {year} {2020})},\ \Eprint
  {http://arxiv.org/abs/2010.00709} {arXiv:2010.00709 [hep-ph]} \BibitemShut
  {NoStop}%
\bibitem [{\citenamefont {Hastings}(2007)}]{Hastings_2007}%
  \BibitemOpen
  \bibfield  {author} {\bibinfo {author} {\bibfnamefont {M.~B.}\ \bibnamefont
  {Hastings}},\ }\href {\doibase 10.1088/1742-5468/2007/08/p08024} {\bibfield
  {journal} {\bibinfo  {journal} {Journal of Statistical Mechanics: Theory and
  Experiment}\ }\textbf {\bibinfo {volume} {2007}},\ \bibinfo {pages} {P08024}
  (\bibinfo {year} {2007})}\BibitemShut {NoStop}%
\bibitem [{\citenamefont {Eisert}\ \emph {et~al.}(2010)\citenamefont {Eisert},
  \citenamefont {Cramer},\ and\ \citenamefont {Plenio}}]{Eisert_2010}%
  \BibitemOpen
  \bibfield  {author} {\bibinfo {author} {\bibfnamefont {J.}~\bibnamefont
  {Eisert}}, \bibinfo {author} {\bibfnamefont {M.}~\bibnamefont {Cramer}}, \
  and\ \bibinfo {author} {\bibfnamefont {M.~B.}\ \bibnamefont {Plenio}},\
  }\href {\doibase 10.1103/revmodphys.82.277} {\bibfield  {journal} {\bibinfo
  {journal} {Reviews of Modern Physics}\ }\textbf {\bibinfo {volume} {82}},\
  \bibinfo {pages} {277–306} (\bibinfo {year} {2010})}\BibitemShut {NoStop}%
\bibitem [{\citenamefont {Dowling}\ and\ \citenamefont
  {Nielsen}(2008)}]{10.5555/2016985.2016986}%
  \BibitemOpen
  \bibfield  {author} {\bibinfo {author} {\bibfnamefont {M.~R.}\ \bibnamefont
  {Dowling}}\ and\ \bibinfo {author} {\bibfnamefont {M.~A.}\ \bibnamefont
  {Nielsen}},\ }\href@noop {} {\bibfield  {journal} {\bibinfo  {journal}
  {Quantum Information and Computation}\ }\textbf {\bibinfo {volume} {8}},\
  \bibinfo {pages} {861–899} (\bibinfo {year} {2008})},\ \Eprint
  {http://arxiv.org/abs/quant-ph/0701004} {arXiv:quant-ph/0701004 [quant-ph]}
  \BibitemShut {NoStop}%
\bibitem [{\citenamefont {T\'oth}\ \emph {et~al.}(2007)\citenamefont {T\'oth},
  \citenamefont {Knapp}, \citenamefont {G\"uhne},\ and\ \citenamefont
  {Briegel}}]{PhysRevLett.99.250405}%
  \BibitemOpen
  \bibfield  {author} {\bibinfo {author} {\bibfnamefont {G.}~\bibnamefont
  {T\'oth}}, \bibinfo {author} {\bibfnamefont {C.}~\bibnamefont {Knapp}},
  \bibinfo {author} {\bibfnamefont {O.}~\bibnamefont {G\"uhne}}, \ and\
  \bibinfo {author} {\bibfnamefont {H.~J.}\ \bibnamefont {Briegel}},\ }\href
  {\doibase 10.1103/PhysRevLett.99.250405} {\bibfield  {journal} {\bibinfo
  {journal} {Phys. Rev. Lett.}\ }\textbf {\bibinfo {volume} {99}},\ \bibinfo
  {pages} {250405} (\bibinfo {year} {2007})}\BibitemShut {NoStop}%
\bibitem [{\citenamefont {Walter}\ \emph {et~al.}(2013)\citenamefont {Walter},
  \citenamefont {Doran}, \citenamefont {Gross},\ and\ \citenamefont
  {Christandl}}]{Walter1205}%
  \BibitemOpen
  \bibfield  {author} {\bibinfo {author} {\bibfnamefont {M.}~\bibnamefont
  {Walter}}, \bibinfo {author} {\bibfnamefont {B.}~\bibnamefont {Doran}},
  \bibinfo {author} {\bibfnamefont {D.}~\bibnamefont {Gross}}, \ and\ \bibinfo
  {author} {\bibfnamefont {M.}~\bibnamefont {Christandl}},\ }\href {\doibase
  10.1126/science.1232957} {\bibfield  {journal} {\bibinfo  {journal}
  {Science}\ }\textbf {\bibinfo {volume} {340}},\ \bibinfo {pages} {1205}
  (\bibinfo {year} {2013})},\ \Eprint
  {http://arxiv.org/abs/https://science.sciencemag.org/content/340/6137/1205.full.pdf}
  {https://science.sciencemag.org/content/340/6137/1205.full.pdf} \BibitemShut
  {NoStop}%
\bibitem [{\citenamefont {Swingle}(2018)}]{SwingleSpacetimeEntanglement}%
  \BibitemOpen
  \bibfield  {author} {\bibinfo {author} {\bibfnamefont {B.}~\bibnamefont
  {Swingle}},\ }\href {\doibase 10.1146/annurev-conmatphys-033117-054219}
  {\bibfield  {journal} {\bibinfo  {journal} {Annual Review of Condensed Matter
  Physics}\ }\textbf {\bibinfo {volume} {9}},\ \bibinfo {pages} {345} (\bibinfo
  {year} {2018})},\ \Eprint
  {http://arxiv.org/abs/https://doi.org/10.1146/annurev-conmatphys-033117-054219}
  {https://doi.org/10.1146/annurev-conmatphys-033117-054219} \BibitemShut
  {NoStop}%
\bibitem [{\citenamefont {'t~Hooft}(1993)}]{tHooft:1993dmi}%
  \BibitemOpen
  \bibfield  {author} {\bibinfo {author} {\bibfnamefont {G.}~\bibnamefont
  {'t~Hooft}},\ }\bibfield  {booktitle} {\emph {\bibinfo {booktitle}
  {Salamfestschrift. A collection of talks from the conference on highlights of
  particle and condensed matter physics}},\ }\href
  {https://arxiv.org/abs/gr-qc/9310026} {\bibfield  {journal} {\bibinfo
  {journal} {Conf. Proc. C}\ }\textbf {\bibinfo {volume} {930308}},\ \bibinfo
  {pages} {284} (\bibinfo {year} {1993})},\ \Eprint
  {http://arxiv.org/abs/gr-qc/9310026} {arXiv:gr-qc/9310026} \BibitemShut
  {NoStop}%
\bibitem [{\citenamefont {Susskind}(1995)}]{Susskind:1994vu}%
  \BibitemOpen
  \bibfield  {author} {\bibinfo {author} {\bibfnamefont {L.}~\bibnamefont
  {Susskind}},\ }\href {\doibase 10.1063/1.531249} {\bibfield  {journal}
  {\bibinfo  {journal} {J. Math. Phys.}\ }\textbf {\bibinfo {volume} {36}},\
  \bibinfo {pages} {6377} (\bibinfo {year} {1995})},\ \Eprint
  {http://arxiv.org/abs/hep-th/9409089} {arXiv:hep-th/9409089} \BibitemShut
  {NoStop}%
\bibitem [{\citenamefont {Bekenstein}(1973)}]{Bekenstein:1973ur}%
  \BibitemOpen
  \bibfield  {author} {\bibinfo {author} {\bibfnamefont {J.~D.}\ \bibnamefont
  {Bekenstein}},\ }\href {\doibase 10.1103/PhysRevD.7.2333} {\bibfield
  {journal} {\bibinfo  {journal} {Phys. Rev. D}\ }\textbf {\bibinfo {volume}
  {7}},\ \bibinfo {pages} {2333} (\bibinfo {year} {1973})}\BibitemShut
  {NoStop}%
\bibitem [{\citenamefont {Hawking}(1975)}]{Hawking:1974sw}%
  \BibitemOpen
  \bibfield  {author} {\bibinfo {author} {\bibfnamefont {S.~W.}\ \bibnamefont
  {Hawking}},\ }\href {\doibase 10.1007/BF02345020} {\bibfield  {journal}
  {\bibinfo  {journal} {Commun. Math. Phys.}\ }\textbf {\bibinfo {volume}
  {43}},\ \bibinfo {pages} {199} (\bibinfo {year} {1975})},\ \bibinfo {note}
  {[Erratum: Commun.Math.Phys. 46, 206 (1976)]}\BibitemShut {NoStop}%
\bibitem [{\citenamefont {Balasubramanian}\ \emph {et~al.}(2012)\citenamefont
  {Balasubramanian}, \citenamefont {McDermott},\ and\ \citenamefont
  {Van~Raamsdonk}}]{Balasubramanian:2011wt}%
  \BibitemOpen
  \bibfield  {author} {\bibinfo {author} {\bibfnamefont {V.}~\bibnamefont
  {Balasubramanian}}, \bibinfo {author} {\bibfnamefont {M.~B.}\ \bibnamefont
  {McDermott}}, \ and\ \bibinfo {author} {\bibfnamefont {M.}~\bibnamefont
  {Van~Raamsdonk}},\ }\href {\doibase 10.1103/PhysRevD.86.045014} {\bibfield
  {journal} {\bibinfo  {journal} {Phys. Rev. D}\ }\textbf {\bibinfo {volume}
  {86}},\ \bibinfo {pages} {045014} (\bibinfo {year} {2012})},\ \Eprint
  {http://arxiv.org/abs/1108.3568} {arXiv:1108.3568 [hep-th]} \BibitemShut
  {NoStop}%
\bibitem [{\citenamefont {Jordan}\ \emph {et~al.}(2014)\citenamefont {Jordan},
  \citenamefont {Lee},\ and\ \citenamefont
  {Preskill}}]{DBLP:journals/qic/JordanLP14}%
  \BibitemOpen
  \bibfield  {author} {\bibinfo {author} {\bibfnamefont {S.~P.}\ \bibnamefont
  {Jordan}}, \bibinfo {author} {\bibfnamefont {K.~S.~M.}\ \bibnamefont {Lee}},
  \ and\ \bibinfo {author} {\bibfnamefont {J.}~\bibnamefont {Preskill}},\
  }\href {\doibase 10.26421/QIC14.11-12-8} {\bibfield  {journal} {\bibinfo
  {journal} {Quantum Inf. Comput.}\ }\textbf {\bibinfo {volume} {14}},\
  \bibinfo {pages} {1014} (\bibinfo {year} {2014})}\BibitemShut {NoStop}%
\bibitem [{\citenamefont {Jordan}\ \emph {et~al.}(2012)\citenamefont {Jordan},
  \citenamefont {Lee},\ and\ \citenamefont {Preskill}}]{Jordan1130}%
  \BibitemOpen
  \bibfield  {author} {\bibinfo {author} {\bibfnamefont {S.~P.}\ \bibnamefont
  {Jordan}}, \bibinfo {author} {\bibfnamefont {K.~S.~M.}\ \bibnamefont {Lee}},
  \ and\ \bibinfo {author} {\bibfnamefont {J.}~\bibnamefont {Preskill}},\
  }\href {\doibase 10.1126/science.1217069} {\bibfield  {journal} {\bibinfo
  {journal} {Science}\ }\textbf {\bibinfo {volume} {336}},\ \bibinfo {pages}
  {1130} (\bibinfo {year} {2012})}\BibitemShut {NoStop}%
\bibitem [{\citenamefont {Somma}(2016)}]{somma2016quantum}%
  \BibitemOpen
  \bibfield  {author} {\bibinfo {author} {\bibfnamefont {R.~D.}\ \bibnamefont
  {Somma}},\ }\href@noop {} {\enquote {\bibinfo {title} {Quantum simulations of
  one dimensional quantum systems},}\ } (\bibinfo {year} {2016}),\ \Eprint
  {http://arxiv.org/abs/1503.06319} {arXiv:1503.06319 [quant-ph]} \BibitemShut
  {NoStop}%
\bibitem [{\citenamefont {Macridin}\ \emph
  {et~al.}(2018{\natexlab{a}})\citenamefont {Macridin}, \citenamefont
  {Spentzouris}, \citenamefont {Amundson},\ and\ \citenamefont
  {Harnik}}]{PhysRevLett.121.110504}%
  \BibitemOpen
  \bibfield  {author} {\bibinfo {author} {\bibfnamefont {A.}~\bibnamefont
  {Macridin}}, \bibinfo {author} {\bibfnamefont {P.}~\bibnamefont
  {Spentzouris}}, \bibinfo {author} {\bibfnamefont {J.}~\bibnamefont
  {Amundson}}, \ and\ \bibinfo {author} {\bibfnamefont {R.}~\bibnamefont
  {Harnik}},\ }\href {\doibase 10.1103/PhysRevLett.121.110504} {\bibfield
  {journal} {\bibinfo  {journal} {Phys. Rev. Lett.}\ }\textbf {\bibinfo
  {volume} {121}},\ \bibinfo {pages} {110504} (\bibinfo {year}
  {2018}{\natexlab{a}})}\BibitemShut {NoStop}%
\bibitem [{\citenamefont {Macridin}\ \emph
  {et~al.}(2018{\natexlab{b}})\citenamefont {Macridin}, \citenamefont
  {Spentzouris}, \citenamefont {Amundson},\ and\ \citenamefont
  {Harnik}}]{PhysRevA.98.042312}%
  \BibitemOpen
  \bibfield  {author} {\bibinfo {author} {\bibfnamefont {A.}~\bibnamefont
  {Macridin}}, \bibinfo {author} {\bibfnamefont {P.}~\bibnamefont
  {Spentzouris}}, \bibinfo {author} {\bibfnamefont {J.}~\bibnamefont
  {Amundson}}, \ and\ \bibinfo {author} {\bibfnamefont {R.}~\bibnamefont
  {Harnik}},\ }\href {\doibase 10.1103/PhysRevA.98.042312} {\bibfield
  {journal} {\bibinfo  {journal} {Phys. Rev. A}\ }\textbf {\bibinfo {volume}
  {98}},\ \bibinfo {pages} {042312} (\bibinfo {year}
  {2018}{\natexlab{b}})}\BibitemShut {NoStop}%
\bibitem [{\citenamefont {Klco}\ and\ \citenamefont
  {Savage}(2019)}]{PhysRevA.99.052335}%
  \BibitemOpen
  \bibfield  {author} {\bibinfo {author} {\bibfnamefont {N.}~\bibnamefont
  {Klco}}\ and\ \bibinfo {author} {\bibfnamefont {M.~J.}\ \bibnamefont
  {Savage}},\ }\href {\doibase 10.1103/PhysRevA.99.052335} {\bibfield
  {journal} {\bibinfo  {journal} {Phys. Rev. A}\ }\textbf {\bibinfo {volume}
  {99}},\ \bibinfo {pages} {052335} (\bibinfo {year} {2019})}\BibitemShut
  {NoStop}%
\bibitem [{\citenamefont {Barata}\ \emph {et~al.}(2021)\citenamefont {Barata},
  \citenamefont {Mueller}, \citenamefont {Tarasov},\ and\ \citenamefont
  {Venugopalan}}]{Barata:2020jtq}%
  \BibitemOpen
  \bibfield  {author} {\bibinfo {author} {\bibfnamefont {J.~a.}\ \bibnamefont
  {Barata}}, \bibinfo {author} {\bibfnamefont {N.}~\bibnamefont {Mueller}},
  \bibinfo {author} {\bibfnamefont {A.}~\bibnamefont {Tarasov}}, \ and\
  \bibinfo {author} {\bibfnamefont {R.}~\bibnamefont {Venugopalan}},\ }\href
  {\doibase 10.1103/PhysRevA.103.042410} {\bibfield  {journal} {\bibinfo
  {journal} {Phys. Rev. A}\ }\textbf {\bibinfo {volume} {103}},\ \bibinfo
  {pages} {042410} (\bibinfo {year} {2021})},\ \Eprint
  {http://arxiv.org/abs/2012.00020} {arXiv:2012.00020 [hep-th]} \BibitemShut
  {NoStop}%
\bibitem [{\citenamefont {Klco}\ and\ \citenamefont
  {Savage}(2020{\natexlab{a}})}]{Klco:2019xro}%
  \BibitemOpen
  \bibfield  {author} {\bibinfo {author} {\bibfnamefont {N.}~\bibnamefont
  {Klco}}\ and\ \bibinfo {author} {\bibfnamefont {M.~J.}\ \bibnamefont
  {Savage}},\ }\href {\doibase 10.1103/PhysRevA.102.012612} {\bibfield
  {journal} {\bibinfo  {journal} {Phys. Rev. A}\ }\textbf {\bibinfo {volume}
  {102}},\ \bibinfo {pages} {012612} (\bibinfo {year} {2020}{\natexlab{a}})},\
  \Eprint {http://arxiv.org/abs/1904.10440} {arXiv:1904.10440 [quant-ph]}
  \BibitemShut {NoStop}%
\bibitem [{\citenamefont {Klco}\ and\ \citenamefont
  {Savage}(2020{\natexlab{b}})}]{Klco:2019yrb}%
  \BibitemOpen
  \bibfield  {author} {\bibinfo {author} {\bibfnamefont {N.}~\bibnamefont
  {Klco}}\ and\ \bibinfo {author} {\bibfnamefont {M.~J.}\ \bibnamefont
  {Savage}},\ }\href {\doibase 10.1103/PhysRevA.102.012619} {\bibfield
  {journal} {\bibinfo  {journal} {Phys. Rev. A}\ }\textbf {\bibinfo {volume}
  {102}},\ \bibinfo {pages} {012619} (\bibinfo {year} {2020}{\natexlab{b}})},\
  \Eprint {http://arxiv.org/abs/1912.03577} {arXiv:1912.03577 [quant-ph]}
  \BibitemShut {NoStop}%
\bibitem [{\citenamefont {Klco}\ and\ \citenamefont
  {Savage}(2020{\natexlab{c}})}]{Klco:2020aud}%
  \BibitemOpen
  \bibfield  {author} {\bibinfo {author} {\bibfnamefont {N.}~\bibnamefont
  {Klco}}\ and\ \bibinfo {author} {\bibfnamefont {M.~J.}\ \bibnamefont
  {Savage}},\ }\href {\doibase 10.1103/PhysRevA.102.052422} {\bibfield
  {journal} {\bibinfo  {journal} {Phys. Rev. A}\ }\textbf {\bibinfo {volume}
  {102}},\ \bibinfo {pages} {052422} (\bibinfo {year} {2020}{\natexlab{c}})},\
  \Eprint {http://arxiv.org/abs/2002.02018} {arXiv:2002.02018 [quant-ph]}
  \BibitemShut {NoStop}%
\bibitem [{\citenamefont {Barenco}\ \emph {et~al.}(1996)\citenamefont
  {Barenco}, \citenamefont {Ekert}, \citenamefont {Suominen},\ and\
  \citenamefont {T\"orm\"a}}]{PhysRevA.54.139}%
  \BibitemOpen
  \bibfield  {author} {\bibinfo {author} {\bibfnamefont {A.}~\bibnamefont
  {Barenco}}, \bibinfo {author} {\bibfnamefont {A.}~\bibnamefont {Ekert}},
  \bibinfo {author} {\bibfnamefont {K.-A.}\ \bibnamefont {Suominen}}, \ and\
  \bibinfo {author} {\bibfnamefont {P.}~\bibnamefont {T\"orm\"a}},\ }\href
  {\doibase 10.1103/PhysRevA.54.139} {\bibfield  {journal} {\bibinfo  {journal}
  {Phys. Rev. A}\ }\textbf {\bibinfo {volume} {54}},\ \bibinfo {pages} {139}
  (\bibinfo {year} {1996})}\BibitemShut {NoStop}%
\bibitem [{\citenamefont {Braunstein}\ and\ \citenamefont
  {Pati}(2002)}]{braunstein2000speed}%
  \BibitemOpen
  \bibfield  {author} {\bibinfo {author} {\bibfnamefont {S.~L.}\ \bibnamefont
  {Braunstein}}\ and\ \bibinfo {author} {\bibfnamefont {A.~K.}\ \bibnamefont
  {Pati}},\ }\href {\doibase 10.26421/QIC2.5-6} {\bibfield  {journal} {\bibinfo
   {journal} {Quantum Information and Computation}\ }\textbf {\bibinfo {volume}
  {2}},\ \bibinfo {pages} {399} (\bibinfo {year} {2002})},\ \Eprint
  {http://arxiv.org/abs/0008018} {arXiv:0008018 [hep-ph]} \BibitemShut
  {NoStop}%
\bibitem [{\citenamefont {Jozsa}\ and\ \citenamefont
  {Linden}(2003)}]{Jozsa_2003}%
  \BibitemOpen
  \bibfield  {author} {\bibinfo {author} {\bibfnamefont {R.}~\bibnamefont
  {Jozsa}}\ and\ \bibinfo {author} {\bibfnamefont {N.}~\bibnamefont {Linden}},\
  }\href {\doibase 10.1098/rspa.2002.1097} {\bibfield  {journal} {\bibinfo
  {journal} {Proceedings of the Royal Society of London. Series A:
  Mathematical, Physical and Engineering Sciences}\ }\textbf {\bibinfo {volume}
  {459}},\ \bibinfo {pages} {2011–2032} (\bibinfo {year} {2003})}\BibitemShut
  {NoStop}%
\bibitem [{\citenamefont {Schuch}\ \emph {et~al.}(2008)\citenamefont {Schuch},
  \citenamefont {Wolf}, \citenamefont {Verstraete},\ and\ \citenamefont
  {Cirac}}]{PhysRevLett.100.030504}%
  \BibitemOpen
  \bibfield  {author} {\bibinfo {author} {\bibfnamefont {N.}~\bibnamefont
  {Schuch}}, \bibinfo {author} {\bibfnamefont {M.~M.}\ \bibnamefont {Wolf}},
  \bibinfo {author} {\bibfnamefont {F.}~\bibnamefont {Verstraete}}, \ and\
  \bibinfo {author} {\bibfnamefont {J.~I.}\ \bibnamefont {Cirac}},\ }\href
  {\doibase 10.1103/PhysRevLett.100.030504} {\bibfield  {journal} {\bibinfo
  {journal} {Phys. Rev. Lett.}\ }\textbf {\bibinfo {volume} {100}},\ \bibinfo
  {pages} {030504} (\bibinfo {year} {2008})}\BibitemShut {NoStop}%
\bibitem [{\citenamefont {Gottesman}(1997)}]{Gottesman:1997zz}%
  \BibitemOpen
  \bibfield  {author} {\bibinfo {author} {\bibfnamefont {D.}~\bibnamefont
  {Gottesman}},\ }\href@noop {} {\bibfield  {journal} {\bibinfo  {journal}
  {CalTech Ph.D. Thesis}\ } (\bibinfo {year} {1997})},\ \Eprint
  {http://arxiv.org/abs/quant-ph/9705052} {arXiv:quant-ph/9705052} \BibitemShut
  {NoStop}%
\bibitem [{\citenamefont {Gottesman}(1998)}]{Gottesman:1998hu}%
  \BibitemOpen
  \bibfield  {author} {\bibinfo {author} {\bibfnamefont {D.}~\bibnamefont
  {Gottesman}},\ }in\ \href@noop {} {\emph {\bibinfo {booktitle} {{22nd
  International Colloquium on Group Theoretical Methods in Physics}}}}\
  (\bibinfo {year} {1998})\ \Eprint {http://arxiv.org/abs/quant-ph/9807006}
  {arXiv:quant-ph/9807006} \BibitemShut {NoStop}%
\bibitem [{\citenamefont {Nielsen}\ and\ \citenamefont
  {Chuang}(2010)}]{9781107002173}%
  \BibitemOpen
  \bibfield  {author} {\bibinfo {author} {\bibfnamefont {M.~A.}\ \bibnamefont
  {Nielsen}}\ and\ \bibinfo {author} {\bibfnamefont {I.~L.}\ \bibnamefont
  {Chuang}},\ }\href {\doibase 10.1017/CBO9780511976667} {\emph {\bibinfo
  {title} {Quantum Computation and Quantum Information}}}\ (\bibinfo {year}
  {2010})\BibitemShut {NoStop}%
\bibitem [{\citenamefont {Kitaev}(1997{\natexlab{a}})}]{Kitaev_1997}%
  \BibitemOpen
  \bibfield  {author} {\bibinfo {author} {\bibfnamefont {A.~Y.}\ \bibnamefont
  {Kitaev}},\ }\href {\doibase 10.1070/rm1997v052n06abeh002155} {\bibfield
  {journal} {\bibinfo  {journal} {Russian Mathematical Surveys}\ }\textbf
  {\bibinfo {volume} {52}},\ \bibinfo {pages} {1191} (\bibinfo {year}
  {1997}{\natexlab{a}})}\BibitemShut {NoStop}%
\bibitem [{\citenamefont {{Dawson}}\ and\ \citenamefont
  {{Nielsen}}(2005)}]{2005quant.ph..5030D}%
  \BibitemOpen
  \bibfield  {author} {\bibinfo {author} {\bibfnamefont {C.~M.}\ \bibnamefont
  {{Dawson}}}\ and\ \bibinfo {author} {\bibfnamefont {M.~A.}\ \bibnamefont
  {{Nielsen}}},\ }\href@noop {} {\bibfield  {journal} {\bibinfo  {journal}
  {arXiv e-prints}\ } (\bibinfo {year} {2005})},\ \Eprint
  {http://arxiv.org/abs/quant-ph/0505030} {arXiv:quant-ph/0505030 [quant-ph]}
  \BibitemShut {NoStop}%
\bibitem [{\citenamefont {Aaronson}\ and\ \citenamefont
  {Gottesman}(2004)}]{Aaronson_2004}%
  \BibitemOpen
  \bibfield  {author} {\bibinfo {author} {\bibfnamefont {S.}~\bibnamefont
  {Aaronson}}\ and\ \bibinfo {author} {\bibfnamefont {D.}~\bibnamefont
  {Gottesman}},\ }\href {\doibase 10.1103/physreva.70.052328} {\bibfield
  {journal} {\bibinfo  {journal} {Physical Review A}\ }\textbf {\bibinfo
  {volume} {70}} (\bibinfo {year} {2004}),\
  10.1103/physreva.70.052328}\BibitemShut {NoStop}%
\bibitem [{\citenamefont {Uno}\ and\ \citenamefont
  {Hayakawa}(2008)}]{10.1143/PTP.120.413}%
  \BibitemOpen
  \bibfield  {author} {\bibinfo {author} {\bibfnamefont {S.}~\bibnamefont
  {Uno}}\ and\ \bibinfo {author} {\bibfnamefont {M.}~\bibnamefont {Hayakawa}},\
  }\href {\doibase 10.1143/PTP.120.413} {\bibfield  {journal} {\bibinfo
  {journal} {Progress of Theoretical Physics}\ }\textbf {\bibinfo {volume}
  {120}},\ \bibinfo {pages} {413} (\bibinfo {year} {2008})},\ \Eprint
  {http://arxiv.org/abs/https://academic.oup.com/ptp/article-pdf/120/3/413/5203205/120-3-413.pdf}
  {https://academic.oup.com/ptp/article-pdf/120/3/413/5203205/120-3-413.pdf}
  \BibitemShut {NoStop}%
\bibitem [{\citenamefont {Borsanyi}\ \emph {et~al.}(2015)\citenamefont
  {Borsanyi}, \citenamefont {Durr}, \citenamefont {Fodor}, \citenamefont
  {Hoelbling}, \citenamefont {Katz}, \citenamefont {Krieg}, \citenamefont
  {Lellouch}, \citenamefont {Lippert}, \citenamefont {Portelli}, \citenamefont
  {Szabo},\ and\ \citenamefont {et~al.}}]{Borsanyi_2015}%
  \BibitemOpen
  \bibfield  {author} {\bibinfo {author} {\bibfnamefont {S.}~\bibnamefont
  {Borsanyi}}, \bibinfo {author} {\bibfnamefont {S.}~\bibnamefont {Durr}},
  \bibinfo {author} {\bibfnamefont {Z.}~\bibnamefont {Fodor}}, \bibinfo
  {author} {\bibfnamefont {C.}~\bibnamefont {Hoelbling}}, \bibinfo {author}
  {\bibfnamefont {S.~D.}\ \bibnamefont {Katz}}, \bibinfo {author}
  {\bibfnamefont {S.}~\bibnamefont {Krieg}}, \bibinfo {author} {\bibfnamefont
  {L.}~\bibnamefont {Lellouch}}, \bibinfo {author} {\bibfnamefont
  {T.}~\bibnamefont {Lippert}}, \bibinfo {author} {\bibfnamefont
  {A.}~\bibnamefont {Portelli}}, \bibinfo {author} {\bibfnamefont {K.~K.}\
  \bibnamefont {Szabo}}, \ and\ \bibinfo {author} {\bibnamefont {et~al.}},\
  }\href {\doibase 10.1126/science.1257050} {\bibfield  {journal} {\bibinfo
  {journal} {Science}\ }\textbf {\bibinfo {volume} {347}},\ \bibinfo {pages}
  {1452–1455} (\bibinfo {year} {2015})}\BibitemShut {NoStop}%
\bibitem [{\citenamefont {Fodor}\ \emph {et~al.}(2016)\citenamefont {Fodor},
  \citenamefont {Hoelbling}, \citenamefont {Katz}, \citenamefont {Lellouch},
  \citenamefont {Portelli}, \citenamefont {Szabo},\ and\ \citenamefont
  {Toth}}]{FODOR2016245}%
  \BibitemOpen
  \bibfield  {author} {\bibinfo {author} {\bibfnamefont {Z.}~\bibnamefont
  {Fodor}}, \bibinfo {author} {\bibfnamefont {C.}~\bibnamefont {Hoelbling}},
  \bibinfo {author} {\bibfnamefont {S.}~\bibnamefont {Katz}}, \bibinfo {author}
  {\bibfnamefont {L.}~\bibnamefont {Lellouch}}, \bibinfo {author}
  {\bibfnamefont {A.}~\bibnamefont {Portelli}}, \bibinfo {author}
  {\bibfnamefont {K.}~\bibnamefont {Szabo}}, \ and\ \bibinfo {author}
  {\bibfnamefont {B.}~\bibnamefont {Toth}},\ }\href {\doibase
  https://doi.org/10.1016/j.physletb.2016.01.047} {\bibfield  {journal}
  {\bibinfo  {journal} {Physics Letters B}\ }\textbf {\bibinfo {volume}
  {755}},\ \bibinfo {pages} {245} (\bibinfo {year} {2016})}\BibitemShut
  {NoStop}%
\bibitem [{\citenamefont {Hansen}\ \emph {et~al.}(2018)\citenamefont {Hansen},
  \citenamefont {Lucini}, \citenamefont {Patella},\ and\ \citenamefont
  {Tantalo}}]{Hansen_2018}%
  \BibitemOpen
  \bibfield  {author} {\bibinfo {author} {\bibfnamefont {M.}~\bibnamefont
  {Hansen}}, \bibinfo {author} {\bibfnamefont {B.}~\bibnamefont {Lucini}},
  \bibinfo {author} {\bibfnamefont {A.}~\bibnamefont {Patella}}, \ and\
  \bibinfo {author} {\bibfnamefont {N.}~\bibnamefont {Tantalo}},\ }\href
  {\doibase 10.1051/epjconf/201817509001} {\bibfield  {journal} {\bibinfo
  {journal} {EPJ Web of Conferences}\ }\textbf {\bibinfo {volume} {175}},\
  \bibinfo {pages} {09001} (\bibinfo {year} {2018})}\BibitemShut {NoStop}%
\bibitem [{\citenamefont {Davoudi}\ \emph {et~al.}(2019)\citenamefont
  {Davoudi}, \citenamefont {Harrison}, \citenamefont {J\"uttner}, \citenamefont
  {Portelli},\ and\ \citenamefont {Savage}}]{Davoudi:2018qpl}%
  \BibitemOpen
  \bibfield  {author} {\bibinfo {author} {\bibfnamefont {Z.}~\bibnamefont
  {Davoudi}}, \bibinfo {author} {\bibfnamefont {J.}~\bibnamefont {Harrison}},
  \bibinfo {author} {\bibfnamefont {A.}~\bibnamefont {J\"uttner}}, \bibinfo
  {author} {\bibfnamefont {A.}~\bibnamefont {Portelli}}, \ and\ \bibinfo
  {author} {\bibfnamefont {M.~J.}\ \bibnamefont {Savage}},\ }\href {\doibase
  10.1103/PhysRevD.99.034510} {\bibfield  {journal} {\bibinfo  {journal} {Phys.
  Rev. D}\ }\textbf {\bibinfo {volume} {99}},\ \bibinfo {pages} {034510}
  (\bibinfo {year} {2019})},\ \Eprint {http://arxiv.org/abs/1810.05923}
  {arXiv:1810.05923 [hep-lat]} \BibitemShut {NoStop}%
\bibitem [{\citenamefont {Osborne}\ and\ \citenamefont
  {Nielsen}(2002)}]{Osborne_2002}%
  \BibitemOpen
  \bibfield  {author} {\bibinfo {author} {\bibfnamefont {T.~J.}\ \bibnamefont
  {Osborne}}\ and\ \bibinfo {author} {\bibfnamefont {M.~A.}\ \bibnamefont
  {Nielsen}},\ }\href {\doibase 10.1023/a:1019601218492} {\bibfield  {journal}
  {\bibinfo  {journal} {Quantum Information Processing}\ }\textbf {\bibinfo
  {volume} {1}},\ \bibinfo {pages} {45–53} (\bibinfo {year}
  {2002})}\BibitemShut {NoStop}%
\bibitem [{\citenamefont {Lloyd}(1996)}]{Lloyd1073}%
  \BibitemOpen
  \bibfield  {author} {\bibinfo {author} {\bibfnamefont {S.}~\bibnamefont
  {Lloyd}},\ }\href {\doibase 10.1126/science.273.5278.1073} {\bibfield
  {journal} {\bibinfo  {journal} {Science}\ }\textbf {\bibinfo {volume}
  {273}},\ \bibinfo {pages} {1073} (\bibinfo {year} {1996})}\BibitemShut
  {NoStop}%
\bibitem [{\citenamefont {Shor}(1994)}]{365700}%
  \BibitemOpen
  \bibfield  {author} {\bibinfo {author} {\bibfnamefont {P.}~\bibnamefont
  {Shor}},\ }in\ \href {\doibase 10.1109/SFCS.1994.365700} {\emph {\bibinfo
  {booktitle} {Proceedings 35th Annual Symposium on Foundations of Computer
  Science}}},\ \bibinfo {series and number} {SFCS '94}\ (\bibinfo  {publisher}
  {IEEE Computer Society},\ \bibinfo {year} {1994})\ pp.\ \bibinfo {pages}
  {124--134}\BibitemShut {NoStop}%
\bibitem [{\citenamefont {Wecker}\ \emph {et~al.}(2014)\citenamefont {Wecker},
  \citenamefont {Bauer}, \citenamefont {Clark}, \citenamefont {Hastings},\ and\
  \citenamefont {Troyer}}]{PhysRevA.90.022305}%
  \BibitemOpen
  \bibfield  {author} {\bibinfo {author} {\bibfnamefont {D.}~\bibnamefont
  {Wecker}}, \bibinfo {author} {\bibfnamefont {B.}~\bibnamefont {Bauer}},
  \bibinfo {author} {\bibfnamefont {B.~K.}\ \bibnamefont {Clark}}, \bibinfo
  {author} {\bibfnamefont {M.~B.}\ \bibnamefont {Hastings}}, \ and\ \bibinfo
  {author} {\bibfnamefont {M.}~\bibnamefont {Troyer}},\ }\href {\doibase
  10.1103/PhysRevA.90.022305} {\bibfield  {journal} {\bibinfo  {journal} {Phys.
  Rev. A}\ }\textbf {\bibinfo {volume} {90}},\ \bibinfo {pages} {022305}
  (\bibinfo {year} {2014})}\BibitemShut {NoStop}%
\bibitem [{\citenamefont {Reiher}\ \emph {et~al.}(2017)\citenamefont {Reiher},
  \citenamefont {Wiebe}, \citenamefont {Svore}, \citenamefont {Wecker},\ and\
  \citenamefont {Troyer}}]{Reiher7555}%
  \BibitemOpen
  \bibfield  {author} {\bibinfo {author} {\bibfnamefont {M.}~\bibnamefont
  {Reiher}}, \bibinfo {author} {\bibfnamefont {N.}~\bibnamefont {Wiebe}},
  \bibinfo {author} {\bibfnamefont {K.~M.}\ \bibnamefont {Svore}}, \bibinfo
  {author} {\bibfnamefont {D.}~\bibnamefont {Wecker}}, \ and\ \bibinfo {author}
  {\bibfnamefont {M.}~\bibnamefont {Troyer}},\ }\href {\doibase
  10.1073/pnas.1619152114} {\bibfield  {journal} {\bibinfo  {journal}
  {Proceedings of the National Academy of Sciences}\ }\textbf {\bibinfo
  {volume} {114}},\ \bibinfo {pages} {7555} (\bibinfo {year}
  {2017})}\BibitemShut {NoStop}%
\bibitem [{\citenamefont {Wecker}\ \emph {et~al.}(2015)\citenamefont {Wecker},
  \citenamefont {Hastings}, \citenamefont {Wiebe}, \citenamefont {Clark},
  \citenamefont {Nayak},\ and\ \citenamefont {Troyer}}]{PhysRevA.92.062318}%
  \BibitemOpen
  \bibfield  {author} {\bibinfo {author} {\bibfnamefont {D.}~\bibnamefont
  {Wecker}}, \bibinfo {author} {\bibfnamefont {M.~B.}\ \bibnamefont
  {Hastings}}, \bibinfo {author} {\bibfnamefont {N.}~\bibnamefont {Wiebe}},
  \bibinfo {author} {\bibfnamefont {B.~K.}\ \bibnamefont {Clark}}, \bibinfo
  {author} {\bibfnamefont {C.}~\bibnamefont {Nayak}}, \ and\ \bibinfo {author}
  {\bibfnamefont {M.}~\bibnamefont {Troyer}},\ }\href {\doibase
  10.1103/PhysRevA.92.062318} {\bibfield  {journal} {\bibinfo  {journal} {Phys.
  Rev. A}\ }\textbf {\bibinfo {volume} {92}},\ \bibinfo {pages} {062318}
  (\bibinfo {year} {2015})}\BibitemShut {NoStop}%
\bibitem [{\citenamefont {Kivlichan}\ \emph {et~al.}(2018)\citenamefont
  {Kivlichan}, \citenamefont {McClean}, \citenamefont {Wiebe}, \citenamefont
  {Gidney}, \citenamefont {Aspuru-Guzik}, \citenamefont {Chan},\ and\
  \citenamefont {Babbush}}]{PhysRevLett.120.110501}%
  \BibitemOpen
  \bibfield  {author} {\bibinfo {author} {\bibfnamefont {I.~D.}\ \bibnamefont
  {Kivlichan}}, \bibinfo {author} {\bibfnamefont {J.}~\bibnamefont {McClean}},
  \bibinfo {author} {\bibfnamefont {N.}~\bibnamefont {Wiebe}}, \bibinfo
  {author} {\bibfnamefont {C.}~\bibnamefont {Gidney}}, \bibinfo {author}
  {\bibfnamefont {A.}~\bibnamefont {Aspuru-Guzik}}, \bibinfo {author}
  {\bibfnamefont {G.~K.-L.}\ \bibnamefont {Chan}}, \ and\ \bibinfo {author}
  {\bibfnamefont {R.}~\bibnamefont {Babbush}},\ }\href {\doibase
  10.1103/PhysRevLett.120.110501} {\bibfield  {journal} {\bibinfo  {journal}
  {Phys. Rev. Lett.}\ }\textbf {\bibinfo {volume} {120}},\ \bibinfo {pages}
  {110501} (\bibinfo {year} {2018})}\BibitemShut {NoStop}%
\bibitem [{\citenamefont {Kivlichan}\ \emph {et~al.}(2020)\citenamefont
  {Kivlichan}, \citenamefont {Gidney}, \citenamefont {Berry}, \citenamefont
  {Wiebe}, \citenamefont {McClean}, \citenamefont {Sun}, \citenamefont {Jiang},
  \citenamefont {Rubin}, \citenamefont {Fowler}, \citenamefont {Aspuru-Guzik},
  \citenamefont {Neven},\ and\ \citenamefont
  {Babbush}}]{Kivlichan2020improvedfault}%
  \BibitemOpen
  \bibfield  {author} {\bibinfo {author} {\bibfnamefont {I.~D.}\ \bibnamefont
  {Kivlichan}}, \bibinfo {author} {\bibfnamefont {C.}~\bibnamefont {Gidney}},
  \bibinfo {author} {\bibfnamefont {D.~W.}\ \bibnamefont {Berry}}, \bibinfo
  {author} {\bibfnamefont {N.}~\bibnamefont {Wiebe}}, \bibinfo {author}
  {\bibfnamefont {J.}~\bibnamefont {McClean}}, \bibinfo {author} {\bibfnamefont
  {W.}~\bibnamefont {Sun}}, \bibinfo {author} {\bibfnamefont {Z.}~\bibnamefont
  {Jiang}}, \bibinfo {author} {\bibfnamefont {N.}~\bibnamefont {Rubin}},
  \bibinfo {author} {\bibfnamefont {A.}~\bibnamefont {Fowler}}, \bibinfo
  {author} {\bibfnamefont {A.}~\bibnamefont {Aspuru-Guzik}}, \bibinfo {author}
  {\bibfnamefont {H.}~\bibnamefont {Neven}}, \ and\ \bibinfo {author}
  {\bibfnamefont {R.}~\bibnamefont {Babbush}},\ }\href {\doibase
  10.22331/q-2020-07-16-296} {\bibfield  {journal} {\bibinfo  {journal}
  {{Quantum}}\ }\textbf {\bibinfo {volume} {4}},\ \bibinfo {pages} {296}
  (\bibinfo {year} {2020})}\BibitemShut {NoStop}%
\bibitem [{\citenamefont {Kaplan}\ \emph
  {et~al.}(1998{\natexlab{a}})\citenamefont {Kaplan}, \citenamefont {Savage},\
  and\ \citenamefont {Wise}}]{Kaplan:1998tg}%
  \BibitemOpen
  \bibfield  {author} {\bibinfo {author} {\bibfnamefont {D.~B.}\ \bibnamefont
  {Kaplan}}, \bibinfo {author} {\bibfnamefont {M.~J.}\ \bibnamefont {Savage}},
  \ and\ \bibinfo {author} {\bibfnamefont {M.~B.}\ \bibnamefont {Wise}},\
  }\href {\doibase 10.1016/S0370-2693(98)00210-X} {\bibfield  {journal}
  {\bibinfo  {journal} {Phys. Lett. B}\ }\textbf {\bibinfo {volume} {424}},\
  \bibinfo {pages} {390} (\bibinfo {year} {1998}{\natexlab{a}})},\ \Eprint
  {http://arxiv.org/abs/nucl-th/9801034} {arXiv:nucl-th/9801034} \BibitemShut
  {NoStop}%
\bibitem [{\citenamefont {Kaplan}\ \emph
  {et~al.}(1998{\natexlab{b}})\citenamefont {Kaplan}, \citenamefont {Savage},\
  and\ \citenamefont {Wise}}]{Kaplan:1998we}%
  \BibitemOpen
  \bibfield  {author} {\bibinfo {author} {\bibfnamefont {D.~B.}\ \bibnamefont
  {Kaplan}}, \bibinfo {author} {\bibfnamefont {M.~J.}\ \bibnamefont {Savage}},
  \ and\ \bibinfo {author} {\bibfnamefont {M.~B.}\ \bibnamefont {Wise}},\
  }\href {\doibase 10.1016/S0550-3213(98)00440-4} {\bibfield  {journal}
  {\bibinfo  {journal} {Nucl. Phys. B}\ }\textbf {\bibinfo {volume} {534}},\
  \bibinfo {pages} {329} (\bibinfo {year} {1998}{\natexlab{b}})},\ \Eprint
  {http://arxiv.org/abs/nucl-th/9802075} {arXiv:nucl-th/9802075} \BibitemShut
  {NoStop}%
\bibitem [{\citenamefont {van Kolck}(1999)}]{vanKolck:1998bw}%
  \BibitemOpen
  \bibfield  {author} {\bibinfo {author} {\bibfnamefont {U.}~\bibnamefont {van
  Kolck}},\ }\href {\doibase 10.1016/S0375-9474(98)00612-5} {\bibfield
  {journal} {\bibinfo  {journal} {Nucl. Phys. A}\ }\textbf {\bibinfo {volume}
  {645}},\ \bibinfo {pages} {273} (\bibinfo {year} {1999})},\ \Eprint
  {http://arxiv.org/abs/nucl-th/9808007} {arXiv:nucl-th/9808007} \BibitemShut
  {NoStop}%
\bibitem [{\citenamefont {Byrnes}\ and\ \citenamefont
  {Yamamoto}(2006)}]{PhysRevA.73.022328}%
  \BibitemOpen
  \bibfield  {author} {\bibinfo {author} {\bibfnamefont {T.}~\bibnamefont
  {Byrnes}}\ and\ \bibinfo {author} {\bibfnamefont {Y.}~\bibnamefont
  {Yamamoto}},\ }\href {\doibase 10.1103/PhysRevA.73.022328} {\bibfield
  {journal} {\bibinfo  {journal} {Phys. Rev. A}\ }\textbf {\bibinfo {volume}
  {73}},\ \bibinfo {pages} {022328} (\bibinfo {year} {2006})}\BibitemShut
  {NoStop}%
\bibitem [{\citenamefont {Shaw}\ \emph {et~al.}(2020)\citenamefont {Shaw},
  \citenamefont {Lougovski}, \citenamefont {Stryker},\ and\ \citenamefont
  {Wiebe}}]{Shaw2020quantumalgorithms}%
  \BibitemOpen
  \bibfield  {author} {\bibinfo {author} {\bibfnamefont {A.~F.}\ \bibnamefont
  {Shaw}}, \bibinfo {author} {\bibfnamefont {P.}~\bibnamefont {Lougovski}},
  \bibinfo {author} {\bibfnamefont {J.~R.}\ \bibnamefont {Stryker}}, \ and\
  \bibinfo {author} {\bibfnamefont {N.}~\bibnamefont {Wiebe}},\ }\href
  {\doibase 10.22331/q-2020-08-10-306} {\bibfield  {journal} {\bibinfo
  {journal} {{Quantum}}\ }\textbf {\bibinfo {volume} {4}},\ \bibinfo {pages}
  {306} (\bibinfo {year} {2020})}\BibitemShut {NoStop}%
\bibitem [{\citenamefont {Mathis}\ \emph {et~al.}(2020)\citenamefont {Mathis},
  \citenamefont {Mazzola},\ and\ \citenamefont
  {Tavernelli}}]{PhysRevD.102.094501}%
  \BibitemOpen
  \bibfield  {author} {\bibinfo {author} {\bibfnamefont {S.~V.}\ \bibnamefont
  {Mathis}}, \bibinfo {author} {\bibfnamefont {G.}~\bibnamefont {Mazzola}}, \
  and\ \bibinfo {author} {\bibfnamefont {I.}~\bibnamefont {Tavernelli}},\
  }\href {\doibase 10.1103/PhysRevD.102.094501} {\bibfield  {journal} {\bibinfo
   {journal} {Phys. Rev. D}\ }\textbf {\bibinfo {volume} {102}},\ \bibinfo
  {pages} {094501} (\bibinfo {year} {2020})}\BibitemShut {NoStop}%
\bibitem [{\citenamefont {Haah}\ \emph {et~al.}(2021)\citenamefont {Haah},
  \citenamefont {Hastings}, \citenamefont {Kothari},\ and\ \citenamefont
  {Low}}]{doi:10.1137/18M1231511}%
  \BibitemOpen
  \bibfield  {author} {\bibinfo {author} {\bibfnamefont {J.}~\bibnamefont
  {Haah}}, \bibinfo {author} {\bibfnamefont {M.~B.}\ \bibnamefont {Hastings}},
  \bibinfo {author} {\bibfnamefont {R.}~\bibnamefont {Kothari}}, \ and\
  \bibinfo {author} {\bibfnamefont {G.~H.}\ \bibnamefont {Low}},\ }\href
  {\doibase 10.1137/18M1231511} {\bibfield  {journal} {\bibinfo  {journal}
  {SIAM Journal on Computing}\ ,\ \bibinfo {pages} {FOCS18}} (\bibinfo {year}
  {2021})}\BibitemShut {NoStop}%
\bibitem [{\citenamefont {Trotter}(1959)}]{10.2307/2033649}%
  \BibitemOpen
  \bibfield  {author} {\bibinfo {author} {\bibfnamefont {H.~F.}\ \bibnamefont
  {Trotter}},\ }\href {http://www.jstor.org/stable/2033649} {\bibfield
  {journal} {\bibinfo  {journal} {Proceedings of the American Mathematical
  Society}\ }\textbf {\bibinfo {volume} {10}},\ \bibinfo {pages} {545}
  (\bibinfo {year} {1959})}\BibitemShut {NoStop}%
\bibitem [{\citenamefont {Suzuki}(1991)}]{doi:10.1063/1.529425}%
  \BibitemOpen
  \bibfield  {author} {\bibinfo {author} {\bibfnamefont {M.}~\bibnamefont
  {Suzuki}},\ }\href {\doibase 10.1063/1.529425} {\bibfield  {journal}
  {\bibinfo  {journal} {Journal of Mathematical Physics}\ }\textbf {\bibinfo
  {volume} {32}},\ \bibinfo {pages} {400} (\bibinfo {year} {1991})}\BibitemShut
  {NoStop}%
\bibitem [{\citenamefont {Childs}\ \emph {et~al.}(2018)\citenamefont {Childs},
  \citenamefont {Maslov}, \citenamefont {Nam}, \citenamefont {Ross},\ and\
  \citenamefont {Su}}]{Childs9456}%
  \BibitemOpen
  \bibfield  {author} {\bibinfo {author} {\bibfnamefont {A.~M.}\ \bibnamefont
  {Childs}}, \bibinfo {author} {\bibfnamefont {D.}~\bibnamefont {Maslov}},
  \bibinfo {author} {\bibfnamefont {Y.}~\bibnamefont {Nam}}, \bibinfo {author}
  {\bibfnamefont {N.~J.}\ \bibnamefont {Ross}}, \ and\ \bibinfo {author}
  {\bibfnamefont {Y.}~\bibnamefont {Su}},\ }\href {\doibase
  10.1073/pnas.1801723115} {\bibfield  {journal} {\bibinfo  {journal}
  {Proceedings of the National Academy of Sciences}\ }\textbf {\bibinfo
  {volume} {115}},\ \bibinfo {pages} {9456} (\bibinfo {year}
  {2018})}\BibitemShut {NoStop}%
\bibitem [{\citenamefont {Childs}\ and\ \citenamefont
  {Wiebe}(2012)}]{Childs2012}%
  \BibitemOpen
  \bibfield  {author} {\bibinfo {author} {\bibfnamefont {A.~M.}\ \bibnamefont
  {Childs}}\ and\ \bibinfo {author} {\bibfnamefont {N.}~\bibnamefont {Wiebe}},\
  }\href {\doibase 10.26421/qic12.11-12} {\bibfield  {journal} {\bibinfo
  {journal} {Quantum Information and Computation}\ }\textbf {\bibinfo {volume}
  {12}} (\bibinfo {year} {2012}),\ 10.26421/qic12.11-12}\BibitemShut {NoStop}%
\bibitem [{\citenamefont {Berry}\ \emph
  {et~al.}(2015{\natexlab{a}})\citenamefont {Berry}, \citenamefont {Childs},
  \citenamefont {Cleve}, \citenamefont {Kothari},\ and\ \citenamefont
  {Somma}}]{PhysRevLett.114.090502}%
  \BibitemOpen
  \bibfield  {author} {\bibinfo {author} {\bibfnamefont {D.~W.}\ \bibnamefont
  {Berry}}, \bibinfo {author} {\bibfnamefont {A.~M.}\ \bibnamefont {Childs}},
  \bibinfo {author} {\bibfnamefont {R.}~\bibnamefont {Cleve}}, \bibinfo
  {author} {\bibfnamefont {R.}~\bibnamefont {Kothari}}, \ and\ \bibinfo
  {author} {\bibfnamefont {R.~D.}\ \bibnamefont {Somma}},\ }\href {\doibase
  10.1103/PhysRevLett.114.090502} {\bibfield  {journal} {\bibinfo  {journal}
  {Phys. Rev. Lett.}\ }\textbf {\bibinfo {volume} {114}},\ \bibinfo {pages}
  {090502} (\bibinfo {year} {2015}{\natexlab{a}})}\BibitemShut {NoStop}%
\bibitem [{\citenamefont {Berry}\ \emph
  {et~al.}(2015{\natexlab{b}})\citenamefont {Berry}, \citenamefont {Childs},\
  and\ \citenamefont {Kothari}}]{7354428}%
  \BibitemOpen
  \bibfield  {author} {\bibinfo {author} {\bibfnamefont {D.~W.}\ \bibnamefont
  {Berry}}, \bibinfo {author} {\bibfnamefont {A.~M.}\ \bibnamefont {Childs}}, \
  and\ \bibinfo {author} {\bibfnamefont {R.}~\bibnamefont {Kothari}},\ }in\
  \href {\doibase 10.1109/FOCS.2015.54} {\emph {\bibinfo {booktitle} {2015 IEEE
  56th Annual Symposium on Foundations of Computer Science}}}\ (\bibinfo {year}
  {2015})\ pp.\ \bibinfo {pages} {792--809}\BibitemShut {NoStop}%
\bibitem [{\citenamefont {Low}\ and\ \citenamefont
  {Chuang}(2017)}]{PhysRevLett.118.010501}%
  \BibitemOpen
  \bibfield  {author} {\bibinfo {author} {\bibfnamefont {G.~H.}\ \bibnamefont
  {Low}}\ and\ \bibinfo {author} {\bibfnamefont {I.~L.}\ \bibnamefont
  {Chuang}},\ }\href {\doibase 10.1103/PhysRevLett.118.010501} {\bibfield
  {journal} {\bibinfo  {journal} {Phys. Rev. Lett.}\ }\textbf {\bibinfo
  {volume} {118}},\ \bibinfo {pages} {010501} (\bibinfo {year}
  {2017})}\BibitemShut {NoStop}%
\bibitem [{\citenamefont {Low}\ and\ \citenamefont
  {Chuang}(2019)}]{Low2019hamiltonian}%
  \BibitemOpen
  \bibfield  {author} {\bibinfo {author} {\bibfnamefont {G.~H.}\ \bibnamefont
  {Low}}\ and\ \bibinfo {author} {\bibfnamefont {I.~L.}\ \bibnamefont
  {Chuang}},\ }\href {\doibase 10.22331/q-2019-07-12-163} {\bibfield  {journal}
  {\bibinfo  {journal} {{Quantum}}\ }\textbf {\bibinfo {volume} {3}},\ \bibinfo
  {pages} {163} (\bibinfo {year} {2019})}\BibitemShut {NoStop}%
\bibitem [{\citenamefont {Berry}\ \emph {et~al.}(2007)\citenamefont {Berry},
  \citenamefont {Ahokas}, \citenamefont {Cleve},\ and\ \citenamefont
  {Sanders}}]{Berry2007}%
  \BibitemOpen
  \bibfield  {author} {\bibinfo {author} {\bibfnamefont {D.~W.}\ \bibnamefont
  {Berry}}, \bibinfo {author} {\bibfnamefont {G.}~\bibnamefont {Ahokas}},
  \bibinfo {author} {\bibfnamefont {R.}~\bibnamefont {Cleve}}, \ and\ \bibinfo
  {author} {\bibfnamefont {B.~C.}\ \bibnamefont {Sanders}},\ }\href {\doibase
  0.1007/s00220-006-0150-x} {\bibfield  {journal} {\bibinfo  {journal}
  {Communications in Mathematical Physics}\ }\textbf {\bibinfo {volume}
  {270}},\ \bibinfo {pages} {359} (\bibinfo {year} {2007})}\BibitemShut
  {NoStop}%
\bibitem [{\citenamefont {Atia}\ and\ \citenamefont
  {Aharonov}(2017)}]{Atia2017}%
  \BibitemOpen
  \bibfield  {author} {\bibinfo {author} {\bibfnamefont {Y.}~\bibnamefont
  {Atia}}\ and\ \bibinfo {author} {\bibfnamefont {D.}~\bibnamefont
  {Aharonov}},\ }\href {\doibase 10.1038/s41467-017-01637-7} {\bibfield
  {journal} {\bibinfo  {journal} {Nature Communications}\ }\textbf {\bibinfo
  {volume} {8}},\ \bibinfo {pages} {1572} (\bibinfo {year} {2017})}\BibitemShut
  {NoStop}%
\bibitem [{\citenamefont {Knill}(1995)}]{Knill:1995kz}%
  \BibitemOpen
  \bibfield  {author} {\bibinfo {author} {\bibfnamefont {E.}~\bibnamefont
  {Knill}},\ }\href@noop {} {\  (\bibinfo {year} {1995})},\ \Eprint
  {http://arxiv.org/abs/quant-ph/9508006} {arXiv:quant-ph/9508006} \BibitemShut
  {NoStop}%
\bibitem [{\citenamefont {Childs}\ and\ \citenamefont
  {Kothari}(2010)}]{Childs2010}%
  \BibitemOpen
  \bibfield  {author} {\bibinfo {author} {\bibfnamefont {A.}~\bibnamefont
  {Childs}}\ and\ \bibinfo {author} {\bibfnamefont {R.}~\bibnamefont
  {Kothari}},\ }\href {\doibase 10.26421/qic10.7-8} {\bibfield  {journal}
  {\bibinfo  {journal} {Quantum Information and Computation}\ }\textbf
  {\bibinfo {volume} {10}} (\bibinfo {year} {2010}),\
  10.26421/qic10.7-8}\BibitemShut {NoStop}%
\bibitem [{\citenamefont {Abrams}\ and\ \citenamefont
  {Lloyd}(1997)}]{PhysRevLett.79.2586}%
  \BibitemOpen
  \bibfield  {author} {\bibinfo {author} {\bibfnamefont {D.~S.}\ \bibnamefont
  {Abrams}}\ and\ \bibinfo {author} {\bibfnamefont {S.}~\bibnamefont {Lloyd}},\
  }\href {\doibase 10.1103/PhysRevLett.79.2586} {\bibfield  {journal} {\bibinfo
   {journal} {Phys. Rev. Lett.}\ }\textbf {\bibinfo {volume} {79}},\ \bibinfo
  {pages} {2586} (\bibinfo {year} {1997})}\BibitemShut {NoStop}%
\bibitem [{\citenamefont {Aspuru-Guzik}\ \emph {et~al.}(2005)\citenamefont
  {Aspuru-Guzik}, \citenamefont {Dutoi}, \citenamefont {Love},\ and\
  \citenamefont {Head-Gordon}}]{Aspuru-Guzik1704}%
  \BibitemOpen
  \bibfield  {author} {\bibinfo {author} {\bibfnamefont {A.}~\bibnamefont
  {Aspuru-Guzik}}, \bibinfo {author} {\bibfnamefont {A.~D.}\ \bibnamefont
  {Dutoi}}, \bibinfo {author} {\bibfnamefont {P.~J.}\ \bibnamefont {Love}}, \
  and\ \bibinfo {author} {\bibfnamefont {M.}~\bibnamefont {Head-Gordon}},\
  }\href {\doibase 10.1126/science.1113479} {\bibfield  {journal} {\bibinfo
  {journal} {Science}\ }\textbf {\bibinfo {volume} {309}},\ \bibinfo {pages}
  {1704} (\bibinfo {year} {2005})},\ \Eprint
  {http://arxiv.org/abs/https://science.sciencemag.org/content/309/5741/1704.full.pdf}
  {https://science.sciencemag.org/content/309/5741/1704.full.pdf} \BibitemShut
  {NoStop}%
\bibitem [{\citenamefont {Babbush}\ \emph {et~al.}(2019)\citenamefont
  {Babbush}, \citenamefont {Berry}, \citenamefont {McClean},\ and\
  \citenamefont {Neven}}]{Babbush2019}%
  \BibitemOpen
  \bibfield  {author} {\bibinfo {author} {\bibfnamefont {R.}~\bibnamefont
  {Babbush}}, \bibinfo {author} {\bibfnamefont {D.~W.}\ \bibnamefont {Berry}},
  \bibinfo {author} {\bibfnamefont {J.~R.}\ \bibnamefont {McClean}}, \ and\
  \bibinfo {author} {\bibfnamefont {H.}~\bibnamefont {Neven}},\ }\href
  {\doibase 10.1038/s41534-019-0199-y} {\bibfield  {journal} {\bibinfo
  {journal} {npj Quantum Information}\ }\textbf {\bibinfo {volume} {5}},\
  \bibinfo {pages} {92} (\bibinfo {year} {2019})}\BibitemShut {NoStop}%
\bibitem [{\citenamefont {Su}\ \emph {et~al.}(2020)\citenamefont {Su},
  \citenamefont {Huang},\ and\ \citenamefont {Campbell}}]{su2020nearly}%
  \BibitemOpen
  \bibfield  {author} {\bibinfo {author} {\bibfnamefont {Y.}~\bibnamefont
  {Su}}, \bibinfo {author} {\bibfnamefont {H.-Y.}\ \bibnamefont {Huang}}, \
  and\ \bibinfo {author} {\bibfnamefont {E.~T.}\ \bibnamefont {Campbell}},\
  }\href@noop {} {\enquote {\bibinfo {title} {Nearly tight trotterization of
  interacting electrons},}\ } (\bibinfo {year} {2020}),\ \Eprint
  {http://arxiv.org/abs/2012.09194} {arXiv:2012.09194 [quant-ph]} \BibitemShut
  {NoStop}%
\bibitem [{\citenamefont {Childs}\ \emph {et~al.}(2019)\citenamefont {Childs},
  \citenamefont {Ostrander},\ and\ \citenamefont
  {Su}}]{Childs2019fasterquantum}%
  \BibitemOpen
  \bibfield  {author} {\bibinfo {author} {\bibfnamefont {A.~M.}\ \bibnamefont
  {Childs}}, \bibinfo {author} {\bibfnamefont {A.}~\bibnamefont {Ostrander}}, \
  and\ \bibinfo {author} {\bibfnamefont {Y.}~\bibnamefont {Su}},\ }\href
  {\doibase 10.22331/q-2019-09-02-182} {\bibfield  {journal} {\bibinfo
  {journal} {{Quantum}}\ }\textbf {\bibinfo {volume} {3}},\ \bibinfo {pages}
  {182} (\bibinfo {year} {2019})}\BibitemShut {NoStop}%
\bibitem [{\citenamefont {Campbell}(2019)}]{PhysRevLett.123.070503}%
  \BibitemOpen
  \bibfield  {author} {\bibinfo {author} {\bibfnamefont {E.}~\bibnamefont
  {Campbell}},\ }\href {\doibase 10.1103/PhysRevLett.123.070503} {\bibfield
  {journal} {\bibinfo  {journal} {Phys. Rev. Lett.}\ }\textbf {\bibinfo
  {volume} {123}},\ \bibinfo {pages} {070503} (\bibinfo {year}
  {2019})}\BibitemShut {NoStop}%
\bibitem [{\citenamefont {Chen}\ \emph {et~al.}(2020)\citenamefont {Chen},
  \citenamefont {Huang}, \citenamefont {Kueng},\ and\ \citenamefont
  {Tropp}}]{chen2020quantum}%
  \BibitemOpen
  \bibfield  {author} {\bibinfo {author} {\bibfnamefont {C.-F.}\ \bibnamefont
  {Chen}}, \bibinfo {author} {\bibfnamefont {H.-Y.}\ \bibnamefont {Huang}},
  \bibinfo {author} {\bibfnamefont {R.}~\bibnamefont {Kueng}}, \ and\ \bibinfo
  {author} {\bibfnamefont {J.~A.}\ \bibnamefont {Tropp}},\ }\href@noop {}
  {\enquote {\bibinfo {title} {Quantum simulation via randomized product
  formulas: Low gate complexity with accuracy guarantees},}\ } (\bibinfo {year}
  {2020}),\ \Eprint {http://arxiv.org/abs/2008.11751} {arXiv:2008.11751
  [quant-ph]} \BibitemShut {NoStop}%
\bibitem [{\citenamefont {Faehrmann}\ \emph {et~al.}(2021)\citenamefont
  {Faehrmann}, \citenamefont {Steudtner}, \citenamefont {Kueng}, \citenamefont
  {Kieferova},\ and\ \citenamefont {Eisert}}]{faehrmann2021randomizing}%
  \BibitemOpen
  \bibfield  {author} {\bibinfo {author} {\bibfnamefont {P.~K.}\ \bibnamefont
  {Faehrmann}}, \bibinfo {author} {\bibfnamefont {M.}~\bibnamefont
  {Steudtner}}, \bibinfo {author} {\bibfnamefont {R.}~\bibnamefont {Kueng}},
  \bibinfo {author} {\bibfnamefont {M.}~\bibnamefont {Kieferova}}, \ and\
  \bibinfo {author} {\bibfnamefont {J.}~\bibnamefont {Eisert}},\ }\href@noop {}
  {\enquote {\bibinfo {title} {Randomizing multi-product formulas for improved
  hamiltonian simulation},}\ } (\bibinfo {year} {2021}),\ \Eprint
  {http://arxiv.org/abs/2101.07808} {arXiv:2101.07808 [quant-ph]} \BibitemShut
  {NoStop}%
\bibitem [{\citenamefont {Berry}\ \emph {et~al.}(2020)\citenamefont {Berry},
  \citenamefont {Childs}, \citenamefont {Su}, \citenamefont {Wang},\ and\
  \citenamefont {Wiebe}}]{Berry2020timedependent}%
  \BibitemOpen
  \bibfield  {author} {\bibinfo {author} {\bibfnamefont {D.~W.}\ \bibnamefont
  {Berry}}, \bibinfo {author} {\bibfnamefont {A.~M.}\ \bibnamefont {Childs}},
  \bibinfo {author} {\bibfnamefont {Y.}~\bibnamefont {Su}}, \bibinfo {author}
  {\bibfnamefont {X.}~\bibnamefont {Wang}}, \ and\ \bibinfo {author}
  {\bibfnamefont {N.}~\bibnamefont {Wiebe}},\ }\href {\doibase
  10.22331/q-2020-04-20-254} {\bibfield  {journal} {\bibinfo  {journal}
  {{Quantum}}\ }\textbf {\bibinfo {volume} {4}},\ \bibinfo {pages} {254}
  (\bibinfo {year} {2020})}\BibitemShut {NoStop}%
\bibitem [{\citenamefont {{{\c{S}}ahino{\u{g}}lu}}\ and\ \citenamefont
  {{Somma}}(2020)}]{2020arXiv200602660S}%
  \BibitemOpen
  \bibfield  {author} {\bibinfo {author} {\bibfnamefont {B.}~\bibnamefont
  {{{\c{S}}ahino{\u{g}}lu}}}\ and\ \bibinfo {author} {\bibfnamefont {R.~D.}\
  \bibnamefont {{Somma}}},\ }\href@noop {} {\bibfield  {journal} {\bibinfo
  {journal} {arXiv e-prints}\ ,\ \bibinfo {eid} {arXiv:2006.02660}} (\bibinfo
  {year} {2020})},\ \Eprint {http://arxiv.org/abs/2006.02660} {arXiv:2006.02660
  [quant-ph]} \BibitemShut {NoStop}%
\bibitem [{\citenamefont {Martonosi}\ and\ \citenamefont
  {Roetteler}(2019)}]{martonosi2019steps}%
  \BibitemOpen
  \bibfield  {author} {\bibinfo {author} {\bibfnamefont {M.}~\bibnamefont
  {Martonosi}}\ and\ \bibinfo {author} {\bibfnamefont {M.}~\bibnamefont
  {Roetteler}},\ }\href@noop {} {\enquote {\bibinfo {title} {Next steps in
  quantum computing: Computer science's role},}\ } (\bibinfo {year} {2019}),\
  \Eprint {http://arxiv.org/abs/1903.10541} {arXiv:1903.10541 [cs.ET]}
  \BibitemShut {NoStop}%
\bibitem [{\citenamefont {Moll}\ \emph {et~al.}(2018)\citenamefont {Moll},
  \citenamefont {Barkoutsos}, \citenamefont {Bishop}, \citenamefont {Chow},
  \citenamefont {Cross}, \citenamefont {Egger}, \citenamefont {Filipp},
  \citenamefont {Fuhrer}, \citenamefont {Gambetta}, \citenamefont {Ganzhorn},\
  and\ \citenamefont {et~al.}}]{Moll_2018}%
  \BibitemOpen
  \bibfield  {author} {\bibinfo {author} {\bibfnamefont {N.}~\bibnamefont
  {Moll}}, \bibinfo {author} {\bibfnamefont {P.}~\bibnamefont {Barkoutsos}},
  \bibinfo {author} {\bibfnamefont {L.~S.}\ \bibnamefont {Bishop}}, \bibinfo
  {author} {\bibfnamefont {J.~M.}\ \bibnamefont {Chow}}, \bibinfo {author}
  {\bibfnamefont {A.}~\bibnamefont {Cross}}, \bibinfo {author} {\bibfnamefont
  {D.~J.}\ \bibnamefont {Egger}}, \bibinfo {author} {\bibfnamefont
  {S.}~\bibnamefont {Filipp}}, \bibinfo {author} {\bibfnamefont
  {A.}~\bibnamefont {Fuhrer}}, \bibinfo {author} {\bibfnamefont {J.~M.}\
  \bibnamefont {Gambetta}}, \bibinfo {author} {\bibfnamefont {M.}~\bibnamefont
  {Ganzhorn}}, \ and\ \bibinfo {author} {\bibnamefont {et~al.}},\ }\href
  {\doibase 10.1088/2058-9565/aab822} {\bibfield  {journal} {\bibinfo
  {journal} {Quantum Science and Technology}\ }\textbf {\bibinfo {volume}
  {3}},\ \bibinfo {pages} {030503} (\bibinfo {year} {2018})}\BibitemShut
  {NoStop}%
\bibitem [{\citenamefont {Kelly}(2018)}]{googlebristlecone}%
  \BibitemOpen
  \bibfield  {author} {\bibinfo {author} {\bibfnamefont {J.}~\bibnamefont
  {Kelly}},\ }\href
  {https://ai.googleblog.com/2018/03/a-preview-of-bristlecone-googles-new.html}
  {\enquote {\bibinfo {title} {A preview of bristlecone, google’s new quantum
  processor},}\ } (\bibinfo {year} {2018})\BibitemShut {NoStop}%
\bibitem [{\citenamefont {Babbush}\ \emph {et~al.}(2018)\citenamefont
  {Babbush}, \citenamefont {Wiebe}, \citenamefont {McClean}, \citenamefont
  {McClain}, \citenamefont {Neven},\ and\ \citenamefont
  {Chan}}]{Babbush2018LowDepthQS}%
  \BibitemOpen
  \bibfield  {author} {\bibinfo {author} {\bibfnamefont {R.}~\bibnamefont
  {Babbush}}, \bibinfo {author} {\bibfnamefont {N.}~\bibnamefont {Wiebe}},
  \bibinfo {author} {\bibfnamefont {J.}~\bibnamefont {McClean}}, \bibinfo
  {author} {\bibfnamefont {J.~D.}\ \bibnamefont {McClain}}, \bibinfo {author}
  {\bibfnamefont {H.}~\bibnamefont {Neven}}, \ and\ \bibinfo {author}
  {\bibfnamefont {G.}~\bibnamefont {Chan}},\ }\href@noop {} {\bibfield
  {journal} {\bibinfo  {journal} {Physical Review X}\ }\textbf {\bibinfo
  {volume} {8}},\ \bibinfo {pages} {011044} (\bibinfo {year}
  {2018})}\BibitemShut {NoStop}%
\bibitem [{\citenamefont {Lee}\ \emph {et~al.}(2020{\natexlab{a}})\citenamefont
  {Lee}, \citenamefont {Berry}, \citenamefont {Gidney}, \citenamefont
  {Huggins}, \citenamefont {McClean}, \citenamefont {Wiebe},\ and\
  \citenamefont {Babbush}}]{lee2020efficient}%
  \BibitemOpen
  \bibfield  {author} {\bibinfo {author} {\bibfnamefont {J.}~\bibnamefont
  {Lee}}, \bibinfo {author} {\bibfnamefont {D.~W.}\ \bibnamefont {Berry}},
  \bibinfo {author} {\bibfnamefont {C.}~\bibnamefont {Gidney}}, \bibinfo
  {author} {\bibfnamefont {W.~J.}\ \bibnamefont {Huggins}}, \bibinfo {author}
  {\bibfnamefont {J.~R.}\ \bibnamefont {McClean}}, \bibinfo {author}
  {\bibfnamefont {N.}~\bibnamefont {Wiebe}}, \ and\ \bibinfo {author}
  {\bibfnamefont {R.}~\bibnamefont {Babbush}},\ }\href@noop {} {\enquote
  {\bibinfo {title} {Even more efficient quantum computations of chemistry
  through tensor hypercontraction},}\ } (\bibinfo {year}
  {2020}{\natexlab{a}}),\ \Eprint {http://arxiv.org/abs/2011.03494}
  {arXiv:2011.03494 [quant-ph]} \BibitemShut {NoStop}%
\bibitem [{\citenamefont {Lee}(2009)}]{LEE2009117}%
  \BibitemOpen
  \bibfield  {author} {\bibinfo {author} {\bibfnamefont {D.}~\bibnamefont
  {Lee}},\ }\href {\doibase https://doi.org/10.1016/j.ppnp.2008.12.001}
  {\bibfield  {journal} {\bibinfo  {journal} {Progress in Particle and Nuclear
  Physics}\ }\textbf {\bibinfo {volume} {63}},\ \bibinfo {pages} {117}
  (\bibinfo {year} {2009})}\BibitemShut {NoStop}%
\bibitem [{\citenamefont {Chakrabarti}\ \emph {et~al.}(2021)\citenamefont
  {Chakrabarti}, \citenamefont {Krishnakumar}, \citenamefont {Mazzola},
  \citenamefont {Stamatopoulos}, \citenamefont {Woerner},\ and\ \citenamefont
  {Zeng}}]{Chakrabarti_2021}%
  \BibitemOpen
  \bibfield  {author} {\bibinfo {author} {\bibfnamefont {S.}~\bibnamefont
  {Chakrabarti}}, \bibinfo {author} {\bibfnamefont {R.}~\bibnamefont
  {Krishnakumar}}, \bibinfo {author} {\bibfnamefont {G.}~\bibnamefont
  {Mazzola}}, \bibinfo {author} {\bibfnamefont {N.}~\bibnamefont
  {Stamatopoulos}}, \bibinfo {author} {\bibfnamefont {S.}~\bibnamefont
  {Woerner}}, \ and\ \bibinfo {author} {\bibfnamefont {W.~J.}\ \bibnamefont
  {Zeng}},\ }\href {\doibase 10.22331/q-2021-06-01-463} {\bibfield  {journal}
  {\bibinfo  {journal} {Quantum}\ }\textbf {\bibinfo {volume} {5}},\ \bibinfo
  {pages} {463} (\bibinfo {year} {2021})}\BibitemShut {NoStop}%
\bibitem [{\citenamefont {Christ}(2011)}]{Christ2011}%
  \BibitemOpen
  \bibfield  {author} {\bibinfo {author} {\bibfnamefont {N.~H.}\ \bibnamefont
  {Christ}},\ }\enquote {\bibinfo {title} {Qcdsp and qcdoc computers},}\ in\
  \href {\doibase 10.1007/978-0-387-09766-4_304} {\emph {\bibinfo {booktitle}
  {Encyclopedia of Parallel Computing}}},\ \bibinfo {editor} {edited by\
  \bibinfo {editor} {\bibfnamefont {D.}~\bibnamefont {Padua}}}\ (\bibinfo
  {publisher} {Springer US},\ \bibinfo {address} {Boston, MA},\ \bibinfo {year}
  {2011})\ pp.\ \bibinfo {pages} {1668--1677}\BibitemShut {NoStop}%
\bibitem [{\citenamefont {Marinari}(1986)}]{Marinari1986}%
  \BibitemOpen
  \bibfield  {author} {\bibinfo {author} {\bibfnamefont {E.}~\bibnamefont
  {Marinari}},\ }\enquote {\bibinfo {title} {The ape computer and lattice gauge
  theories},}\ in\ \href {\doibase 10.1007/978-1-4613-2231-3_28} {\emph
  {\bibinfo {booktitle} {Lattice Gauge Theory: A Challenge in Large-Scale
  Computing}}},\ \bibinfo {editor} {edited by\ \bibinfo {editor} {\bibfnamefont
  {B.}~\bibnamefont {Bunk}}, \bibinfo {editor} {\bibfnamefont {K.~H.}\
  \bibnamefont {M{\"u}tter}}, \ and\ \bibinfo {editor} {\bibfnamefont
  {K.}~\bibnamefont {Schilling}}}\ (\bibinfo  {publisher} {Springer US},\
  \bibinfo {address} {Boston, MA},\ \bibinfo {year} {1986})\ pp.\ \bibinfo
  {pages} {295--304}\BibitemShut {NoStop}%
\bibitem [{\citenamefont {Pogorelov}\ \emph {et~al.}(2021)\citenamefont
  {Pogorelov}, \citenamefont {Feldker}, \citenamefont {Marciniak},
  \citenamefont {Postler}, \citenamefont {Jacob}, \citenamefont
  {Krieglsteiner}, \citenamefont {Podlesnic}, \citenamefont {Meth},
  \citenamefont {Negnevitsky}, \citenamefont {Stadler}, \citenamefont
  {H\"ofer}, \citenamefont {W\"achter}, \citenamefont {Lakhmanskiy},
  \citenamefont {Blatt}, \citenamefont {Schindler},\ and\ \citenamefont
  {Monz}}]{PRXQuantum.2.020343}%
  \BibitemOpen
  \bibfield  {author} {\bibinfo {author} {\bibfnamefont {I.}~\bibnamefont
  {Pogorelov}}, \bibinfo {author} {\bibfnamefont {T.}~\bibnamefont {Feldker}},
  \bibinfo {author} {\bibfnamefont {C.~D.}\ \bibnamefont {Marciniak}}, \bibinfo
  {author} {\bibfnamefont {L.}~\bibnamefont {Postler}}, \bibinfo {author}
  {\bibfnamefont {G.}~\bibnamefont {Jacob}}, \bibinfo {author} {\bibfnamefont
  {O.}~\bibnamefont {Krieglsteiner}}, \bibinfo {author} {\bibfnamefont
  {V.}~\bibnamefont {Podlesnic}}, \bibinfo {author} {\bibfnamefont
  {M.}~\bibnamefont {Meth}}, \bibinfo {author} {\bibfnamefont {V.}~\bibnamefont
  {Negnevitsky}}, \bibinfo {author} {\bibfnamefont {M.}~\bibnamefont
  {Stadler}}, \bibinfo {author} {\bibfnamefont {B.}~\bibnamefont {H\"ofer}},
  \bibinfo {author} {\bibfnamefont {C.}~\bibnamefont {W\"achter}}, \bibinfo
  {author} {\bibfnamefont {K.}~\bibnamefont {Lakhmanskiy}}, \bibinfo {author}
  {\bibfnamefont {R.}~\bibnamefont {Blatt}}, \bibinfo {author} {\bibfnamefont
  {P.}~\bibnamefont {Schindler}}, \ and\ \bibinfo {author} {\bibfnamefont
  {T.}~\bibnamefont {Monz}},\ }\href {\doibase 10.1103/PRXQuantum.2.020343}
  {\bibfield  {journal} {\bibinfo  {journal} {PRX Quantum}\ }\textbf {\bibinfo
  {volume} {2}},\ \bibinfo {pages} {020343} (\bibinfo {year}
  {2021})}\BibitemShut {NoStop}%
\bibitem [{\citenamefont {team}\ and\ \citenamefont
  {collaborators}(2020)}]{Cirq}%
  \BibitemOpen
  \bibfield  {author} {\bibinfo {author} {\bibfnamefont {Q.~A.}\ \bibnamefont
  {team}}\ and\ \bibinfo {author} {\bibnamefont {collaborators}},\ }\href
  {\doibase 10.5281/zenodo.4062499} {\enquote {\bibinfo {title} {Cirq},}\ }
  (\bibinfo {year} {2020})\BibitemShut {NoStop}%
\bibitem [{\citenamefont {Aleksandrowicz}\ \emph {et~al.}(2019)\citenamefont
  {Aleksandrowicz}, \citenamefont {Alexander}, \citenamefont {Barkoutsos},
  \citenamefont {Bello}, \citenamefont {Ben-Haim}, \citenamefont {Bucher},
  \citenamefont {Cabrera-Hernández}, \citenamefont {Carballo-Franquis},
  \citenamefont {Chen}, \citenamefont {Chen}, \citenamefont {Chow},
  \citenamefont {Córcoles-Gonzales}, \citenamefont {Cross}, \citenamefont
  {Cross}, \citenamefont {Cruz-Benito}, \citenamefont {Culver}, \citenamefont
  {González}, \citenamefont {Torre}, \citenamefont {Ding}, \citenamefont
  {Dumitrescu}, \citenamefont {Duran}, \citenamefont {Eendebak}, \citenamefont
  {Everitt}, \citenamefont {Sertage}, \citenamefont {Frisch}, \citenamefont
  {Fuhrer}, \citenamefont {Gambetta}, \citenamefont {Gago}, \citenamefont
  {Gomez-Mosquera}, \citenamefont {Greenberg}, \citenamefont {Hamamura},
  \citenamefont {Havlicek}, \citenamefont {Hellmers}, \citenamefont {Łukasz
  Herok}, \citenamefont {Horii}, \citenamefont {Hu}, \citenamefont {Imamichi},
  \citenamefont {Itoko}, \citenamefont {Javadi-Abhari}, \citenamefont
  {Kanazawa}, \citenamefont {Karazeev}, \citenamefont {Krsulich}, \citenamefont
  {Liu}, \citenamefont {Luh}, \citenamefont {Maeng}, \citenamefont {Marques},
  \citenamefont {Martín-Fernández}, \citenamefont {McClure}, \citenamefont
  {McKay}, \citenamefont {Meesala}, \citenamefont {Mezzacapo}, \citenamefont
  {Moll}, \citenamefont {Rodríguez}, \citenamefont {Nannicini}, \citenamefont
  {Nation}, \citenamefont {Ollitrault}, \citenamefont {O'Riordan},
  \citenamefont {Paik}, \citenamefont {Pérez}, \citenamefont {Phan},
  \citenamefont {Pistoia}, \citenamefont {Prutyanov}, \citenamefont {Reuter},
  \citenamefont {Rice}, \citenamefont {Davila}, \citenamefont {Rudy},
  \citenamefont {Ryu}, \citenamefont {Sathaye}, \citenamefont {Schnabel},
  \citenamefont {Schoute}, \citenamefont {Setia}, \citenamefont {Shi},
  \citenamefont {Silva}, \citenamefont {Siraichi}, \citenamefont {Sivarajah},
  \citenamefont {Smolin}, \citenamefont {Soeken}, \citenamefont {Takahashi},
  \citenamefont {Tavernelli}, \citenamefont {Taylor}, \citenamefont {Taylour},
  \citenamefont {Trabing}, \citenamefont {Treinish}, \citenamefont {Turner},
  \citenamefont {Vogt-Lee}, \citenamefont {Vuillot}, \citenamefont {Wildstrom},
  \citenamefont {Wilson}, \citenamefont {Winston}, \citenamefont {Wood},
  \citenamefont {Wood}, \citenamefont {Wörner}, \citenamefont {Akhalwaya},\
  and\ \citenamefont {Zoufal}}]{gadi_aleksandrowicz_2019_2562111}%
  \BibitemOpen
  \bibfield  {author} {\bibinfo {author} {\bibfnamefont {G.}~\bibnamefont
  {Aleksandrowicz}}, \bibinfo {author} {\bibfnamefont {T.}~\bibnamefont
  {Alexander}}, \bibinfo {author} {\bibfnamefont {P.}~\bibnamefont
  {Barkoutsos}}, \bibinfo {author} {\bibfnamefont {L.}~\bibnamefont {Bello}},
  \bibinfo {author} {\bibfnamefont {Y.}~\bibnamefont {Ben-Haim}}, \bibinfo
  {author} {\bibfnamefont {D.}~\bibnamefont {Bucher}}, \bibinfo {author}
  {\bibfnamefont {F.~J.}\ \bibnamefont {Cabrera-Hernández}}, \bibinfo {author}
  {\bibfnamefont {J.}~\bibnamefont {Carballo-Franquis}}, \bibinfo {author}
  {\bibfnamefont {A.}~\bibnamefont {Chen}}, \bibinfo {author} {\bibfnamefont
  {C.-F.}\ \bibnamefont {Chen}}, \bibinfo {author} {\bibfnamefont {J.~M.}\
  \bibnamefont {Chow}}, \bibinfo {author} {\bibfnamefont {A.~D.}\ \bibnamefont
  {Córcoles-Gonzales}}, \bibinfo {author} {\bibfnamefont {A.~J.}\ \bibnamefont
  {Cross}}, \bibinfo {author} {\bibfnamefont {A.}~\bibnamefont {Cross}},
  \bibinfo {author} {\bibfnamefont {J.}~\bibnamefont {Cruz-Benito}}, \bibinfo
  {author} {\bibfnamefont {C.}~\bibnamefont {Culver}}, \bibinfo {author}
  {\bibfnamefont {S.~D. L.~P.}\ \bibnamefont {González}}, \bibinfo {author}
  {\bibfnamefont {E.~D.~L.}\ \bibnamefont {Torre}}, \bibinfo {author}
  {\bibfnamefont {D.}~\bibnamefont {Ding}}, \bibinfo {author} {\bibfnamefont
  {E.}~\bibnamefont {Dumitrescu}}, \bibinfo {author} {\bibfnamefont
  {I.}~\bibnamefont {Duran}}, \bibinfo {author} {\bibfnamefont
  {P.}~\bibnamefont {Eendebak}}, \bibinfo {author} {\bibfnamefont
  {M.}~\bibnamefont {Everitt}}, \bibinfo {author} {\bibfnamefont {I.~F.}\
  \bibnamefont {Sertage}}, \bibinfo {author} {\bibfnamefont {A.}~\bibnamefont
  {Frisch}}, \bibinfo {author} {\bibfnamefont {A.}~\bibnamefont {Fuhrer}},
  \bibinfo {author} {\bibfnamefont {J.}~\bibnamefont {Gambetta}}, \bibinfo
  {author} {\bibfnamefont {B.~G.}\ \bibnamefont {Gago}}, \bibinfo {author}
  {\bibfnamefont {J.}~\bibnamefont {Gomez-Mosquera}}, \bibinfo {author}
  {\bibfnamefont {D.}~\bibnamefont {Greenberg}}, \bibinfo {author}
  {\bibfnamefont {I.}~\bibnamefont {Hamamura}}, \bibinfo {author}
  {\bibfnamefont {V.}~\bibnamefont {Havlicek}}, \bibinfo {author}
  {\bibfnamefont {J.}~\bibnamefont {Hellmers}}, \bibinfo {author} {\bibnamefont
  {Łukasz Herok}}, \bibinfo {author} {\bibfnamefont {H.}~\bibnamefont
  {Horii}}, \bibinfo {author} {\bibfnamefont {S.}~\bibnamefont {Hu}}, \bibinfo
  {author} {\bibfnamefont {T.}~\bibnamefont {Imamichi}}, \bibinfo {author}
  {\bibfnamefont {T.}~\bibnamefont {Itoko}}, \bibinfo {author} {\bibfnamefont
  {A.}~\bibnamefont {Javadi-Abhari}}, \bibinfo {author} {\bibfnamefont
  {N.}~\bibnamefont {Kanazawa}}, \bibinfo {author} {\bibfnamefont
  {A.}~\bibnamefont {Karazeev}}, \bibinfo {author} {\bibfnamefont
  {K.}~\bibnamefont {Krsulich}}, \bibinfo {author} {\bibfnamefont
  {P.}~\bibnamefont {Liu}}, \bibinfo {author} {\bibfnamefont {Y.}~\bibnamefont
  {Luh}}, \bibinfo {author} {\bibfnamefont {Y.}~\bibnamefont {Maeng}}, \bibinfo
  {author} {\bibfnamefont {M.}~\bibnamefont {Marques}}, \bibinfo {author}
  {\bibfnamefont {F.~J.}\ \bibnamefont {Martín-Fernández}}, \bibinfo {author}
  {\bibfnamefont {D.~T.}\ \bibnamefont {McClure}}, \bibinfo {author}
  {\bibfnamefont {D.}~\bibnamefont {McKay}}, \bibinfo {author} {\bibfnamefont
  {S.}~\bibnamefont {Meesala}}, \bibinfo {author} {\bibfnamefont
  {A.}~\bibnamefont {Mezzacapo}}, \bibinfo {author} {\bibfnamefont
  {N.}~\bibnamefont {Moll}}, \bibinfo {author} {\bibfnamefont {D.~M.}\
  \bibnamefont {Rodríguez}}, \bibinfo {author} {\bibfnamefont
  {G.}~\bibnamefont {Nannicini}}, \bibinfo {author} {\bibfnamefont
  {P.}~\bibnamefont {Nation}}, \bibinfo {author} {\bibfnamefont
  {P.}~\bibnamefont {Ollitrault}}, \bibinfo {author} {\bibfnamefont {L.~J.}\
  \bibnamefont {O'Riordan}}, \bibinfo {author} {\bibfnamefont {H.}~\bibnamefont
  {Paik}}, \bibinfo {author} {\bibfnamefont {J.}~\bibnamefont {Pérez}},
  \bibinfo {author} {\bibfnamefont {A.}~\bibnamefont {Phan}}, \bibinfo {author}
  {\bibfnamefont {M.}~\bibnamefont {Pistoia}}, \bibinfo {author} {\bibfnamefont
  {V.}~\bibnamefont {Prutyanov}}, \bibinfo {author} {\bibfnamefont
  {M.}~\bibnamefont {Reuter}}, \bibinfo {author} {\bibfnamefont
  {J.}~\bibnamefont {Rice}}, \bibinfo {author} {\bibfnamefont {A.~R.}\
  \bibnamefont {Davila}}, \bibinfo {author} {\bibfnamefont {R.~H.~P.}\
  \bibnamefont {Rudy}}, \bibinfo {author} {\bibfnamefont {M.}~\bibnamefont
  {Ryu}}, \bibinfo {author} {\bibfnamefont {N.}~\bibnamefont {Sathaye}},
  \bibinfo {author} {\bibfnamefont {C.}~\bibnamefont {Schnabel}}, \bibinfo
  {author} {\bibfnamefont {E.}~\bibnamefont {Schoute}}, \bibinfo {author}
  {\bibfnamefont {K.}~\bibnamefont {Setia}}, \bibinfo {author} {\bibfnamefont
  {Y.}~\bibnamefont {Shi}}, \bibinfo {author} {\bibfnamefont {A.}~\bibnamefont
  {Silva}}, \bibinfo {author} {\bibfnamefont {Y.}~\bibnamefont {Siraichi}},
  \bibinfo {author} {\bibfnamefont {S.}~\bibnamefont {Sivarajah}}, \bibinfo
  {author} {\bibfnamefont {J.~A.}\ \bibnamefont {Smolin}}, \bibinfo {author}
  {\bibfnamefont {M.}~\bibnamefont {Soeken}}, \bibinfo {author} {\bibfnamefont
  {H.}~\bibnamefont {Takahashi}}, \bibinfo {author} {\bibfnamefont
  {I.}~\bibnamefont {Tavernelli}}, \bibinfo {author} {\bibfnamefont
  {C.}~\bibnamefont {Taylor}}, \bibinfo {author} {\bibfnamefont
  {P.}~\bibnamefont {Taylour}}, \bibinfo {author} {\bibfnamefont
  {K.}~\bibnamefont {Trabing}}, \bibinfo {author} {\bibfnamefont
  {M.}~\bibnamefont {Treinish}}, \bibinfo {author} {\bibfnamefont
  {W.}~\bibnamefont {Turner}}, \bibinfo {author} {\bibfnamefont
  {D.}~\bibnamefont {Vogt-Lee}}, \bibinfo {author} {\bibfnamefont
  {C.}~\bibnamefont {Vuillot}}, \bibinfo {author} {\bibfnamefont {J.~A.}\
  \bibnamefont {Wildstrom}}, \bibinfo {author} {\bibfnamefont {J.}~\bibnamefont
  {Wilson}}, \bibinfo {author} {\bibfnamefont {E.}~\bibnamefont {Winston}},
  \bibinfo {author} {\bibfnamefont {C.}~\bibnamefont {Wood}}, \bibinfo {author}
  {\bibfnamefont {S.}~\bibnamefont {Wood}}, \bibinfo {author} {\bibfnamefont
  {S.}~\bibnamefont {Wörner}}, \bibinfo {author} {\bibfnamefont {I.~Y.}\
  \bibnamefont {Akhalwaya}}, \ and\ \bibinfo {author} {\bibfnamefont
  {C.}~\bibnamefont {Zoufal}},\ }\href {\doibase 10.5281/zenodo.2562111}
  {\enquote {\bibinfo {title} {{Qiskit: An Open-source Framework for Quantum
  Computing}},}\ } (\bibinfo {year} {2019})\BibitemShut {NoStop}%
\bibitem [{\citenamefont {Alexander}\ \emph {et~al.}(2020)\citenamefont
  {Alexander}, \citenamefont {Kanazawa}, \citenamefont {Egger}, \citenamefont
  {Capelluto}, \citenamefont {Wood}, \citenamefont {Javadi-Abhari},\ and\
  \citenamefont {C~McKay}}]{Alexander_2020}%
  \BibitemOpen
  \bibfield  {author} {\bibinfo {author} {\bibfnamefont {T.}~\bibnamefont
  {Alexander}}, \bibinfo {author} {\bibfnamefont {N.}~\bibnamefont {Kanazawa}},
  \bibinfo {author} {\bibfnamefont {D.~J.}\ \bibnamefont {Egger}}, \bibinfo
  {author} {\bibfnamefont {L.}~\bibnamefont {Capelluto}}, \bibinfo {author}
  {\bibfnamefont {C.~J.}\ \bibnamefont {Wood}}, \bibinfo {author}
  {\bibfnamefont {A.}~\bibnamefont {Javadi-Abhari}}, \ and\ \bibinfo {author}
  {\bibfnamefont {D.}~\bibnamefont {C~McKay}},\ }\href {\doibase
  10.1088/2058-9565/aba404} {\bibfield  {journal} {\bibinfo  {journal} {Quantum
  Science and Technology}\ }\textbf {\bibinfo {volume} {5}},\ \bibinfo {pages}
  {044006} (\bibinfo {year} {2020})}\BibitemShut {NoStop}%
\bibitem [{\citenamefont {Cross}\ \emph {et~al.}(2017)\citenamefont {Cross},
  \citenamefont {Bishop}, \citenamefont {Smolin},\ and\ \citenamefont
  {Gambetta}}]{cross2017open}%
  \BibitemOpen
  \bibfield  {author} {\bibinfo {author} {\bibfnamefont {A.~W.}\ \bibnamefont
  {Cross}}, \bibinfo {author} {\bibfnamefont {L.~S.}\ \bibnamefont {Bishop}},
  \bibinfo {author} {\bibfnamefont {J.~A.}\ \bibnamefont {Smolin}}, \ and\
  \bibinfo {author} {\bibfnamefont {J.~M.}\ \bibnamefont {Gambetta}},\
  }\href@noop {} {\enquote {\bibinfo {title} {Open quantum assembly
  language},}\ } (\bibinfo {year} {2017}),\ \Eprint
  {http://arxiv.org/abs/1707.03429} {arXiv:1707.03429 [quant-ph]} \BibitemShut
  {NoStop}%
\bibitem [{\citenamefont {McCaskey}\ \emph {et~al.}(2020)\citenamefont
  {McCaskey}, \citenamefont {Lyakh}, \citenamefont {Dumitrescu}, \citenamefont
  {Powers},\ and\ \citenamefont {Humble}}]{McCaskey_2020}%
  \BibitemOpen
  \bibfield  {author} {\bibinfo {author} {\bibfnamefont {A.~J.}\ \bibnamefont
  {McCaskey}}, \bibinfo {author} {\bibfnamefont {D.~I.}\ \bibnamefont {Lyakh}},
  \bibinfo {author} {\bibfnamefont {E.~F.}\ \bibnamefont {Dumitrescu}},
  \bibinfo {author} {\bibfnamefont {S.~S.}\ \bibnamefont {Powers}}, \ and\
  \bibinfo {author} {\bibfnamefont {T.~S.}\ \bibnamefont {Humble}},\ }\href
  {\doibase 10.1088/2058-9565/ab6bf6} {\bibfield  {journal} {\bibinfo
  {journal} {Quantum Science and Technology}\ }\textbf {\bibinfo {volume}
  {5}},\ \bibinfo {pages} {024002} (\bibinfo {year} {2020})}\BibitemShut
  {NoStop}%
\bibitem [{\citenamefont {{Microsoft Q\# Language Design Team}}()}]{Qsharp}%
  \BibitemOpen
  \bibfield  {author} {\bibinfo {author} {\bibnamefont {{Microsoft Q\# Language
  Design Team}}},\ }\href@noop {} {\enquote {\bibinfo {title} {Q\#},}\
  }\bibinfo {howpublished}
  {\url{https://github.com/microsoft/qsharp-language}}\BibitemShut {NoStop}%
\bibitem [{\citenamefont {Li}\ and\ \citenamefont
  {Benjamin}(2017)}]{PhysRevX.7.021050}%
  \BibitemOpen
  \bibfield  {author} {\bibinfo {author} {\bibfnamefont {Y.}~\bibnamefont
  {Li}}\ and\ \bibinfo {author} {\bibfnamefont {S.~C.}\ \bibnamefont
  {Benjamin}},\ }\href {\doibase 10.1103/PhysRevX.7.021050} {\bibfield
  {journal} {\bibinfo  {journal} {Phys. Rev. X}\ }\textbf {\bibinfo {volume}
  {7}},\ \bibinfo {pages} {021050} (\bibinfo {year} {2017})}\BibitemShut
  {NoStop}%
\bibitem [{\citenamefont {Temme}\ \emph {et~al.}(2017)\citenamefont {Temme},
  \citenamefont {Bravyi},\ and\ \citenamefont
  {Gambetta}}]{PhysRevLett.119.180509}%
  \BibitemOpen
  \bibfield  {author} {\bibinfo {author} {\bibfnamefont {K.}~\bibnamefont
  {Temme}}, \bibinfo {author} {\bibfnamefont {S.}~\bibnamefont {Bravyi}}, \
  and\ \bibinfo {author} {\bibfnamefont {J.~M.}\ \bibnamefont {Gambetta}},\
  }\href {\doibase 10.1103/PhysRevLett.119.180509} {\bibfield  {journal}
  {\bibinfo  {journal} {Phys. Rev. Lett.}\ }\textbf {\bibinfo {volume} {119}},\
  \bibinfo {pages} {180509} (\bibinfo {year} {2017})}\BibitemShut {NoStop}%
\bibitem [{\citenamefont {Kandala}\ \emph {et~al.}(2019)\citenamefont
  {Kandala}, \citenamefont {Temme}, \citenamefont {Córcoles}, \citenamefont
  {Mezzacapo}, \citenamefont {Chow},\ and\ \citenamefont
  {Gambetta}}]{Kandala_2019}%
  \BibitemOpen
  \bibfield  {author} {\bibinfo {author} {\bibfnamefont {A.}~\bibnamefont
  {Kandala}}, \bibinfo {author} {\bibfnamefont {K.}~\bibnamefont {Temme}},
  \bibinfo {author} {\bibfnamefont {A.~D.}\ \bibnamefont {Córcoles}}, \bibinfo
  {author} {\bibfnamefont {A.}~\bibnamefont {Mezzacapo}}, \bibinfo {author}
  {\bibfnamefont {J.~M.}\ \bibnamefont {Chow}}, \ and\ \bibinfo {author}
  {\bibfnamefont {J.~M.}\ \bibnamefont {Gambetta}},\ }\href {\doibase
  10.1038/s41586-019-1040-7} {\bibfield  {journal} {\bibinfo  {journal}
  {Nature}\ }\textbf {\bibinfo {volume} {567}},\ \bibinfo {pages} {491–495}
  (\bibinfo {year} {2019})},\ \Eprint {http://arxiv.org/abs/1805.04492}
  {arXiv:1805.04492 [quant-ph]} \BibitemShut {NoStop}%
\bibitem [{\citenamefont {Endo}\ \emph {et~al.}(2018)\citenamefont {Endo},
  \citenamefont {Benjamin},\ and\ \citenamefont {Li}}]{PhysRevX.8.031027}%
  \BibitemOpen
  \bibfield  {author} {\bibinfo {author} {\bibfnamefont {S.}~\bibnamefont
  {Endo}}, \bibinfo {author} {\bibfnamefont {S.~C.}\ \bibnamefont {Benjamin}},
  \ and\ \bibinfo {author} {\bibfnamefont {Y.}~\bibnamefont {Li}},\ }\href
  {\doibase 10.1103/PhysRevX.8.031027} {\bibfield  {journal} {\bibinfo
  {journal} {Phys. Rev. X}\ }\textbf {\bibinfo {volume} {8}},\ \bibinfo {pages}
  {031027} (\bibinfo {year} {2018})}\BibitemShut {NoStop}%
\bibitem [{\citenamefont {Watrous}(2008)}]{watrous2008quantum}%
  \BibitemOpen
  \bibfield  {author} {\bibinfo {author} {\bibfnamefont {J.}~\bibnamefont
  {Watrous}},\ }\href@noop {} {\enquote {\bibinfo {title} {Quantum
  computational complexity},}\ } (\bibinfo {year} {2008}),\ \Eprint
  {http://arxiv.org/abs/0804.3401} {arXiv:0804.3401 [quant-ph]} \BibitemShut
  {NoStop}%
\bibitem [{\citenamefont {Bernstein}\ and\ \citenamefont
  {Vazirani}(1993)}]{10.1145/167088.167097}%
  \BibitemOpen
  \bibfield  {author} {\bibinfo {author} {\bibfnamefont {E.}~\bibnamefont
  {Bernstein}}\ and\ \bibinfo {author} {\bibfnamefont {U.}~\bibnamefont
  {Vazirani}},\ }in\ \href {\doibase 10.1145/167088.167097} {\emph {\bibinfo
  {booktitle} {Proceedings of the Twenty-Fifth Annual ACM Symposium on Theory
  of Computing}}},\ \bibinfo {series and number} {STOC '93}\ (\bibinfo
  {publisher} {Association for Computing Machinery},\ \bibinfo {address} {New
  York, NY, USA},\ \bibinfo {year} {1993})\ p.\ \bibinfo {pages}
  {11–20}\BibitemShut {NoStop}%
\bibitem [{\citenamefont {Cleve}()}]{doi:10.1142/9789810248185_0004}%
  \BibitemOpen
  \bibfield  {author} {\bibinfo {author} {\bibfnamefont {R.}~\bibnamefont
  {Cleve}},\ }\enquote {\bibinfo {title} {An introduction to quantum complexity
  theory},}\ in\ \href {\doibase 10.1142/9789810248185_0004} {\emph {\bibinfo
  {booktitle} {Quantum Computation and Quantum Information Theory}}},\ pp.\
  \bibinfo {pages} {103--127},\ \Eprint
  {http://arxiv.org/abs/https://www.worldscientific.com/doi/pdf/10.1142/9789810248185\_0004}
  {https://www.worldscientific.com/doi/pdf/10.1142/9789810248185\_0004}
  \BibitemShut {NoStop}%
\bibitem [{com(2013)}]{complexityhighlevel}%
  \BibitemOpen
  \href {https://dblp.org/db/journals/eccc/eccc20.html} {\enquote {\bibinfo
  {title} {{Electronic Colloquium on Computational Complexity}},}\ } (\bibinfo
  {year} {2013})\BibitemShut {NoStop}%
\bibitem [{\citenamefont {Jordan}\ \emph {et~al.}(2018)\citenamefont {Jordan},
  \citenamefont {Krovi}, \citenamefont {Lee},\ and\ \citenamefont
  {Preskill}}]{Jordan_2018}%
  \BibitemOpen
  \bibfield  {author} {\bibinfo {author} {\bibfnamefont {S.~P.}\ \bibnamefont
  {Jordan}}, \bibinfo {author} {\bibfnamefont {H.}~\bibnamefont {Krovi}},
  \bibinfo {author} {\bibfnamefont {K.~S.~M.}\ \bibnamefont {Lee}}, \ and\
  \bibinfo {author} {\bibfnamefont {J.}~\bibnamefont {Preskill}},\ }\href
  {\doibase 10.22331/q-2018-01-08-44} {\bibfield  {journal} {\bibinfo
  {journal} {Quantum}\ }\textbf {\bibinfo {volume} {2}},\ \bibinfo {pages} {44}
  (\bibinfo {year} {2018})}\BibitemShut {NoStop}%
\bibitem [{\citenamefont {Barahona}(1982)}]{Barahona_1982}%
  \BibitemOpen
  \bibfield  {author} {\bibinfo {author} {\bibfnamefont {F.}~\bibnamefont
  {Barahona}},\ }\href {\doibase 10.1088/0305-4470/15/10/028} {\bibfield
  {journal} {\bibinfo  {journal} {Journal of Physics A: Mathematical and
  General}\ }\textbf {\bibinfo {volume} {15}},\ \bibinfo {pages} {3241}
  (\bibinfo {year} {1982})}\BibitemShut {NoStop}%
\bibitem [{\citenamefont {Istrail}(2000)}]{10.1145/335305.335316}%
  \BibitemOpen
  \bibfield  {author} {\bibinfo {author} {\bibfnamefont {S.}~\bibnamefont
  {Istrail}},\ }in\ \href {\doibase 10.1145/335305.335316} {\emph {\bibinfo
  {booktitle} {Proceedings of the Thirty-Second Annual ACM Symposium on Theory
  of Computing}}},\ \bibinfo {series and number} {STOC '00}\ (\bibinfo
  {publisher} {Association for Computing Machinery},\ \bibinfo {address} {New
  York, NY, USA},\ \bibinfo {year} {2000})\ p.\ \bibinfo {pages}
  {87–96}\BibitemShut {NoStop}%
\bibitem [{\citenamefont {Troyer}\ and\ \citenamefont
  {Wiese}(2005)}]{Troyer_2005}%
  \BibitemOpen
  \bibfield  {author} {\bibinfo {author} {\bibfnamefont {M.}~\bibnamefont
  {Troyer}}\ and\ \bibinfo {author} {\bibfnamefont {U.-J.}\ \bibnamefont
  {Wiese}},\ }\href {\doibase 10.1103/physrevlett.94.170201} {\bibfield
  {journal} {\bibinfo  {journal} {Physical Review Letters}\ }\textbf {\bibinfo
  {volume} {94}} (\bibinfo {year} {2005}),\
  10.1103/physrevlett.94.170201}\BibitemShut {NoStop}%
\bibitem [{\citenamefont {Kempe}\ \emph {et~al.}(2005)\citenamefont {Kempe},
  \citenamefont {Kitaev},\ and\ \citenamefont
  {Regev}}]{10.1007/978-3-540-30538-5_31}%
  \BibitemOpen
  \bibfield  {author} {\bibinfo {author} {\bibfnamefont {J.}~\bibnamefont
  {Kempe}}, \bibinfo {author} {\bibfnamefont {A.}~\bibnamefont {Kitaev}}, \
  and\ \bibinfo {author} {\bibfnamefont {O.}~\bibnamefont {Regev}},\ }in\
  \href@noop {} {\emph {\bibinfo {booktitle} {FSTTCS 2004: Foundations of
  Software Technology and Theoretical Computer Science}}},\ \bibinfo {editor}
  {edited by\ \bibinfo {editor} {\bibfnamefont {K.}~\bibnamefont {Lodaya}}\
  and\ \bibinfo {editor} {\bibfnamefont {M.}~\bibnamefont {Mahajan}}}\
  (\bibinfo  {publisher} {Springer Berlin Heidelberg},\ \bibinfo {address}
  {Berlin, Heidelberg},\ \bibinfo {year} {2005})\ pp.\ \bibinfo {pages}
  {372--383}\BibitemShut {NoStop}%
\bibitem [{\citenamefont {Kaye}(2000)}]{Kaye2000}%
  \BibitemOpen
  \bibfield  {author} {\bibinfo {author} {\bibfnamefont {R.}~\bibnamefont
  {Kaye}},\ }\href {\doibase 10.1007/BF03025367} {\bibfield  {journal}
  {\bibinfo  {journal} {The Mathematical Intelligencer}\ }\textbf {\bibinfo
  {volume} {22}},\ \bibinfo {pages} {9} (\bibinfo {year} {2000})}\BibitemShut
  {NoStop}%
\bibitem [{\citenamefont {Scott}\ \emph {et~al.}(2011)\citenamefont {Scott},
  \citenamefont {Stege},\ and\ \citenamefont {van Rooij}}]{Scott2011}%
  \BibitemOpen
  \bibfield  {author} {\bibinfo {author} {\bibfnamefont {A.}~\bibnamefont
  {Scott}}, \bibinfo {author} {\bibfnamefont {U.}~\bibnamefont {Stege}}, \ and\
  \bibinfo {author} {\bibfnamefont {I.}~\bibnamefont {van Rooij}},\ }\href
  {\doibase 10.1007/s00283-011-9256-x} {\bibfield  {journal} {\bibinfo
  {journal} {The Mathematical Intelligencer}\ }\textbf {\bibinfo {volume}
  {33}},\ \bibinfo {pages} {5} (\bibinfo {year} {2011})}\BibitemShut {NoStop}%
\bibitem [{\citenamefont {Dempsey}\ and\ \citenamefont
  {Guinn}(2020)}]{dempsey_et_al:LIPIcs:2020:12773}%
  \BibitemOpen
  \bibfield  {author} {\bibinfo {author} {\bibfnamefont {R.}~\bibnamefont
  {Dempsey}}\ and\ \bibinfo {author} {\bibfnamefont {C.}~\bibnamefont
  {Guinn}},\ }in\ \href {\doibase 10.4230/LIPIcs.FUN.2021.12} {\emph {\bibinfo
  {booktitle} {10th International Conference on Fun with Algorithms (FUN
  2021)}}},\ \bibinfo {series} {Leibniz International Proceedings in
  Informatics (LIPIcs)}, Vol.\ \bibinfo {volume} {157},\ \bibinfo {editor}
  {edited by\ \bibinfo {editor} {\bibfnamefont {M.}~\bibnamefont
  {Farach-Colton}}, \bibinfo {editor} {\bibfnamefont {G.}~\bibnamefont
  {Prencipe}}, \ and\ \bibinfo {editor} {\bibfnamefont {R.}~\bibnamefont
  {Uehara}}}\ (\bibinfo  {publisher} {Schloss Dagstuhl--Leibniz-Zentrum f{\"u}r
  Informatik},\ \bibinfo {address} {Dagstuhl, Germany},\ \bibinfo {year}
  {2020})\ pp.\ \bibinfo {pages} {12:1--12:10}\BibitemShut {NoStop}%
\bibitem [{\citenamefont {{Monasson}}\ \emph {et~al.}(1999)\citenamefont
  {{Monasson}}, \citenamefont {{Zecchina}}, \citenamefont {{Kirkpatrick}},
  \citenamefont {{Selman}},\ and\ \citenamefont
  {{Troyansky}}}]{1999Natur.400..133M}%
  \BibitemOpen
  \bibfield  {author} {\bibinfo {author} {\bibfnamefont {R.}~\bibnamefont
  {{Monasson}}}, \bibinfo {author} {\bibfnamefont {R.}~\bibnamefont
  {{Zecchina}}}, \bibinfo {author} {\bibfnamefont {S.}~\bibnamefont
  {{Kirkpatrick}}}, \bibinfo {author} {\bibfnamefont {B.}~\bibnamefont
  {{Selman}}}, \ and\ \bibinfo {author} {\bibfnamefont {L.}~\bibnamefont
  {{Troyansky}}},\ }\href {\doibase 10.1038/22055} {\bibfield  {journal}
  {\bibinfo  {journal} {Nature}\ }\textbf {\bibinfo {volume} {400}},\ \bibinfo
  {pages} {133} (\bibinfo {year} {1999})}\BibitemShut {NoStop}%
\bibitem [{\citenamefont {Beane}\ \emph {et~al.}(2015)\citenamefont {Beane},
  \citenamefont {Detmold}, \citenamefont {Orginos},\ and\ \citenamefont
  {Savage}}]{Beane:2014oea}%
  \BibitemOpen
  \bibfield  {author} {\bibinfo {author} {\bibfnamefont {S.~R.}\ \bibnamefont
  {Beane}}, \bibinfo {author} {\bibfnamefont {W.}~\bibnamefont {Detmold}},
  \bibinfo {author} {\bibfnamefont {K.}~\bibnamefont {Orginos}}, \ and\
  \bibinfo {author} {\bibfnamefont {M.~J.}\ \bibnamefont {Savage}},\ }\href
  {\doibase 10.1088/0954-3899/42/3/034022} {\bibfield  {journal} {\bibinfo
  {journal} {J. Phys. G}\ }\textbf {\bibinfo {volume} {42}},\ \bibinfo {pages}
  {034022} (\bibinfo {year} {2015})},\ \Eprint {http://arxiv.org/abs/1410.2937}
  {arXiv:1410.2937 [nucl-th]} \BibitemShut {NoStop}%
\bibitem [{\citenamefont {Golterman}()}]{Golterman}%
  \BibitemOpen
  \bibfield  {author} {\bibinfo {author} {\bibfnamefont {M.}~\bibnamefont
  {Golterman}},\ }\href {\doibase 10.1007/bfb0104897} {\bibinfo  {journal}
  {Lecture Notes in Physics}\ ,\ \bibinfo {pages} {46–61}}\BibitemShut
  {NoStop}%
\bibitem [{\citenamefont {Savage}(2006)}]{Savage:2006et}%
  \BibitemOpen
\bibfield  {journal} {  }\bibfield  {author} {\bibinfo {author} {\bibfnamefont
  {M.~J.}\ \bibnamefont {Savage}},\ }\href {\doibase 10.1063/1.2220225}
  {\bibfield  {journal} {\bibinfo  {journal} {AIP Conf. Proc.}\ }\textbf
  {\bibinfo {volume} {842}},\ \bibinfo {pages} {177} (\bibinfo {year}
  {2006})},\ \Eprint {http://arxiv.org/abs/nucl-th/0601001}
  {arXiv:nucl-th/0601001} \BibitemShut {NoStop}%
\bibitem [{\citenamefont {Symanzik}(1983{\natexlab{a}})}]{Symanzik:1983dc}%
  \BibitemOpen
  \bibfield  {author} {\bibinfo {author} {\bibfnamefont {K.}~\bibnamefont
  {Symanzik}},\ }\href {\doibase 10.1016/0550-3213(83)90468-6} {\bibfield
  {journal} {\bibinfo  {journal} {Nucl. Phys.}\ }\textbf {\bibinfo {volume}
  {B226}},\ \bibinfo {pages} {187} (\bibinfo {year}
  {1983}{\natexlab{a}})}\BibitemShut {NoStop}%
\bibitem [{\citenamefont {Symanzik}(1983{\natexlab{b}})}]{Symanzik:1983gh}%
  \BibitemOpen
  \bibfield  {author} {\bibinfo {author} {\bibfnamefont {K.}~\bibnamefont
  {Symanzik}},\ }\href {\doibase 10.1016/0550-3213(83)90469-8} {\bibfield
  {journal} {\bibinfo  {journal} {Nucl. Phys.}\ }\textbf {\bibinfo {volume}
  {B226}},\ \bibinfo {pages} {205} (\bibinfo {year}
  {1983}{\natexlab{b}})}\BibitemShut {NoStop}%
\bibitem [{\citenamefont {Roggero}\ and\ \citenamefont
  {Baroni}(2020)}]{PhysRevA.101.022328}%
  \BibitemOpen
  \bibfield  {author} {\bibinfo {author} {\bibfnamefont {A.}~\bibnamefont
  {Roggero}}\ and\ \bibinfo {author} {\bibfnamefont {A.}~\bibnamefont
  {Baroni}},\ }\href {\doibase 10.1103/PhysRevA.101.022328} {\bibfield
  {journal} {\bibinfo  {journal} {Phys. Rev. A}\ }\textbf {\bibinfo {volume}
  {101}},\ \bibinfo {pages} {022328} (\bibinfo {year} {2020})}\BibitemShut
  {NoStop}%
\bibitem [{\citenamefont {Colò}(2020)}]{doi:10.1080/23746149.2020.1740061}%
  \BibitemOpen
  \bibfield  {author} {\bibinfo {author} {\bibfnamefont {G.}~\bibnamefont
  {Colò}},\ }\href {\doibase 10.1080/23746149.2020.1740061} {\bibfield
  {journal} {\bibinfo  {journal} {Advances in Physics: X}\ }\textbf {\bibinfo
  {volume} {5}},\ \bibinfo {pages} {1740061} (\bibinfo {year}
  {2020})}\BibitemShut {NoStop}%
\bibitem [{\citenamefont {Hohenberg}\ and\ \citenamefont
  {Kohn}(1964)}]{PhysRev.136.B864}%
  \BibitemOpen
  \bibfield  {author} {\bibinfo {author} {\bibfnamefont {P.}~\bibnamefont
  {Hohenberg}}\ and\ \bibinfo {author} {\bibfnamefont {W.}~\bibnamefont
  {Kohn}},\ }\href {\doibase 10.1103/PhysRev.136.B864} {\bibfield  {journal}
  {\bibinfo  {journal} {Phys. Rev.}\ }\textbf {\bibinfo {volume} {136}},\
  \bibinfo {pages} {B864} (\bibinfo {year} {1964})}\BibitemShut {NoStop}%
\bibitem [{\citenamefont {Schuch}\ and\ \citenamefont
  {Verstraete}(2009)}]{Schuch_2009}%
  \BibitemOpen
  \bibfield  {author} {\bibinfo {author} {\bibfnamefont {N.}~\bibnamefont
  {Schuch}}\ and\ \bibinfo {author} {\bibfnamefont {F.}~\bibnamefont
  {Verstraete}},\ }\href {\doibase 10.1038/nphys1370} {\bibfield  {journal}
  {\bibinfo  {journal} {Nature Physics}\ }\textbf {\bibinfo {volume} {5}},\
  \bibinfo {pages} {732–735} (\bibinfo {year} {2009})}\BibitemShut {NoStop}%
\bibitem [{\citenamefont {Wigner}(1937{\natexlab{a}})}]{PhysRev.51.106}%
  \BibitemOpen
  \bibfield  {author} {\bibinfo {author} {\bibfnamefont {E.}~\bibnamefont
  {Wigner}},\ }\href {\doibase 10.1103/PhysRev.51.106} {\bibfield  {journal}
  {\bibinfo  {journal} {Phys. Rev.}\ }\textbf {\bibinfo {volume} {51}},\
  \bibinfo {pages} {106} (\bibinfo {year} {1937}{\natexlab{a}})}\BibitemShut
  {NoStop}%
\bibitem [{\citenamefont {Wigner}(1937{\natexlab{b}})}]{PhysRev.51.947}%
  \BibitemOpen
  \bibfield  {author} {\bibinfo {author} {\bibfnamefont {E.}~\bibnamefont
  {Wigner}},\ }\href {\doibase 10.1103/PhysRev.51.947} {\bibfield  {journal}
  {\bibinfo  {journal} {Phys. Rev.}\ }\textbf {\bibinfo {volume} {51}},\
  \bibinfo {pages} {947} (\bibinfo {year} {1937}{\natexlab{b}})}\BibitemShut
  {NoStop}%
\bibitem [{\citenamefont {Kaplan}\ and\ \citenamefont
  {Savage}(1996)}]{Kaplan:1995yg}%
  \BibitemOpen
  \bibfield  {author} {\bibinfo {author} {\bibfnamefont {D.~B.}\ \bibnamefont
  {Kaplan}}\ and\ \bibinfo {author} {\bibfnamefont {M.~J.}\ \bibnamefont
  {Savage}},\ }\href {\doibase 10.1016/0370-2693(95)01277-X} {\bibfield
  {journal} {\bibinfo  {journal} {Phys. Lett. B}\ }\textbf {\bibinfo {volume}
  {365}},\ \bibinfo {pages} {244} (\bibinfo {year} {1996})},\ \Eprint
  {http://arxiv.org/abs/hep-ph/9509371} {arXiv:hep-ph/9509371} \BibitemShut
  {NoStop}%
\bibitem [{\citenamefont {Lee}(2016)}]{Lee:2016sxr}%
  \BibitemOpen
  \bibfield  {author} {\bibinfo {author} {\bibfnamefont {D.}~\bibnamefont
  {Lee}},\ }\href {\doibase 10.1142/S021830131641010X} {\bibfield  {journal}
  {\bibinfo  {journal} {Int. J. Mod. Phys. E}\ }\textbf {\bibinfo {volume}
  {25}},\ \bibinfo {pages} {1641010} (\bibinfo {year} {2016})}\BibitemShut
  {NoStop}%
\bibitem [{\citenamefont {Carlson}\ \emph {et~al.}(2015)\citenamefont
  {Carlson}, \citenamefont {Gandolfi}, \citenamefont {Pederiva}, \citenamefont
  {Pieper}, \citenamefont {Schiavilla}, \citenamefont {Schmidt},\ and\
  \citenamefont {Wiringa}}]{RevModPhys.87.1067}%
  \BibitemOpen
  \bibfield  {author} {\bibinfo {author} {\bibfnamefont {J.}~\bibnamefont
  {Carlson}}, \bibinfo {author} {\bibfnamefont {S.}~\bibnamefont {Gandolfi}},
  \bibinfo {author} {\bibfnamefont {F.}~\bibnamefont {Pederiva}}, \bibinfo
  {author} {\bibfnamefont {S.~C.}\ \bibnamefont {Pieper}}, \bibinfo {author}
  {\bibfnamefont {R.}~\bibnamefont {Schiavilla}}, \bibinfo {author}
  {\bibfnamefont {K.~E.}\ \bibnamefont {Schmidt}}, \ and\ \bibinfo {author}
  {\bibfnamefont {R.~B.}\ \bibnamefont {Wiringa}},\ }\href {\doibase
  10.1103/RevModPhys.87.1067} {\bibfield  {journal} {\bibinfo  {journal} {Rev.
  Mod. Phys.}\ }\textbf {\bibinfo {volume} {87}},\ \bibinfo {pages} {1067}
  (\bibinfo {year} {2015})}\BibitemShut {NoStop}%
\bibitem [{\citenamefont {Weinberg}(1990)}]{Weinberg:1990rz}%
  \BibitemOpen
  \bibfield  {author} {\bibinfo {author} {\bibfnamefont {S.}~\bibnamefont
  {Weinberg}},\ }\href {\doibase 10.1016/0370-2693(90)90938-3} {\bibfield
  {journal} {\bibinfo  {journal} {Phys. Lett. B}\ }\textbf {\bibinfo {volume}
  {251}},\ \bibinfo {pages} {288} (\bibinfo {year} {1990})}\BibitemShut
  {NoStop}%
\bibitem [{\citenamefont {Weinberg}(1991)}]{Weinberg:1991um}%
  \BibitemOpen
  \bibfield  {author} {\bibinfo {author} {\bibfnamefont {S.}~\bibnamefont
  {Weinberg}},\ }\href {\doibase 10.1016/0550-3213(91)90231-L} {\bibfield
  {journal} {\bibinfo  {journal} {Nucl. Phys. B}\ }\textbf {\bibinfo {volume}
  {363}},\ \bibinfo {pages} {3} (\bibinfo {year} {1991})}\BibitemShut {NoStop}%
\bibitem [{\citenamefont {Ordonez}\ \emph {et~al.}(1996)\citenamefont
  {Ordonez}, \citenamefont {Ray},\ and\ \citenamefont {van
  Kolck}}]{Ordonez:1995rz}%
  \BibitemOpen
  \bibfield  {author} {\bibinfo {author} {\bibfnamefont {C.}~\bibnamefont
  {Ordonez}}, \bibinfo {author} {\bibfnamefont {L.}~\bibnamefont {Ray}}, \ and\
  \bibinfo {author} {\bibfnamefont {U.}~\bibnamefont {van Kolck}},\ }\href
  {\doibase 10.1103/PhysRevC.53.2086} {\bibfield  {journal} {\bibinfo
  {journal} {Phys. Rev. C}\ }\textbf {\bibinfo {volume} {53}},\ \bibinfo
  {pages} {2086} (\bibinfo {year} {1996})},\ \Eprint
  {http://arxiv.org/abs/hep-ph/9511380} {arXiv:hep-ph/9511380} \BibitemShut
  {NoStop}%
\bibitem [{\citenamefont {Kaplan}\ \emph {et~al.}(1996)\citenamefont {Kaplan},
  \citenamefont {Savage},\ and\ \citenamefont {Wise}}]{Kaplan:1996xu}%
  \BibitemOpen
  \bibfield  {author} {\bibinfo {author} {\bibfnamefont {D.~B.}\ \bibnamefont
  {Kaplan}}, \bibinfo {author} {\bibfnamefont {M.~J.}\ \bibnamefont {Savage}},
  \ and\ \bibinfo {author} {\bibfnamefont {M.~B.}\ \bibnamefont {Wise}},\
  }\href {\doibase 10.1016/0550-3213(96)00357-4} {\bibfield  {journal}
  {\bibinfo  {journal} {Nucl. Phys. B}\ }\textbf {\bibinfo {volume} {478}},\
  \bibinfo {pages} {629} (\bibinfo {year} {1996})},\ \Eprint
  {http://arxiv.org/abs/nucl-th/9605002} {arXiv:nucl-th/9605002} \BibitemShut
  {NoStop}%
\bibitem [{\citenamefont {Epelbaum}\ \emph {et~al.}(1998)\citenamefont
  {Epelbaum}, \citenamefont {Gloeckle},\ and\ \citenamefont
  {Meissner}}]{Epelbaum:1998hg}%
  \BibitemOpen
  \bibfield  {author} {\bibinfo {author} {\bibfnamefont {E.}~\bibnamefont
  {Epelbaum}}, \bibinfo {author} {\bibfnamefont {W.}~\bibnamefont {Gloeckle}},
  \ and\ \bibinfo {author} {\bibfnamefont {U.-G.}\ \bibnamefont {Meissner}},\
  }\href {\doibase 10.1016/S0370-2693(98)01045-4} {\bibfield  {journal}
  {\bibinfo  {journal} {Phys. Lett. B}\ }\textbf {\bibinfo {volume} {439}},\
  \bibinfo {pages} {1} (\bibinfo {year} {1998})},\ \Eprint
  {http://arxiv.org/abs/nucl-th/9804005} {arXiv:nucl-th/9804005} \BibitemShut
  {NoStop}%
\bibitem [{\citenamefont {Machleidt}\ and\ \citenamefont
  {Entem}(2011)}]{Machleidt_2011}%
  \BibitemOpen
  \bibfield  {author} {\bibinfo {author} {\bibfnamefont {R.}~\bibnamefont
  {Machleidt}}\ and\ \bibinfo {author} {\bibfnamefont {D.}~\bibnamefont
  {Entem}},\ }\href {\doibase 10.1016/j.physrep.2011.02.001} {\bibfield
  {journal} {\bibinfo  {journal} {Physics Reports}\ }\textbf {\bibinfo {volume}
  {503}},\ \bibinfo {pages} {1–75} (\bibinfo {year} {2011})}\BibitemShut
  {NoStop}%
\bibitem [{\citenamefont {Bogner}\ \emph {et~al.}(2003)\citenamefont {Bogner},
  \citenamefont {Kuo},\ and\ \citenamefont {Schwenk}}]{Bogner:2003wn}%
  \BibitemOpen
  \bibfield  {author} {\bibinfo {author} {\bibfnamefont {S.~K.}\ \bibnamefont
  {Bogner}}, \bibinfo {author} {\bibfnamefont {T.~T.~S.}\ \bibnamefont {Kuo}},
  \ and\ \bibinfo {author} {\bibfnamefont {A.}~\bibnamefont {Schwenk}},\ }\href
  {\doibase 10.1016/j.physrep.2003.07.001} {\bibfield  {journal} {\bibinfo
  {journal} {Phys. Rept.}\ }\textbf {\bibinfo {volume} {386}},\ \bibinfo
  {pages} {1} (\bibinfo {year} {2003})},\ \Eprint
  {http://arxiv.org/abs/nucl-th/0305035} {arXiv:nucl-th/0305035} \BibitemShut
  {NoStop}%
\bibitem [{\citenamefont {Hergert}(2020)}]{Hergert_2020}%
  \BibitemOpen
  \bibfield  {author} {\bibinfo {author} {\bibfnamefont {H.}~\bibnamefont
  {Hergert}},\ }\href {\doibase 10.3389/fphy.2020.00379} {\bibfield  {journal}
  {\bibinfo  {journal} {Frontiers in Physics}\ }\textbf {\bibinfo {volume} {8}}
  (\bibinfo {year} {2020}),\ 10.3389/fphy.2020.00379}\BibitemShut {NoStop}%
\bibitem [{\citenamefont {Parisi}(1984)}]{Parisi:1983ae}%
  \BibitemOpen
  \bibfield  {author} {\bibinfo {author} {\bibfnamefont {G.}~\bibnamefont
  {Parisi}},\ }\href {\doibase 10.1016/0370-1573(84)90081-4} {\bibfield
  {journal} {\bibinfo  {journal} {Phys. Rept.}\ }\textbf {\bibinfo {volume}
  {103}},\ \bibinfo {pages} {203} (\bibinfo {year} {1984})}\BibitemShut
  {NoStop}%
\bibitem [{\citenamefont {Lepage}(1989)}]{Lepage:1989hd}%
  \BibitemOpen
  \bibfield  {author} {\bibinfo {author} {\bibfnamefont {G.~P.}\ \bibnamefont
  {Lepage}},\ }in\ \href@noop {} {\emph {\bibinfo {booktitle} {{Theoretical
  Advanced Study Institute in Elementary Particle Physics}}}}\ (\bibinfo {year}
  {1989})\BibitemShut {NoStop}%
\bibitem [{\citenamefont {Beane}\ \emph {et~al.}(2008)\citenamefont {Beane},
  \citenamefont {Orginos},\ and\ \citenamefont {Savage}}]{Beane:2008dv}%
  \BibitemOpen
  \bibfield  {author} {\bibinfo {author} {\bibfnamefont {S.~R.}\ \bibnamefont
  {Beane}}, \bibinfo {author} {\bibfnamefont {K.}~\bibnamefont {Orginos}}, \
  and\ \bibinfo {author} {\bibfnamefont {M.~J.}\ \bibnamefont {Savage}},\
  }\href {\doibase 10.1142/S0218301308010404} {\bibfield  {journal} {\bibinfo
  {journal} {Int. J. Mod. Phys. E}\ }\textbf {\bibinfo {volume} {17}},\
  \bibinfo {pages} {1157} (\bibinfo {year} {2008})},\ \Eprint
  {http://arxiv.org/abs/0805.4629} {arXiv:0805.4629 [hep-lat]} \BibitemShut
  {NoStop}%
\bibitem [{\citenamefont {O'Gorman}\ \emph {et~al.}(2021)\citenamefont
  {O'Gorman}, \citenamefont {Irani}, \citenamefont {Whitfield},\ and\
  \citenamefont {Fefferman}}]{ogorman2021electronic}%
  \BibitemOpen
  \bibfield  {author} {\bibinfo {author} {\bibfnamefont {B.}~\bibnamefont
  {O'Gorman}}, \bibinfo {author} {\bibfnamefont {S.}~\bibnamefont {Irani}},
  \bibinfo {author} {\bibfnamefont {J.}~\bibnamefont {Whitfield}}, \ and\
  \bibinfo {author} {\bibfnamefont {B.}~\bibnamefont {Fefferman}},\ }\href@noop
  {} {\enquote {\bibinfo {title} {Electronic structure in a fixed basis is
  qma-complete},}\ } (\bibinfo {year} {2021}),\ \Eprint
  {http://arxiv.org/abs/2103.08215} {arXiv:2103.08215 [quant-ph]} \BibitemShut
  {NoStop}%
\bibitem [{\citenamefont {Beane}\ \emph {et~al.}(2013)\citenamefont {Beane},
  \citenamefont {Chang}, \citenamefont {Cohen}, \citenamefont {Detmold},
  \citenamefont {Lin}, \citenamefont {Luu}, \citenamefont {Orginos},
  \citenamefont {Parreno}, \citenamefont {Savage},\ and\ \citenamefont
  {Walker-Loud}}]{Beane:2012vq}%
  \BibitemOpen
  \bibfield  {author} {\bibinfo {author} {\bibfnamefont {S.~R.}\ \bibnamefont
  {Beane}}, \bibinfo {author} {\bibfnamefont {E.}~\bibnamefont {Chang}},
  \bibinfo {author} {\bibfnamefont {S.~D.}\ \bibnamefont {Cohen}}, \bibinfo
  {author} {\bibfnamefont {W.}~\bibnamefont {Detmold}}, \bibinfo {author}
  {\bibfnamefont {H.~W.}\ \bibnamefont {Lin}}, \bibinfo {author} {\bibfnamefont
  {T.~C.}\ \bibnamefont {Luu}}, \bibinfo {author} {\bibfnamefont
  {K.}~\bibnamefont {Orginos}}, \bibinfo {author} {\bibfnamefont
  {A.}~\bibnamefont {Parreno}}, \bibinfo {author} {\bibfnamefont {M.~J.}\
  \bibnamefont {Savage}}, \ and\ \bibinfo {author} {\bibfnamefont
  {A.}~\bibnamefont {Walker-Loud}} (\bibinfo {collaboration} {NPLQCD}),\ }\href
  {\doibase 10.1103/PhysRevD.87.034506} {\bibfield  {journal} {\bibinfo
  {journal} {Phys. Rev. D}\ }\textbf {\bibinfo {volume} {87}},\ \bibinfo
  {pages} {034506} (\bibinfo {year} {2013})},\ \Eprint
  {http://arxiv.org/abs/1206.5219} {arXiv:1206.5219 [hep-lat]} \BibitemShut
  {NoStop}%
\bibitem [{\citenamefont {Barnea}\ \emph {et~al.}(2015)\citenamefont {Barnea},
  \citenamefont {Contessi}, \citenamefont {Gazit}, \citenamefont {Pederiva},\
  and\ \citenamefont {van Kolck}}]{Barnea:2013uqa}%
  \BibitemOpen
  \bibfield  {author} {\bibinfo {author} {\bibfnamefont {N.}~\bibnamefont
  {Barnea}}, \bibinfo {author} {\bibfnamefont {L.}~\bibnamefont {Contessi}},
  \bibinfo {author} {\bibfnamefont {D.}~\bibnamefont {Gazit}}, \bibinfo
  {author} {\bibfnamefont {F.}~\bibnamefont {Pederiva}}, \ and\ \bibinfo
  {author} {\bibfnamefont {U.}~\bibnamefont {van Kolck}},\ }\href {\doibase
  10.1103/PhysRevLett.114.052501} {\bibfield  {journal} {\bibinfo  {journal}
  {Phys. Rev. Lett.}\ }\textbf {\bibinfo {volume} {114}},\ \bibinfo {pages}
  {052501} (\bibinfo {year} {2015})},\ \Eprint {http://arxiv.org/abs/1311.4966}
  {arXiv:1311.4966 [nucl-th]} \BibitemShut {NoStop}%
\bibitem [{\citenamefont {Kirscher}\ \emph {et~al.}(2015)\citenamefont
  {Kirscher}, \citenamefont {Barnea}, \citenamefont {Gazit}, \citenamefont
  {Pederiva},\ and\ \citenamefont {van Kolck}}]{Kirscher:2015yda}%
  \BibitemOpen
  \bibfield  {author} {\bibinfo {author} {\bibfnamefont {J.}~\bibnamefont
  {Kirscher}}, \bibinfo {author} {\bibfnamefont {N.}~\bibnamefont {Barnea}},
  \bibinfo {author} {\bibfnamefont {D.}~\bibnamefont {Gazit}}, \bibinfo
  {author} {\bibfnamefont {F.}~\bibnamefont {Pederiva}}, \ and\ \bibinfo
  {author} {\bibfnamefont {U.}~\bibnamefont {van Kolck}},\ }\href {\doibase
  10.1103/PhysRevC.92.054002} {\bibfield  {journal} {\bibinfo  {journal} {Phys.
  Rev. C}\ }\textbf {\bibinfo {volume} {92}},\ \bibinfo {pages} {054002}
  (\bibinfo {year} {2015})},\ \Eprint {http://arxiv.org/abs/1506.09048}
  {arXiv:1506.09048 [nucl-th]} \BibitemShut {NoStop}%
\bibitem [{\citenamefont {Contessi}\ \emph {et~al.}(2017)\citenamefont
  {Contessi}, \citenamefont {Lovato}, \citenamefont {Pederiva}, \citenamefont
  {Roggero}, \citenamefont {Kirscher},\ and\ \citenamefont {van
  Kolck}}]{Contessi:2017rww}%
  \BibitemOpen
  \bibfield  {author} {\bibinfo {author} {\bibfnamefont {L.}~\bibnamefont
  {Contessi}}, \bibinfo {author} {\bibfnamefont {A.}~\bibnamefont {Lovato}},
  \bibinfo {author} {\bibfnamefont {F.}~\bibnamefont {Pederiva}}, \bibinfo
  {author} {\bibfnamefont {A.}~\bibnamefont {Roggero}}, \bibinfo {author}
  {\bibfnamefont {J.}~\bibnamefont {Kirscher}}, \ and\ \bibinfo {author}
  {\bibfnamefont {U.}~\bibnamefont {van Kolck}},\ }\href {\doibase
  10.1016/j.physletb.2017.07.048} {\bibfield  {journal} {\bibinfo  {journal}
  {Phys. Lett. B}\ }\textbf {\bibinfo {volume} {772}},\ \bibinfo {pages} {839}
  (\bibinfo {year} {2017})},\ \Eprint {http://arxiv.org/abs/1701.06516}
  {arXiv:1701.06516 [nucl-th]} \BibitemShut {NoStop}%
\bibitem [{\citenamefont {Bansal}\ \emph {et~al.}(2018)\citenamefont {Bansal},
  \citenamefont {Binder}, \citenamefont {Ekstr\"om}, \citenamefont {Hagen},
  \citenamefont {Jansen},\ and\ \citenamefont {Papenbrock}}]{Bansal:2017pwn}%
  \BibitemOpen
  \bibfield  {author} {\bibinfo {author} {\bibfnamefont {A.}~\bibnamefont
  {Bansal}}, \bibinfo {author} {\bibfnamefont {S.}~\bibnamefont {Binder}},
  \bibinfo {author} {\bibfnamefont {A.}~\bibnamefont {Ekstr\"om}}, \bibinfo
  {author} {\bibfnamefont {G.}~\bibnamefont {Hagen}}, \bibinfo {author}
  {\bibfnamefont {G.~R.}\ \bibnamefont {Jansen}}, \ and\ \bibinfo {author}
  {\bibfnamefont {T.}~\bibnamefont {Papenbrock}},\ }\href {\doibase
  10.1103/PhysRevC.98.054301} {\bibfield  {journal} {\bibinfo  {journal} {Phys.
  Rev. C}\ }\textbf {\bibinfo {volume} {98}},\ \bibinfo {pages} {054301}
  (\bibinfo {year} {2018})},\ \Eprint {http://arxiv.org/abs/1712.10246}
  {arXiv:1712.10246 [nucl-th]} \BibitemShut {NoStop}%
\bibitem [{\citenamefont {Beane}\ \emph {et~al.}(2011)\citenamefont {Beane},
  \citenamefont {Detmold}, \citenamefont {Orginos},\ and\ \citenamefont
  {Savage}}]{Beane:2010em}%
  \BibitemOpen
  \bibfield  {author} {\bibinfo {author} {\bibfnamefont {S.~R.}\ \bibnamefont
  {Beane}}, \bibinfo {author} {\bibfnamefont {W.}~\bibnamefont {Detmold}},
  \bibinfo {author} {\bibfnamefont {K.}~\bibnamefont {Orginos}}, \ and\
  \bibinfo {author} {\bibfnamefont {M.~J.}\ \bibnamefont {Savage}},\ }\href
  {\doibase 10.1016/j.ppnp.2010.08.002} {\bibfield  {journal} {\bibinfo
  {journal} {Prog. Part. Nucl. Phys.}\ }\textbf {\bibinfo {volume} {66}},\
  \bibinfo {pages} {1} (\bibinfo {year} {2011})},\ \Eprint
  {http://arxiv.org/abs/1004.2935} {arXiv:1004.2935 [hep-lat]} \BibitemShut
  {NoStop}%
\bibitem [{\citenamefont {Savage}(2016)}]{Savage:2016egr}%
  \BibitemOpen
  \bibfield  {author} {\bibinfo {author} {\bibfnamefont {M.~J.}\ \bibnamefont
  {Savage}},\ }\href {\doibase 10.22323/1.256.0021} {\bibfield  {journal}
  {\bibinfo  {journal} {PoS}\ }\textbf {\bibinfo {volume} {LATTICE2016}},\
  \bibinfo {pages} {021} (\bibinfo {year} {2016})},\ \Eprint
  {http://arxiv.org/abs/1611.02078} {arXiv:1611.02078 [hep-lat]} \BibitemShut
  {NoStop}%
\bibitem [{\citenamefont {Drischler}\ \emph {et~al.}(2019)\citenamefont
  {Drischler}, \citenamefont {Haxton}, \citenamefont {McElvain}, \citenamefont
  {Mereghetti}, \citenamefont {Nicholson}, \citenamefont {Vranas},\ and\
  \citenamefont {Walker-Loud}}]{Drischler:2019xuo}%
  \BibitemOpen
  \bibfield  {author} {\bibinfo {author} {\bibfnamefont {C.}~\bibnamefont
  {Drischler}}, \bibinfo {author} {\bibfnamefont {W.}~\bibnamefont {Haxton}},
  \bibinfo {author} {\bibfnamefont {K.}~\bibnamefont {McElvain}}, \bibinfo
  {author} {\bibfnamefont {E.}~\bibnamefont {Mereghetti}}, \bibinfo {author}
  {\bibfnamefont {A.}~\bibnamefont {Nicholson}}, \bibinfo {author}
  {\bibfnamefont {P.}~\bibnamefont {Vranas}}, \ and\ \bibinfo {author}
  {\bibfnamefont {A.}~\bibnamefont {Walker-Loud}},\ }\href@noop {} {\
  (\bibinfo {year} {2019})},\ \Eprint {http://arxiv.org/abs/1910.07961}
  {arXiv:1910.07961 [nucl-th]} \BibitemShut {NoStop}%
\bibitem [{\citenamefont {Davoudi}\ \emph
  {et~al.}(2021{\natexlab{a}})\citenamefont {Davoudi}, \citenamefont {Detmold},
  \citenamefont {Orginos}, \citenamefont {Parre\~no}, \citenamefont {Savage},
  \citenamefont {Shanahan},\ and\ \citenamefont {Wagman}}]{Davoudi:2020ngi}%
  \BibitemOpen
  \bibfield  {author} {\bibinfo {author} {\bibfnamefont {Z.}~\bibnamefont
  {Davoudi}}, \bibinfo {author} {\bibfnamefont {W.}~\bibnamefont {Detmold}},
  \bibinfo {author} {\bibfnamefont {K.}~\bibnamefont {Orginos}}, \bibinfo
  {author} {\bibfnamefont {A.}~\bibnamefont {Parre\~no}}, \bibinfo {author}
  {\bibfnamefont {M.~J.}\ \bibnamefont {Savage}}, \bibinfo {author}
  {\bibfnamefont {P.}~\bibnamefont {Shanahan}}, \ and\ \bibinfo {author}
  {\bibfnamefont {M.~L.}\ \bibnamefont {Wagman}},\ }\href {\doibase
  10.1016/j.physrep.2020.10.004} {\bibfield  {journal} {\bibinfo  {journal}
  {Phys. Rept.}\ }\textbf {\bibinfo {volume} {900}},\ \bibinfo {pages} {1}
  (\bibinfo {year} {2021}{\natexlab{a}})},\ \Eprint
  {http://arxiv.org/abs/2008.11160} {arXiv:2008.11160 [hep-lat]} \BibitemShut
  {NoStop}%
\bibitem [{\citenamefont {Aoki}\ and\ \citenamefont
  {Doi}(2020)}]{Aoki:2020bew}%
  \BibitemOpen
  \bibfield  {author} {\bibinfo {author} {\bibfnamefont {S.}~\bibnamefont
  {Aoki}}\ and\ \bibinfo {author} {\bibfnamefont {T.}~\bibnamefont {Doi}},\
  }\href {\doibase 10.3389/fphy.2020.00307} {\bibfield  {journal} {\bibinfo
  {journal} {Front. in Phys.}\ }\textbf {\bibinfo {volume} {8}},\ \bibinfo
  {pages} {307} (\bibinfo {year} {2020})},\ \Eprint
  {http://arxiv.org/abs/2003.10730} {arXiv:2003.10730 [hep-lat]} \BibitemShut
  {NoStop}%
\bibitem [{\citenamefont {Ademollo}\ and\ \citenamefont
  {Gatto}(1964)}]{Ademollo:1964sr}%
  \BibitemOpen
  \bibfield  {author} {\bibinfo {author} {\bibfnamefont {M.}~\bibnamefont
  {Ademollo}}\ and\ \bibinfo {author} {\bibfnamefont {R.}~\bibnamefont
  {Gatto}},\ }\href {\doibase 10.1103/PhysRevLett.13.264} {\bibfield  {journal}
  {\bibinfo  {journal} {Phys. Rev. Lett.}\ }\textbf {\bibinfo {volume} {13}},\
  \bibinfo {pages} {264} (\bibinfo {year} {1964})}\BibitemShut {NoStop}%
\bibitem [{\citenamefont {Luke}(1990)}]{Luke:1990eg}%
  \BibitemOpen
  \bibfield  {author} {\bibinfo {author} {\bibfnamefont {M.~E.}\ \bibnamefont
  {Luke}},\ }\href {\doibase 10.1016/0370-2693(90)90568-Q} {\bibfield
  {journal} {\bibinfo  {journal} {Phys. Lett. B}\ }\textbf {\bibinfo {volume}
  {252}},\ \bibinfo {pages} {447} (\bibinfo {year} {1990})}\BibitemShut
  {NoStop}%
\bibitem [{\citenamefont {Bauer}\ \emph {et~al.}(2000)\citenamefont {Bauer},
  \citenamefont {Fleming},\ and\ \citenamefont {Luke}}]{Bauer:2000ew}%
  \BibitemOpen
  \bibfield  {author} {\bibinfo {author} {\bibfnamefont {C.~W.}\ \bibnamefont
  {Bauer}}, \bibinfo {author} {\bibfnamefont {S.}~\bibnamefont {Fleming}}, \
  and\ \bibinfo {author} {\bibfnamefont {M.~E.}\ \bibnamefont {Luke}},\ }\href
  {\doibase 10.1103/PhysRevD.63.014006} {\bibfield  {journal} {\bibinfo
  {journal} {Phys. Rev. D}\ }\textbf {\bibinfo {volume} {63}},\ \bibinfo
  {pages} {014006} (\bibinfo {year} {2000})},\ \Eprint
  {http://arxiv.org/abs/hep-ph/0005275} {arXiv:hep-ph/0005275} \BibitemShut
  {NoStop}%
\bibitem [{\citenamefont {Bauer}\ \emph {et~al.}(2001)\citenamefont {Bauer},
  \citenamefont {Fleming}, \citenamefont {Pirjol},\ and\ \citenamefont
  {Stewart}}]{Bauer:2000yr}%
  \BibitemOpen
  \bibfield  {author} {\bibinfo {author} {\bibfnamefont {C.~W.}\ \bibnamefont
  {Bauer}}, \bibinfo {author} {\bibfnamefont {S.}~\bibnamefont {Fleming}},
  \bibinfo {author} {\bibfnamefont {D.}~\bibnamefont {Pirjol}}, \ and\ \bibinfo
  {author} {\bibfnamefont {I.~W.}\ \bibnamefont {Stewart}},\ }\href {\doibase
  10.1103/PhysRevD.63.114020} {\bibfield  {journal} {\bibinfo  {journal} {Phys.
  Rev. D}\ }\textbf {\bibinfo {volume} {63}},\ \bibinfo {pages} {114020}
  (\bibinfo {year} {2001})},\ \Eprint {http://arxiv.org/abs/hep-ph/0011336}
  {arXiv:hep-ph/0011336} \BibitemShut {NoStop}%
\bibitem [{\citenamefont {Bauer}\ and\ \citenamefont
  {Stewart}(2001)}]{Bauer:2001ct}%
  \BibitemOpen
  \bibfield  {author} {\bibinfo {author} {\bibfnamefont {C.~W.}\ \bibnamefont
  {Bauer}}\ and\ \bibinfo {author} {\bibfnamefont {I.~W.}\ \bibnamefont
  {Stewart}},\ }\href {\doibase 10.1016/S0370-2693(01)00902-9} {\bibfield
  {journal} {\bibinfo  {journal} {Phys. Lett. B}\ }\textbf {\bibinfo {volume}
  {516}},\ \bibinfo {pages} {134} (\bibinfo {year} {2001})},\ \Eprint
  {http://arxiv.org/abs/hep-ph/0107001} {arXiv:hep-ph/0107001} \BibitemShut
  {NoStop}%
\bibitem [{\citenamefont {Bauer}\ \emph
  {et~al.}(2021{\natexlab{a}})\citenamefont {Bauer}, \citenamefont {Freytsis},\
  and\ \citenamefont {Nachman}}]{Bauer:2021gup}%
  \BibitemOpen
  \bibfield  {author} {\bibinfo {author} {\bibfnamefont {C.~W.}\ \bibnamefont
  {Bauer}}, \bibinfo {author} {\bibfnamefont {M.}~\bibnamefont {Freytsis}}, \
  and\ \bibinfo {author} {\bibfnamefont {B.}~\bibnamefont {Nachman}},\
  }\href@noop {} {\  (\bibinfo {year} {2021}{\natexlab{a}})},\ \Eprint
  {http://arxiv.org/abs/2102.05044} {arXiv:2102.05044 [hep-ph]} \BibitemShut
  {NoStop}%
\bibitem [{\citenamefont {Avkhadiev}\ \emph {et~al.}(2020)\citenamefont
  {Avkhadiev}, \citenamefont {Shanahan},\ and\ \citenamefont
  {Young}}]{Avkhadiev:2019niu}%
  \BibitemOpen
  \bibfield  {author} {\bibinfo {author} {\bibfnamefont {A.}~\bibnamefont
  {Avkhadiev}}, \bibinfo {author} {\bibfnamefont {P.~E.}\ \bibnamefont
  {Shanahan}}, \ and\ \bibinfo {author} {\bibfnamefont {R.~D.}\ \bibnamefont
  {Young}},\ }\href {\doibase 10.1103/PhysRevLett.124.080501} {\bibfield
  {journal} {\bibinfo  {journal} {Phys. Rev. Lett.}\ }\textbf {\bibinfo
  {volume} {124}},\ \bibinfo {pages} {080501} (\bibinfo {year} {2020})},\
  \Eprint {http://arxiv.org/abs/1908.04194} {arXiv:1908.04194 [hep-lat]}
  \BibitemShut {NoStop}%
\bibitem [{\citenamefont {Moosavian}\ \emph {et~al.}(2019)\citenamefont
  {Moosavian}, \citenamefont {Garrison},\ and\ \citenamefont
  {Jordan}}]{Moosavian:2019rxg}%
  \BibitemOpen
  \bibfield  {author} {\bibinfo {author} {\bibfnamefont {A.~H.}\ \bibnamefont
  {Moosavian}}, \bibinfo {author} {\bibfnamefont {J.~R.}\ \bibnamefont
  {Garrison}}, \ and\ \bibinfo {author} {\bibfnamefont {S.~P.}\ \bibnamefont
  {Jordan}},\ }\href@noop {} {\  (\bibinfo {year} {2019})},\ \Eprint
  {http://arxiv.org/abs/1911.03505} {arXiv:1911.03505 [quant-ph]} \BibitemShut
  {NoStop}%
\bibitem [{\citenamefont {Provasoli}\ \emph {et~al.}(2019)\citenamefont
  {Provasoli}, \citenamefont {Nachman}, \citenamefont {de~Jong},\ and\
  \citenamefont {Bauer}}]{Provasoli:2019tzz}%
  \BibitemOpen
  \bibfield  {author} {\bibinfo {author} {\bibfnamefont {D.}~\bibnamefont
  {Provasoli}}, \bibinfo {author} {\bibfnamefont {B.}~\bibnamefont {Nachman}},
  \bibinfo {author} {\bibfnamefont {W.~A.}\ \bibnamefont {de~Jong}}, \ and\
  \bibinfo {author} {\bibfnamefont {C.~W.}\ \bibnamefont {Bauer}},\ }\href@noop
  {} {\  (\bibinfo {year} {2019})},\ \Eprint {http://arxiv.org/abs/1901.08148}
  {arXiv:1901.08148 [quant-ph]} \BibitemShut {NoStop}%
\bibitem [{\citenamefont {Bauer}\ \emph
  {et~al.}(2021{\natexlab{b}})\citenamefont {Bauer}, \citenamefont {de~Jong},
  \citenamefont {Nachman},\ and\ \citenamefont {Provasoli}}]{Bauer:2019qxa}%
  \BibitemOpen
  \bibfield  {author} {\bibinfo {author} {\bibfnamefont {C.~W.}\ \bibnamefont
  {Bauer}}, \bibinfo {author} {\bibfnamefont {W.~A.}\ \bibnamefont {de~Jong}},
  \bibinfo {author} {\bibfnamefont {B.}~\bibnamefont {Nachman}}, \ and\
  \bibinfo {author} {\bibfnamefont {D.}~\bibnamefont {Provasoli}},\ }\href
  {\doibase 10.1103/PhysRevLett.126.062001} {\bibfield  {journal} {\bibinfo
  {journal} {Phys. Rev. Lett.}\ }\textbf {\bibinfo {volume} {126}},\ \bibinfo
  {pages} {062001} (\bibinfo {year} {2021}{\natexlab{b}})},\ \Eprint
  {http://arxiv.org/abs/1904.03196} {arXiv:1904.03196 [hep-ph]} \BibitemShut
  {NoStop}%
\bibitem [{\citenamefont {Jordan}\ and\ \citenamefont
  {Wigner}(1928)}]{Jordan1928}%
  \BibitemOpen
  \bibfield  {author} {\bibinfo {author} {\bibfnamefont {P.}~\bibnamefont
  {Jordan}}\ and\ \bibinfo {author} {\bibfnamefont {E.}~\bibnamefont
  {Wigner}},\ }\href {\doibase 10.1007/BF01331938} {\bibfield  {journal}
  {\bibinfo  {journal} {Zeitschrift für Physik}\ }\textbf {\bibinfo {volume}
  {47}},\ \bibinfo {pages} {631} (\bibinfo {year} {1928})}\BibitemShut
  {NoStop}%
\bibitem [{\citenamefont {Ortiz}\ \emph {et~al.}(2001)\citenamefont {Ortiz},
  \citenamefont {Gubernatis}, \citenamefont {Knill},\ and\ \citenamefont
  {Laflamme}}]{PhysRevA.64.022319}%
  \BibitemOpen
  \bibfield  {author} {\bibinfo {author} {\bibfnamefont {G.}~\bibnamefont
  {Ortiz}}, \bibinfo {author} {\bibfnamefont {J.~E.}\ \bibnamefont
  {Gubernatis}}, \bibinfo {author} {\bibfnamefont {E.}~\bibnamefont {Knill}}, \
  and\ \bibinfo {author} {\bibfnamefont {R.}~\bibnamefont {Laflamme}},\ }\href
  {\doibase 10.1103/PhysRevA.64.022319} {\bibfield  {journal} {\bibinfo
  {journal} {Phys. Rev. A}\ }\textbf {\bibinfo {volume} {64}},\ \bibinfo
  {pages} {022319} (\bibinfo {year} {2001})}\BibitemShut {NoStop}%
\bibitem [{\citenamefont {Somma}\ \emph {et~al.}(2002)\citenamefont {Somma},
  \citenamefont {Ortiz}, \citenamefont {Gubernatis}, \citenamefont {Knill},\
  and\ \citenamefont {Laflamme}}]{PhysRevA.65.042323}%
  \BibitemOpen
  \bibfield  {author} {\bibinfo {author} {\bibfnamefont {R.}~\bibnamefont
  {Somma}}, \bibinfo {author} {\bibfnamefont {G.}~\bibnamefont {Ortiz}},
  \bibinfo {author} {\bibfnamefont {J.~E.}\ \bibnamefont {Gubernatis}},
  \bibinfo {author} {\bibfnamefont {E.}~\bibnamefont {Knill}}, \ and\ \bibinfo
  {author} {\bibfnamefont {R.}~\bibnamefont {Laflamme}},\ }\href {\doibase
  10.1103/PhysRevA.65.042323} {\bibfield  {journal} {\bibinfo  {journal} {Phys.
  Rev. A}\ }\textbf {\bibinfo {volume} {65}},\ \bibinfo {pages} {042323}
  (\bibinfo {year} {2002})}\BibitemShut {NoStop}%
\bibitem [{\citenamefont {Bravyi}\ and\ \citenamefont
  {Kitaev}(2002)}]{BRAVYI2002210}%
  \BibitemOpen
  \bibfield  {author} {\bibinfo {author} {\bibfnamefont {S.~B.}\ \bibnamefont
  {Bravyi}}\ and\ \bibinfo {author} {\bibfnamefont {A.~Y.}\ \bibnamefont
  {Kitaev}},\ }\href {\doibase https://doi.org/10.1006/aphy.2002.6254}
  {\bibfield  {journal} {\bibinfo  {journal} {Annals of Physics}\ }\textbf
  {\bibinfo {volume} {298}},\ \bibinfo {pages} {210} (\bibinfo {year}
  {2002})}\BibitemShut {NoStop}%
\bibitem [{\citenamefont {Seeley}\ \emph {et~al.}(2012)\citenamefont {Seeley},
  \citenamefont {Richard},\ and\ \citenamefont {Love}}]{doi:10.1063/1.4768229}%
  \BibitemOpen
  \bibfield  {author} {\bibinfo {author} {\bibfnamefont {J.~T.}\ \bibnamefont
  {Seeley}}, \bibinfo {author} {\bibfnamefont {M.~J.}\ \bibnamefont {Richard}},
  \ and\ \bibinfo {author} {\bibfnamefont {P.~J.}\ \bibnamefont {Love}},\
  }\href {\doibase 10.1063/1.4768229} {\bibfield  {journal} {\bibinfo
  {journal} {The Journal of Chemical Physics}\ }\textbf {\bibinfo {volume}
  {137}},\ \bibinfo {pages} {224109} (\bibinfo {year} {2012})}\BibitemShut
  {NoStop}%
\bibitem [{\citenamefont {Verstraete}\ and\ \citenamefont
  {Cirac}(2005)}]{Verstraete_2005}%
  \BibitemOpen
  \bibfield  {author} {\bibinfo {author} {\bibfnamefont {F.}~\bibnamefont
  {Verstraete}}\ and\ \bibinfo {author} {\bibfnamefont {J.~I.}\ \bibnamefont
  {Cirac}},\ }\href {\doibase 10.1088/1742-5468/2005/09/p09012} {\bibfield
  {journal} {\bibinfo  {journal} {Journal of Statistical Mechanics: Theory and
  Experiment}\ }\textbf {\bibinfo {volume} {2005}},\ \bibinfo {pages} {P09012}
  (\bibinfo {year} {2005})}\BibitemShut {NoStop}%
\bibitem [{\citenamefont {Ball}(2005)}]{PhysRevLett.95.176407}%
  \BibitemOpen
  \bibfield  {author} {\bibinfo {author} {\bibfnamefont {R.~C.}\ \bibnamefont
  {Ball}},\ }\href {\doibase 10.1103/PhysRevLett.95.176407} {\bibfield
  {journal} {\bibinfo  {journal} {Phys. Rev. Lett.}\ }\textbf {\bibinfo
  {volume} {95}},\ \bibinfo {pages} {176407} (\bibinfo {year}
  {2005})}\BibitemShut {NoStop}%
\bibitem [{\citenamefont {Havl\'{\i}\ifmmode~\check{c}\else \v{c}\fi{}ek}\
  \emph {et~al.}(2017)\citenamefont {Havl\'{\i}\ifmmode~\check{c}\else
  \v{c}\fi{}ek}, \citenamefont {Troyer},\ and\ \citenamefont
  {Whitfield}}]{PhysRevA.95.032332}%
  \BibitemOpen
  \bibfield  {author} {\bibinfo {author} {\bibfnamefont {V.~c.~v.}\
  \bibnamefont {Havl\'{\i}\ifmmode~\check{c}\else \v{c}\fi{}ek}}, \bibinfo
  {author} {\bibfnamefont {M.}~\bibnamefont {Troyer}}, \ and\ \bibinfo {author}
  {\bibfnamefont {J.~D.}\ \bibnamefont {Whitfield}},\ }\href {\doibase
  10.1103/PhysRevA.95.032332} {\bibfield  {journal} {\bibinfo  {journal} {Phys.
  Rev. A}\ }\textbf {\bibinfo {volume} {95}},\ \bibinfo {pages} {032332}
  (\bibinfo {year} {2017})}\BibitemShut {NoStop}%
\bibitem [{\citenamefont {Steudtner}\ and\ \citenamefont
  {Wehner}(2018)}]{Steudtner_2018}%
  \BibitemOpen
  \bibfield  {author} {\bibinfo {author} {\bibfnamefont {M.}~\bibnamefont
  {Steudtner}}\ and\ \bibinfo {author} {\bibfnamefont {S.}~\bibnamefont
  {Wehner}},\ }\href {\doibase 10.1088/1367-2630/aac54f} {\bibfield  {journal}
  {\bibinfo  {journal} {New Journal of Physics}\ }\textbf {\bibinfo {volume}
  {20}},\ \bibinfo {pages} {063010} (\bibinfo {year} {2018})}\BibitemShut
  {NoStop}%
\bibitem [{\citenamefont {Derby}\ and\ \citenamefont
  {Klassen}(2020)}]{derby2020compact}%
  \BibitemOpen
  \bibfield  {author} {\bibinfo {author} {\bibfnamefont {C.}~\bibnamefont
  {Derby}}\ and\ \bibinfo {author} {\bibfnamefont {J.}~\bibnamefont
  {Klassen}},\ }\href@noop {} {\enquote {\bibinfo {title} {A compact fermion to
  qubit mapping},}\ } (\bibinfo {year} {2020}),\ \Eprint
  {http://arxiv.org/abs/2003.06939} {arXiv:2003.06939 [quant-ph]} \BibitemShut
  {NoStop}%
\bibitem [{\citenamefont {Jiang}\ \emph {et~al.}(2020)\citenamefont {Jiang},
  \citenamefont {Kalev}, \citenamefont {Mruczkiewicz},\ and\ \citenamefont
  {Neven}}]{Jiang2020optimalfermionto}%
  \BibitemOpen
  \bibfield  {author} {\bibinfo {author} {\bibfnamefont {Z.}~\bibnamefont
  {Jiang}}, \bibinfo {author} {\bibfnamefont {A.}~\bibnamefont {Kalev}},
  \bibinfo {author} {\bibfnamefont {W.}~\bibnamefont {Mruczkiewicz}}, \ and\
  \bibinfo {author} {\bibfnamefont {H.}~\bibnamefont {Neven}},\ }\href
  {\doibase 10.22331/q-2020-06-04-276} {\bibfield  {journal} {\bibinfo
  {journal} {{Quantum}}\ }\textbf {\bibinfo {volume} {4}},\ \bibinfo {pages}
  {276} (\bibinfo {year} {2020})}\BibitemShut {NoStop}%
\bibitem [{\citenamefont {Setia}\ and\ \citenamefont
  {Whitfield}(2018)}]{doi:10.1063/1.5019371}%
  \BibitemOpen
  \bibfield  {author} {\bibinfo {author} {\bibfnamefont {K.}~\bibnamefont
  {Setia}}\ and\ \bibinfo {author} {\bibfnamefont {J.~D.}\ \bibnamefont
  {Whitfield}},\ }\href {\doibase 10.1063/1.5019371} {\bibfield  {journal}
  {\bibinfo  {journal} {The Journal of Chemical Physics}\ }\textbf {\bibinfo
  {volume} {148}},\ \bibinfo {pages} {164104} (\bibinfo {year}
  {2018})}\BibitemShut {NoStop}%
\bibitem [{\citenamefont {Jiang}\ \emph {et~al.}(2019)\citenamefont {Jiang},
  \citenamefont {McClean}, \citenamefont {Babbush},\ and\ \citenamefont
  {Neven}}]{PhysRevApplied.12.064041}%
  \BibitemOpen
  \bibfield  {author} {\bibinfo {author} {\bibfnamefont {Z.}~\bibnamefont
  {Jiang}}, \bibinfo {author} {\bibfnamefont {J.}~\bibnamefont {McClean}},
  \bibinfo {author} {\bibfnamefont {R.}~\bibnamefont {Babbush}}, \ and\
  \bibinfo {author} {\bibfnamefont {H.}~\bibnamefont {Neven}},\ }\href
  {\doibase 10.1103/PhysRevApplied.12.064041} {\bibfield  {journal} {\bibinfo
  {journal} {Phys. Rev. Applied}\ }\textbf {\bibinfo {volume} {12}},\ \bibinfo
  {pages} {064041} (\bibinfo {year} {2019})}\BibitemShut {NoStop}%
\bibitem [{\citenamefont {Chandrasekharan}\ and\ \citenamefont
  {Wiese}(1997{\natexlab{a}})}]{Chandrasekharan:1996ih}%
  \BibitemOpen
  \bibfield  {author} {\bibinfo {author} {\bibfnamefont {S.}~\bibnamefont
  {Chandrasekharan}}\ and\ \bibinfo {author} {\bibfnamefont {U.~J.}\
  \bibnamefont {Wiese}},\ }\href {\doibase 10.1016/S0550-3213(97)00006-0}
  {\bibfield  {journal} {\bibinfo  {journal} {Nucl. Phys. B}\ }\textbf
  {\bibinfo {volume} {492}},\ \bibinfo {pages} {455} (\bibinfo {year}
  {1997}{\natexlab{a}})},\ \Eprint {http://arxiv.org/abs/hep-lat/9609042}
  {arXiv:hep-lat/9609042} \BibitemShut {NoStop}%
\bibitem [{\citenamefont {Brower}\ \emph {et~al.}(1999)\citenamefont {Brower},
  \citenamefont {Chandrasekharan},\ and\ \citenamefont
  {Wiese}}]{Brower:1997ha}%
  \BibitemOpen
  \bibfield  {author} {\bibinfo {author} {\bibfnamefont {R.}~\bibnamefont
  {Brower}}, \bibinfo {author} {\bibfnamefont {S.}~\bibnamefont
  {Chandrasekharan}}, \ and\ \bibinfo {author} {\bibfnamefont {U.~J.}\
  \bibnamefont {Wiese}},\ }\href {\doibase 10.1103/PhysRevD.60.094502}
  {\bibfield  {journal} {\bibinfo  {journal} {Phys. Rev. D}\ }\textbf {\bibinfo
  {volume} {60}},\ \bibinfo {pages} {094502} (\bibinfo {year} {1999})},\
  \Eprint {http://arxiv.org/abs/hep-th/9704106} {arXiv:hep-th/9704106}
  \BibitemShut {NoStop}%
\bibitem [{\citenamefont {Zache}\ \emph {et~al.}(2021)\citenamefont {Zache},
  \citenamefont {Van~Damme}, \citenamefont {Halimeh}, \citenamefont {Hauke},\
  and\ \citenamefont {Banerjee}}]{Zache:2021lrh}%
  \BibitemOpen
  \bibfield  {author} {\bibinfo {author} {\bibfnamefont {T.~V.}\ \bibnamefont
  {Zache}}, \bibinfo {author} {\bibfnamefont {M.}~\bibnamefont {Van~Damme}},
  \bibinfo {author} {\bibfnamefont {J.~C.}\ \bibnamefont {Halimeh}}, \bibinfo
  {author} {\bibfnamefont {P.}~\bibnamefont {Hauke}}, \ and\ \bibinfo {author}
  {\bibfnamefont {D.}~\bibnamefont {Banerjee}},\ }\href@noop {} {\  (\bibinfo
  {year} {2021})},\ \Eprint {http://arxiv.org/abs/2104.00025} {arXiv:2104.00025
  [hep-lat]} \BibitemShut {NoStop}%
\bibitem [{\citenamefont {Singh}\ and\ \citenamefont
  {Chandrasekharan}(2019)}]{Singh:2019uwd}%
  \BibitemOpen
  \bibfield  {author} {\bibinfo {author} {\bibfnamefont {H.}~\bibnamefont
  {Singh}}\ and\ \bibinfo {author} {\bibfnamefont {S.}~\bibnamefont
  {Chandrasekharan}},\ }\href {\doibase 10.1103/PhysRevD.100.054505} {\bibfield
   {journal} {\bibinfo  {journal} {Phys. Rev. D}\ }\textbf {\bibinfo {volume}
  {100}},\ \bibinfo {pages} {054505} (\bibinfo {year} {2019})},\ \Eprint
  {http://arxiv.org/abs/1905.13204} {arXiv:1905.13204 [hep-lat]} \BibitemShut
  {NoStop}%
\bibitem [{\citenamefont {Bhattacharya}\ \emph {et~al.}(2021)\citenamefont
  {Bhattacharya}, \citenamefont {Buser}, \citenamefont {Chandrasekharan},
  \citenamefont {Gupta},\ and\ \citenamefont {Singh}}]{Bhattacharya:2020gpm}%
  \BibitemOpen
  \bibfield  {author} {\bibinfo {author} {\bibfnamefont {T.}~\bibnamefont
  {Bhattacharya}}, \bibinfo {author} {\bibfnamefont {A.~J.}\ \bibnamefont
  {Buser}}, \bibinfo {author} {\bibfnamefont {S.}~\bibnamefont
  {Chandrasekharan}}, \bibinfo {author} {\bibfnamefont {R.}~\bibnamefont
  {Gupta}}, \ and\ \bibinfo {author} {\bibfnamefont {H.}~\bibnamefont
  {Singh}},\ }\href {\doibase 10.1103/PhysRevLett.126.172001} {\bibfield
  {journal} {\bibinfo  {journal} {Phys. Rev. Lett.}\ }\textbf {\bibinfo
  {volume} {126}},\ \bibinfo {pages} {172001} (\bibinfo {year}
  {2021})}\BibitemShut {NoStop}%
\bibitem [{\citenamefont {Ba\~nuls}\ \emph {et~al.}(2020)\citenamefont
  {Ba\~nuls} \emph {et~al.}}]{Banuls:2019bmf}%
  \BibitemOpen
  \bibfield  {author} {\bibinfo {author} {\bibfnamefont {M.~C.}\ \bibnamefont
  {Ba\~nuls}} \emph {et~al.},\ }\href {\doibase 10.1140/epjd/e2020-100571-8}
  {\bibfield  {journal} {\bibinfo  {journal} {Eur. Phys. J. D}\ }\textbf
  {\bibinfo {volume} {74}},\ \bibinfo {pages} {165} (\bibinfo {year} {2020})},\
  \Eprint {http://arxiv.org/abs/1911.00003} {arXiv:1911.00003 [quant-ph]}
  \BibitemShut {NoStop}%
\bibitem [{\citenamefont {Zohar}(2021)}]{Zohar:2021nyc}%
  \BibitemOpen
  \bibfield  {author} {\bibinfo {author} {\bibfnamefont {E.}~\bibnamefont
  {Zohar}},\ }\href@noop {} {\  (\bibinfo {year} {2021})},\ \Eprint
  {http://arxiv.org/abs/2106.04609} {arXiv:2106.04609 [quant-ph]} \BibitemShut
  {NoStop}%
\bibitem [{\citenamefont {Petcher}\ and\ \citenamefont
  {Weingarten}(1980)}]{PhysRevD.22.2465}%
  \BibitemOpen
  \bibfield  {author} {\bibinfo {author} {\bibfnamefont {D.}~\bibnamefont
  {Petcher}}\ and\ \bibinfo {author} {\bibfnamefont {D.~H.}\ \bibnamefont
  {Weingarten}},\ }\href {\doibase 10.1103/PhysRevD.22.2465} {\bibfield
  {journal} {\bibinfo  {journal} {Phys. Rev. D}\ }\textbf {\bibinfo {volume}
  {22}},\ \bibinfo {pages} {2465} (\bibinfo {year} {1980})}\BibitemShut
  {NoStop}%
\bibitem [{\citenamefont {Bhanot}\ and\ \citenamefont
  {Rebbi}(1981)}]{PhysRevD.24.3319}%
  \BibitemOpen
  \bibfield  {author} {\bibinfo {author} {\bibfnamefont {G.}~\bibnamefont
  {Bhanot}}\ and\ \bibinfo {author} {\bibfnamefont {C.}~\bibnamefont {Rebbi}},\
  }\href {\doibase 10.1103/PhysRevD.24.3319} {\bibfield  {journal} {\bibinfo
  {journal} {Phys. Rev. D}\ }\textbf {\bibinfo {volume} {24}},\ \bibinfo
  {pages} {3319} (\bibinfo {year} {1981})}\BibitemShut {NoStop}%
\bibitem [{\citenamefont {Grosse}\ and\ \citenamefont
  {Kühnelt}(1981)}]{GROSSE198177}%
  \BibitemOpen
  \bibfield  {author} {\bibinfo {author} {\bibfnamefont {H.}~\bibnamefont
  {Grosse}}\ and\ \bibinfo {author} {\bibfnamefont {H.}~\bibnamefont
  {Kühnelt}},\ }\href {\doibase https://doi.org/10.1016/0370-2693(81)90494-9}
  {\bibfield  {journal} {\bibinfo  {journal} {Physics Letters B}\ }\textbf
  {\bibinfo {volume} {101}},\ \bibinfo {pages} {77} (\bibinfo {year}
  {1981})}\BibitemShut {NoStop}%
\bibitem [{\citenamefont {Alexandru}\ \emph {et~al.}(2019)\citenamefont
  {Alexandru}, \citenamefont {Bedaque}, \citenamefont {Harmalkar},
  \citenamefont {Lamm}, \citenamefont {Lawrence},\ and\ \citenamefont
  {Warrington}}]{Alexandru:2019nsa}%
  \BibitemOpen
  \bibfield  {author} {\bibinfo {author} {\bibfnamefont {A.}~\bibnamefont
  {Alexandru}}, \bibinfo {author} {\bibfnamefont {P.~F.}\ \bibnamefont
  {Bedaque}}, \bibinfo {author} {\bibfnamefont {S.}~\bibnamefont {Harmalkar}},
  \bibinfo {author} {\bibfnamefont {H.}~\bibnamefont {Lamm}}, \bibinfo {author}
  {\bibfnamefont {S.}~\bibnamefont {Lawrence}}, \ and\ \bibinfo {author}
  {\bibfnamefont {N.~C.}\ \bibnamefont {Warrington}} (\bibinfo {collaboration}
  {NuQS}),\ }\href {\doibase 10.1103/PhysRevD.100.114501} {\bibfield  {journal}
  {\bibinfo  {journal} {Phys. Rev. D}\ }\textbf {\bibinfo {volume} {100}},\
  \bibinfo {pages} {114501} (\bibinfo {year} {2019})},\ \Eprint
  {http://arxiv.org/abs/1906.11213} {arXiv:1906.11213 [hep-lat]} \BibitemShut
  {NoStop}%
\bibitem [{\citenamefont {Kogut}\ and\ \citenamefont
  {Susskind}(1975)}]{PhysRevD.11.395}%
  \BibitemOpen
  \bibfield  {author} {\bibinfo {author} {\bibfnamefont {J.}~\bibnamefont
  {Kogut}}\ and\ \bibinfo {author} {\bibfnamefont {L.}~\bibnamefont
  {Susskind}},\ }\href {\doibase 10.1103/PhysRevD.11.395} {\bibfield  {journal}
  {\bibinfo  {journal} {Phys. Rev. D}\ }\textbf {\bibinfo {volume} {11}},\
  \bibinfo {pages} {395} (\bibinfo {year} {1975})}\BibitemShut {NoStop}%
\bibitem [{\citenamefont {Kogut}(1979)}]{RevModPhys.51.659}%
  \BibitemOpen
  \bibfield  {author} {\bibinfo {author} {\bibfnamefont {J.~B.}\ \bibnamefont
  {Kogut}},\ }\href {\doibase 10.1103/RevModPhys.51.659} {\bibfield  {journal}
  {\bibinfo  {journal} {Rev. Mod. Phys.}\ }\textbf {\bibinfo {volume} {51}},\
  \bibinfo {pages} {659} (\bibinfo {year} {1979})}\BibitemShut {NoStop}%
\bibitem [{\citenamefont {Bañuls}\ \emph {et~al.}(2020)\citenamefont
  {Bañuls}, \citenamefont {Blatt}, \citenamefont {Catani}, \citenamefont
  {Celi}, \citenamefont {Cirac}, \citenamefont {Dalmonte}, \citenamefont
  {Fallani}, \citenamefont {Jansen}, \citenamefont {Lewenstein}, \citenamefont
  {Montangero},\ and\ \citenamefont {et~al.}}]{Banuls_2020}%
  \BibitemOpen
  \bibfield  {author} {\bibinfo {author} {\bibfnamefont {M.~C.}\ \bibnamefont
  {Bañuls}}, \bibinfo {author} {\bibfnamefont {R.}~\bibnamefont {Blatt}},
  \bibinfo {author} {\bibfnamefont {J.}~\bibnamefont {Catani}}, \bibinfo
  {author} {\bibfnamefont {A.}~\bibnamefont {Celi}}, \bibinfo {author}
  {\bibfnamefont {J.~I.}\ \bibnamefont {Cirac}}, \bibinfo {author}
  {\bibfnamefont {M.}~\bibnamefont {Dalmonte}}, \bibinfo {author}
  {\bibfnamefont {L.}~\bibnamefont {Fallani}}, \bibinfo {author} {\bibfnamefont
  {K.}~\bibnamefont {Jansen}}, \bibinfo {author} {\bibfnamefont
  {M.}~\bibnamefont {Lewenstein}}, \bibinfo {author} {\bibfnamefont
  {S.}~\bibnamefont {Montangero}}, \ and\ \bibinfo {author} {\bibnamefont
  {et~al.}},\ }\href {\doibase 10.1140/epjd/e2020-100571-8} {\bibfield
  {journal} {\bibinfo  {journal} {The European Physical Journal D}\ }\textbf
  {\bibinfo {volume} {74}} (\bibinfo {year} {2020}),\
  10.1140/epjd/e2020-100571-8}\BibitemShut {NoStop}%
\bibitem [{\citenamefont {Klco}\ \emph {et~al.}(2020)\citenamefont {Klco},
  \citenamefont {Savage},\ and\ \citenamefont {Stryker}}]{PhysRevD.101.074512}%
  \BibitemOpen
  \bibfield  {author} {\bibinfo {author} {\bibfnamefont {N.}~\bibnamefont
  {Klco}}, \bibinfo {author} {\bibfnamefont {M.~J.}\ \bibnamefont {Savage}}, \
  and\ \bibinfo {author} {\bibfnamefont {J.~R.}\ \bibnamefont {Stryker}},\
  }\href {\doibase 10.1103/PhysRevD.101.074512} {\bibfield  {journal} {\bibinfo
   {journal} {Phys. Rev. D}\ }\textbf {\bibinfo {volume} {101}},\ \bibinfo
  {pages} {074512} (\bibinfo {year} {2020})}\BibitemShut {NoStop}%
\bibitem [{\citenamefont {Ciavarella}\ \emph {et~al.}(2021)\citenamefont
  {Ciavarella}, \citenamefont {Klco},\ and\ \citenamefont
  {Savage}}]{PhysRevD.103.094501}%
  \BibitemOpen
  \bibfield  {author} {\bibinfo {author} {\bibfnamefont {A.}~\bibnamefont
  {Ciavarella}}, \bibinfo {author} {\bibfnamefont {N.}~\bibnamefont {Klco}}, \
  and\ \bibinfo {author} {\bibfnamefont {M.~J.}\ \bibnamefont {Savage}},\
  }\href {\doibase 10.1103/PhysRevD.103.094501} {\bibfield  {journal} {\bibinfo
   {journal} {Phys. Rev. D}\ }\textbf {\bibinfo {volume} {103}},\ \bibinfo
  {pages} {094501} (\bibinfo {year} {2021})}\BibitemShut {NoStop}%
\bibitem [{\citenamefont {Klimov}\ \emph {et~al.}(2003)\citenamefont {Klimov},
  \citenamefont {Guzman}, \citenamefont {Retamal},\ and\ \citenamefont
  {Saavedra}}]{osti_20636540}%
  \BibitemOpen
  \bibfield  {author} {\bibinfo {author} {\bibfnamefont {A.~B.}\ \bibnamefont
  {Klimov}}, \bibinfo {author} {\bibfnamefont {R.}~\bibnamefont {Guzman}},
  \bibinfo {author} {\bibfnamefont {J.~C.}\ \bibnamefont {Retamal}}, \ and\
  \bibinfo {author} {\bibfnamefont {C.}~\bibnamefont {Saavedra}},\ }\href
  {\doibase 10.1103/PhysRevA.67.062313} {\bibfield  {journal} {\bibinfo
  {journal} {Physical Review. A}\ }\textbf {\bibinfo {volume} {67}} (\bibinfo
  {year} {2003}),\ 10.1103/PhysRevA.67.062313}\BibitemShut {NoStop}%
\bibitem [{\citenamefont {Low}\ \emph {et~al.}(2020)\citenamefont {Low},
  \citenamefont {White}, \citenamefont {Cox}, \citenamefont {Day},\ and\
  \citenamefont {Senko}}]{Low_2020}%
  \BibitemOpen
  \bibfield  {author} {\bibinfo {author} {\bibfnamefont {P.~J.}\ \bibnamefont
  {Low}}, \bibinfo {author} {\bibfnamefont {B.~M.}\ \bibnamefont {White}},
  \bibinfo {author} {\bibfnamefont {A.~A.}\ \bibnamefont {Cox}}, \bibinfo
  {author} {\bibfnamefont {M.~L.}\ \bibnamefont {Day}}, \ and\ \bibinfo
  {author} {\bibfnamefont {C.}~\bibnamefont {Senko}},\ }\href {\doibase
  10.1103/physrevresearch.2.033128} {\bibfield  {journal} {\bibinfo  {journal}
  {Physical Review Research}\ }\textbf {\bibinfo {volume} {2}} (\bibinfo {year}
  {2020}),\ 10.1103/physrevresearch.2.033128}\BibitemShut {NoStop}%
\bibitem [{\citenamefont {Lu}\ \emph {et~al.}(2020)\citenamefont {Lu},
  \citenamefont {Hu}, \citenamefont {Alshaykh}, \citenamefont {Moore},
  \citenamefont {Wang}, \citenamefont {Imany}, \citenamefont {Weiner},\ and\
  \citenamefont {Kais}}]{lu2020quantum}%
  \BibitemOpen
  \bibfield  {author} {\bibinfo {author} {\bibfnamefont {H.-H.}\ \bibnamefont
  {Lu}}, \bibinfo {author} {\bibfnamefont {Z.}~\bibnamefont {Hu}}, \bibinfo
  {author} {\bibfnamefont {M.~S.}\ \bibnamefont {Alshaykh}}, \bibinfo {author}
  {\bibfnamefont {A.~J.}\ \bibnamefont {Moore}}, \bibinfo {author}
  {\bibfnamefont {Y.}~\bibnamefont {Wang}}, \bibinfo {author} {\bibfnamefont
  {P.}~\bibnamefont {Imany}}, \bibinfo {author} {\bibfnamefont {A.~M.}\
  \bibnamefont {Weiner}}, \ and\ \bibinfo {author} {\bibfnamefont
  {S.}~\bibnamefont {Kais}},\ }\href {\doibase 10.1002/qute.201900074}
  {\bibfield  {journal} {\bibinfo  {journal} {Advanced Quantum Technologies}\
  }\textbf {\bibinfo {volume} {3}},\ \bibinfo {pages} {1900074} (\bibinfo
  {year} {2020})},\ \Eprint {http://arxiv.org/abs/1906.11401} {arXiv:1906.11401
  [quant-ph]} \BibitemShut {NoStop}%
\bibitem [{\citenamefont {Sawant}\ \emph {et~al.}(2020)\citenamefont {Sawant},
  \citenamefont {Blackmore}, \citenamefont {Gregory}, \citenamefont
  {Mur-Petit}, \citenamefont {Jaksch}, \citenamefont {Aldegunde}, \citenamefont
  {Hutson}, \citenamefont {Tarbutt},\ and\ \citenamefont
  {Cornish}}]{Sawant_2020}%
  \BibitemOpen
  \bibfield  {author} {\bibinfo {author} {\bibfnamefont {R.}~\bibnamefont
  {Sawant}}, \bibinfo {author} {\bibfnamefont {J.~A.}\ \bibnamefont
  {Blackmore}}, \bibinfo {author} {\bibfnamefont {P.~D.}\ \bibnamefont
  {Gregory}}, \bibinfo {author} {\bibfnamefont {J.}~\bibnamefont {Mur-Petit}},
  \bibinfo {author} {\bibfnamefont {D.}~\bibnamefont {Jaksch}}, \bibinfo
  {author} {\bibfnamefont {J.}~\bibnamefont {Aldegunde}}, \bibinfo {author}
  {\bibfnamefont {J.~M.}\ \bibnamefont {Hutson}}, \bibinfo {author}
  {\bibfnamefont {M.~R.}\ \bibnamefont {Tarbutt}}, \ and\ \bibinfo {author}
  {\bibfnamefont {S.~L.}\ \bibnamefont {Cornish}},\ }\href {\doibase
  10.1088/1367-2630/ab60f4} {\bibfield  {journal} {\bibinfo  {journal} {New
  Journal of Physics}\ }\textbf {\bibinfo {volume} {22}},\ \bibinfo {pages}
  {013027} (\bibinfo {year} {2020})}\BibitemShut {NoStop}%
\bibitem [{\citenamefont {Holland}\ \emph {et~al.}(2020)\citenamefont
  {Holland}, \citenamefont {Wendt}, \citenamefont {Kravvaris}, \citenamefont
  {Wu}, \citenamefont {Erich~Ormand}, \citenamefont {DuBois}, \citenamefont
  {Quaglioni},\ and\ \citenamefont {Pederiva}}]{Holland:2019zju}%
  \BibitemOpen
  \bibfield  {author} {\bibinfo {author} {\bibfnamefont {E.~T.}\ \bibnamefont
  {Holland}}, \bibinfo {author} {\bibfnamefont {K.~A.}\ \bibnamefont {Wendt}},
  \bibinfo {author} {\bibfnamefont {K.}~\bibnamefont {Kravvaris}}, \bibinfo
  {author} {\bibfnamefont {X.}~\bibnamefont {Wu}}, \bibinfo {author}
  {\bibfnamefont {W.}~\bibnamefont {Erich~Ormand}}, \bibinfo {author}
  {\bibfnamefont {J.~L.}\ \bibnamefont {DuBois}}, \bibinfo {author}
  {\bibfnamefont {S.}~\bibnamefont {Quaglioni}}, \ and\ \bibinfo {author}
  {\bibfnamefont {F.}~\bibnamefont {Pederiva}},\ }\href {\doibase
  10.1103/PhysRevA.101.062307} {\bibfield  {journal} {\bibinfo  {journal}
  {Phys. Rev. A}\ }\textbf {\bibinfo {volume} {101}},\ \bibinfo {pages}
  {062307} (\bibinfo {year} {2020})},\ \Eprint
  {http://arxiv.org/abs/1908.08222} {arXiv:1908.08222 [quant-ph]} \BibitemShut
  {NoStop}%
\bibitem [{SRF()}]{SRFFNAL}%
  \BibitemOpen
  \href@noop {} {\enquote {\bibinfo {title} {Fermilab quantum technologies},}\
  }\bibinfo {howpublished} {\url{https://td.fnal.gov/quantum-technologies/}},\
  \bibinfo {note} {accessed: 2021-05-31}\BibitemShut {NoStop}%
\bibitem [{\citenamefont {Gokhale}\ \emph {et~al.}(2019)\citenamefont
  {Gokhale}, \citenamefont {Baker}, \citenamefont {Duckering}, \citenamefont
  {Brown}, \citenamefont {Brown},\ and\ \citenamefont
  {Chong}}]{10.1145/3307650.3322253}%
  \BibitemOpen
  \bibfield  {author} {\bibinfo {author} {\bibfnamefont {P.}~\bibnamefont
  {Gokhale}}, \bibinfo {author} {\bibfnamefont {J.~M.}\ \bibnamefont {Baker}},
  \bibinfo {author} {\bibfnamefont {C.}~\bibnamefont {Duckering}}, \bibinfo
  {author} {\bibfnamefont {N.~C.}\ \bibnamefont {Brown}}, \bibinfo {author}
  {\bibfnamefont {K.~R.}\ \bibnamefont {Brown}}, \ and\ \bibinfo {author}
  {\bibfnamefont {F.~T.}\ \bibnamefont {Chong}},\ }in\ \href {\doibase
  10.1145/3307650.3322253} {\emph {\bibinfo {booktitle} {Proceedings of the
  46th International Symposium on Computer Architecture}}},\ \bibinfo {series
  and number} {ISCA '19}\ (\bibinfo  {publisher} {Association for Computing
  Machinery},\ \bibinfo {address} {New York, NY, USA},\ \bibinfo {year}
  {2019})\ p.\ \bibinfo {pages} {554–566}\BibitemShut {NoStop}%
\bibitem [{\citenamefont {Horn}(1981)}]{HORN1981149}%
  \BibitemOpen
  \bibfield  {author} {\bibinfo {author} {\bibfnamefont {D.}~\bibnamefont
  {Horn}},\ }\href {\doibase https://doi.org/10.1016/0370-2693(81)90763-2}
  {\bibfield  {journal} {\bibinfo  {journal} {Physics Letters B}\ }\textbf
  {\bibinfo {volume} {100}},\ \bibinfo {pages} {149 } (\bibinfo {year}
  {1981})}\BibitemShut {NoStop}%
\bibitem [{\citenamefont {Orland}\ and\ \citenamefont
  {Rohrlich}(1990)}]{ORLAND1990647}%
  \BibitemOpen
  \bibfield  {author} {\bibinfo {author} {\bibfnamefont {P.}~\bibnamefont
  {Orland}}\ and\ \bibinfo {author} {\bibfnamefont {D.}~\bibnamefont
  {Rohrlich}},\ }\href {\doibase https://doi.org/10.1016/0550-3213(90)90646-U}
  {\bibfield  {journal} {\bibinfo  {journal} {Nuclear Physics B}\ }\textbf
  {\bibinfo {volume} {338}},\ \bibinfo {pages} {647 } (\bibinfo {year}
  {1990})}\BibitemShut {NoStop}%
\bibitem [{\citenamefont {Chandrasekharan}\ and\ \citenamefont
  {Wiese}(1997{\natexlab{b}})}]{Chandrasekharan_1997}%
  \BibitemOpen
  \bibfield  {author} {\bibinfo {author} {\bibfnamefont {S.}~\bibnamefont
  {Chandrasekharan}}\ and\ \bibinfo {author} {\bibfnamefont {U.-J.}\
  \bibnamefont {Wiese}},\ }\href {\doibase 10.1016/s0550-3213(97)80041-7}
  {\bibfield  {journal} {\bibinfo  {journal} {Nuclear Physics B}\ }\textbf
  {\bibinfo {volume} {492}},\ \bibinfo {pages} {455–471} (\bibinfo {year}
  {1997}{\natexlab{b}})},\ \Eprint {http://arxiv.org/abs/hep-lat/9609042}
  {arXiv:hep-lat/9609042} \BibitemShut {NoStop}%
\bibitem [{\citenamefont {Brower}\ \emph {et~al.}(2004)\citenamefont {Brower},
  \citenamefont {Chandrasekharan}, \citenamefont {Riederer},\ and\
  \citenamefont {Wiese}}]{Brower:2003vy}%
  \BibitemOpen
  \bibfield  {author} {\bibinfo {author} {\bibfnamefont {R.}~\bibnamefont
  {Brower}}, \bibinfo {author} {\bibfnamefont {S.}~\bibnamefont
  {Chandrasekharan}}, \bibinfo {author} {\bibfnamefont {S.}~\bibnamefont
  {Riederer}}, \ and\ \bibinfo {author} {\bibfnamefont {U.}~\bibnamefont
  {Wiese}},\ }\href {\doibase 10.1016/j.nuclphysb.2004.06.007} {\bibfield
  {journal} {\bibinfo  {journal} {Nucl. Phys. B}\ }\textbf {\bibinfo {volume}
  {693}},\ \bibinfo {pages} {149} (\bibinfo {year} {2004})},\ \Eprint
  {http://arxiv.org/abs/hep-lat/0309182} {arXiv:hep-lat/0309182} \BibitemShut
  {NoStop}%
\bibitem [{\citenamefont {Wiese}(2006)}]{Wiese:2006kp}%
  \BibitemOpen
  \bibfield  {author} {\bibinfo {author} {\bibfnamefont {U.}~\bibnamefont
  {Wiese}},\ }\href {\doibase 10.1016/j.nuclphysbps.2006.01.027} {\bibfield
  {journal} {\bibinfo  {journal} {Nucl. Phys. B Proc. Suppl.}\ }\textbf
  {\bibinfo {volume} {153}},\ \bibinfo {pages} {336} (\bibinfo {year}
  {2006})},\ \bibinfo {note} {proceedings of the Workshop on Computational
  Hadron Physics}\BibitemShut {NoStop}%
\bibitem [{\citenamefont {Wiese}(2020)}]{Uwe:CERN_2020}%
  \BibitemOpen
  \bibfield  {author} {\bibinfo {author} {\bibfnamefont {U.-J.}\ \bibnamefont
  {Wiese}},\ }\href@noop {} {\enquote {\bibinfo {title} {{Quantum Link Models
  for the Quantum Simulation of Gauge Theories}},}\ }\bibinfo {howpublished}
  {\url{https://indico.cern.ch/event/940556/attachments/2078878/3491481/CERN2020new.pdf}}
  (\bibinfo {year} {2020}),\ \bibinfo {note} {accessed: 2021-05-31}\BibitemShut
  {NoStop}%
\bibitem [{\citenamefont {Mezzacapo}\ \emph
  {et~al.}(2015{\natexlab{a}})\citenamefont {Mezzacapo}, \citenamefont {Rico},
  \citenamefont {Sab\'{\i}n}, \citenamefont {Egusquiza}, \citenamefont
  {Lamata},\ and\ \citenamefont {Solano}}]{PhysRevLett.115.240502}%
  \BibitemOpen
  \bibfield  {author} {\bibinfo {author} {\bibfnamefont {A.}~\bibnamefont
  {Mezzacapo}}, \bibinfo {author} {\bibfnamefont {E.}~\bibnamefont {Rico}},
  \bibinfo {author} {\bibfnamefont {C.}~\bibnamefont {Sab\'{\i}n}}, \bibinfo
  {author} {\bibfnamefont {I.~L.}\ \bibnamefont {Egusquiza}}, \bibinfo {author}
  {\bibfnamefont {L.}~\bibnamefont {Lamata}}, \ and\ \bibinfo {author}
  {\bibfnamefont {E.}~\bibnamefont {Solano}},\ }\href {\doibase
  10.1103/PhysRevLett.115.240502} {\bibfield  {journal} {\bibinfo  {journal}
  {Phys. Rev. Lett.}\ }\textbf {\bibinfo {volume} {115}},\ \bibinfo {pages}
  {240502} (\bibinfo {year} {2015}{\natexlab{a}})}\BibitemShut {NoStop}%
\bibitem [{\citenamefont {Bernien}\ \emph {et~al.}(2017)\citenamefont
  {Bernien}, \citenamefont {Schwartz}, \citenamefont {Keesling}, \citenamefont
  {Levine}, \citenamefont {Omran}, \citenamefont {Pichler}, \citenamefont
  {Choi}, \citenamefont {Zibrov}, \citenamefont {Endres}, \citenamefont
  {Greiner}, \citenamefont {Vuleti{\'{c}}},\ and\ \citenamefont
  {Lukin}}]{Bernien2017}%
  \BibitemOpen
  \bibfield  {author} {\bibinfo {author} {\bibfnamefont {H.}~\bibnamefont
  {Bernien}}, \bibinfo {author} {\bibfnamefont {S.}~\bibnamefont {Schwartz}},
  \bibinfo {author} {\bibfnamefont {A.}~\bibnamefont {Keesling}}, \bibinfo
  {author} {\bibfnamefont {H.}~\bibnamefont {Levine}}, \bibinfo {author}
  {\bibfnamefont {A.}~\bibnamefont {Omran}}, \bibinfo {author} {\bibfnamefont
  {H.}~\bibnamefont {Pichler}}, \bibinfo {author} {\bibfnamefont
  {S.}~\bibnamefont {Choi}}, \bibinfo {author} {\bibfnamefont {A.~S.}\
  \bibnamefont {Zibrov}}, \bibinfo {author} {\bibfnamefont {M.}~\bibnamefont
  {Endres}}, \bibinfo {author} {\bibfnamefont {M.}~\bibnamefont {Greiner}},
  \bibinfo {author} {\bibfnamefont {V.}~\bibnamefont {Vuleti{\'{c}}}}, \ and\
  \bibinfo {author} {\bibfnamefont {M.~D.}\ \bibnamefont {Lukin}},\ }\href
  {\doibase 10.1038/nature24622} {\bibfield  {journal} {\bibinfo  {journal}
  {Nature}\ }\textbf {\bibinfo {volume} {551}},\ \bibinfo {pages} {579}
  (\bibinfo {year} {2017})}\BibitemShut {NoStop}%
\bibitem [{\citenamefont {Shen}\ \emph {et~al.}(2020)\citenamefont {Shen},
  \citenamefont {Luo}, \citenamefont {Highman}, \citenamefont {Clark},
  \citenamefont {DeMarco}, \citenamefont {El-Khadra},\ and\ \citenamefont
  {Gadway}}]{Shen:2020coq}%
  \BibitemOpen
  \bibfield  {author} {\bibinfo {author} {\bibfnamefont {J.}~\bibnamefont
  {Shen}}, \bibinfo {author} {\bibfnamefont {D.}~\bibnamefont {Luo}}, \bibinfo
  {author} {\bibfnamefont {M.}~\bibnamefont {Highman}}, \bibinfo {author}
  {\bibfnamefont {B.~K.}\ \bibnamefont {Clark}}, \bibinfo {author}
  {\bibfnamefont {B.}~\bibnamefont {DeMarco}}, \bibinfo {author} {\bibfnamefont
  {A.~X.}\ \bibnamefont {El-Khadra}}, \ and\ \bibinfo {author} {\bibfnamefont
  {B.}~\bibnamefont {Gadway}},\ }\href {\doibase 10.22323/1.363.0125}
  {\bibfield  {journal} {\bibinfo  {journal} {PoS}\ }\textbf {\bibinfo {volume}
  {LATTICE2019}},\ \bibinfo {pages} {125} (\bibinfo {year} {2020})},\ \Eprint
  {http://arxiv.org/abs/2001.10002} {arXiv:2001.10002 [hep-lat]} \BibitemShut
  {NoStop}%
\bibitem [{\citenamefont {Surace}\ \emph
  {et~al.}(2020{\natexlab{a}})\citenamefont {Surace}, \citenamefont {Mazza},
  \citenamefont {Giudici}, \citenamefont {Lerose}, \citenamefont {Gambassi},\
  and\ \citenamefont {Dalmonte}}]{Surace_2020}%
  \BibitemOpen
  \bibfield  {author} {\bibinfo {author} {\bibfnamefont {F.~M.}\ \bibnamefont
  {Surace}}, \bibinfo {author} {\bibfnamefont {P.~P.}\ \bibnamefont {Mazza}},
  \bibinfo {author} {\bibfnamefont {G.}~\bibnamefont {Giudici}}, \bibinfo
  {author} {\bibfnamefont {A.}~\bibnamefont {Lerose}}, \bibinfo {author}
  {\bibfnamefont {A.}~\bibnamefont {Gambassi}}, \ and\ \bibinfo {author}
  {\bibfnamefont {M.}~\bibnamefont {Dalmonte}},\ }\href {\doibase
  10.1103/physrevx.10.021041} {\bibfield  {journal} {\bibinfo  {journal}
  {Physical Review X}\ }\textbf {\bibinfo {volume} {10}} (\bibinfo {year}
  {2020}{\natexlab{a}}),\ 10.1103/physrevx.10.021041}\BibitemShut {NoStop}%
\bibitem [{\citenamefont {Stannigel}\ \emph {et~al.}(2014)\citenamefont
  {Stannigel}, \citenamefont {Hauke}, \citenamefont {Marcos}, \citenamefont
  {Hafezi}, \citenamefont {Diehl}, \citenamefont {Dalmonte},\ and\
  \citenamefont {Zoller}}]{Stannigel:2013zka}%
  \BibitemOpen
  \bibfield  {author} {\bibinfo {author} {\bibfnamefont {K.}~\bibnamefont
  {Stannigel}}, \bibinfo {author} {\bibfnamefont {P.}~\bibnamefont {Hauke}},
  \bibinfo {author} {\bibfnamefont {D.}~\bibnamefont {Marcos}}, \bibinfo
  {author} {\bibfnamefont {M.}~\bibnamefont {Hafezi}}, \bibinfo {author}
  {\bibfnamefont {S.}~\bibnamefont {Diehl}}, \bibinfo {author} {\bibfnamefont
  {M.}~\bibnamefont {Dalmonte}}, \ and\ \bibinfo {author} {\bibfnamefont
  {P.}~\bibnamefont {Zoller}},\ }\href {\doibase
  10.1103/PhysRevLett.112.120406} {\bibfield  {journal} {\bibinfo  {journal}
  {Phys. Rev. Lett.}\ }\textbf {\bibinfo {volume} {112}},\ \bibinfo {pages}
  {120406} (\bibinfo {year} {2014})},\ \Eprint {http://arxiv.org/abs/1308.0528}
  {arXiv:1308.0528 [quant-ph]} \BibitemShut {NoStop}%
\bibitem [{\citenamefont {Mezzacapo}\ \emph
  {et~al.}(2015{\natexlab{b}})\citenamefont {Mezzacapo}, \citenamefont {Rico},
  \citenamefont {Sab\'\i{}n}, \citenamefont {Egusquiza}, \citenamefont
  {Lamata},\ and\ \citenamefont {Solano}}]{Mezzacapo:2015bra}%
  \BibitemOpen
  \bibfield  {author} {\bibinfo {author} {\bibfnamefont {A.}~\bibnamefont
  {Mezzacapo}}, \bibinfo {author} {\bibfnamefont {E.}~\bibnamefont {Rico}},
  \bibinfo {author} {\bibfnamefont {C.}~\bibnamefont {Sab\'\i{}n}}, \bibinfo
  {author} {\bibfnamefont {I.~L.}\ \bibnamefont {Egusquiza}}, \bibinfo {author}
  {\bibfnamefont {L.}~\bibnamefont {Lamata}}, \ and\ \bibinfo {author}
  {\bibfnamefont {E.}~\bibnamefont {Solano}},\ }\href {\doibase
  10.1103/PhysRevLett.115.240502} {\bibfield  {journal} {\bibinfo  {journal}
  {Phys. Rev. Lett.}\ }\textbf {\bibinfo {volume} {115}},\ \bibinfo {pages}
  {240502} (\bibinfo {year} {2015}{\natexlab{b}})},\ \Eprint
  {http://arxiv.org/abs/1505.04720} {arXiv:1505.04720 [quant-ph]} \BibitemShut
  {NoStop}%
\bibitem [{\citenamefont {Singh}(2019)}]{singh2019qubit}%
  \BibitemOpen
  \bibfield  {author} {\bibinfo {author} {\bibfnamefont {H.}~\bibnamefont
  {Singh}},\ }\href {https://arxiv.org/abs/1911.12353} {\enquote {\bibinfo
  {title} {Qubit $o(n)$ nonlinear sigma models},}\ } (\bibinfo {year} {2019}),\
  \Eprint {http://arxiv.org/abs/1911.12353} {arXiv:1911.12353 [hep-lat]}
  \BibitemShut {NoStop}%
\bibitem [{\citenamefont {Surace}\ \emph
  {et~al.}(2020{\natexlab{b}})\citenamefont {Surace}, \citenamefont {Mazza},
  \citenamefont {Giudici}, \citenamefont {Lerose}, \citenamefont {Gambassi},\
  and\ \citenamefont {Dalmonte}}]{Surace:2019dtp}%
  \BibitemOpen
  \bibfield  {author} {\bibinfo {author} {\bibfnamefont {F.~M.}\ \bibnamefont
  {Surace}}, \bibinfo {author} {\bibfnamefont {P.~P.}\ \bibnamefont {Mazza}},
  \bibinfo {author} {\bibfnamefont {G.}~\bibnamefont {Giudici}}, \bibinfo
  {author} {\bibfnamefont {A.}~\bibnamefont {Lerose}}, \bibinfo {author}
  {\bibfnamefont {A.}~\bibnamefont {Gambassi}}, \ and\ \bibinfo {author}
  {\bibfnamefont {M.}~\bibnamefont {Dalmonte}},\ }\href {\doibase
  10.1103/PhysRevX.10.021041} {\bibfield  {journal} {\bibinfo  {journal} {Phys.
  Rev. X}\ }\textbf {\bibinfo {volume} {10}},\ \bibinfo {pages} {021041}
  (\bibinfo {year} {2020}{\natexlab{b}})},\ \Eprint
  {http://arxiv.org/abs/1902.09551} {arXiv:1902.09551 [cond-mat.quant-gas]}
  \BibitemShut {NoStop}%
\bibitem [{\citenamefont {Van~Damme}\ \emph {et~al.}(2020)\citenamefont
  {Van~Damme}, \citenamefont {Halimeh},\ and\ \citenamefont
  {Hauke}}]{VanDamme:2020rur}%
  \BibitemOpen
  \bibfield  {author} {\bibinfo {author} {\bibfnamefont {M.}~\bibnamefont
  {Van~Damme}}, \bibinfo {author} {\bibfnamefont {J.~C.}\ \bibnamefont
  {Halimeh}}, \ and\ \bibinfo {author} {\bibfnamefont {P.}~\bibnamefont
  {Hauke}},\ }\href@noop {} {\  (\bibinfo {year} {2020})},\ \Eprint
  {http://arxiv.org/abs/2010.07338} {arXiv:2010.07338 [cond-mat.quant-gas]}
  \BibitemShut {NoStop}%
\bibitem [{\citenamefont {Tan}\ \emph {et~al.}(2021)\citenamefont {Tan} \emph
  {et~al.}}]{Tan:2019kya}%
  \BibitemOpen
  \bibfield  {author} {\bibinfo {author} {\bibfnamefont {W.~L.}\ \bibnamefont
  {Tan}} \emph {et~al.},\ }\href {\doibase 10.1038/s41567-021-01194-3}
  {\bibfield  {journal} {\bibinfo  {journal} {Nature Phys.}\ }\textbf {\bibinfo
  {volume} {17}},\ \bibinfo {pages} {742} (\bibinfo {year} {2021})},\ \Eprint
  {http://arxiv.org/abs/1912.11117} {arXiv:1912.11117 [quant-ph]} \BibitemShut
  {NoStop}%
\bibitem [{\citenamefont {Verdel}\ \emph {et~al.}(2020)\citenamefont {Verdel},
  \citenamefont {Liu}, \citenamefont {Whitsitt}, \citenamefont {Gorshkov},\
  and\ \citenamefont {Heyl}}]{Verdel:2019chj}%
  \BibitemOpen
  \bibfield  {author} {\bibinfo {author} {\bibfnamefont {R.}~\bibnamefont
  {Verdel}}, \bibinfo {author} {\bibfnamefont {F.}~\bibnamefont {Liu}},
  \bibinfo {author} {\bibfnamefont {S.}~\bibnamefont {Whitsitt}}, \bibinfo
  {author} {\bibfnamefont {A.~V.}\ \bibnamefont {Gorshkov}}, \ and\ \bibinfo
  {author} {\bibfnamefont {M.}~\bibnamefont {Heyl}},\ }\href {\doibase
  10.1103/PhysRevB.102.014308} {\bibfield  {journal} {\bibinfo  {journal}
  {Phys. Rev. B}\ }\textbf {\bibinfo {volume} {102}},\ \bibinfo {pages}
  {014308} (\bibinfo {year} {2020})},\ \Eprint
  {http://arxiv.org/abs/1911.11382} {arXiv:1911.11382 [cond-mat.stat-mech]}
  \BibitemShut {NoStop}%
\bibitem [{\citenamefont {Gustafson}\ \emph {et~al.}(2021)\citenamefont
  {Gustafson}, \citenamefont {Zhu}, \citenamefont {Dreher}, \citenamefont
  {Linke},\ and\ \citenamefont {Meurice}}]{Gustafson:2021imb}%
  \BibitemOpen
  \bibfield  {author} {\bibinfo {author} {\bibfnamefont {E.}~\bibnamefont
  {Gustafson}}, \bibinfo {author} {\bibfnamefont {Y.}~\bibnamefont {Zhu}},
  \bibinfo {author} {\bibfnamefont {P.}~\bibnamefont {Dreher}}, \bibinfo
  {author} {\bibfnamefont {N.~M.}\ \bibnamefont {Linke}}, \ and\ \bibinfo
  {author} {\bibfnamefont {Y.}~\bibnamefont {Meurice}},\ }\href@noop {} {\
  (\bibinfo {year} {2021})},\ \Eprint {http://arxiv.org/abs/2103.06848}
  {arXiv:2103.06848 [hep-lat]} \BibitemShut {NoStop}%
\bibitem [{\citenamefont {Zohar}\ and\ \citenamefont
  {Cirac}(2018)}]{Zohar:2018cwb}%
  \BibitemOpen
  \bibfield  {author} {\bibinfo {author} {\bibfnamefont {E.}~\bibnamefont
  {Zohar}}\ and\ \bibinfo {author} {\bibfnamefont {J.~I.}\ \bibnamefont
  {Cirac}},\ }\href {\doibase 10.1103/PhysRevB.98.075119} {\bibfield  {journal}
  {\bibinfo  {journal} {Phys. Rev. B}\ }\textbf {\bibinfo {volume} {98}},\
  \bibinfo {pages} {075119} (\bibinfo {year} {2018})},\ \Eprint
  {http://arxiv.org/abs/1805.05347} {arXiv:1805.05347 [quant-ph]} \BibitemShut
  {NoStop}%
\bibitem [{\citenamefont {Zohar}\ and\ \citenamefont
  {Cirac}(2019)}]{Zohar:2019ygc}%
  \BibitemOpen
  \bibfield  {author} {\bibinfo {author} {\bibfnamefont {E.}~\bibnamefont
  {Zohar}}\ and\ \bibinfo {author} {\bibfnamefont {J.~I.}\ \bibnamefont
  {Cirac}},\ }\href {\doibase 10.1103/PhysRevD.99.114511} {\bibfield  {journal}
  {\bibinfo  {journal} {Phys. Rev. D}\ }\textbf {\bibinfo {volume} {99}},\
  \bibinfo {pages} {114511} (\bibinfo {year} {2019})},\ \Eprint
  {http://arxiv.org/abs/1905.00652} {arXiv:1905.00652 [quant-ph]} \BibitemShut
  {NoStop}%
\bibitem [{\citenamefont {Aidelsburger}\ \emph {et~al.}(2021)\citenamefont
  {Aidelsburger}, \citenamefont {Barbiero}, \citenamefont {Bermudez},
  \citenamefont {Chanda}, \citenamefont {Dauphin}, \citenamefont
  {González-Cuadra}, \citenamefont {Grzybowski}, \citenamefont {Hands},
  \citenamefont {Jendrzejewski}, \citenamefont {Jünemann}, \citenamefont
  {Juzeliunas}, \citenamefont {Kasper}, \citenamefont {Piga}, \citenamefont
  {Ran}, \citenamefont {Rizzi}, \citenamefont {Sierra}, \citenamefont
  {Tagliacozzo}, \citenamefont {Tirrito}, \citenamefont {Zache}, \citenamefont
  {Zakrzewski}, \citenamefont {Zohar},\ and\ \citenamefont
  {Lewenstein}}]{aidelsburger2021cold}%
  \BibitemOpen
  \bibfield  {author} {\bibinfo {author} {\bibfnamefont {M.}~\bibnamefont
  {Aidelsburger}}, \bibinfo {author} {\bibfnamefont {L.}~\bibnamefont
  {Barbiero}}, \bibinfo {author} {\bibfnamefont {A.}~\bibnamefont {Bermudez}},
  \bibinfo {author} {\bibfnamefont {T.}~\bibnamefont {Chanda}}, \bibinfo
  {author} {\bibfnamefont {A.}~\bibnamefont {Dauphin}}, \bibinfo {author}
  {\bibfnamefont {D.}~\bibnamefont {González-Cuadra}}, \bibinfo {author}
  {\bibfnamefont {P.~R.}\ \bibnamefont {Grzybowski}}, \bibinfo {author}
  {\bibfnamefont {S.}~\bibnamefont {Hands}}, \bibinfo {author} {\bibfnamefont
  {F.}~\bibnamefont {Jendrzejewski}}, \bibinfo {author} {\bibfnamefont
  {J.}~\bibnamefont {Jünemann}}, \bibinfo {author} {\bibfnamefont
  {G.}~\bibnamefont {Juzeliunas}}, \bibinfo {author} {\bibfnamefont
  {V.}~\bibnamefont {Kasper}}, \bibinfo {author} {\bibfnamefont
  {A.}~\bibnamefont {Piga}}, \bibinfo {author} {\bibfnamefont {S.-J.}\
  \bibnamefont {Ran}}, \bibinfo {author} {\bibfnamefont {M.}~\bibnamefont
  {Rizzi}}, \bibinfo {author} {\bibfnamefont {G.}~\bibnamefont {Sierra}},
  \bibinfo {author} {\bibfnamefont {L.}~\bibnamefont {Tagliacozzo}}, \bibinfo
  {author} {\bibfnamefont {E.}~\bibnamefont {Tirrito}}, \bibinfo {author}
  {\bibfnamefont {T.~V.}\ \bibnamefont {Zache}}, \bibinfo {author}
  {\bibfnamefont {J.}~\bibnamefont {Zakrzewski}}, \bibinfo {author}
  {\bibfnamefont {E.}~\bibnamefont {Zohar}}, \ and\ \bibinfo {author}
  {\bibfnamefont {M.}~\bibnamefont {Lewenstein}},\ }\href@noop {} {\bibfield
  {journal} {\bibinfo  {journal} {arXiv preprint arXiv:2106.03063}\ } (\bibinfo
  {year} {2021})}\BibitemShut {NoStop}%
\bibitem [{\citenamefont {Buser}\ \emph {et~al.}(2020)\citenamefont {Buser},
  \citenamefont {Gharibyan}, \citenamefont {Hanada}, \citenamefont {Honda},\
  and\ \citenamefont {Liu}}]{Buser:2020cvn}%
  \BibitemOpen
  \bibfield  {author} {\bibinfo {author} {\bibfnamefont {A.}~\bibnamefont
  {Buser}}, \bibinfo {author} {\bibfnamefont {H.}~\bibnamefont {Gharibyan}},
  \bibinfo {author} {\bibfnamefont {M.}~\bibnamefont {Hanada}}, \bibinfo
  {author} {\bibfnamefont {M.}~\bibnamefont {Honda}}, \ and\ \bibinfo {author}
  {\bibfnamefont {J.}~\bibnamefont {Liu}},\ }\href@noop {} {\  (\bibinfo {year}
  {2020})},\ \Eprint {http://arxiv.org/abs/2011.06576} {arXiv:2011.06576
  [hep-th]} \BibitemShut {NoStop}%
\bibitem [{\citenamefont {Schwinger}(1965)}]{Schwinger:1965}%
  \BibitemOpen
  \bibfield  {author} {\bibinfo {author} {\bibfnamefont {J.}~\bibnamefont
  {Schwinger}},\ }in\ \href@noop {} {\emph {\bibinfo {booktitle} {Quantum
  Theory of Angular Momentum}}},\ \bibinfo {editor} {edited by\ \bibinfo
  {editor} {\bibfnamefont {L.~C.}\ \bibnamefont {Biedenharn}}\ and\ \bibinfo
  {editor} {\bibfnamefont {H.~J.~V.}\ \bibnamefont {Dam}}}\ (\bibinfo
  {publisher} {Academic},\ \bibinfo {address} {New York},\ \bibinfo {year}
  {1965})\BibitemShut {NoStop}%
\bibitem [{\citenamefont {Mathur}\ \emph {et~al.}(2010)\citenamefont {Mathur},
  \citenamefont {Raychowdhury},\ and\ \citenamefont
  {Anishetty}}]{Mathur:2010wc}%
  \BibitemOpen
  \bibfield  {author} {\bibinfo {author} {\bibfnamefont {M.}~\bibnamefont
  {Mathur}}, \bibinfo {author} {\bibfnamefont {I.}~\bibnamefont
  {Raychowdhury}}, \ and\ \bibinfo {author} {\bibfnamefont {R.}~\bibnamefont
  {Anishetty}},\ }\href {\doibase 10.1063/1.3464267} {\bibfield  {journal}
  {\bibinfo  {journal} {J. Math. Phys.}\ }\textbf {\bibinfo {volume} {51}},\
  \bibinfo {pages} {093504} (\bibinfo {year} {2010})},\ \Eprint
  {http://arxiv.org/abs/1003.5487} {arXiv:1003.5487 [math-ph]} \BibitemShut
  {NoStop}%
\bibitem [{\citenamefont {Raychowdhury}\ and\ \citenamefont
  {Stryker}(2018)}]{Raychowdhury:2018osk}%
  \BibitemOpen
  \bibfield  {author} {\bibinfo {author} {\bibfnamefont {I.}~\bibnamefont
  {Raychowdhury}}\ and\ \bibinfo {author} {\bibfnamefont {J.~R.}\ \bibnamefont
  {Stryker}},\ }\href@noop {} {\  (\bibinfo {year} {2018})},\ \Eprint
  {http://arxiv.org/abs/1812.07554} {arXiv:1812.07554 [hep-lat]} \BibitemShut
  {NoStop}%
\bibitem [{\citenamefont {Raychowdhury}\ and\ \citenamefont
  {Stryker}(2020)}]{Raychowdhury_2020}%
  \BibitemOpen
  \bibfield  {author} {\bibinfo {author} {\bibfnamefont {I.}~\bibnamefont
  {Raychowdhury}}\ and\ \bibinfo {author} {\bibfnamefont {J.~R.}\ \bibnamefont
  {Stryker}},\ }\href {\doibase 10.1103/physrevd.101.114502} {\bibfield
  {journal} {\bibinfo  {journal} {Physical Review D}\ }\textbf {\bibinfo
  {volume} {101}} (\bibinfo {year} {2020}),\
  10.1103/physrevd.101.114502}\BibitemShut {NoStop}%
\bibitem [{\citenamefont {Gonz\'alez-Cuadra}\ \emph {et~al.}(2017)\citenamefont
  {Gonz\'alez-Cuadra}, \citenamefont {Zohar},\ and\ \citenamefont
  {Cirac}}]{Gonzalez-Cuadra:2017lvz}%
  \BibitemOpen
  \bibfield  {author} {\bibinfo {author} {\bibfnamefont {D.}~\bibnamefont
  {Gonz\'alez-Cuadra}}, \bibinfo {author} {\bibfnamefont {E.}~\bibnamefont
  {Zohar}}, \ and\ \bibinfo {author} {\bibfnamefont {J.~I.}\ \bibnamefont
  {Cirac}},\ }\href {\doibase 10.1088/1367-2630/aa6f37} {\bibfield  {journal}
  {\bibinfo  {journal} {New J. Phys.}\ }\textbf {\bibinfo {volume} {19}},\
  \bibinfo {pages} {063038} (\bibinfo {year} {2017})},\ \Eprint
  {http://arxiv.org/abs/1702.05492} {arXiv:1702.05492 [quant-ph]} \BibitemShut
  {NoStop}%
\bibitem [{\citenamefont {Mil}\ \emph {et~al.}(2020)\citenamefont {Mil},
  \citenamefont {Zache}, \citenamefont {Hegde}, \citenamefont {Xia},
  \citenamefont {Bhatt}, \citenamefont {Oberthaler}, \citenamefont {Hauke},
  \citenamefont {Berges},\ and\ \citenamefont {Jendrzejewski}}]{Mil:2019pbt}%
  \BibitemOpen
  \bibfield  {author} {\bibinfo {author} {\bibfnamefont {A.}~\bibnamefont
  {Mil}}, \bibinfo {author} {\bibfnamefont {T.~V.}\ \bibnamefont {Zache}},
  \bibinfo {author} {\bibfnamefont {A.}~\bibnamefont {Hegde}}, \bibinfo
  {author} {\bibfnamefont {A.}~\bibnamefont {Xia}}, \bibinfo {author}
  {\bibfnamefont {R.~P.}\ \bibnamefont {Bhatt}}, \bibinfo {author}
  {\bibfnamefont {M.~K.}\ \bibnamefont {Oberthaler}}, \bibinfo {author}
  {\bibfnamefont {P.}~\bibnamefont {Hauke}}, \bibinfo {author} {\bibfnamefont
  {J.}~\bibnamefont {Berges}}, \ and\ \bibinfo {author} {\bibfnamefont
  {F.}~\bibnamefont {Jendrzejewski}},\ }\href {\doibase
  10.1126/science.aaz5312} {\bibfield  {journal} {\bibinfo  {journal}
  {Science}\ }\textbf {\bibinfo {volume} {367}},\ \bibinfo {pages} {1128}
  (\bibinfo {year} {2020})},\ \Eprint {http://arxiv.org/abs/1909.07641}
  {arXiv:1909.07641 [cond-mat.quant-gas]} \BibitemShut {NoStop}%
\bibitem [{\citenamefont {Ott}\ \emph {et~al.}(2020)\citenamefont {Ott},
  \citenamefont {Zache}, \citenamefont {Jendrzejewski},\ and\ \citenamefont
  {Berges}}]{Ott:2020ycj}%
  \BibitemOpen
  \bibfield  {author} {\bibinfo {author} {\bibfnamefont {R.}~\bibnamefont
  {Ott}}, \bibinfo {author} {\bibfnamefont {T.~V.}\ \bibnamefont {Zache}},
  \bibinfo {author} {\bibfnamefont {F.}~\bibnamefont {Jendrzejewski}}, \ and\
  \bibinfo {author} {\bibfnamefont {J.}~\bibnamefont {Berges}},\ }\href@noop {}
  {\  (\bibinfo {year} {2020})},\ \Eprint {http://arxiv.org/abs/2012.10432}
  {arXiv:2012.10432 [cond-mat.quant-gas]} \BibitemShut {NoStop}%
\bibitem [{\citenamefont {Dasgupta}\ and\ \citenamefont
  {Raychowdhury}(2020)}]{Dasgupta:2020itb}%
  \BibitemOpen
  \bibfield  {author} {\bibinfo {author} {\bibfnamefont {R.}~\bibnamefont
  {Dasgupta}}\ and\ \bibinfo {author} {\bibfnamefont {I.}~\bibnamefont
  {Raychowdhury}},\ }\href@noop {} {\  (\bibinfo {year} {2020})},\ \Eprint
  {http://arxiv.org/abs/2009.13969} {arXiv:2009.13969 [hep-lat]} \BibitemShut
  {NoStop}%
\bibitem [{\citenamefont {Kreshchuk}\ \emph {et~al.}(2020)\citenamefont
  {Kreshchuk}, \citenamefont {Kirby}, \citenamefont {Goldstein}, \citenamefont
  {Beauchemin},\ and\ \citenamefont {Love}}]{Kreshchuk:2020dla}%
  \BibitemOpen
  \bibfield  {author} {\bibinfo {author} {\bibfnamefont {M.}~\bibnamefont
  {Kreshchuk}}, \bibinfo {author} {\bibfnamefont {W.~M.}\ \bibnamefont
  {Kirby}}, \bibinfo {author} {\bibfnamefont {G.}~\bibnamefont {Goldstein}},
  \bibinfo {author} {\bibfnamefont {H.}~\bibnamefont {Beauchemin}}, \ and\
  \bibinfo {author} {\bibfnamefont {P.~J.}\ \bibnamefont {Love}},\ }\href@noop
  {} {\  (\bibinfo {year} {2020})},\ \Eprint {http://arxiv.org/abs/2002.04016}
  {arXiv:2002.04016 [quant-ph]} \BibitemShut {NoStop}%
\bibitem [{\citenamefont {Echevarria}\ \emph {et~al.}(2020)\citenamefont
  {Echevarria}, \citenamefont {Egusquiza}, \citenamefont {Rico},\ and\
  \citenamefont {Schnell}}]{Echevarria:2020wct}%
  \BibitemOpen
  \bibfield  {author} {\bibinfo {author} {\bibfnamefont {M.~G.}\ \bibnamefont
  {Echevarria}}, \bibinfo {author} {\bibfnamefont {I.~L.}\ \bibnamefont
  {Egusquiza}}, \bibinfo {author} {\bibfnamefont {E.}~\bibnamefont {Rico}}, \
  and\ \bibinfo {author} {\bibfnamefont {G.}~\bibnamefont {Schnell}},\
  }\href@noop {} {\  (\bibinfo {year} {2020})},\ \Eprint
  {http://arxiv.org/abs/2011.01275} {arXiv:2011.01275 [quant-ph]} \BibitemShut
  {NoStop}%
\bibitem [{\citenamefont {Kreshchuk}\ \emph
  {et~al.}(2021{\natexlab{a}})\citenamefont {Kreshchuk}, \citenamefont {Jia},
  \citenamefont {Kirby}, \citenamefont {Goldstein}, \citenamefont {Vary},\ and\
  \citenamefont {Love}}]{Kreshchuk:2020aiq}%
  \BibitemOpen
  \bibfield  {author} {\bibinfo {author} {\bibfnamefont {M.}~\bibnamefont
  {Kreshchuk}}, \bibinfo {author} {\bibfnamefont {S.}~\bibnamefont {Jia}},
  \bibinfo {author} {\bibfnamefont {W.~M.}\ \bibnamefont {Kirby}}, \bibinfo
  {author} {\bibfnamefont {G.}~\bibnamefont {Goldstein}}, \bibinfo {author}
  {\bibfnamefont {J.~P.}\ \bibnamefont {Vary}}, \ and\ \bibinfo {author}
  {\bibfnamefont {P.~J.}\ \bibnamefont {Love}},\ }\href {\doibase
  10.1103/PhysRevA.103.062601} {\bibfield  {journal} {\bibinfo  {journal}
  {Phys. Rev. A}\ }\textbf {\bibinfo {volume} {103}},\ \bibinfo {pages}
  {062601} (\bibinfo {year} {2021}{\natexlab{a}})},\ \Eprint
  {http://arxiv.org/abs/2011.13443} {arXiv:2011.13443 [quant-ph]} \BibitemShut
  {NoStop}%
\bibitem [{\citenamefont {Kreshchuk}\ \emph
  {et~al.}(2021{\natexlab{b}})\citenamefont {Kreshchuk}, \citenamefont {Jia},
  \citenamefont {Kirby}, \citenamefont {Goldstein}, \citenamefont {Vary},\ and\
  \citenamefont {Love}}]{Kreshchuk:2020kcz}%
  \BibitemOpen
  \bibfield  {author} {\bibinfo {author} {\bibfnamefont {M.}~\bibnamefont
  {Kreshchuk}}, \bibinfo {author} {\bibfnamefont {S.}~\bibnamefont {Jia}},
  \bibinfo {author} {\bibfnamefont {W.~M.}\ \bibnamefont {Kirby}}, \bibinfo
  {author} {\bibfnamefont {G.}~\bibnamefont {Goldstein}}, \bibinfo {author}
  {\bibfnamefont {J.~P.}\ \bibnamefont {Vary}}, \ and\ \bibinfo {author}
  {\bibfnamefont {P.~J.}\ \bibnamefont {Love}},\ }\href {\doibase
  10.3390/e23050597} {\bibfield  {journal} {\bibinfo  {journal} {Entropy}\
  }\textbf {\bibinfo {volume} {23}},\ \bibinfo {pages} {597} (\bibinfo {year}
  {2021}{\natexlab{b}})},\ \Eprint {http://arxiv.org/abs/2009.07885}
  {arXiv:2009.07885 [quant-ph]} \BibitemShut {NoStop}%
\bibitem [{\citenamefont {Zache}\ \emph {et~al.}(2020)\citenamefont {Zache},
  \citenamefont {Schweigler}, \citenamefont {Erne}, \citenamefont
  {Schmiedmayer},\ and\ \citenamefont {Berges}}]{Zache:2019xkx}%
  \BibitemOpen
  \bibfield  {author} {\bibinfo {author} {\bibfnamefont {T.~V.}\ \bibnamefont
  {Zache}}, \bibinfo {author} {\bibfnamefont {T.}~\bibnamefont {Schweigler}},
  \bibinfo {author} {\bibfnamefont {S.}~\bibnamefont {Erne}}, \bibinfo {author}
  {\bibfnamefont {J.}~\bibnamefont {Schmiedmayer}}, \ and\ \bibinfo {author}
  {\bibfnamefont {J.}~\bibnamefont {Berges}},\ }\href {\doibase
  10.1103/PhysRevX.10.011020} {\bibfield  {journal} {\bibinfo  {journal} {Phys.
  Rev. X}\ }\textbf {\bibinfo {volume} {10}},\ \bibinfo {pages} {011020}
  (\bibinfo {year} {2020})},\ \Eprint {http://arxiv.org/abs/1909.12815}
  {arXiv:1909.12815 [cond-mat.quant-gas]} \BibitemShut {NoStop}%
\bibitem [{\citenamefont {Bermudez}\ \emph {et~al.}(2017)\citenamefont
  {Bermudez}, \citenamefont {Aarts},\ and\ \citenamefont
  {M\"uller}}]{Bermudez:2017yrq}%
  \BibitemOpen
  \bibfield  {author} {\bibinfo {author} {\bibfnamefont {A.}~\bibnamefont
  {Bermudez}}, \bibinfo {author} {\bibfnamefont {G.}~\bibnamefont {Aarts}}, \
  and\ \bibinfo {author} {\bibfnamefont {M.}~\bibnamefont {M\"uller}},\ }\href
  {\doibase 10.1103/PhysRevX.7.041012} {\bibfield  {journal} {\bibinfo
  {journal} {Phys. Rev. X}\ }\textbf {\bibinfo {volume} {7}},\ \bibinfo {pages}
  {041012} (\bibinfo {year} {2017})},\ \Eprint
  {http://arxiv.org/abs/1704.02877} {arXiv:1704.02877 [quant-ph]} \BibitemShut
  {NoStop}%
\bibitem [{\citenamefont {Mart\'\i{}n-V\'azquez}\ \emph
  {et~al.}(2021)\citenamefont {Mart\'\i{}n-V\'azquez}, \citenamefont {Aarts},
  \citenamefont {M\"uller},\ and\ \citenamefont
  {Bermudez}}]{Martin-Vazquez:2021mxc}%
  \BibitemOpen
  \bibfield  {author} {\bibinfo {author} {\bibfnamefont {G.}~\bibnamefont
  {Mart\'\i{}n-V\'azquez}}, \bibinfo {author} {\bibfnamefont {G.}~\bibnamefont
  {Aarts}}, \bibinfo {author} {\bibfnamefont {M.}~\bibnamefont {M\"uller}}, \
  and\ \bibinfo {author} {\bibfnamefont {A.}~\bibnamefont {Bermudez}},\
  }\href@noop {} {\  (\bibinfo {year} {2021})},\ \Eprint
  {http://arxiv.org/abs/2105.06886} {arXiv:2105.06886 [quant-ph]} \BibitemShut
  {NoStop}%
\bibitem [{\citenamefont {Zohar}\ \emph
  {et~al.}(2013{\natexlab{a}})\citenamefont {Zohar}, \citenamefont {Cirac},\
  and\ \citenamefont {Reznik}}]{Zohar:2013zla}%
  \BibitemOpen
  \bibfield  {author} {\bibinfo {author} {\bibfnamefont {E.}~\bibnamefont
  {Zohar}}, \bibinfo {author} {\bibfnamefont {J.~I.}\ \bibnamefont {Cirac}}, \
  and\ \bibinfo {author} {\bibfnamefont {B.}~\bibnamefont {Reznik}},\ }\href
  {\doibase 10.1103/PhysRevA.88.023617} {\bibfield  {journal} {\bibinfo
  {journal} {Phys. Rev. A}\ }\textbf {\bibinfo {volume} {88}},\ \bibinfo
  {pages} {023617} (\bibinfo {year} {2013}{\natexlab{a}})},\ \Eprint
  {http://arxiv.org/abs/1303.5040} {arXiv:1303.5040 [quant-ph]} \BibitemShut
  {NoStop}%
\bibitem [{\citenamefont {Zohar}\ \emph
  {et~al.}(2013{\natexlab{b}})\citenamefont {Zohar}, \citenamefont {Cirac},\
  and\ \citenamefont {Reznik}}]{Zohar:2012xf}%
  \BibitemOpen
  \bibfield  {author} {\bibinfo {author} {\bibfnamefont {E.}~\bibnamefont
  {Zohar}}, \bibinfo {author} {\bibfnamefont {J.~I.}\ \bibnamefont {Cirac}}, \
  and\ \bibinfo {author} {\bibfnamefont {B.}~\bibnamefont {Reznik}},\ }\href
  {\doibase 10.1103/PhysRevLett.110.125304} {\bibfield  {journal} {\bibinfo
  {journal} {Phys. Rev. Lett.}\ }\textbf {\bibinfo {volume} {110}},\ \bibinfo
  {pages} {125304} (\bibinfo {year} {2013}{\natexlab{b}})},\ \Eprint
  {http://arxiv.org/abs/1211.2241} {arXiv:1211.2241 [quant-ph]} \BibitemShut
  {NoStop}%
\bibitem [{\citenamefont {Davoudi}\ \emph
  {et~al.}(2021{\natexlab{b}})\citenamefont {Davoudi}, \citenamefont {Linke},\
  and\ \citenamefont {Pagano}}]{Davoudi:2021ney}%
  \BibitemOpen
  \bibfield  {author} {\bibinfo {author} {\bibfnamefont {Z.}~\bibnamefont
  {Davoudi}}, \bibinfo {author} {\bibfnamefont {N.~M.}\ \bibnamefont {Linke}},
  \ and\ \bibinfo {author} {\bibfnamefont {G.}~\bibnamefont {Pagano}},\
  }\href@noop {} {\  (\bibinfo {year} {2021}{\natexlab{b}})},\ \Eprint
  {http://arxiv.org/abs/2104.09346} {arXiv:2104.09346 [quant-ph]} \BibitemShut
  {NoStop}%
\bibitem [{\citenamefont {Stryker}(2019)}]{Stryker:2018efp}%
  \BibitemOpen
  \bibfield  {author} {\bibinfo {author} {\bibfnamefont {J.~R.}\ \bibnamefont
  {Stryker}},\ }\href {\doibase 10.1103/PhysRevA.99.042301} {\bibfield
  {journal} {\bibinfo  {journal} {Phys. Rev. A}\ }\textbf {\bibinfo {volume}
  {99}},\ \bibinfo {pages} {042301} (\bibinfo {year} {2019})},\ \Eprint
  {http://arxiv.org/abs/1812.01617} {arXiv:1812.01617 [quant-ph]} \BibitemShut
  {NoStop}%
\bibitem [{\citenamefont {Goldman}\ and\ \citenamefont
  {Dalibard}(2014)}]{Goldman:2014xja}%
  \BibitemOpen
  \bibfield  {author} {\bibinfo {author} {\bibfnamefont {N.}~\bibnamefont
  {Goldman}}\ and\ \bibinfo {author} {\bibfnamefont {J.}~\bibnamefont
  {Dalibard}},\ }\href {\doibase 10.1103/PhysRevX.4.031027} {\bibfield
  {journal} {\bibinfo  {journal} {Phys. Rev. X}\ }\textbf {\bibinfo {volume}
  {4}},\ \bibinfo {pages} {031027} (\bibinfo {year} {2014})},\ \bibinfo {note}
  {[Erratum: Phys.Rev.X 5, 029902 (2015)]},\ \Eprint
  {http://arxiv.org/abs/1404.4373} {arXiv:1404.4373 [cond-mat.quant-gas]}
  \BibitemShut {NoStop}%
\bibitem [{\citenamefont {Halimeh}\ and\ \citenamefont
  {Hauke}(2020)}]{Halimeh:2019svu}%
  \BibitemOpen
  \bibfield  {author} {\bibinfo {author} {\bibfnamefont {J.~C.}\ \bibnamefont
  {Halimeh}}\ and\ \bibinfo {author} {\bibfnamefont {P.}~\bibnamefont
  {Hauke}},\ }\href {\doibase 10.1103/PhysRevLett.125.030503} {\bibfield
  {journal} {\bibinfo  {journal} {Phys. Rev. Lett.}\ }\textbf {\bibinfo
  {volume} {125}},\ \bibinfo {pages} {030503} (\bibinfo {year} {2020})},\
  \Eprint {http://arxiv.org/abs/2001.00024} {arXiv:2001.00024
  [cond-mat.quant-gas]} \BibitemShut {NoStop}%
\bibitem [{\citenamefont {Kasper}\ \emph {et~al.}(2020)\citenamefont {Kasper},
  \citenamefont {Zache}, \citenamefont {Jendrzejewski}, \citenamefont
  {Lewenstein},\ and\ \citenamefont {Zohar}}]{Kasper:2020owz}%
  \BibitemOpen
  \bibfield  {author} {\bibinfo {author} {\bibfnamefont {V.}~\bibnamefont
  {Kasper}}, \bibinfo {author} {\bibfnamefont {T.~V.}\ \bibnamefont {Zache}},
  \bibinfo {author} {\bibfnamefont {F.}~\bibnamefont {Jendrzejewski}}, \bibinfo
  {author} {\bibfnamefont {M.}~\bibnamefont {Lewenstein}}, \ and\ \bibinfo
  {author} {\bibfnamefont {E.}~\bibnamefont {Zohar}},\ }\href@noop {} {\
  (\bibinfo {year} {2020})},\ \Eprint {http://arxiv.org/abs/2012.08620}
  {arXiv:2012.08620 [quant-ph]} \BibitemShut {NoStop}%
\bibitem [{\citenamefont {Lamm}\ \emph {et~al.}(2020)\citenamefont {Lamm},
  \citenamefont {Lawrence},\ and\ \citenamefont {Yamauchi}}]{Lamm:2020jwv}%
  \BibitemOpen
  \bibfield  {author} {\bibinfo {author} {\bibfnamefont {H.}~\bibnamefont
  {Lamm}}, \bibinfo {author} {\bibfnamefont {S.}~\bibnamefont {Lawrence}}, \
  and\ \bibinfo {author} {\bibfnamefont {Y.}~\bibnamefont {Yamauchi}} (\bibinfo
  {collaboration} {NuQS}),\ }\href@noop {} {\  (\bibinfo {year} {2020})},\
  \Eprint {http://arxiv.org/abs/2005.12688} {arXiv:2005.12688 [quant-ph]}
  \BibitemShut {NoStop}%
\bibitem [{\citenamefont {Halimeh}\ \emph
  {et~al.}(2020{\natexlab{a}})\citenamefont {Halimeh}, \citenamefont {Lang},
  \citenamefont {Mildenberger}, \citenamefont {Jiang},\ and\ \citenamefont
  {Hauke}}]{Halimeh:2020ecg}%
  \BibitemOpen
  \bibfield  {author} {\bibinfo {author} {\bibfnamefont {J.~C.}\ \bibnamefont
  {Halimeh}}, \bibinfo {author} {\bibfnamefont {H.}~\bibnamefont {Lang}},
  \bibinfo {author} {\bibfnamefont {J.}~\bibnamefont {Mildenberger}}, \bibinfo
  {author} {\bibfnamefont {Z.}~\bibnamefont {Jiang}}, \ and\ \bibinfo {author}
  {\bibfnamefont {P.}~\bibnamefont {Hauke}},\ }\href@noop {} {\  (\bibinfo
  {year} {2020}{\natexlab{a}})},\ \Eprint {http://arxiv.org/abs/2007.00668}
  {arXiv:2007.00668 [quant-ph]} \BibitemShut {NoStop}%
\bibitem [{\citenamefont {Halimeh}\ \emph
  {et~al.}(2020{\natexlab{b}})\citenamefont {Halimeh}, \citenamefont {Kasper},\
  and\ \citenamefont {Hauke}}]{Halimeh:2020djb}%
  \BibitemOpen
  \bibfield  {author} {\bibinfo {author} {\bibfnamefont {J.~C.}\ \bibnamefont
  {Halimeh}}, \bibinfo {author} {\bibfnamefont {V.}~\bibnamefont {Kasper}}, \
  and\ \bibinfo {author} {\bibfnamefont {P.}~\bibnamefont {Hauke}},\
  }\href@noop {} {\  (\bibinfo {year} {2020}{\natexlab{b}})},\ \Eprint
  {http://arxiv.org/abs/2009.07848} {arXiv:2009.07848 [cond-mat.quant-gas]}
  \BibitemShut {NoStop}%
\bibitem [{\citenamefont {Van~Damme}\ \emph {et~al.}(2021)\citenamefont
  {Van~Damme}, \citenamefont {Lang}, \citenamefont {Hauke},\ and\ \citenamefont
  {Halimeh}}]{VanDamme:2021teo}%
  \BibitemOpen
  \bibfield  {author} {\bibinfo {author} {\bibfnamefont {M.}~\bibnamefont
  {Van~Damme}}, \bibinfo {author} {\bibfnamefont {H.}~\bibnamefont {Lang}},
  \bibinfo {author} {\bibfnamefont {P.}~\bibnamefont {Hauke}}, \ and\ \bibinfo
  {author} {\bibfnamefont {J.~C.}\ \bibnamefont {Halimeh}},\ }\href@noop {} {\
  (\bibinfo {year} {2021})},\ \Eprint {http://arxiv.org/abs/2104.07040}
  {arXiv:2104.07040 [cond-mat.quant-gas]} \BibitemShut {NoStop}%
\bibitem [{\citenamefont {Almheiri}\ \emph {et~al.}(2015)\citenamefont
  {Almheiri}, \citenamefont {Dong},\ and\ \citenamefont
  {Harlow}}]{Almheiri:2014lwa}%
  \BibitemOpen
  \bibfield  {author} {\bibinfo {author} {\bibfnamefont {A.}~\bibnamefont
  {Almheiri}}, \bibinfo {author} {\bibfnamefont {X.}~\bibnamefont {Dong}}, \
  and\ \bibinfo {author} {\bibfnamefont {D.}~\bibnamefont {Harlow}},\ }\href
  {\doibase 10.1007/JHEP04(2015)163} {\bibfield  {journal} {\bibinfo  {journal}
  {JHEP}\ }\textbf {\bibinfo {volume} {04}},\ \bibinfo {pages} {163} (\bibinfo
  {year} {2015})},\ \Eprint {http://arxiv.org/abs/1411.7041} {arXiv:1411.7041
  [hep-th]} \BibitemShut {NoStop}%
\bibitem [{\citenamefont {Pastawski}\ \emph {et~al.}(2015)\citenamefont
  {Pastawski}, \citenamefont {Yoshida}, \citenamefont {Harlow},\ and\
  \citenamefont {Preskill}}]{Pastawski_2015}%
  \BibitemOpen
  \bibfield  {author} {\bibinfo {author} {\bibfnamefont {F.}~\bibnamefont
  {Pastawski}}, \bibinfo {author} {\bibfnamefont {B.}~\bibnamefont {Yoshida}},
  \bibinfo {author} {\bibfnamefont {D.}~\bibnamefont {Harlow}}, \ and\ \bibinfo
  {author} {\bibfnamefont {J.}~\bibnamefont {Preskill}},\ }\href {\doibase
  10.1007/jhep06(2015)149} {\bibfield  {journal} {\bibinfo  {journal} {Journal
  of High Energy Physics}\ }\textbf {\bibinfo {volume} {2015}} (\bibinfo {year}
  {2015}),\ 10.1007/jhep06(2015)149}\BibitemShut {NoStop}%
\bibitem [{\citenamefont {Savary}\ and\ \citenamefont
  {Balents}(2016)}]{Savary_2016}%
  \BibitemOpen
  \bibfield  {author} {\bibinfo {author} {\bibfnamefont {L.}~\bibnamefont
  {Savary}}\ and\ \bibinfo {author} {\bibfnamefont {L.}~\bibnamefont
  {Balents}},\ }\href {\doibase 10.1088/0034-4885/80/1/016502} {\bibfield
  {journal} {\bibinfo  {journal} {Reports on Progress in Physics}\ }\textbf
  {\bibinfo {volume} {80}},\ \bibinfo {pages} {016502} (\bibinfo {year}
  {2016})}\BibitemShut {NoStop}%
\bibitem [{\citenamefont {Campbell}\ \emph {et~al.}(2017)\citenamefont
  {Campbell}, \citenamefont {Terhal},\ and\ \citenamefont
  {Vuillot}}]{Campbell_2017}%
  \BibitemOpen
  \bibfield  {author} {\bibinfo {author} {\bibfnamefont {E.~T.}\ \bibnamefont
  {Campbell}}, \bibinfo {author} {\bibfnamefont {B.~M.}\ \bibnamefont
  {Terhal}}, \ and\ \bibinfo {author} {\bibfnamefont {C.}~\bibnamefont
  {Vuillot}},\ }\href {\doibase 10.1038/nature23460} {\bibfield  {journal}
  {\bibinfo  {journal} {Nature}\ }\textbf {\bibinfo {volume} {549}},\ \bibinfo
  {pages} {172–179} (\bibinfo {year} {2017})}\BibitemShut {NoStop}%
\bibitem [{\citenamefont {Kitaev}(1997{\natexlab{b}})}]{kitaev1997quantum}%
  \BibitemOpen
  \bibfield  {author} {\bibinfo {author} {\bibfnamefont {A.~Y.}\ \bibnamefont
  {Kitaev}},\ }in\ \href@noop {} {\emph {\bibinfo {booktitle} {Proceedings of
  the Third International Conference on Quantum Communication and
  Measurement}}}\ (\bibinfo  {publisher} {Springer},\ \bibinfo {year} {1997})\
  pp.\ \bibinfo {pages} {181--188}\BibitemShut {NoStop}%
\bibitem [{\citenamefont {Bravyi}\ and\ \citenamefont
  {Kitaev}(1998)}]{Bravyi:1998sy}%
  \BibitemOpen
  \bibfield  {author} {\bibinfo {author} {\bibfnamefont {S.~B.}\ \bibnamefont
  {Bravyi}}\ and\ \bibinfo {author} {\bibfnamefont {A.~Y.}\ \bibnamefont
  {Kitaev}},\ }\href@noop {} {\  (\bibinfo {year} {1998})},\ \Eprint
  {http://arxiv.org/abs/quant-ph/9811052} {arXiv:quant-ph/9811052} \BibitemShut
  {NoStop}%
\bibitem [{\citenamefont {Dennis}\ \emph {et~al.}(2002)\citenamefont {Dennis},
  \citenamefont {Kitaev}, \citenamefont {Landahl},\ and\ \citenamefont
  {Preskill}}]{Dennis_2002}%
  \BibitemOpen
  \bibfield  {author} {\bibinfo {author} {\bibfnamefont {E.}~\bibnamefont
  {Dennis}}, \bibinfo {author} {\bibfnamefont {A.}~\bibnamefont {Kitaev}},
  \bibinfo {author} {\bibfnamefont {A.}~\bibnamefont {Landahl}}, \ and\
  \bibinfo {author} {\bibfnamefont {J.}~\bibnamefont {Preskill}},\ }\href
  {\doibase 10.1063/1.1499754} {\bibfield  {journal} {\bibinfo  {journal}
  {Journal of Mathematical Physics}\ }\textbf {\bibinfo {volume} {43}},\
  \bibinfo {pages} {4452–4505} (\bibinfo {year} {2002})}\BibitemShut
  {NoStop}%
\bibitem [{\citenamefont {Fowler}\ \emph {et~al.}(2012)\citenamefont {Fowler},
  \citenamefont {Mariantoni}, \citenamefont {Martinis},\ and\ \citenamefont
  {Cleland}}]{PhysRevA.86.032324}%
  \BibitemOpen
  \bibfield  {author} {\bibinfo {author} {\bibfnamefont {A.~G.}\ \bibnamefont
  {Fowler}}, \bibinfo {author} {\bibfnamefont {M.}~\bibnamefont {Mariantoni}},
  \bibinfo {author} {\bibfnamefont {J.~M.}\ \bibnamefont {Martinis}}, \ and\
  \bibinfo {author} {\bibfnamefont {A.~N.}\ \bibnamefont {Cleland}},\ }\href
  {\doibase 10.1103/PhysRevA.86.032324} {\bibfield  {journal} {\bibinfo
  {journal} {Phys. Rev. A}\ }\textbf {\bibinfo {volume} {86}},\ \bibinfo
  {pages} {032324} (\bibinfo {year} {2012})}\BibitemShut {NoStop}%
\bibitem [{\citenamefont {Kitaev}\ and\ \citenamefont
  {Laumann}(2009)}]{kitaev2009topological}%
  \BibitemOpen
  \bibfield  {author} {\bibinfo {author} {\bibfnamefont {A.}~\bibnamefont
  {Kitaev}}\ and\ \bibinfo {author} {\bibfnamefont {C.}~\bibnamefont
  {Laumann}},\ }\href@noop {} {\enquote {\bibinfo {title} {Topological phases
  and quantum computation},}\ } (\bibinfo {year} {2009}),\ \Eprint
  {http://arxiv.org/abs/0904.2771} {arXiv:0904.2771 [cond-mat.mes-hall]}
  \BibitemShut {NoStop}%
\bibitem [{\citenamefont {Mathur}\ and\ \citenamefont
  {Sreeraj}(2016)}]{Mathur:2016cko}%
  \BibitemOpen
  \bibfield  {author} {\bibinfo {author} {\bibfnamefont {M.}~\bibnamefont
  {Mathur}}\ and\ \bibinfo {author} {\bibfnamefont {T.~P.}\ \bibnamefont
  {Sreeraj}},\ }\href {\doibase 10.1103/PhysRevD.94.085029} {\bibfield
  {journal} {\bibinfo  {journal} {Phys. Rev. D}\ }\textbf {\bibinfo {volume}
  {94}},\ \bibinfo {pages} {085029} (\bibinfo {year} {2016})},\ \Eprint
  {http://arxiv.org/abs/1604.00315} {arXiv:1604.00315 [hep-lat]} \BibitemShut
  {NoStop}%
\bibitem [{\citenamefont {Jochym-O’Connor}\ and\ \citenamefont
  {Yoder}(2021)}]{Jochym_O_Connor_2021}%
  \BibitemOpen
  \bibfield  {author} {\bibinfo {author} {\bibfnamefont {T.}~\bibnamefont
  {Jochym-O’Connor}}\ and\ \bibinfo {author} {\bibfnamefont {T.~J.}\
  \bibnamefont {Yoder}},\ }\href {\doibase 10.1103/physrevresearch.3.013118}
  {\bibfield  {journal} {\bibinfo  {journal} {Physical Review Research}\
  }\textbf {\bibinfo {volume} {3}} (\bibinfo {year} {2021}),\
  10.1103/physrevresearch.3.013118}\BibitemShut {NoStop}%
\bibitem [{\citenamefont {Bombin}\ and\ \citenamefont
  {Martin-Delgado}(2006)}]{PhysRevLett.97.180501}%
  \BibitemOpen
  \bibfield  {author} {\bibinfo {author} {\bibfnamefont {H.}~\bibnamefont
  {Bombin}}\ and\ \bibinfo {author} {\bibfnamefont {M.~A.}\ \bibnamefont
  {Martin-Delgado}},\ }\href {\doibase 10.1103/PhysRevLett.97.180501}
  {\bibfield  {journal} {\bibinfo  {journal} {Phys. Rev. Lett.}\ }\textbf
  {\bibinfo {volume} {97}},\ \bibinfo {pages} {180501} (\bibinfo {year}
  {2006})}\BibitemShut {NoStop}%
\bibitem [{\citenamefont {Bomb\'{\i}n}(2015)}]{PhysRevX.5.031043}%
  \BibitemOpen
  \bibfield  {author} {\bibinfo {author} {\bibfnamefont {H.}~\bibnamefont
  {Bomb\'{\i}n}},\ }\href {\doibase 10.1103/PhysRevX.5.031043} {\bibfield
  {journal} {\bibinfo  {journal} {Phys. Rev. X}\ }\textbf {\bibinfo {volume}
  {5}},\ \bibinfo {pages} {031043} (\bibinfo {year} {2015})}\BibitemShut
  {NoStop}%
\bibitem [{\citenamefont {Brown}\ \emph
  {et~al.}(2016{\natexlab{a}})\citenamefont {Brown}, \citenamefont
  {Nickerson},\ and\ \citenamefont {Browne}}]{Brown_2016}%
  \BibitemOpen
  \bibfield  {author} {\bibinfo {author} {\bibfnamefont {B.~J.}\ \bibnamefont
  {Brown}}, \bibinfo {author} {\bibfnamefont {N.~H.}\ \bibnamefont
  {Nickerson}}, \ and\ \bibinfo {author} {\bibfnamefont {D.~E.}\ \bibnamefont
  {Browne}},\ }\href {\doibase 10.1038/ncomms12302} {\bibfield  {journal}
  {\bibinfo  {journal} {Nature Communications}\ }\textbf {\bibinfo {volume}
  {7}} (\bibinfo {year} {2016}{\natexlab{a}}),\
  10.1038/ncomms12302}\BibitemShut {NoStop}%
\bibitem [{\citenamefont {Wootton}\ \emph {et~al.}(2010)\citenamefont
  {Wootton}, \citenamefont {Heath},\ and\ \citenamefont
  {Pachos}}]{Wootton2010LocalizationAQ}%
  \BibitemOpen
  \bibfield  {author} {\bibinfo {author} {\bibfnamefont {J.~R.}\ \bibnamefont
  {Wootton}}, \bibinfo {author} {\bibfnamefont {R.}~\bibnamefont {Heath}}, \
  and\ \bibinfo {author} {\bibfnamefont {J.}~\bibnamefont {Pachos}}\ }(\bibinfo
  {year} {2010})\BibitemShut {NoStop}%
\bibitem [{\citenamefont {Andrist}\ \emph {et~al.}(2011)\citenamefont
  {Andrist}, \citenamefont {Katzgraber}, \citenamefont {Bombin},\ and\
  \citenamefont {Martin-Delgado}}]{Andrist_2011}%
  \BibitemOpen
  \bibfield  {author} {\bibinfo {author} {\bibfnamefont {R.~S.}\ \bibnamefont
  {Andrist}}, \bibinfo {author} {\bibfnamefont {H.~G.}\ \bibnamefont
  {Katzgraber}}, \bibinfo {author} {\bibfnamefont {H.}~\bibnamefont {Bombin}},
  \ and\ \bibinfo {author} {\bibfnamefont {M.~A.}\ \bibnamefont
  {Martin-Delgado}},\ }\href {\doibase 10.1088/1367-2630/13/8/083006}
  {\bibfield  {journal} {\bibinfo  {journal} {New Journal of Physics}\ }\textbf
  {\bibinfo {volume} {13}},\ \bibinfo {pages} {083006} (\bibinfo {year}
  {2011})}\BibitemShut {NoStop}%
\bibitem [{\citenamefont {Brown}\ \emph
  {et~al.}(2016{\natexlab{b}})\citenamefont {Brown}, \citenamefont
  {Nickerson},\ and\ \citenamefont {Browne}}]{Brown2016}%
  \BibitemOpen
  \bibfield  {author} {\bibinfo {author} {\bibfnamefont {B.~J.}\ \bibnamefont
  {Brown}}, \bibinfo {author} {\bibfnamefont {N.~H.}\ \bibnamefont
  {Nickerson}}, \ and\ \bibinfo {author} {\bibfnamefont {D.~E.}\ \bibnamefont
  {Browne}},\ }\href {\doibase 10.1038/ncomms12302} {\bibfield  {journal}
  {\bibinfo  {journal} {Nature Communications}\ }\textbf {\bibinfo {volume}
  {7}},\ \bibinfo {pages} {12302} (\bibinfo {year}
  {2016}{\natexlab{b}})}\BibitemShut {NoStop}%
\bibitem [{\citenamefont {Kitaev}(2006)}]{kitaev2006anyons}%
  \BibitemOpen
  \bibfield  {author} {\bibinfo {author} {\bibfnamefont {A.}~\bibnamefont
  {Kitaev}},\ }\href {\doibase 10.1016/j.aop.2005.10.005} {\bibfield  {journal}
  {\bibinfo  {journal} {Annals of Physics}\ }\textbf {\bibinfo {volume}
  {321}},\ \bibinfo {pages} {2} (\bibinfo {year} {2006})}\BibitemShut {NoStop}%
\bibitem [{\citenamefont {Kaplan}(1992)}]{Kaplan_1992}%
  \BibitemOpen
  \bibfield  {author} {\bibinfo {author} {\bibfnamefont {D.~B.}\ \bibnamefont
  {Kaplan}},\ }\href {\doibase 10.1016/0370-2693(92)91112-m} {\bibfield
  {journal} {\bibinfo  {journal} {Physics Letters B}\ }\textbf {\bibinfo
  {volume} {288}},\ \bibinfo {pages} {342–347} (\bibinfo {year}
  {1992})}\BibitemShut {NoStop}%
\bibitem [{\citenamefont {Kaplan}()}]{Kaplan_youtube_Harvard2020}%
  \BibitemOpen
  \bibfield  {author} {\bibinfo {author} {\bibfnamefont {D.~B.}\ \bibnamefont
  {Kaplan}},\ }\href
  {https://www.youtube.com/watch?v=1Ua33eZMh78&list=PL0NRmB0fnLJQAnYwkpt9PN2PBKx4rvdup&index=3}
  {\enquote {\bibinfo {title} {Domain wall fermions and chiral gauge
  theories:topological insulators},}\ }\BibitemShut {NoStop}%
\bibitem [{\citenamefont {Albash}\ and\ \citenamefont
  {Lidar}(2018)}]{RevModPhys.90.015002}%
  \BibitemOpen
  \bibfield  {author} {\bibinfo {author} {\bibfnamefont {T.}~\bibnamefont
  {Albash}}\ and\ \bibinfo {author} {\bibfnamefont {D.~A.}\ \bibnamefont
  {Lidar}},\ }\href {\doibase 10.1103/RevModPhys.90.015002} {\bibfield
  {journal} {\bibinfo  {journal} {Rev. Mod. Phys.}\ }\textbf {\bibinfo {volume}
  {90}},\ \bibinfo {pages} {015002} (\bibinfo {year} {2018})}\BibitemShut
  {NoStop}%
\bibitem [{\citenamefont {{Kaplan}}\ \emph {et~al.}(2017)\citenamefont
  {{Kaplan}}, \citenamefont {{Klco}},\ and\ \citenamefont
  {{Roggero}}}]{2017arXiv170908250K}%
  \BibitemOpen
  \bibfield  {author} {\bibinfo {author} {\bibfnamefont {D.~B.}\ \bibnamefont
  {{Kaplan}}}, \bibinfo {author} {\bibfnamefont {N.}~\bibnamefont {{Klco}}}, \
  and\ \bibinfo {author} {\bibfnamefont {A.}~\bibnamefont {{Roggero}}},\
  }\href@noop {} {\bibfield  {journal} {\bibinfo  {journal} {arXiv e-prints}\
  ,\ \bibinfo {eid} {arXiv:1709.08250}} (\bibinfo {year} {2017})},\ \Eprint
  {http://arxiv.org/abs/1709.08250} {arXiv:1709.08250 [quant-ph]} \BibitemShut
  {NoStop}%
\bibitem [{\citenamefont {Lee}\ \emph {et~al.}(2020{\natexlab{b}})\citenamefont
  {Lee}, \citenamefont {Bonitati}, \citenamefont {Given}, \citenamefont
  {Hicks}, \citenamefont {Li}, \citenamefont {Lu}, \citenamefont {Rai},
  \citenamefont {Sarkar},\ and\ \citenamefont {Watkins}}]{LEE2020135536}%
  \BibitemOpen
  \bibfield  {author} {\bibinfo {author} {\bibfnamefont {D.}~\bibnamefont
  {Lee}}, \bibinfo {author} {\bibfnamefont {J.}~\bibnamefont {Bonitati}},
  \bibinfo {author} {\bibfnamefont {G.}~\bibnamefont {Given}}, \bibinfo
  {author} {\bibfnamefont {C.}~\bibnamefont {Hicks}}, \bibinfo {author}
  {\bibfnamefont {N.}~\bibnamefont {Li}}, \bibinfo {author} {\bibfnamefont
  {B.-N.}\ \bibnamefont {Lu}}, \bibinfo {author} {\bibfnamefont
  {A.}~\bibnamefont {Rai}}, \bibinfo {author} {\bibfnamefont {A.}~\bibnamefont
  {Sarkar}}, \ and\ \bibinfo {author} {\bibfnamefont {J.}~\bibnamefont
  {Watkins}},\ }\href {\doibase https://doi.org/10.1016/j.physletb.2020.135536}
  {\bibfield  {journal} {\bibinfo  {journal} {Physics Letters B}\ }\textbf
  {\bibinfo {volume} {807}},\ \bibinfo {pages} {135536} (\bibinfo {year}
  {2020}{\natexlab{b}})}\BibitemShut {NoStop}%
\bibitem [{\citenamefont {Metcalf}\ \emph {et~al.}(2020)\citenamefont
  {Metcalf}, \citenamefont {Moussa}, \citenamefont {de~Jong},\ and\
  \citenamefont {Sarovar}}]{PhysRevResearch.2.023214}%
  \BibitemOpen
  \bibfield  {author} {\bibinfo {author} {\bibfnamefont {M.}~\bibnamefont
  {Metcalf}}, \bibinfo {author} {\bibfnamefont {J.~E.}\ \bibnamefont {Moussa}},
  \bibinfo {author} {\bibfnamefont {W.~A.}\ \bibnamefont {de~Jong}}, \ and\
  \bibinfo {author} {\bibfnamefont {M.}~\bibnamefont {Sarovar}},\ }\href
  {\doibase 10.1103/PhysRevResearch.2.023214} {\bibfield  {journal} {\bibinfo
  {journal} {Phys. Rev. Research}\ }\textbf {\bibinfo {volume} {2}},\ \bibinfo
  {pages} {023214} (\bibinfo {year} {2020})}\BibitemShut {NoStop}%
\bibitem [{\citenamefont {Somma}\ and\ \citenamefont
  {Boixo}(2013)}]{doi:10.1137/120871997}%
  \BibitemOpen
  \bibfield  {author} {\bibinfo {author} {\bibfnamefont {R.~D.}\ \bibnamefont
  {Somma}}\ and\ \bibinfo {author} {\bibfnamefont {S.}~\bibnamefont {Boixo}},\
  }\href {\doibase 10.1137/120871997} {\bibfield  {journal} {\bibinfo
  {journal} {SIAM Journal on Computing}\ }\textbf {\bibinfo {volume} {42}},\
  \bibinfo {pages} {593} (\bibinfo {year} {2013})}\BibitemShut {NoStop}%
\bibitem [{\citenamefont {Du}\ \emph {et~al.}(2021)\citenamefont {Du},
  \citenamefont {Vary}, \citenamefont {Zhao},\ and\ \citenamefont
  {Zuo}}]{du2021ab}%
  \BibitemOpen
  \bibfield  {author} {\bibinfo {author} {\bibfnamefont {W.}~\bibnamefont
  {Du}}, \bibinfo {author} {\bibfnamefont {J.~P.}\ \bibnamefont {Vary}},
  \bibinfo {author} {\bibfnamefont {X.}~\bibnamefont {Zhao}}, \ and\ \bibinfo
  {author} {\bibfnamefont {W.}~\bibnamefont {Zuo}},\ }\href@noop {} {\enquote
  {\bibinfo {title} {Ab initio nuclear structure via quantum adiabatic
  algorithm},}\ } (\bibinfo {year} {2021}),\ \Eprint
  {http://arxiv.org/abs/2105.08910} {arXiv:2105.08910 [nucl-th]} \BibitemShut
  {NoStop}%
\bibitem [{\citenamefont {{Kitaev}}(1995)}]{1995quant.ph.11026K}%
  \BibitemOpen
  \bibfield  {author} {\bibinfo {author} {\bibfnamefont {A.~Y.}\ \bibnamefont
  {{Kitaev}}},\ }\href@noop {} {\bibfield  {journal} {\bibinfo  {journal}
  {arXiv e-prints}\ ,\ \bibinfo {eid} {quant-ph/9511026}} (\bibinfo {year}
  {1995})},\ \Eprint {http://arxiv.org/abs/quant-ph/9511026}
  {arXiv:quant-ph/9511026 [quant-ph]} \BibitemShut {NoStop}%
\bibitem [{\citenamefont {Abrams}\ and\ \citenamefont
  {Lloyd}(1999)}]{PhysRevLett.83.5162}%
  \BibitemOpen
  \bibfield  {author} {\bibinfo {author} {\bibfnamefont {D.~S.}\ \bibnamefont
  {Abrams}}\ and\ \bibinfo {author} {\bibfnamefont {S.}~\bibnamefont {Lloyd}},\
  }\href {\doibase 10.1103/PhysRevLett.83.5162} {\bibfield  {journal} {\bibinfo
   {journal} {Phys. Rev. Lett.}\ }\textbf {\bibinfo {volume} {83}},\ \bibinfo
  {pages} {5162} (\bibinfo {year} {1999})}\BibitemShut {NoStop}%
\bibitem [{\citenamefont {Svore}\ \emph {et~al.}(2014)\citenamefont {Svore},
  \citenamefont {Hastings},\ and\ \citenamefont
  {Freedman}}]{10.5555/2600508.2600515}%
  \BibitemOpen
  \bibfield  {author} {\bibinfo {author} {\bibfnamefont {K.~M.}\ \bibnamefont
  {Svore}}, \bibinfo {author} {\bibfnamefont {M.~B.}\ \bibnamefont {Hastings}},
  \ and\ \bibinfo {author} {\bibfnamefont {M.}~\bibnamefont {Freedman}},\
  }\href@noop {} {\bibfield  {journal} {\bibinfo  {journal} {Quantum Info.
  Comput.}\ }\textbf {\bibinfo {volume} {14}},\ \bibinfo {pages} {306–328}
  (\bibinfo {year} {2014})}\BibitemShut {NoStop}%
\bibitem [{\citenamefont {Wiebe}\ and\ \citenamefont
  {Granade}(2016)}]{PhysRevLett.117.010503}%
  \BibitemOpen
  \bibfield  {author} {\bibinfo {author} {\bibfnamefont {N.}~\bibnamefont
  {Wiebe}}\ and\ \bibinfo {author} {\bibfnamefont {C.}~\bibnamefont
  {Granade}},\ }\href {\doibase 10.1103/PhysRevLett.117.010503} {\bibfield
  {journal} {\bibinfo  {journal} {Phys. Rev. Lett.}\ }\textbf {\bibinfo
  {volume} {117}},\ \bibinfo {pages} {010503} (\bibinfo {year}
  {2016})}\BibitemShut {NoStop}%
\bibitem [{\citenamefont {Ovrum}\ and\ \citenamefont
  {Hjorth-Jensen}(2007)}]{ovrum2007quantum}%
  \BibitemOpen
  \bibfield  {author} {\bibinfo {author} {\bibfnamefont {E.}~\bibnamefont
  {Ovrum}}\ and\ \bibinfo {author} {\bibfnamefont {M.}~\bibnamefont
  {Hjorth-Jensen}},\ }\href@noop {} {\enquote {\bibinfo {title} {Quantum
  computation algorithm for many-body studies},}\ } (\bibinfo {year} {2007}),\
  \Eprint {http://arxiv.org/abs/0705.1928} {arXiv:0705.1928 [quant-ph]}
  \BibitemShut {NoStop}%
\bibitem [{\citenamefont {Poulin}\ \emph {et~al.}(2018)\citenamefont {Poulin},
  \citenamefont {Kitaev}, \citenamefont {Steiger}, \citenamefont {Hastings},\
  and\ \citenamefont {Troyer}}]{PhysRevLett.121.010501}%
  \BibitemOpen
  \bibfield  {author} {\bibinfo {author} {\bibfnamefont {D.}~\bibnamefont
  {Poulin}}, \bibinfo {author} {\bibfnamefont {A.}~\bibnamefont {Kitaev}},
  \bibinfo {author} {\bibfnamefont {D.~S.}\ \bibnamefont {Steiger}}, \bibinfo
  {author} {\bibfnamefont {M.~B.}\ \bibnamefont {Hastings}}, \ and\ \bibinfo
  {author} {\bibfnamefont {M.}~\bibnamefont {Troyer}},\ }\href {\doibase
  10.1103/PhysRevLett.121.010501} {\bibfield  {journal} {\bibinfo  {journal}
  {Phys. Rev. Lett.}\ }\textbf {\bibinfo {volume} {121}},\ \bibinfo {pages}
  {010501} (\bibinfo {year} {2018})}\BibitemShut {NoStop}%
\bibitem [{\citenamefont {Lin}\ and\ \citenamefont
  {Tong}(2020)}]{Lin2020nearoptimalground}%
  \BibitemOpen
  \bibfield  {author} {\bibinfo {author} {\bibfnamefont {L.}~\bibnamefont
  {Lin}}\ and\ \bibinfo {author} {\bibfnamefont {Y.}~\bibnamefont {Tong}},\
  }\href {\doibase 10.22331/q-2020-12-14-372} {\bibfield  {journal} {\bibinfo
  {journal} {{Quantum}}\ }\textbf {\bibinfo {volume} {4}},\ \bibinfo {pages}
  {372} (\bibinfo {year} {2020})}\BibitemShut {NoStop}%
\bibitem [{\citenamefont {Choi}\ \emph {et~al.}(2021)\citenamefont {Choi},
  \citenamefont {Lee}, \citenamefont {Bonitati}, \citenamefont {Qian},\ and\
  \citenamefont {Watkins}}]{choi2021rodeo}%
  \BibitemOpen
  \bibfield  {author} {\bibinfo {author} {\bibfnamefont {K.}~\bibnamefont
  {Choi}}, \bibinfo {author} {\bibfnamefont {D.}~\bibnamefont {Lee}}, \bibinfo
  {author} {\bibfnamefont {J.}~\bibnamefont {Bonitati}}, \bibinfo {author}
  {\bibfnamefont {Z.}~\bibnamefont {Qian}}, \ and\ \bibinfo {author}
  {\bibfnamefont {J.}~\bibnamefont {Watkins}},\ }\href@noop {} {\enquote
  {\bibinfo {title} {Rodeo algorithm for quantum computing},}\ } (\bibinfo
  {year} {2021}),\ \Eprint {http://arxiv.org/abs/2009.04092} {arXiv:2009.04092
  [quant-ph]} \BibitemShut {NoStop}%
\bibitem [{\citenamefont {Turro}\ \emph {et~al.}(2021)\citenamefont {Turro},
  \citenamefont {Amitrano}, \citenamefont {Luchi}, \citenamefont {Wendt},
  \citenamefont {DuBois}, \citenamefont {Quaglioni},\ and\ \citenamefont
  {Pederiva}}]{turro2021imaginary}%
  \BibitemOpen
  \bibfield  {author} {\bibinfo {author} {\bibfnamefont {F.}~\bibnamefont
  {Turro}}, \bibinfo {author} {\bibfnamefont {V.}~\bibnamefont {Amitrano}},
  \bibinfo {author} {\bibfnamefont {P.}~\bibnamefont {Luchi}}, \bibinfo
  {author} {\bibfnamefont {K.~A.}\ \bibnamefont {Wendt}}, \bibinfo {author}
  {\bibfnamefont {J.~L.}\ \bibnamefont {DuBois}}, \bibinfo {author}
  {\bibfnamefont {S.}~\bibnamefont {Quaglioni}}, \ and\ \bibinfo {author}
  {\bibfnamefont {F.}~\bibnamefont {Pederiva}},\ }\href@noop {} {\enquote
  {\bibinfo {title} {Imaginary time propagation on a quantum chip},}\ }
  (\bibinfo {year} {2021}),\ \Eprint {http://arxiv.org/abs/2102.12260}
  {arXiv:2102.12260 [quant-ph]} \BibitemShut {NoStop}%
\bibitem [{\citenamefont {Peruzzo}\ \emph {et~al.}(2014)\citenamefont
  {Peruzzo}, \citenamefont {McClean}, \citenamefont {Shadbolt}, \citenamefont
  {Yung}, \citenamefont {Zhou}, \citenamefont {Love}, \citenamefont
  {Aspuru-Guzik},\ and\ \citenamefont {O’Brien}}]{Peruzzo_2014}%
  \BibitemOpen
  \bibfield  {author} {\bibinfo {author} {\bibfnamefont {A.}~\bibnamefont
  {Peruzzo}}, \bibinfo {author} {\bibfnamefont {J.}~\bibnamefont {McClean}},
  \bibinfo {author} {\bibfnamefont {P.}~\bibnamefont {Shadbolt}}, \bibinfo
  {author} {\bibfnamefont {M.-H.}\ \bibnamefont {Yung}}, \bibinfo {author}
  {\bibfnamefont {X.-Q.}\ \bibnamefont {Zhou}}, \bibinfo {author}
  {\bibfnamefont {P.~J.}\ \bibnamefont {Love}}, \bibinfo {author}
  {\bibfnamefont {A.}~\bibnamefont {Aspuru-Guzik}}, \ and\ \bibinfo {author}
  {\bibfnamefont {J.~L.}\ \bibnamefont {O’Brien}},\ }\href {\doibase
  10.1038/ncomms5213} {\bibfield  {journal} {\bibinfo  {journal} {Nature
  Communications}\ }\textbf {\bibinfo {volume} {5}} (\bibinfo {year} {2014}),\
  10.1038/ncomms5213}\BibitemShut {NoStop}%
\bibitem [{\citenamefont {McClean}\ \emph {et~al.}(2016)\citenamefont
  {McClean}, \citenamefont {Romero}, \citenamefont {Babbush},\ and\
  \citenamefont {Aspuru-Guzik}}]{McClean_2016}%
  \BibitemOpen
  \bibfield  {author} {\bibinfo {author} {\bibfnamefont {J.~R.}\ \bibnamefont
  {McClean}}, \bibinfo {author} {\bibfnamefont {J.}~\bibnamefont {Romero}},
  \bibinfo {author} {\bibfnamefont {R.}~\bibnamefont {Babbush}}, \ and\
  \bibinfo {author} {\bibfnamefont {A.}~\bibnamefont {Aspuru-Guzik}},\ }\href
  {\doibase 10.1088/1367-2630/18/2/023023} {\bibfield  {journal} {\bibinfo
  {journal} {New Journal of Physics}\ }\textbf {\bibinfo {volume} {18}},\
  \bibinfo {pages} {023023} (\bibinfo {year} {2016})}\BibitemShut {NoStop}%
\bibitem [{\citenamefont {Sim}\ \emph {et~al.}(2019)\citenamefont {Sim},
  \citenamefont {Johnson},\ and\ \citenamefont
  {Aspuru‐Guzik}}]{AspuruGuzikSim2019}%
  \BibitemOpen
  \bibfield  {author} {\bibinfo {author} {\bibfnamefont {S.}~\bibnamefont
  {Sim}}, \bibinfo {author} {\bibfnamefont {P.~D.}\ \bibnamefont {Johnson}}, \
  and\ \bibinfo {author} {\bibfnamefont {A.}~\bibnamefont {Aspuru‐Guzik}},\
  }\href {\doibase 10.1002/qute.201900070} {\bibfield  {journal} {\bibinfo
  {journal} {Advanced Quantum Technologies}\ }\textbf {\bibinfo {volume} {2}},\
  \bibinfo {pages} {1900070} (\bibinfo {year} {2019})}\BibitemShut {NoStop}%
\bibitem [{\citenamefont {Farhi}\ \emph {et~al.}(2014)\citenamefont {Farhi},
  \citenamefont {Goldstone},\ and\ \citenamefont {Gutmann}}]{farhi2014quantum}%
  \BibitemOpen
  \bibfield  {author} {\bibinfo {author} {\bibfnamefont {E.}~\bibnamefont
  {Farhi}}, \bibinfo {author} {\bibfnamefont {J.}~\bibnamefont {Goldstone}}, \
  and\ \bibinfo {author} {\bibfnamefont {S.}~\bibnamefont {Gutmann}},\
  }\href@noop {} {\enquote {\bibinfo {title} {A quantum approximate
  optimization algorithm},}\ } (\bibinfo {year} {2014}),\ \Eprint
  {http://arxiv.org/abs/1411.4028} {arXiv:1411.4028 [quant-ph]} \BibitemShut
  {NoStop}%
\bibitem [{\citenamefont {McArdle}\ \emph {et~al.}(2019)\citenamefont
  {McArdle}, \citenamefont {Jones}, \citenamefont {Endo}, \citenamefont {Li},
  \citenamefont {Benjamin},\ and\ \citenamefont {Yuan}}]{McArdle_2019}%
  \BibitemOpen
  \bibfield  {author} {\bibinfo {author} {\bibfnamefont {S.}~\bibnamefont
  {McArdle}}, \bibinfo {author} {\bibfnamefont {T.}~\bibnamefont {Jones}},
  \bibinfo {author} {\bibfnamefont {S.}~\bibnamefont {Endo}}, \bibinfo {author}
  {\bibfnamefont {Y.}~\bibnamefont {Li}}, \bibinfo {author} {\bibfnamefont
  {S.~C.}\ \bibnamefont {Benjamin}}, \ and\ \bibinfo {author} {\bibfnamefont
  {X.}~\bibnamefont {Yuan}},\ }\href {\doibase 10.1038/s41534-019-0187-2}
  {\bibfield  {journal} {\bibinfo  {journal} {npj Quantum Information}\
  }\textbf {\bibinfo {volume} {5}} (\bibinfo {year} {2019}),\
  10.1038/s41534-019-0187-2}\BibitemShut {NoStop}%
\bibitem [{\citenamefont {Motta}\ \emph {et~al.}(2019)\citenamefont {Motta},
  \citenamefont {Sun}, \citenamefont {Tan}, \citenamefont {O’Rourke},
  \citenamefont {Ye}, \citenamefont {Minnich}, \citenamefont {Brandão},\ and\
  \citenamefont {Chan}}]{Motta_2019}%
  \BibitemOpen
  \bibfield  {author} {\bibinfo {author} {\bibfnamefont {M.}~\bibnamefont
  {Motta}}, \bibinfo {author} {\bibfnamefont {C.}~\bibnamefont {Sun}}, \bibinfo
  {author} {\bibfnamefont {A.~T.~K.}\ \bibnamefont {Tan}}, \bibinfo {author}
  {\bibfnamefont {M.~J.}\ \bibnamefont {O’Rourke}}, \bibinfo {author}
  {\bibfnamefont {E.}~\bibnamefont {Ye}}, \bibinfo {author} {\bibfnamefont
  {A.~J.}\ \bibnamefont {Minnich}}, \bibinfo {author} {\bibfnamefont {F.~G.
  S.~L.}\ \bibnamefont {Brandão}}, \ and\ \bibinfo {author} {\bibfnamefont
  {G.~K.-L.}\ \bibnamefont {Chan}},\ }\href {\doibase
  10.1038/s41567-019-0704-4} {\bibfield  {journal} {\bibinfo  {journal} {Nature
  Physics}\ }\textbf {\bibinfo {volume} {16}},\ \bibinfo {pages} {205–210}
  (\bibinfo {year} {2019})}\BibitemShut {NoStop}%
\bibitem [{\citenamefont {O'Malley}\ \emph {et~al.}(2016)\citenamefont
  {O'Malley}, \citenamefont {Babbush}, \citenamefont {Kivlichan}, \citenamefont
  {Romero}, \citenamefont {McClean}, \citenamefont {Barends}, \citenamefont
  {Kelly}, \citenamefont {Roushan}, \citenamefont {Tranter}, \citenamefont
  {Ding}, \citenamefont {Campbell}, \citenamefont {Chen}, \citenamefont {Chen},
  \citenamefont {Chiaro}, \citenamefont {Dunsworth}, \citenamefont {Fowler},
  \citenamefont {Jeffrey}, \citenamefont {Lucero}, \citenamefont {Megrant},
  \citenamefont {Mutus}, \citenamefont {Neeley}, \citenamefont {Neill},
  \citenamefont {Quintana}, \citenamefont {Sank}, \citenamefont {Vainsencher},
  \citenamefont {Wenner}, \citenamefont {White}, \citenamefont {Coveney},
  \citenamefont {Love}, \citenamefont {Neven}, \citenamefont {Aspuru-Guzik},\
  and\ \citenamefont {Martinis}}]{PhysRevX.6.031007}%
  \BibitemOpen
  \bibfield  {author} {\bibinfo {author} {\bibfnamefont {P.~J.~J.}\
  \bibnamefont {O'Malley}}, \bibinfo {author} {\bibfnamefont {R.}~\bibnamefont
  {Babbush}}, \bibinfo {author} {\bibfnamefont {I.~D.}\ \bibnamefont
  {Kivlichan}}, \bibinfo {author} {\bibfnamefont {J.}~\bibnamefont {Romero}},
  \bibinfo {author} {\bibfnamefont {J.~R.}\ \bibnamefont {McClean}}, \bibinfo
  {author} {\bibfnamefont {R.}~\bibnamefont {Barends}}, \bibinfo {author}
  {\bibfnamefont {J.}~\bibnamefont {Kelly}}, \bibinfo {author} {\bibfnamefont
  {P.}~\bibnamefont {Roushan}}, \bibinfo {author} {\bibfnamefont
  {A.}~\bibnamefont {Tranter}}, \bibinfo {author} {\bibfnamefont
  {N.}~\bibnamefont {Ding}}, \bibinfo {author} {\bibfnamefont {B.}~\bibnamefont
  {Campbell}}, \bibinfo {author} {\bibfnamefont {Y.}~\bibnamefont {Chen}},
  \bibinfo {author} {\bibfnamefont {Z.}~\bibnamefont {Chen}}, \bibinfo {author}
  {\bibfnamefont {B.}~\bibnamefont {Chiaro}}, \bibinfo {author} {\bibfnamefont
  {A.}~\bibnamefont {Dunsworth}}, \bibinfo {author} {\bibfnamefont {A.~G.}\
  \bibnamefont {Fowler}}, \bibinfo {author} {\bibfnamefont {E.}~\bibnamefont
  {Jeffrey}}, \bibinfo {author} {\bibfnamefont {E.}~\bibnamefont {Lucero}},
  \bibinfo {author} {\bibfnamefont {A.}~\bibnamefont {Megrant}}, \bibinfo
  {author} {\bibfnamefont {J.~Y.}\ \bibnamefont {Mutus}}, \bibinfo {author}
  {\bibfnamefont {M.}~\bibnamefont {Neeley}}, \bibinfo {author} {\bibfnamefont
  {C.}~\bibnamefont {Neill}}, \bibinfo {author} {\bibfnamefont
  {C.}~\bibnamefont {Quintana}}, \bibinfo {author} {\bibfnamefont
  {D.}~\bibnamefont {Sank}}, \bibinfo {author} {\bibfnamefont {A.}~\bibnamefont
  {Vainsencher}}, \bibinfo {author} {\bibfnamefont {J.}~\bibnamefont {Wenner}},
  \bibinfo {author} {\bibfnamefont {T.~C.}\ \bibnamefont {White}}, \bibinfo
  {author} {\bibfnamefont {P.~V.}\ \bibnamefont {Coveney}}, \bibinfo {author}
  {\bibfnamefont {P.~J.}\ \bibnamefont {Love}}, \bibinfo {author}
  {\bibfnamefont {H.}~\bibnamefont {Neven}}, \bibinfo {author} {\bibfnamefont
  {A.}~\bibnamefont {Aspuru-Guzik}}, \ and\ \bibinfo {author} {\bibfnamefont
  {J.~M.}\ \bibnamefont {Martinis}},\ }\href {\doibase
  10.1103/PhysRevX.6.031007} {\bibfield  {journal} {\bibinfo  {journal} {Phys.
  Rev. X}\ }\textbf {\bibinfo {volume} {6}},\ \bibinfo {pages} {031007}
  (\bibinfo {year} {2016})}\BibitemShut {NoStop}%
\bibitem [{\citenamefont {Bittel}\ and\ \citenamefont
  {Kliesch}(2021)}]{bittel2021training}%
  \BibitemOpen
  \bibfield  {author} {\bibinfo {author} {\bibfnamefont {L.}~\bibnamefont
  {Bittel}}\ and\ \bibinfo {author} {\bibfnamefont {M.}~\bibnamefont
  {Kliesch}},\ }\href@noop {} {\enquote {\bibinfo {title} {Training variational
  quantum algorithms is np-hard -- even for logarithmically many qubits and
  free fermionic systems},}\ } (\bibinfo {year} {2021}),\ \Eprint
  {http://arxiv.org/abs/2101.07267} {arXiv:2101.07267 [quant-ph]} \BibitemShut
  {NoStop}%
\bibitem [{\citenamefont {Cerezo}\ \emph {et~al.}(2020)\citenamefont {Cerezo},
  \citenamefont {Arrasmith}, \citenamefont {Babbush}, \citenamefont {Benjamin},
  \citenamefont {Endo}, \citenamefont {Fujii}, \citenamefont {McClean},
  \citenamefont {Mitarai}, \citenamefont {Yuan}, \citenamefont {Cincio},\ and\
  \citenamefont {Coles}}]{cerezo2020variational}%
  \BibitemOpen
  \bibfield  {author} {\bibinfo {author} {\bibfnamefont {M.}~\bibnamefont
  {Cerezo}}, \bibinfo {author} {\bibfnamefont {A.}~\bibnamefont {Arrasmith}},
  \bibinfo {author} {\bibfnamefont {R.}~\bibnamefont {Babbush}}, \bibinfo
  {author} {\bibfnamefont {S.~C.}\ \bibnamefont {Benjamin}}, \bibinfo {author}
  {\bibfnamefont {S.}~\bibnamefont {Endo}}, \bibinfo {author} {\bibfnamefont
  {K.}~\bibnamefont {Fujii}}, \bibinfo {author} {\bibfnamefont {J.~R.}\
  \bibnamefont {McClean}}, \bibinfo {author} {\bibfnamefont {K.}~\bibnamefont
  {Mitarai}}, \bibinfo {author} {\bibfnamefont {X.}~\bibnamefont {Yuan}},
  \bibinfo {author} {\bibfnamefont {L.}~\bibnamefont {Cincio}}, \ and\ \bibinfo
  {author} {\bibfnamefont {P.~J.}\ \bibnamefont {Coles}},\ }\href@noop {}
  {\enquote {\bibinfo {title} {Variational quantum algorithms},}\ } (\bibinfo
  {year} {2020}),\ \Eprint {http://arxiv.org/abs/2012.09265} {arXiv:2012.09265
  [quant-ph]} \BibitemShut {NoStop}%
\bibitem [{\citenamefont {Biamonte}\ \emph {et~al.}(2017)\citenamefont
  {Biamonte}, \citenamefont {Wittek}, \citenamefont {Pancotti}, \citenamefont
  {Rebentrost}, \citenamefont {Wiebe},\ and\ \citenamefont
  {Lloyd}}]{Biamonte2017}%
  \BibitemOpen
  \bibfield  {author} {\bibinfo {author} {\bibfnamefont {J.}~\bibnamefont
  {Biamonte}}, \bibinfo {author} {\bibfnamefont {P.}~\bibnamefont {Wittek}},
  \bibinfo {author} {\bibfnamefont {N.}~\bibnamefont {Pancotti}}, \bibinfo
  {author} {\bibfnamefont {P.}~\bibnamefont {Rebentrost}}, \bibinfo {author}
  {\bibfnamefont {N.}~\bibnamefont {Wiebe}}, \ and\ \bibinfo {author}
  {\bibfnamefont {S.}~\bibnamefont {Lloyd}},\ }\href {\doibase
  10.1038/nature23474} {\bibfield  {journal} {\bibinfo  {journal} {Nature}\
  }\textbf {\bibinfo {volume} {549}},\ \bibinfo {pages} {195} (\bibinfo {year}
  {2017})}\BibitemShut {NoStop}%
\bibitem [{\citenamefont {Filipek}\ \emph {et~al.}(2021)\citenamefont
  {Filipek}, \citenamefont {Hsu}, \citenamefont {Roggero},\ and\ \citenamefont
  {Wiebe}}]{filipek2021quantum}%
  \BibitemOpen
  \bibfield  {author} {\bibinfo {author} {\bibfnamefont {J.}~\bibnamefont
  {Filipek}}, \bibinfo {author} {\bibfnamefont {S.-C.}\ \bibnamefont {Hsu}},
  \bibinfo {author} {\bibfnamefont {A.}~\bibnamefont {Roggero}}, \ and\
  \bibinfo {author} {\bibfnamefont {N.}~\bibnamefont {Wiebe}},\ }\href@noop {}
  {\enquote {\bibinfo {title} {Quantum machine learning with squid},}\ }
  (\bibinfo {year} {2021}),\ \Eprint {http://arxiv.org/abs/2105.00098}
  {arXiv:2105.00098 [quant-ph]} \BibitemShut {NoStop}%
\bibitem [{\citenamefont {Guan}\ \emph {et~al.}(2021)\citenamefont {Guan},
  \citenamefont {Perdue}, \citenamefont {Pesah}, \citenamefont {Schuld},
  \citenamefont {Terashi}, \citenamefont {Vallecorsa},\ and\ \citenamefont
  {Vlimant}}]{Guan_2021}%
  \BibitemOpen
  \bibfield  {author} {\bibinfo {author} {\bibfnamefont {W.}~\bibnamefont
  {Guan}}, \bibinfo {author} {\bibfnamefont {G.}~\bibnamefont {Perdue}},
  \bibinfo {author} {\bibfnamefont {A.}~\bibnamefont {Pesah}}, \bibinfo
  {author} {\bibfnamefont {M.}~\bibnamefont {Schuld}}, \bibinfo {author}
  {\bibfnamefont {K.}~\bibnamefont {Terashi}}, \bibinfo {author} {\bibfnamefont
  {S.}~\bibnamefont {Vallecorsa}}, \ and\ \bibinfo {author} {\bibfnamefont
  {J.-R.}\ \bibnamefont {Vlimant}},\ }\href {\doibase 10.1088/2632-2153/abc17d}
  {\bibfield  {journal} {\bibinfo  {journal} {Machine Learning: Science and
  Technology}\ }\textbf {\bibinfo {volume} {2}},\ \bibinfo {pages} {011003}
  (\bibinfo {year} {2021})}\BibitemShut {NoStop}%
\bibitem [{\citenamefont {{Aaronson}}(2015)}]{2015NatPh..11..291A}%
  \BibitemOpen
  \bibfield  {author} {\bibinfo {author} {\bibfnamefont {S.}~\bibnamefont
  {{Aaronson}}},\ }\href {\doibase 10.1038/nphys3272} {\bibfield  {journal}
  {\bibinfo  {journal} {Nature Physics}\ }\textbf {\bibinfo {volume} {11}},\
  \bibinfo {pages} {291} (\bibinfo {year} {2015})}\BibitemShut {NoStop}%
\bibitem [{\citenamefont {Wiebe}(2020)}]{Wiebe2020}%
  \BibitemOpen
  \bibfield  {author} {\bibinfo {author} {\bibfnamefont {N.}~\bibnamefont
  {Wiebe}},\ }\href {\doibase 10.1088/1367-2630/abac39} {\bibfield  {journal}
  {\bibinfo  {journal} {New Journal of Physics}\ }\textbf {\bibinfo {volume}
  {22}},\ \bibinfo {pages} {091001} (\bibinfo {year} {2020})}\BibitemShut
  {NoStop}%
\bibitem [{\citenamefont {Verteletskyi}\ \emph {et~al.}(2020)\citenamefont
  {Verteletskyi}, \citenamefont {Yen},\ and\ \citenamefont
  {Izmaylov}}]{doi:10.1063/1.5141458}%
  \BibitemOpen
  \bibfield  {author} {\bibinfo {author} {\bibfnamefont {V.}~\bibnamefont
  {Verteletskyi}}, \bibinfo {author} {\bibfnamefont {T.-C.}\ \bibnamefont
  {Yen}}, \ and\ \bibinfo {author} {\bibfnamefont {A.~F.}\ \bibnamefont
  {Izmaylov}},\ }\href {\doibase 10.1063/1.5141458} {\bibfield  {journal}
  {\bibinfo  {journal} {The Journal of Chemical Physics}\ }\textbf {\bibinfo
  {volume} {152}},\ \bibinfo {pages} {124114} (\bibinfo {year}
  {2020})}\BibitemShut {NoStop}%
\bibitem [{\citenamefont {O'Brien}\ \emph {et~al.}(2019)\citenamefont
  {O'Brien}, \citenamefont {Tarasinski},\ and\ \citenamefont
  {Terhal}}]{O_Brien_2019}%
  \BibitemOpen
  \bibfield  {author} {\bibinfo {author} {\bibfnamefont {T.~E.}\ \bibnamefont
  {O'Brien}}, \bibinfo {author} {\bibfnamefont {B.}~\bibnamefont {Tarasinski}},
  \ and\ \bibinfo {author} {\bibfnamefont {B.~M.}\ \bibnamefont {Terhal}},\
  }\href {\doibase 10.1088/1367-2630/aafb8e} {\bibfield  {journal} {\bibinfo
  {journal} {New Journal of Physics}\ }\textbf {\bibinfo {volume} {21}},\
  \bibinfo {pages} {023022} (\bibinfo {year} {2019})}\BibitemShut {NoStop}%
\bibitem [{\citenamefont {O'Brien}\ \emph {et~al.}(2021)\citenamefont
  {O'Brien}, \citenamefont {Polla}, \citenamefont {Rubin}, \citenamefont
  {Huggins}, \citenamefont {McArdle}, \citenamefont {Boixo}, \citenamefont
  {McClean},\ and\ \citenamefont {Babbush}}]{PRXQuantum.2.020317}%
  \BibitemOpen
  \bibfield  {author} {\bibinfo {author} {\bibfnamefont {T.~E.}\ \bibnamefont
  {O'Brien}}, \bibinfo {author} {\bibfnamefont {S.}~\bibnamefont {Polla}},
  \bibinfo {author} {\bibfnamefont {N.~C.}\ \bibnamefont {Rubin}}, \bibinfo
  {author} {\bibfnamefont {W.~J.}\ \bibnamefont {Huggins}}, \bibinfo {author}
  {\bibfnamefont {S.}~\bibnamefont {McArdle}}, \bibinfo {author} {\bibfnamefont
  {S.}~\bibnamefont {Boixo}}, \bibinfo {author} {\bibfnamefont {J.~R.}\
  \bibnamefont {McClean}}, \ and\ \bibinfo {author} {\bibfnamefont
  {R.}~\bibnamefont {Babbush}},\ }\href {\doibase 10.1103/PRXQuantum.2.020317}
  {\bibfield  {journal} {\bibinfo  {journal} {PRX Quantum}\ }\textbf {\bibinfo
  {volume} {2}},\ \bibinfo {pages} {020317} (\bibinfo {year}
  {2021})}\BibitemShut {NoStop}%
\bibitem [{\citenamefont {Funcke}\ \emph {et~al.}(2021)\citenamefont {Funcke},
  \citenamefont {Hartung}, \citenamefont {Jansen}, \citenamefont {K\"uhn},\
  and\ \citenamefont {Stornati}}]{Funcke:2020vkw}%
  \BibitemOpen
  \bibfield  {author} {\bibinfo {author} {\bibfnamefont {L.}~\bibnamefont
  {Funcke}}, \bibinfo {author} {\bibfnamefont {T.}~\bibnamefont {Hartung}},
  \bibinfo {author} {\bibfnamefont {K.}~\bibnamefont {Jansen}}, \bibinfo
  {author} {\bibfnamefont {S.}~\bibnamefont {K\"uhn}}, \ and\ \bibinfo {author}
  {\bibfnamefont {P.}~\bibnamefont {Stornati}},\ }\href {\doibase
  10.22331/q-2021-03-29-422} {\bibfield  {journal} {\bibinfo  {journal}
  {Quantum}\ }\textbf {\bibinfo {volume} {5}},\ \bibinfo {pages} {422}
  (\bibinfo {year} {2021})},\ \Eprint {http://arxiv.org/abs/2011.03532}
  {arXiv:2011.03532 [quant-ph]} \BibitemShut {NoStop}%
\bibitem [{\citenamefont {Huggins}\ \emph {et~al.}(2021)\citenamefont
  {Huggins}, \citenamefont {McClean}, \citenamefont {Rubin}, \citenamefont
  {Jiang}, \citenamefont {Wiebe}, \citenamefont {Whaley},\ and\ \citenamefont
  {Babbush}}]{Huggins_2021}%
  \BibitemOpen
  \bibfield  {author} {\bibinfo {author} {\bibfnamefont {W.~J.}\ \bibnamefont
  {Huggins}}, \bibinfo {author} {\bibfnamefont {J.~R.}\ \bibnamefont
  {McClean}}, \bibinfo {author} {\bibfnamefont {N.~C.}\ \bibnamefont {Rubin}},
  \bibinfo {author} {\bibfnamefont {Z.}~\bibnamefont {Jiang}}, \bibinfo
  {author} {\bibfnamefont {N.}~\bibnamefont {Wiebe}}, \bibinfo {author}
  {\bibfnamefont {K.~B.}\ \bibnamefont {Whaley}}, \ and\ \bibinfo {author}
  {\bibfnamefont {R.}~\bibnamefont {Babbush}},\ }\href {\doibase
  10.1038/s41534-020-00341-7} {\bibfield  {journal} {\bibinfo  {journal} {npj
  Quantum Information}\ }\textbf {\bibinfo {volume} {7}} (\bibinfo {year}
  {2021}),\ 10.1038/s41534-020-00341-7}\BibitemShut {NoStop}%
\bibitem [{\citenamefont {Guzman}\ and\ \citenamefont
  {Lacroix}(2021)}]{guzman2021predicting}%
  \BibitemOpen
  \bibfield  {author} {\bibinfo {author} {\bibfnamefont {E.~A.~R.}\
  \bibnamefont {Guzman}}\ and\ \bibinfo {author} {\bibfnamefont
  {D.}~\bibnamefont {Lacroix}},\ }\href@noop {} {\enquote {\bibinfo {title}
  {Predicting ground state, excited states and long-time evolution of many-body
  systems from short-time evolution on a quantum computer},}\ } (\bibinfo
  {year} {2021}),\ \Eprint {http://arxiv.org/abs/2104.08181} {arXiv:2104.08181
  [quant-ph]} \BibitemShut {NoStop}%
\bibitem [{\citenamefont {Brydges}\ \emph {et~al.}(2019)\citenamefont
  {Brydges}, \citenamefont {Elben}, \citenamefont {Jurcevic}, \citenamefont
  {Vermersch}, \citenamefont {Maier}, \citenamefont {Lanyon}, \citenamefont
  {Zoller}, \citenamefont {Blatt},\ and\ \citenamefont {Roos}}]{Brydges260}%
  \BibitemOpen
  \bibfield  {author} {\bibinfo {author} {\bibfnamefont {T.}~\bibnamefont
  {Brydges}}, \bibinfo {author} {\bibfnamefont {A.}~\bibnamefont {Elben}},
  \bibinfo {author} {\bibfnamefont {P.}~\bibnamefont {Jurcevic}}, \bibinfo
  {author} {\bibfnamefont {B.}~\bibnamefont {Vermersch}}, \bibinfo {author}
  {\bibfnamefont {C.}~\bibnamefont {Maier}}, \bibinfo {author} {\bibfnamefont
  {B.~P.}\ \bibnamefont {Lanyon}}, \bibinfo {author} {\bibfnamefont
  {P.}~\bibnamefont {Zoller}}, \bibinfo {author} {\bibfnamefont
  {R.}~\bibnamefont {Blatt}}, \ and\ \bibinfo {author} {\bibfnamefont {C.~F.}\
  \bibnamefont {Roos}},\ }\href {\doibase 10.1126/science.aau4963} {\bibfield
  {journal} {\bibinfo  {journal} {Science}\ }\textbf {\bibinfo {volume}
  {364}},\ \bibinfo {pages} {260} (\bibinfo {year} {2019})}\BibitemShut
  {NoStop}%
\bibitem [{\citenamefont {Elben}\ \emph {et~al.}(2019)\citenamefont {Elben},
  \citenamefont {Vermersch}, \citenamefont {Roos},\ and\ \citenamefont
  {Zoller}}]{PhysRevA.99.052323}%
  \BibitemOpen
  \bibfield  {author} {\bibinfo {author} {\bibfnamefont {A.}~\bibnamefont
  {Elben}}, \bibinfo {author} {\bibfnamefont {B.}~\bibnamefont {Vermersch}},
  \bibinfo {author} {\bibfnamefont {C.~F.}\ \bibnamefont {Roos}}, \ and\
  \bibinfo {author} {\bibfnamefont {P.}~\bibnamefont {Zoller}},\ }\href
  {\doibase 10.1103/PhysRevA.99.052323} {\bibfield  {journal} {\bibinfo
  {journal} {Phys. Rev. A}\ }\textbf {\bibinfo {volume} {99}},\ \bibinfo
  {pages} {052323} (\bibinfo {year} {2019})}\BibitemShut {NoStop}%
\bibitem [{\citenamefont {Aaronson}(2018)}]{10.1145/3188745.3188802}%
  \BibitemOpen
  \bibfield  {author} {\bibinfo {author} {\bibfnamefont {S.}~\bibnamefont
  {Aaronson}},\ }in\ \href {\doibase 10.1145/3188745.3188802} {\emph {\bibinfo
  {booktitle} {Proceedings of the 50th Annual ACM SIGACT Symposium on Theory of
  Computing}}},\ \bibinfo {series and number} {STOC 2018}\ (\bibinfo
  {publisher} {Association for Computing Machinery},\ \bibinfo {address} {New
  York, NY, USA},\ \bibinfo {year} {2018})\ p.\ \bibinfo {pages}
  {325–338}\BibitemShut {NoStop}%
\bibitem [{\citenamefont {Huang}\ \emph {et~al.}(2020)\citenamefont {Huang},
  \citenamefont {Kueng},\ and\ \citenamefont {Preskill}}]{Huang_2020}%
  \BibitemOpen
  \bibfield  {author} {\bibinfo {author} {\bibfnamefont {H.-Y.}\ \bibnamefont
  {Huang}}, \bibinfo {author} {\bibfnamefont {R.}~\bibnamefont {Kueng}}, \ and\
  \bibinfo {author} {\bibfnamefont {J.}~\bibnamefont {Preskill}},\ }\href
  {\doibase 10.1038/s41567-020-0932-7} {\bibfield  {journal} {\bibinfo
  {journal} {Nature Physics}\ }\textbf {\bibinfo {volume} {16}},\ \bibinfo
  {pages} {1050–1057} (\bibinfo {year} {2020})}\BibitemShut {NoStop}%
\bibitem [{\citenamefont {Magierski}\ \emph {et~al.}(2009)\citenamefont
  {Magierski}, \citenamefont {Wlaz\l{}owski}, \citenamefont {Bulgac},\ and\
  \citenamefont {Drut}}]{PhysRevLett.103.210403}%
  \BibitemOpen
  \bibfield  {author} {\bibinfo {author} {\bibfnamefont {P.}~\bibnamefont
  {Magierski}}, \bibinfo {author} {\bibfnamefont {G.}~\bibnamefont
  {Wlaz\l{}owski}}, \bibinfo {author} {\bibfnamefont {A.}~\bibnamefont
  {Bulgac}}, \ and\ \bibinfo {author} {\bibfnamefont {J.~E.}\ \bibnamefont
  {Drut}},\ }\href {\doibase 10.1103/PhysRevLett.103.210403} {\bibfield
  {journal} {\bibinfo  {journal} {Phys. Rev. Lett.}\ }\textbf {\bibinfo
  {volume} {103}},\ \bibinfo {pages} {210403} (\bibinfo {year}
  {2009})}\BibitemShut {NoStop}%
\bibitem [{\citenamefont {Burnier}\ and\ \citenamefont
  {Rothkopf}(2013)}]{PhysRevLett.111.182003}%
  \BibitemOpen
  \bibfield  {author} {\bibinfo {author} {\bibfnamefont {Y.}~\bibnamefont
  {Burnier}}\ and\ \bibinfo {author} {\bibfnamefont {A.}~\bibnamefont
  {Rothkopf}},\ }\href {\doibase 10.1103/PhysRevLett.111.182003} {\bibfield
  {journal} {\bibinfo  {journal} {Phys. Rev. Lett.}\ }\textbf {\bibinfo
  {volume} {111}},\ \bibinfo {pages} {182003} (\bibinfo {year}
  {2013})}\BibitemShut {NoStop}%
\bibitem [{\citenamefont {Lovato}\ \emph {et~al.}(2016)\citenamefont {Lovato},
  \citenamefont {Gandolfi}, \citenamefont {Carlson}, \citenamefont {Pieper},\
  and\ \citenamefont {Schiavilla}}]{PhysRevLett.117.082501}%
  \BibitemOpen
  \bibfield  {author} {\bibinfo {author} {\bibfnamefont {A.}~\bibnamefont
  {Lovato}}, \bibinfo {author} {\bibfnamefont {S.}~\bibnamefont {Gandolfi}},
  \bibinfo {author} {\bibfnamefont {J.}~\bibnamefont {Carlson}}, \bibinfo
  {author} {\bibfnamefont {S.~C.}\ \bibnamefont {Pieper}}, \ and\ \bibinfo
  {author} {\bibfnamefont {R.}~\bibnamefont {Schiavilla}},\ }\href {\doibase
  10.1103/PhysRevLett.117.082501} {\bibfield  {journal} {\bibinfo  {journal}
  {Phys. Rev. Lett.}\ }\textbf {\bibinfo {volume} {117}},\ \bibinfo {pages}
  {082501} (\bibinfo {year} {2016})}\BibitemShut {NoStop}%
\bibitem [{\citenamefont {Barnea}\ \emph {et~al.}(2010)\citenamefont {Barnea},
  \citenamefont {Efros}, \citenamefont {Leidemann},\ and\ \citenamefont
  {Orlandini}}]{Barnea_2010}%
  \BibitemOpen
  \bibfield  {author} {\bibinfo {author} {\bibfnamefont {N.}~\bibnamefont
  {Barnea}}, \bibinfo {author} {\bibfnamefont {V.~D.}\ \bibnamefont {Efros}},
  \bibinfo {author} {\bibfnamefont {W.}~\bibnamefont {Leidemann}}, \ and\
  \bibinfo {author} {\bibfnamefont {G.}~\bibnamefont {Orlandini}},\ }\href
  {\doibase 10.1007/s00601-009-0081-0} {\bibfield  {journal} {\bibinfo
  {journal} {Few-Body Systems}\ }\textbf {\bibinfo {volume} {47}},\ \bibinfo
  {pages} {201–206} (\bibinfo {year} {2010})}\BibitemShut {NoStop}%
\bibitem [{\citenamefont {Roggero}\ \emph {et~al.}(2013)\citenamefont
  {Roggero}, \citenamefont {Pederiva},\ and\ \citenamefont
  {Orlandini}}]{PhysRevB.88.094302}%
  \BibitemOpen
  \bibfield  {author} {\bibinfo {author} {\bibfnamefont {A.}~\bibnamefont
  {Roggero}}, \bibinfo {author} {\bibfnamefont {F.}~\bibnamefont {Pederiva}}, \
  and\ \bibinfo {author} {\bibfnamefont {G.}~\bibnamefont {Orlandini}},\ }\href
  {\doibase 10.1103/PhysRevB.88.094302} {\bibfield  {journal} {\bibinfo
  {journal} {Phys. Rev. B}\ }\textbf {\bibinfo {volume} {88}},\ \bibinfo
  {pages} {094302} (\bibinfo {year} {2013})}\BibitemShut {NoStop}%
\bibitem [{\citenamefont {Klco}\ \emph {et~al.}(2018)\citenamefont {Klco},
  \citenamefont {Dumitrescu}, \citenamefont {McCaskey}, \citenamefont {Morris},
  \citenamefont {Pooser}, \citenamefont {Sanz}, \citenamefont {Solano},
  \citenamefont {Lougovski},\ and\ \citenamefont
  {Savage}}]{PhysRevA.98.032331}%
  \BibitemOpen
  \bibfield  {author} {\bibinfo {author} {\bibfnamefont {N.}~\bibnamefont
  {Klco}}, \bibinfo {author} {\bibfnamefont {E.~F.}\ \bibnamefont
  {Dumitrescu}}, \bibinfo {author} {\bibfnamefont {A.~J.}\ \bibnamefont
  {McCaskey}}, \bibinfo {author} {\bibfnamefont {T.~D.}\ \bibnamefont
  {Morris}}, \bibinfo {author} {\bibfnamefont {R.~C.}\ \bibnamefont {Pooser}},
  \bibinfo {author} {\bibfnamefont {M.}~\bibnamefont {Sanz}}, \bibinfo {author}
  {\bibfnamefont {E.}~\bibnamefont {Solano}}, \bibinfo {author} {\bibfnamefont
  {P.}~\bibnamefont {Lougovski}}, \ and\ \bibinfo {author} {\bibfnamefont
  {M.~J.}\ \bibnamefont {Savage}},\ }\href {\doibase
  10.1103/PhysRevA.98.032331} {\bibfield  {journal} {\bibinfo  {journal} {Phys.
  Rev. A}\ }\textbf {\bibinfo {volume} {98}},\ \bibinfo {pages} {032331}
  (\bibinfo {year} {2018})}\BibitemShut {NoStop}%
\bibitem [{\citenamefont {Kharzeev}\ and\ \citenamefont
  {Kikuchi}(2020)}]{Kharzeev:2020kgc}%
  \BibitemOpen
  \bibfield  {author} {\bibinfo {author} {\bibfnamefont {D.~E.}\ \bibnamefont
  {Kharzeev}}\ and\ \bibinfo {author} {\bibfnamefont {Y.}~\bibnamefont
  {Kikuchi}},\ }\href {\doibase 10.1103/PhysRevResearch.2.023342} {\bibfield
  {journal} {\bibinfo  {journal} {Phys. Rev. Research}\ }\textbf {\bibinfo
  {volume} {2}},\ \bibinfo {pages} {023342} (\bibinfo {year} {2020})},\ \Eprint
  {http://arxiv.org/abs/2001.00698} {arXiv:2001.00698 [hep-ph]} \BibitemShut
  {NoStop}%
\bibitem [{\citenamefont {Du}\ \emph {et~al.}(2020)\citenamefont {Du},
  \citenamefont {Vary}, \citenamefont {Zhao},\ and\ \citenamefont
  {Zuo}}]{du2020quantum}%
  \BibitemOpen
  \bibfield  {author} {\bibinfo {author} {\bibfnamefont {W.}~\bibnamefont
  {Du}}, \bibinfo {author} {\bibfnamefont {J.~P.}\ \bibnamefont {Vary}},
  \bibinfo {author} {\bibfnamefont {X.}~\bibnamefont {Zhao}}, \ and\ \bibinfo
  {author} {\bibfnamefont {W.}~\bibnamefont {Zuo}},\ }\href@noop {} {\enquote
  {\bibinfo {title} {Quantum simulation of nuclear inelastic scattering},}\ }
  (\bibinfo {year} {2020}),\ \Eprint {http://arxiv.org/abs/2006.01369}
  {arXiv:2006.01369 [nucl-th]} \BibitemShut {NoStop}%
\bibitem [{\citenamefont {Yeter-Aydeniz}\ \emph {et~al.}(2021)\citenamefont
  {Yeter-Aydeniz}, \citenamefont {Bangar}, \citenamefont {Siopsis},\ and\
  \citenamefont {Pooser}}]{yeteraydeniz2021collective}%
  \BibitemOpen
  \bibfield  {author} {\bibinfo {author} {\bibfnamefont {K.}~\bibnamefont
  {Yeter-Aydeniz}}, \bibinfo {author} {\bibfnamefont {S.}~\bibnamefont
  {Bangar}}, \bibinfo {author} {\bibfnamefont {G.}~\bibnamefont {Siopsis}}, \
  and\ \bibinfo {author} {\bibfnamefont {R.~C.}\ \bibnamefont {Pooser}},\
  }\href@noop {} {\enquote {\bibinfo {title} {Collective neutrino oscillations
  on a quantum computer},}\ } (\bibinfo {year} {2021}),\ \Eprint
  {http://arxiv.org/abs/2104.03273} {arXiv:2104.03273 [quant-ph]} \BibitemShut
  {NoStop}%
\bibitem [{\citenamefont {Hall}\ \emph {et~al.}(2021)\citenamefont {Hall},
  \citenamefont {Roggero}, \citenamefont {Baroni},\ and\ \citenamefont
  {Carlson}}]{hall2021simulation}%
  \BibitemOpen
  \bibfield  {author} {\bibinfo {author} {\bibfnamefont {B.}~\bibnamefont
  {Hall}}, \bibinfo {author} {\bibfnamefont {A.}~\bibnamefont {Roggero}},
  \bibinfo {author} {\bibfnamefont {A.}~\bibnamefont {Baroni}}, \ and\ \bibinfo
  {author} {\bibfnamefont {J.}~\bibnamefont {Carlson}},\ }\href@noop {}
  {\enquote {\bibinfo {title} {Simulation of collective neutrino oscillations
  on a quantum computer},}\ } (\bibinfo {year} {2021}),\ \Eprint
  {http://arxiv.org/abs/2102.12556} {arXiv:2102.12556 [quant-ph]} \BibitemShut
  {NoStop}%
\bibitem [{\citenamefont {Baez}\ \emph {et~al.}(2020)\citenamefont {Baez},
  \citenamefont {Goihl}, \citenamefont {Haferkamp}, \citenamefont
  {Bermejo-Vega}, \citenamefont {Gluza},\ and\ \citenamefont
  {Eisert}}]{Baez_2020}%
  \BibitemOpen
  \bibfield  {author} {\bibinfo {author} {\bibfnamefont {M.~L.}\ \bibnamefont
  {Baez}}, \bibinfo {author} {\bibfnamefont {M.}~\bibnamefont {Goihl}},
  \bibinfo {author} {\bibfnamefont {J.}~\bibnamefont {Haferkamp}}, \bibinfo
  {author} {\bibfnamefont {J.}~\bibnamefont {Bermejo-Vega}}, \bibinfo {author}
  {\bibfnamefont {M.}~\bibnamefont {Gluza}}, \ and\ \bibinfo {author}
  {\bibfnamefont {J.}~\bibnamefont {Eisert}},\ }\href {\doibase
  10.1073/pnas.2006103117} {\bibfield  {journal} {\bibinfo  {journal}
  {Proceedings of the National Academy of Sciences}\ }\textbf {\bibinfo
  {volume} {117}},\ \bibinfo {pages} {26123–26134} (\bibinfo {year}
  {2020})}\BibitemShut {NoStop}%
\bibitem [{\citenamefont {Lüscher}(1991)}]{LUSCHER1991531}%
  \BibitemOpen
  \bibfield  {author} {\bibinfo {author} {\bibfnamefont {M.}~\bibnamefont
  {Lüscher}},\ }\href {\doibase https://doi.org/10.1016/0550-3213(91)90366-6}
  {\bibfield  {journal} {\bibinfo  {journal} {Nuclear Physics B}\ }\textbf
  {\bibinfo {volume} {354}},\ \bibinfo {pages} {531} (\bibinfo {year}
  {1991})}\BibitemShut {NoStop}%
\bibitem [{\citenamefont {L{\"u}scher}(1986{\natexlab{a}})}]{Luscher1986}%
  \BibitemOpen
  \bibfield  {author} {\bibinfo {author} {\bibfnamefont {M.}~\bibnamefont
  {L{\"u}scher}},\ }\href {\doibase 10.1007/BF01211589} {\bibfield  {journal}
  {\bibinfo  {journal} {Communications in Mathematical Physics}\ }\textbf
  {\bibinfo {volume} {104}},\ \bibinfo {pages} {177} (\bibinfo {year}
  {1986}{\natexlab{a}})}\BibitemShut {NoStop}%
\bibitem [{\citenamefont {L{\"u}scher}(1986{\natexlab{b}})}]{Luscher1986b}%
  \BibitemOpen
  \bibfield  {author} {\bibinfo {author} {\bibfnamefont {M.}~\bibnamefont
  {L{\"u}scher}},\ }\href {\doibase 10.1007/BF01211097} {\bibfield  {journal}
  {\bibinfo  {journal} {Communications in Mathematical Physics}\ }\textbf
  {\bibinfo {volume} {105}},\ \bibinfo {pages} {153} (\bibinfo {year}
  {1986}{\natexlab{b}})}\BibitemShut {NoStop}%
\bibitem [{\citenamefont {Brice\~no}\ \emph {et~al.}(2021)\citenamefont
  {Brice\~no}, \citenamefont {Guerrero}, \citenamefont {Hansen},\ and\
  \citenamefont {Sturzu}}]{PhysRevD.103.014506}%
  \BibitemOpen
  \bibfield  {author} {\bibinfo {author} {\bibfnamefont {R.~A.}\ \bibnamefont
  {Brice\~no}}, \bibinfo {author} {\bibfnamefont {J.~V.}\ \bibnamefont
  {Guerrero}}, \bibinfo {author} {\bibfnamefont {M.~T.}\ \bibnamefont
  {Hansen}}, \ and\ \bibinfo {author} {\bibfnamefont {A.~M.}\ \bibnamefont
  {Sturzu}},\ }\href {\doibase 10.1103/PhysRevD.103.014506} {\bibfield
  {journal} {\bibinfo  {journal} {Phys. Rev. D}\ }\textbf {\bibinfo {volume}
  {103}},\ \bibinfo {pages} {014506} (\bibinfo {year} {2021})}\BibitemShut
  {NoStop}%
\bibitem [{\citenamefont {Roggero}\ and\ \citenamefont
  {Carlson}(2019)}]{PhysRevC.100.034610}%
  \BibitemOpen
  \bibfield  {author} {\bibinfo {author} {\bibfnamefont {A.}~\bibnamefont
  {Roggero}}\ and\ \bibinfo {author} {\bibfnamefont {J.}~\bibnamefont
  {Carlson}},\ }\href {\doibase 10.1103/PhysRevC.100.034610} {\bibfield
  {journal} {\bibinfo  {journal} {Phys. Rev. C}\ }\textbf {\bibinfo {volume}
  {100}},\ \bibinfo {pages} {034610} (\bibinfo {year} {2019})}\BibitemShut
  {NoStop}%
\bibitem [{\citenamefont {Somma}(2019)}]{Somma_2019}%
  \BibitemOpen
  \bibfield  {author} {\bibinfo {author} {\bibfnamefont {R.~D.}\ \bibnamefont
  {Somma}},\ }\href {\doibase 10.1088/1367-2630/ab5c60} {\bibfield  {journal}
  {\bibinfo  {journal} {New Journal of Physics}\ }\textbf {\bibinfo {volume}
  {21}},\ \bibinfo {pages} {123025} (\bibinfo {year} {2019})}\BibitemShut
  {NoStop}%
\bibitem [{\citenamefont {Lu}\ \emph {et~al.}(2021)\citenamefont {Lu},
  \citenamefont {Ba\~nuls},\ and\ \citenamefont {Cirac}}]{PRXQuantum.2.020321}%
  \BibitemOpen
  \bibfield  {author} {\bibinfo {author} {\bibfnamefont {S.}~\bibnamefont
  {Lu}}, \bibinfo {author} {\bibfnamefont {M.~C.}\ \bibnamefont {Ba\~nuls}}, \
  and\ \bibinfo {author} {\bibfnamefont {J.~I.}\ \bibnamefont {Cirac}},\ }\href
  {\doibase 10.1103/PRXQuantum.2.020321} {\bibfield  {journal} {\bibinfo
  {journal} {PRX Quantum}\ }\textbf {\bibinfo {volume} {2}},\ \bibinfo {pages}
  {020321} (\bibinfo {year} {2021})}\BibitemShut {NoStop}%
\bibitem [{\citenamefont {Roggero}(2020)}]{PhysRevA.102.022409}%
  \BibitemOpen
  \bibfield  {author} {\bibinfo {author} {\bibfnamefont {A.}~\bibnamefont
  {Roggero}},\ }\href {\doibase 10.1103/PhysRevA.102.022409} {\bibfield
  {journal} {\bibinfo  {journal} {Phys. Rev. A}\ }\textbf {\bibinfo {volume}
  {102}},\ \bibinfo {pages} {022409} (\bibinfo {year} {2020})}\BibitemShut
  {NoStop}%
\bibitem [{\citenamefont {Rall}(2020)}]{PhysRevA.102.022408}%
  \BibitemOpen
  \bibfield  {author} {\bibinfo {author} {\bibfnamefont {P.}~\bibnamefont
  {Rall}},\ }\href {\doibase 10.1103/PhysRevA.102.022408} {\bibfield  {journal}
  {\bibinfo  {journal} {Phys. Rev. A}\ }\textbf {\bibinfo {volume} {102}},\
  \bibinfo {pages} {022408} (\bibinfo {year} {2020})}\BibitemShut {NoStop}%
\bibitem [{\citenamefont {Ba\~nuls}\ \emph {et~al.}(2013)\citenamefont
  {Ba\~nuls}, \citenamefont {Cichy}, \citenamefont {Jansen},\ and\
  \citenamefont {Cirac}}]{Banuls:2013jaa}%
  \BibitemOpen
  \bibfield  {author} {\bibinfo {author} {\bibfnamefont {M.~C.}\ \bibnamefont
  {Ba\~nuls}}, \bibinfo {author} {\bibfnamefont {K.}~\bibnamefont {Cichy}},
  \bibinfo {author} {\bibfnamefont {K.}~\bibnamefont {Jansen}}, \ and\ \bibinfo
  {author} {\bibfnamefont {J.~I.}\ \bibnamefont {Cirac}},\ }\href {\doibase
  10.1007/JHEP11(2013)158} {\bibfield  {journal} {\bibinfo  {journal} {JHEP}\
  }\textbf {\bibinfo {volume} {11}},\ \bibinfo {pages} {158} (\bibinfo {year}
  {2013})},\ \Eprint {http://arxiv.org/abs/1305.3765} {arXiv:1305.3765
  [hep-lat]} \BibitemShut {NoStop}%
\bibitem [{\citenamefont {Buyens}\ \emph {et~al.}(2017)\citenamefont {Buyens},
  \citenamefont {Haegeman}, \citenamefont {Hebenstreit}, \citenamefont
  {Verstraete},\ and\ \citenamefont {Van~Acoleyen}}]{Buyens:2016hhu}%
  \BibitemOpen
  \bibfield  {author} {\bibinfo {author} {\bibfnamefont {B.}~\bibnamefont
  {Buyens}}, \bibinfo {author} {\bibfnamefont {J.}~\bibnamefont {Haegeman}},
  \bibinfo {author} {\bibfnamefont {F.}~\bibnamefont {Hebenstreit}}, \bibinfo
  {author} {\bibfnamefont {F.}~\bibnamefont {Verstraete}}, \ and\ \bibinfo
  {author} {\bibfnamefont {K.}~\bibnamefont {Van~Acoleyen}},\ }\href {\doibase
  10.1103/PhysRevD.96.114501} {\bibfield  {journal} {\bibinfo  {journal} {Phys.
  Rev. D}\ }\textbf {\bibinfo {volume} {96}},\ \bibinfo {pages} {114501}
  (\bibinfo {year} {2017})},\ \Eprint {http://arxiv.org/abs/1612.00739}
  {arXiv:1612.00739 [hep-lat]} \BibitemShut {NoStop}%
\bibitem [{\citenamefont {Ba\~nuls}\ \emph {et~al.}(2016)\citenamefont
  {Ba\~nuls}, \citenamefont {Cichy}, \citenamefont {Jansen},\ and\
  \citenamefont {Saito}}]{Banuls:2016lkq}%
  \BibitemOpen
  \bibfield  {author} {\bibinfo {author} {\bibfnamefont {M.~C.}\ \bibnamefont
  {Ba\~nuls}}, \bibinfo {author} {\bibfnamefont {K.}~\bibnamefont {Cichy}},
  \bibinfo {author} {\bibfnamefont {K.}~\bibnamefont {Jansen}}, \ and\ \bibinfo
  {author} {\bibfnamefont {H.}~\bibnamefont {Saito}},\ }\href {\doibase
  10.1103/PhysRevD.93.094512} {\bibfield  {journal} {\bibinfo  {journal} {Phys.
  Rev. D}\ }\textbf {\bibinfo {volume} {93}},\ \bibinfo {pages} {094512}
  (\bibinfo {year} {2016})},\ \Eprint {http://arxiv.org/abs/1603.05002}
  {arXiv:1603.05002 [hep-lat]} \BibitemShut {NoStop}%
\bibitem [{\citenamefont {Funcke}\ \emph {et~al.}(2020)\citenamefont {Funcke},
  \citenamefont {Jansen},\ and\ \citenamefont {K\"uhn}}]{Funcke:2019zna}%
  \BibitemOpen
  \bibfield  {author} {\bibinfo {author} {\bibfnamefont {L.}~\bibnamefont
  {Funcke}}, \bibinfo {author} {\bibfnamefont {K.}~\bibnamefont {Jansen}}, \
  and\ \bibinfo {author} {\bibfnamefont {S.}~\bibnamefont {K\"uhn}},\ }\href
  {\doibase 10.1103/PhysRevD.101.054507} {\bibfield  {journal} {\bibinfo
  {journal} {Phys. Rev. D}\ }\textbf {\bibinfo {volume} {101}},\ \bibinfo
  {pages} {054507} (\bibinfo {year} {2020})},\ \Eprint
  {http://arxiv.org/abs/1908.00551} {arXiv:1908.00551 [hep-lat]} \BibitemShut
  {NoStop}%
\bibitem [{\citenamefont {Yang}\ \emph {et~al.}(2020)\citenamefont {Yang},
  \citenamefont {Sun}, \citenamefont {Ott}, \citenamefont {Wang}, \citenamefont
  {Zache}, \citenamefont {Halimeh}, \citenamefont {Yuan}, \citenamefont
  {Hauke},\ and\ \citenamefont {Pan}}]{Yang_2020}%
  \BibitemOpen
  \bibfield  {author} {\bibinfo {author} {\bibfnamefont {B.}~\bibnamefont
  {Yang}}, \bibinfo {author} {\bibfnamefont {H.}~\bibnamefont {Sun}}, \bibinfo
  {author} {\bibfnamefont {R.}~\bibnamefont {Ott}}, \bibinfo {author}
  {\bibfnamefont {H.-Y.}\ \bibnamefont {Wang}}, \bibinfo {author}
  {\bibfnamefont {T.~V.}\ \bibnamefont {Zache}}, \bibinfo {author}
  {\bibfnamefont {J.~C.}\ \bibnamefont {Halimeh}}, \bibinfo {author}
  {\bibfnamefont {Z.-S.}\ \bibnamefont {Yuan}}, \bibinfo {author}
  {\bibfnamefont {P.}~\bibnamefont {Hauke}}, \ and\ \bibinfo {author}
  {\bibfnamefont {J.-W.}\ \bibnamefont {Pan}},\ }\href {\doibase
  10.1038/s41586-020-2910-8} {\bibfield  {journal} {\bibinfo  {journal}
  {Nature}\ }\textbf {\bibinfo {volume} {587}},\ \bibinfo {pages} {392–396}
  (\bibinfo {year} {2020})}\BibitemShut {NoStop}%
\bibitem [{\citenamefont {Hauke}\ \emph {et~al.}(2013)\citenamefont {Hauke},
  \citenamefont {Marcos}, \citenamefont {Dalmonte},\ and\ \citenamefont
  {Zoller}}]{Hauke:2013jga}%
  \BibitemOpen
  \bibfield  {author} {\bibinfo {author} {\bibfnamefont {P.}~\bibnamefont
  {Hauke}}, \bibinfo {author} {\bibfnamefont {D.}~\bibnamefont {Marcos}},
  \bibinfo {author} {\bibfnamefont {M.}~\bibnamefont {Dalmonte}}, \ and\
  \bibinfo {author} {\bibfnamefont {P.}~\bibnamefont {Zoller}},\ }\href
  {\doibase 10.1103/PhysRevX.3.041018} {\bibfield  {journal} {\bibinfo
  {journal} {Phys. Rev. X}\ }\textbf {\bibinfo {volume} {3}},\ \bibinfo {pages}
  {041018} (\bibinfo {year} {2013})},\ \Eprint {http://arxiv.org/abs/1306.2162}
  {arXiv:1306.2162 [cond-mat.quant-gas]} \BibitemShut {NoStop}%
\bibitem [{\citenamefont {A~Rahman}\ \emph {et~al.}(2021)\citenamefont
  {A~Rahman}, \citenamefont {Lewis}, \citenamefont {Mendicelli},\ and\
  \citenamefont {Powell}}]{Rahman:2021yse}%
  \BibitemOpen
  \bibfield  {author} {\bibinfo {author} {\bibfnamefont {S.}~\bibnamefont
  {A~Rahman}}, \bibinfo {author} {\bibfnamefont {R.}~\bibnamefont {Lewis}},
  \bibinfo {author} {\bibfnamefont {E.}~\bibnamefont {Mendicelli}}, \ and\
  \bibinfo {author} {\bibfnamefont {S.}~\bibnamefont {Powell}},\ }\href@noop {}
  {\  (\bibinfo {year} {2021})},\ \Eprint {http://arxiv.org/abs/2103.08661}
  {arXiv:2103.08661 [hep-lat]} \BibitemShut {NoStop}%
\bibitem [{\citenamefont {Buyens}\ \emph {et~al.}(2016)\citenamefont {Buyens},
  \citenamefont {Verstraete},\ and\ \citenamefont
  {Van~Acoleyen}}]{Buyens:2016ecr}%
  \BibitemOpen
  \bibfield  {author} {\bibinfo {author} {\bibfnamefont {B.}~\bibnamefont
  {Buyens}}, \bibinfo {author} {\bibfnamefont {F.}~\bibnamefont {Verstraete}},
  \ and\ \bibinfo {author} {\bibfnamefont {K.}~\bibnamefont {Van~Acoleyen}},\
  }\href {\doibase 10.1103/PhysRevD.94.085018} {\bibfield  {journal} {\bibinfo
  {journal} {Phys. Rev. D}\ }\textbf {\bibinfo {volume} {94}},\ \bibinfo
  {pages} {085018} (\bibinfo {year} {2016})},\ \Eprint
  {http://arxiv.org/abs/1606.03385} {arXiv:1606.03385 [hep-lat]} \BibitemShut
  {NoStop}%
\bibitem [{\citenamefont {Pichler}\ \emph {et~al.}(2016)\citenamefont
  {Pichler}, \citenamefont {Dalmonte}, \citenamefont {Rico}, \citenamefont
  {Zoller},\ and\ \citenamefont {Montangero}}]{Pichler:2015yqa}%
  \BibitemOpen
  \bibfield  {author} {\bibinfo {author} {\bibfnamefont {T.}~\bibnamefont
  {Pichler}}, \bibinfo {author} {\bibfnamefont {M.}~\bibnamefont {Dalmonte}},
  \bibinfo {author} {\bibfnamefont {E.}~\bibnamefont {Rico}}, \bibinfo {author}
  {\bibfnamefont {P.}~\bibnamefont {Zoller}}, \ and\ \bibinfo {author}
  {\bibfnamefont {S.}~\bibnamefont {Montangero}},\ }\href {\doibase
  10.1103/PhysRevX.6.011023} {\bibfield  {journal} {\bibinfo  {journal} {Phys.
  Rev. X}\ }\textbf {\bibinfo {volume} {6}},\ \bibinfo {pages} {011023}
  (\bibinfo {year} {2016})},\ \Eprint {http://arxiv.org/abs/1505.04440}
  {arXiv:1505.04440 [cond-mat.quant-gas]} \BibitemShut {NoStop}%
\bibitem [{\citenamefont {Butt}\ \emph {et~al.}(2020)\citenamefont {Butt},
  \citenamefont {Catterall}, \citenamefont {Meurice}, \citenamefont {Sakai},\
  and\ \citenamefont {Unmuth-Yockey}}]{Butt:2019uul}%
  \BibitemOpen
  \bibfield  {author} {\bibinfo {author} {\bibfnamefont {N.}~\bibnamefont
  {Butt}}, \bibinfo {author} {\bibfnamefont {S.}~\bibnamefont {Catterall}},
  \bibinfo {author} {\bibfnamefont {Y.}~\bibnamefont {Meurice}}, \bibinfo
  {author} {\bibfnamefont {R.}~\bibnamefont {Sakai}}, \ and\ \bibinfo {author}
  {\bibfnamefont {J.}~\bibnamefont {Unmuth-Yockey}},\ }\href {\doibase
  10.1103/PhysRevD.101.094509} {\bibfield  {journal} {\bibinfo  {journal}
  {Phys. Rev. D}\ }\textbf {\bibinfo {volume} {101}},\ \bibinfo {pages}
  {094509} (\bibinfo {year} {2020})},\ \Eprint
  {http://arxiv.org/abs/1911.01285} {arXiv:1911.01285 [hep-lat]} \BibitemShut
  {NoStop}%
\bibitem [{\citenamefont {Zache}\ \emph {et~al.}(2019)\citenamefont {Zache},
  \citenamefont {Mueller}, \citenamefont {Schneider}, \citenamefont
  {Jendrzejewski}, \citenamefont {Berges},\ and\ \citenamefont
  {Hauke}}]{PhysRevLett.122.050403}%
  \BibitemOpen
  \bibfield  {author} {\bibinfo {author} {\bibfnamefont {T.~V.}\ \bibnamefont
  {Zache}}, \bibinfo {author} {\bibfnamefont {N.}~\bibnamefont {Mueller}},
  \bibinfo {author} {\bibfnamefont {J.~T.}\ \bibnamefont {Schneider}}, \bibinfo
  {author} {\bibfnamefont {F.}~\bibnamefont {Jendrzejewski}}, \bibinfo {author}
  {\bibfnamefont {J.}~\bibnamefont {Berges}}, \ and\ \bibinfo {author}
  {\bibfnamefont {P.}~\bibnamefont {Hauke}},\ }\href {\doibase
  10.1103/PhysRevLett.122.050403} {\bibfield  {journal} {\bibinfo  {journal}
  {Phys. Rev. Lett.}\ }\textbf {\bibinfo {volume} {122}},\ \bibinfo {pages}
  {050403} (\bibinfo {year} {2019})}\BibitemShut {NoStop}%
\bibitem [{\citenamefont {Huang}\ \emph {et~al.}(2019)\citenamefont {Huang},
  \citenamefont {Banerjee},\ and\ \citenamefont
  {Heyl}}]{PhysRevLett.122.250401}%
  \BibitemOpen
  \bibfield  {author} {\bibinfo {author} {\bibfnamefont {Y.-P.}\ \bibnamefont
  {Huang}}, \bibinfo {author} {\bibfnamefont {D.}~\bibnamefont {Banerjee}}, \
  and\ \bibinfo {author} {\bibfnamefont {M.}~\bibnamefont {Heyl}},\ }\href
  {\doibase 10.1103/PhysRevLett.122.250401} {\bibfield  {journal} {\bibinfo
  {journal} {Phys. Rev. Lett.}\ }\textbf {\bibinfo {volume} {122}},\ \bibinfo
  {pages} {250401} (\bibinfo {year} {2019})}\BibitemShut {NoStop}%
\bibitem [{\citenamefont {Zohar}\ \emph {et~al.}(2017)\citenamefont {Zohar},
  \citenamefont {Farace}, \citenamefont {Reznik},\ and\ \citenamefont
  {Cirac}}]{Zohar:2016iic}%
  \BibitemOpen
  \bibfield  {author} {\bibinfo {author} {\bibfnamefont {E.}~\bibnamefont
  {Zohar}}, \bibinfo {author} {\bibfnamefont {A.}~\bibnamefont {Farace}},
  \bibinfo {author} {\bibfnamefont {B.}~\bibnamefont {Reznik}}, \ and\ \bibinfo
  {author} {\bibfnamefont {J.~I.}\ \bibnamefont {Cirac}},\ }\href {\doibase
  10.1103/PhysRevA.95.023604} {\bibfield  {journal} {\bibinfo  {journal} {Phys.
  Rev. A}\ }\textbf {\bibinfo {volume} {95}},\ \bibinfo {pages} {023604}
  (\bibinfo {year} {2017})},\ \Eprint {http://arxiv.org/abs/1607.08121}
  {arXiv:1607.08121 [quant-ph]} \BibitemShut {NoStop}%
\bibitem [{\citenamefont {Muschik}\ \emph {et~al.}(2017)\citenamefont
  {Muschik}, \citenamefont {Heyl}, \citenamefont {Martinez}, \citenamefont
  {Monz}, \citenamefont {Schindler}, \citenamefont {Vogell}, \citenamefont
  {Dalmonte}, \citenamefont {Hauke}, \citenamefont {Blatt},\ and\ \citenamefont
  {Zoller}}]{Muschik:2016tws}%
  \BibitemOpen
  \bibfield  {author} {\bibinfo {author} {\bibfnamefont {C.}~\bibnamefont
  {Muschik}}, \bibinfo {author} {\bibfnamefont {M.}~\bibnamefont {Heyl}},
  \bibinfo {author} {\bibfnamefont {E.}~\bibnamefont {Martinez}}, \bibinfo
  {author} {\bibfnamefont {T.}~\bibnamefont {Monz}}, \bibinfo {author}
  {\bibfnamefont {P.}~\bibnamefont {Schindler}}, \bibinfo {author}
  {\bibfnamefont {B.}~\bibnamefont {Vogell}}, \bibinfo {author} {\bibfnamefont
  {M.}~\bibnamefont {Dalmonte}}, \bibinfo {author} {\bibfnamefont
  {P.}~\bibnamefont {Hauke}}, \bibinfo {author} {\bibfnamefont
  {R.}~\bibnamefont {Blatt}}, \ and\ \bibinfo {author} {\bibfnamefont
  {P.}~\bibnamefont {Zoller}},\ }\href {\doibase 10.1088/1367-2630/aa89ab}
  {\bibfield  {journal} {\bibinfo  {journal} {New J. Phys.}\ }\textbf {\bibinfo
  {volume} {19}},\ \bibinfo {pages} {103020} (\bibinfo {year} {2017})},\
  \Eprint {http://arxiv.org/abs/1612.08653} {arXiv:1612.08653 [quant-ph]}
  \BibitemShut {NoStop}%
\bibitem [{\citenamefont {Luo}\ \emph {et~al.}(2020)\citenamefont {Luo},
  \citenamefont {Shen}, \citenamefont {Highman}, \citenamefont {Clark},
  \citenamefont {DeMarco}, \citenamefont {El-Khadra},\ and\ \citenamefont
  {Gadway}}]{Luo:2019vmi}%
  \BibitemOpen
  \bibfield  {author} {\bibinfo {author} {\bibfnamefont {D.}~\bibnamefont
  {Luo}}, \bibinfo {author} {\bibfnamefont {J.}~\bibnamefont {Shen}}, \bibinfo
  {author} {\bibfnamefont {M.}~\bibnamefont {Highman}}, \bibinfo {author}
  {\bibfnamefont {B.~K.}\ \bibnamefont {Clark}}, \bibinfo {author}
  {\bibfnamefont {B.}~\bibnamefont {DeMarco}}, \bibinfo {author} {\bibfnamefont
  {A.~X.}\ \bibnamefont {El-Khadra}}, \ and\ \bibinfo {author} {\bibfnamefont
  {B.}~\bibnamefont {Gadway}},\ }\href {\doibase 10.1103/PhysRevA.102.032617}
  {\bibfield  {journal} {\bibinfo  {journal} {Phys. Rev. A}\ }\textbf {\bibinfo
  {volume} {102}},\ \bibinfo {pages} {032617} (\bibinfo {year} {2020})},\
  \Eprint {http://arxiv.org/abs/1912.11488} {arXiv:1912.11488 [quant-ph]}
  \BibitemShut {NoStop}%
\bibitem [{\citenamefont {Davoudi}\ \emph {et~al.}(2020)\citenamefont
  {Davoudi}, \citenamefont {Hafezi}, \citenamefont {Monroe}, \citenamefont
  {Pagano}, \citenamefont {Seif},\ and\ \citenamefont
  {Shaw}}]{Davoudi:2019bhy}%
  \BibitemOpen
  \bibfield  {author} {\bibinfo {author} {\bibfnamefont {Z.}~\bibnamefont
  {Davoudi}}, \bibinfo {author} {\bibfnamefont {M.}~\bibnamefont {Hafezi}},
  \bibinfo {author} {\bibfnamefont {C.}~\bibnamefont {Monroe}}, \bibinfo
  {author} {\bibfnamefont {G.}~\bibnamefont {Pagano}}, \bibinfo {author}
  {\bibfnamefont {A.}~\bibnamefont {Seif}}, \ and\ \bibinfo {author}
  {\bibfnamefont {A.}~\bibnamefont {Shaw}},\ }\href {\doibase
  10.1103/PhysRevResearch.2.023015} {\bibfield  {journal} {\bibinfo  {journal}
  {Phys. Rev. Res.}\ }\textbf {\bibinfo {volume} {2}},\ \bibinfo {pages}
  {023015} (\bibinfo {year} {2020})},\ \Eprint
  {http://arxiv.org/abs/1908.03210} {arXiv:1908.03210 [quant-ph]} \BibitemShut
  {NoStop}%
\bibitem [{\citenamefont {Kaplan}\ and\ \citenamefont
  {Stryker}(2020)}]{Kaplan:2018vnj}%
  \BibitemOpen
  \bibfield  {author} {\bibinfo {author} {\bibfnamefont {D.~B.}\ \bibnamefont
  {Kaplan}}\ and\ \bibinfo {author} {\bibfnamefont {J.~R.}\ \bibnamefont
  {Stryker}},\ }\href {\doibase 10.1103/PhysRevD.102.094515} {\bibfield
  {journal} {\bibinfo  {journal} {Phys. Rev. D}\ }\textbf {\bibinfo {volume}
  {102}},\ \bibinfo {pages} {094515} (\bibinfo {year} {2020})},\ \Eprint
  {http://arxiv.org/abs/1806.08797} {arXiv:1806.08797 [hep-lat]} \BibitemShut
  {NoStop}%
\bibitem [{\citenamefont {Magnifico}\ \emph {et~al.}(2020)\citenamefont
  {Magnifico}, \citenamefont {Dalmonte}, \citenamefont {Facchi}, \citenamefont
  {Pascazio}, \citenamefont {Pepe},\ and\ \citenamefont
  {Ercolessi}}]{Magnifico:2019kyj}%
  \BibitemOpen
  \bibfield  {author} {\bibinfo {author} {\bibfnamefont {G.}~\bibnamefont
  {Magnifico}}, \bibinfo {author} {\bibfnamefont {M.}~\bibnamefont {Dalmonte}},
  \bibinfo {author} {\bibfnamefont {P.}~\bibnamefont {Facchi}}, \bibinfo
  {author} {\bibfnamefont {S.}~\bibnamefont {Pascazio}}, \bibinfo {author}
  {\bibfnamefont {F.~V.}\ \bibnamefont {Pepe}}, \ and\ \bibinfo {author}
  {\bibfnamefont {E.}~\bibnamefont {Ercolessi}},\ }\href {\doibase
  10.22331/q-2020-06-15-281} {\bibfield  {journal} {\bibinfo  {journal}
  {Quantum}\ }\textbf {\bibinfo {volume} {4}},\ \bibinfo {pages} {281}
  (\bibinfo {year} {2020})},\ \Eprint {http://arxiv.org/abs/1909.04821}
  {arXiv:1909.04821 [quant-ph]} \BibitemShut {NoStop}%
\bibitem [{\citenamefont {Stryker}(2021)}]{Stryker:2021asy}%
  \BibitemOpen
  \bibfield  {author} {\bibinfo {author} {\bibfnamefont {J.~R.}\ \bibnamefont
  {Stryker}},\ }\href@noop {} {\  (\bibinfo {year} {2021})},\ \Eprint
  {http://arxiv.org/abs/2105.11548} {arXiv:2105.11548 [hep-lat]} \BibitemShut
  {NoStop}%
\bibitem [{\citenamefont {Zohar}\ and\ \citenamefont
  {Reznik}(2011)}]{Zohar:2011cw}%
  \BibitemOpen
  \bibfield  {author} {\bibinfo {author} {\bibfnamefont {E.}~\bibnamefont
  {Zohar}}\ and\ \bibinfo {author} {\bibfnamefont {B.}~\bibnamefont {Reznik}},\
  }\href {\doibase 10.1103/PhysRevLett.107.275301} {\bibfield  {journal}
  {\bibinfo  {journal} {Phys. Rev. Lett.}\ }\textbf {\bibinfo {volume} {107}},\
  \bibinfo {pages} {275301} (\bibinfo {year} {2011})},\ \Eprint
  {http://arxiv.org/abs/1108.1562} {arXiv:1108.1562 [quant-ph]} \BibitemShut
  {NoStop}%
\bibitem [{\citenamefont {Zohar}\ \emph {et~al.}(2012)\citenamefont {Zohar},
  \citenamefont {Cirac},\ and\ \citenamefont {Reznik}}]{Zohar:2012ay}%
  \BibitemOpen
  \bibfield  {author} {\bibinfo {author} {\bibfnamefont {E.}~\bibnamefont
  {Zohar}}, \bibinfo {author} {\bibfnamefont {J.~I.}\ \bibnamefont {Cirac}}, \
  and\ \bibinfo {author} {\bibfnamefont {B.}~\bibnamefont {Reznik}},\ }\href
  {\doibase 10.1103/PhysRevLett.109.125302} {\bibfield  {journal} {\bibinfo
  {journal} {Phys. Rev. Lett.}\ }\textbf {\bibinfo {volume} {109}},\ \bibinfo
  {pages} {125302} (\bibinfo {year} {2012})},\ \Eprint
  {http://arxiv.org/abs/1204.6574} {arXiv:1204.6574 [quant-ph]} \BibitemShut
  {NoStop}%
\bibitem [{\citenamefont {Tagliacozzo}\ \emph {et~al.}(2013)\citenamefont
  {Tagliacozzo}, \citenamefont {Celi}, \citenamefont {Zamora},\ and\
  \citenamefont {Lewenstein}}]{Tagliacozzo:2012vg}%
  \BibitemOpen
  \bibfield  {author} {\bibinfo {author} {\bibfnamefont {L.}~\bibnamefont
  {Tagliacozzo}}, \bibinfo {author} {\bibfnamefont {A.}~\bibnamefont {Celi}},
  \bibinfo {author} {\bibfnamefont {A.}~\bibnamefont {Zamora}}, \ and\ \bibinfo
  {author} {\bibfnamefont {M.}~\bibnamefont {Lewenstein}},\ }\href {\doibase
  10.1016/j.aop.2012.11.009} {\bibfield  {journal} {\bibinfo  {journal} {Annals
  Phys.}\ }\textbf {\bibinfo {volume} {330}},\ \bibinfo {pages} {160} (\bibinfo
  {year} {2013})},\ \Eprint {http://arxiv.org/abs/1205.0496} {arXiv:1205.0496
  [cond-mat.quant-gas]} \BibitemShut {NoStop}%
\bibitem [{\citenamefont {Zohar}\ \emph
  {et~al.}(2013{\natexlab{c}})\citenamefont {Zohar}, \citenamefont {Cirac},\
  and\ \citenamefont {Reznik}}]{Zohar:2012ts}%
  \BibitemOpen
  \bibfield  {author} {\bibinfo {author} {\bibfnamefont {E.}~\bibnamefont
  {Zohar}}, \bibinfo {author} {\bibfnamefont {J.~I.}\ \bibnamefont {Cirac}}, \
  and\ \bibinfo {author} {\bibfnamefont {B.}~\bibnamefont {Reznik}},\ }\href
  {\doibase 10.1103/PhysRevLett.110.055302} {\bibfield  {journal} {\bibinfo
  {journal} {Phys. Rev. Lett.}\ }\textbf {\bibinfo {volume} {110}},\ \bibinfo
  {pages} {055302} (\bibinfo {year} {2013}{\natexlab{c}})},\ \Eprint
  {http://arxiv.org/abs/1208.4299} {arXiv:1208.4299 [quant-ph]} \BibitemShut
  {NoStop}%
\bibitem [{\citenamefont {Wiese}(2013)}]{Wiese:2013uua}%
  \BibitemOpen
  \bibfield  {author} {\bibinfo {author} {\bibfnamefont {U.-J.}\ \bibnamefont
  {Wiese}},\ }\href {\doibase 10.1002/andp.201300104} {\bibfield  {journal}
  {\bibinfo  {journal} {Annalen Phys.}\ }\textbf {\bibinfo {volume} {525}},\
  \bibinfo {pages} {777} (\bibinfo {year} {2013})},\ \Eprint
  {http://arxiv.org/abs/1305.1602} {arXiv:1305.1602 [quant-ph]} \BibitemShut
  {NoStop}%
\bibitem [{\citenamefont {Marcos}\ \emph {et~al.}(2014)\citenamefont {Marcos},
  \citenamefont {Widmer}, \citenamefont {Rico}, \citenamefont {Hafezi},
  \citenamefont {Rabl}, \citenamefont {Wiese},\ and\ \citenamefont
  {Zoller}}]{Marcos:2014lda}%
  \BibitemOpen
  \bibfield  {author} {\bibinfo {author} {\bibfnamefont {D.}~\bibnamefont
  {Marcos}}, \bibinfo {author} {\bibfnamefont {P.}~\bibnamefont {Widmer}},
  \bibinfo {author} {\bibfnamefont {E.}~\bibnamefont {Rico}}, \bibinfo {author}
  {\bibfnamefont {M.}~\bibnamefont {Hafezi}}, \bibinfo {author} {\bibfnamefont
  {P.}~\bibnamefont {Rabl}}, \bibinfo {author} {\bibfnamefont {U.~J.}\
  \bibnamefont {Wiese}}, \ and\ \bibinfo {author} {\bibfnamefont
  {P.}~\bibnamefont {Zoller}},\ }\href {\doibase 10.1016/j.aop.2014.09.011}
  {\bibfield  {journal} {\bibinfo  {journal} {Annals Phys.}\ }\textbf {\bibinfo
  {volume} {351}},\ \bibinfo {pages} {634} (\bibinfo {year} {2014})},\ \Eprint
  {http://arxiv.org/abs/1407.6066} {arXiv:1407.6066 [quant-ph]} \BibitemShut
  {NoStop}%
\bibitem [{\citenamefont {Kuno}\ \emph {et~al.}(2015)\citenamefont {Kuno},
  \citenamefont {Kasamatsu}, \citenamefont {Takahashi}, \citenamefont
  {Ichinose},\ and\ \citenamefont {Matsui}}]{Kuno:2014npa}%
  \BibitemOpen
  \bibfield  {author} {\bibinfo {author} {\bibfnamefont {Y.}~\bibnamefont
  {Kuno}}, \bibinfo {author} {\bibfnamefont {K.}~\bibnamefont {Kasamatsu}},
  \bibinfo {author} {\bibfnamefont {Y.}~\bibnamefont {Takahashi}}, \bibinfo
  {author} {\bibfnamefont {I.}~\bibnamefont {Ichinose}}, \ and\ \bibinfo
  {author} {\bibfnamefont {T.}~\bibnamefont {Matsui}},\ }\href {\doibase
  10.1088/1367-2630/17/6/063005} {\bibfield  {journal} {\bibinfo  {journal}
  {New J. Phys.}\ }\textbf {\bibinfo {volume} {17}},\ \bibinfo {pages} {063005}
  (\bibinfo {year} {2015})},\ \Eprint {http://arxiv.org/abs/1412.7605}
  {arXiv:1412.7605 [cond-mat.quant-gas]} \BibitemShut {NoStop}%
\bibitem [{\citenamefont {Bazavov}\ \emph {et~al.}(2015)\citenamefont
  {Bazavov}, \citenamefont {Meurice}, \citenamefont {Tsai}, \citenamefont
  {Unmuth-Yockey},\ and\ \citenamefont {Zhang}}]{Bazavov:2015kka}%
  \BibitemOpen
  \bibfield  {author} {\bibinfo {author} {\bibfnamefont {A.}~\bibnamefont
  {Bazavov}}, \bibinfo {author} {\bibfnamefont {Y.}~\bibnamefont {Meurice}},
  \bibinfo {author} {\bibfnamefont {S.-W.}\ \bibnamefont {Tsai}}, \bibinfo
  {author} {\bibfnamefont {J.}~\bibnamefont {Unmuth-Yockey}}, \ and\ \bibinfo
  {author} {\bibfnamefont {J.}~\bibnamefont {Zhang}},\ }\href {\doibase
  10.1103/PhysRevD.92.076003} {\bibfield  {journal} {\bibinfo  {journal} {Phys.
  Rev. D}\ }\textbf {\bibinfo {volume} {92}},\ \bibinfo {pages} {076003}
  (\bibinfo {year} {2015})},\ \Eprint {http://arxiv.org/abs/1503.08354}
  {arXiv:1503.08354 [hep-lat]} \BibitemShut {NoStop}%
\bibitem [{\citenamefont {Kasper}\ \emph {et~al.}(2016)\citenamefont {Kasper},
  \citenamefont {Hebenstreit}, \citenamefont {Oberthaler},\ and\ \citenamefont
  {Berges}}]{Kasper:2015cca}%
  \BibitemOpen
  \bibfield  {author} {\bibinfo {author} {\bibfnamefont {V.}~\bibnamefont
  {Kasper}}, \bibinfo {author} {\bibfnamefont {F.}~\bibnamefont {Hebenstreit}},
  \bibinfo {author} {\bibfnamefont {M.}~\bibnamefont {Oberthaler}}, \ and\
  \bibinfo {author} {\bibfnamefont {J.}~\bibnamefont {Berges}},\ }\href
  {\doibase 10.1016/j.physletb.2016.07.036} {\bibfield  {journal} {\bibinfo
  {journal} {Phys. Lett. B}\ }\textbf {\bibinfo {volume} {760}},\ \bibinfo
  {pages} {742} (\bibinfo {year} {2016})},\ \Eprint
  {http://arxiv.org/abs/1506.01238} {arXiv:1506.01238 [cond-mat.quant-gas]}
  \BibitemShut {NoStop}%
\bibitem [{\citenamefont {Brennen}\ \emph {et~al.}(2016)\citenamefont
  {Brennen}, \citenamefont {Pupillo}, \citenamefont {Rico}, \citenamefont
  {Stace},\ and\ \citenamefont {Vodola}}]{Brennen:2015pgn}%
  \BibitemOpen
  \bibfield  {author} {\bibinfo {author} {\bibfnamefont {G.~K.}\ \bibnamefont
  {Brennen}}, \bibinfo {author} {\bibfnamefont {G.}~\bibnamefont {Pupillo}},
  \bibinfo {author} {\bibfnamefont {E.}~\bibnamefont {Rico}}, \bibinfo {author}
  {\bibfnamefont {T.~M.}\ \bibnamefont {Stace}}, \ and\ \bibinfo {author}
  {\bibfnamefont {D.}~\bibnamefont {Vodola}},\ }\href {\doibase
  10.1103/PhysRevLett.117.240504} {\bibfield  {journal} {\bibinfo  {journal}
  {Phys. Rev. Lett.}\ }\textbf {\bibinfo {volume} {117}},\ \bibinfo {pages}
  {240504} (\bibinfo {year} {2016})},\ \Eprint
  {http://arxiv.org/abs/1512.06565} {arXiv:1512.06565 [quant-ph]} \BibitemShut
  {NoStop}%
\bibitem [{\citenamefont {Kuno}\ \emph {et~al.}(2016)\citenamefont {Kuno},
  \citenamefont {Sakane}, \citenamefont {Kasamatsu}, \citenamefont {Ichinose},\
  and\ \citenamefont {Matsui}}]{Kuno:2016xbf}%
  \BibitemOpen
  \bibfield  {author} {\bibinfo {author} {\bibfnamefont {Y.}~\bibnamefont
  {Kuno}}, \bibinfo {author} {\bibfnamefont {S.}~\bibnamefont {Sakane}},
  \bibinfo {author} {\bibfnamefont {K.}~\bibnamefont {Kasamatsu}}, \bibinfo
  {author} {\bibfnamefont {I.}~\bibnamefont {Ichinose}}, \ and\ \bibinfo
  {author} {\bibfnamefont {T.}~\bibnamefont {Matsui}},\ }\href {\doibase
  10.1103/PhysRevA.94.063641} {\bibfield  {journal} {\bibinfo  {journal} {Phys.
  Rev. A}\ }\textbf {\bibinfo {volume} {94}},\ \bibinfo {pages} {063641}
  (\bibinfo {year} {2016})},\ \Eprint {http://arxiv.org/abs/1605.02502}
  {arXiv:1605.02502 [cond-mat.quant-gas]} \BibitemShut {NoStop}%
\bibitem [{\citenamefont {Kasper}\ \emph {et~al.}(2017)\citenamefont {Kasper},
  \citenamefont {Hebenstreit}, \citenamefont {Jendrzejewski}, \citenamefont
  {Oberthaler},\ and\ \citenamefont {Berges}}]{Kasper:2016mzj}%
  \BibitemOpen
  \bibfield  {author} {\bibinfo {author} {\bibfnamefont {V.}~\bibnamefont
  {Kasper}}, \bibinfo {author} {\bibfnamefont {F.}~\bibnamefont {Hebenstreit}},
  \bibinfo {author} {\bibfnamefont {F.}~\bibnamefont {Jendrzejewski}}, \bibinfo
  {author} {\bibfnamefont {M.~K.}\ \bibnamefont {Oberthaler}}, \ and\ \bibinfo
  {author} {\bibfnamefont {J.}~\bibnamefont {Berges}},\ }\href {\doibase
  10.1088/1367-2630/aa54e0} {\bibfield  {journal} {\bibinfo  {journal} {New J.
  Phys.}\ }\textbf {\bibinfo {volume} {19}},\ \bibinfo {pages} {023030}
  (\bibinfo {year} {2017})},\ \Eprint {http://arxiv.org/abs/1608.03480}
  {arXiv:1608.03480 [cond-mat.quant-gas]} \BibitemShut {NoStop}%
\bibitem [{\citenamefont {Paulson}\ \emph
  {et~al.}(2020{\natexlab{a}})\citenamefont {Paulson} \emph
  {et~al.}}]{Paulson:2020zjd}%
  \BibitemOpen
  \bibfield  {author} {\bibinfo {author} {\bibfnamefont {D.}~\bibnamefont
  {Paulson}} \emph {et~al.},\ }\href@noop {} {\  (\bibinfo {year}
  {2020}{\natexlab{a}})},\ \Eprint {http://arxiv.org/abs/2008.09252}
  {arXiv:2008.09252 [quant-ph]} \BibitemShut {NoStop}%
\bibitem [{\citenamefont {Kan}\ \emph {et~al.}(2021)\citenamefont {Kan},
  \citenamefont {Funcke}, \citenamefont {K\"uhn}, \citenamefont {Dellantonio},
  \citenamefont {Zhang}, \citenamefont {Haase}, \citenamefont {Muschik},\ and\
  \citenamefont {Jansen}}]{Kan:2021nyu}%
  \BibitemOpen
  \bibfield  {author} {\bibinfo {author} {\bibfnamefont {A.}~\bibnamefont
  {Kan}}, \bibinfo {author} {\bibfnamefont {L.}~\bibnamefont {Funcke}},
  \bibinfo {author} {\bibfnamefont {S.}~\bibnamefont {K\"uhn}}, \bibinfo
  {author} {\bibfnamefont {L.}~\bibnamefont {Dellantonio}}, \bibinfo {author}
  {\bibfnamefont {J.}~\bibnamefont {Zhang}}, \bibinfo {author} {\bibfnamefont
  {J.~F.}\ \bibnamefont {Haase}}, \bibinfo {author} {\bibfnamefont {C.~A.}\
  \bibnamefont {Muschik}}, \ and\ \bibinfo {author} {\bibfnamefont
  {K.}~\bibnamefont {Jansen}},\ }\href@noop {} {\  (\bibinfo {year} {2021})},\
  \Eprint {http://arxiv.org/abs/2105.06019} {arXiv:2105.06019 [hep-lat]}
  \BibitemShut {NoStop}%
\bibitem [{\citenamefont {Wilson}\ \emph {et~al.}(1994)\citenamefont {Wilson},
  \citenamefont {Walhout}, \citenamefont {Harindranath}, \citenamefont {Zhang},
  \citenamefont {Perry},\ and\ \citenamefont {Glazek}}]{Wilson:1994fk}%
  \BibitemOpen
  \bibfield  {author} {\bibinfo {author} {\bibfnamefont {K.~G.}\ \bibnamefont
  {Wilson}}, \bibinfo {author} {\bibfnamefont {T.~S.}\ \bibnamefont {Walhout}},
  \bibinfo {author} {\bibfnamefont {A.}~\bibnamefont {Harindranath}}, \bibinfo
  {author} {\bibfnamefont {W.-M.}\ \bibnamefont {Zhang}}, \bibinfo {author}
  {\bibfnamefont {R.~J.}\ \bibnamefont {Perry}}, \ and\ \bibinfo {author}
  {\bibfnamefont {S.~D.}\ \bibnamefont {Glazek}},\ }\href {\doibase
  10.1103/PhysRevD.49.6720} {\bibfield  {journal} {\bibinfo  {journal} {Phys.
  Rev. D}\ }\textbf {\bibinfo {volume} {49}},\ \bibinfo {pages} {6720}
  (\bibinfo {year} {1994})},\ \Eprint {http://arxiv.org/abs/hep-th/9401153}
  {arXiv:hep-th/9401153} \BibitemShut {NoStop}%
\bibitem [{\citenamefont {Heinzl}(1995)}]{Heinzl:1995jn}%
  \BibitemOpen
  \bibfield  {author} {\bibinfo {author} {\bibfnamefont {T.}~\bibnamefont
  {Heinzl}},\ }in\ \href@noop {} {\emph {\bibinfo {booktitle} {{5th Meeting on
  Light Cone Quantization and Nonperturbative QCD}}}}\ (\bibinfo {year}
  {1995})\ \Eprint {http://arxiv.org/abs/hep-th/9604018} {arXiv:hep-th/9604018}
  \BibitemShut {NoStop}%
\bibitem [{\citenamefont {Bronzan}(1985)}]{PhysRevD.31.2020}%
  \BibitemOpen
  \bibfield  {author} {\bibinfo {author} {\bibfnamefont {J.~B.}\ \bibnamefont
  {Bronzan}},\ }\href {\doibase 10.1103/PhysRevD.31.2020} {\bibfield  {journal}
  {\bibinfo  {journal} {Phys. Rev. D}\ }\textbf {\bibinfo {volume} {31}},\
  \bibinfo {pages} {2020} (\bibinfo {year} {1985})}\BibitemShut {NoStop}%
\bibitem [{\citenamefont {Ligterink}\ \emph
  {et~al.}(2000{\natexlab{a}})\citenamefont {Ligterink}, \citenamefont
  {Walet},\ and\ \citenamefont {Bishop}}]{LIGTERINK2000983c}%
  \BibitemOpen
  \bibfield  {author} {\bibinfo {author} {\bibfnamefont {N.}~\bibnamefont
  {Ligterink}}, \bibinfo {author} {\bibfnamefont {N.}~\bibnamefont {Walet}}, \
  and\ \bibinfo {author} {\bibfnamefont {R.}~\bibnamefont {Bishop}},\ }\href
  {\doibase https://doi.org/10.1016/S0375-9474(99)00748-4} {\bibfield
  {journal} {\bibinfo  {journal} {Nuclear Physics A}\ }\textbf {\bibinfo
  {volume} {663-664}},\ \bibinfo {pages} {983c} (\bibinfo {year}
  {2000}{\natexlab{a}})}\BibitemShut {NoStop}%
\bibitem [{\citenamefont {Ligterink}\ \emph
  {et~al.}(2000{\natexlab{b}})\citenamefont {Ligterink}, \citenamefont
  {Walet},\ and\ \citenamefont {Bishop}}]{LIGTERINK2000215}%
  \BibitemOpen
  \bibfield  {author} {\bibinfo {author} {\bibfnamefont {N.}~\bibnamefont
  {Ligterink}}, \bibinfo {author} {\bibfnamefont {N.}~\bibnamefont {Walet}}, \
  and\ \bibinfo {author} {\bibfnamefont {R.}~\bibnamefont {Bishop}},\ }\href
  {\doibase https://doi.org/10.1006/aphy.2000.6070} {\bibfield  {journal}
  {\bibinfo  {journal} {Annals of Physics}\ }\textbf {\bibinfo {volume}
  {284}},\ \bibinfo {pages} {215} (\bibinfo {year}
  {2000}{\natexlab{b}})}\BibitemShut {NoStop}%
\bibitem [{\citenamefont {Ba\~nuls}\ \emph {et~al.}(2017)\citenamefont
  {Ba\~nuls}, \citenamefont {Cichy}, \citenamefont {Cirac}, \citenamefont
  {Jansen},\ and\ \citenamefont {K\"uhn}}]{Banuls:2017ena}%
  \BibitemOpen
  \bibfield  {author} {\bibinfo {author} {\bibfnamefont {M.~C.}\ \bibnamefont
  {Ba\~nuls}}, \bibinfo {author} {\bibfnamefont {K.}~\bibnamefont {Cichy}},
  \bibinfo {author} {\bibfnamefont {J.~I.}\ \bibnamefont {Cirac}}, \bibinfo
  {author} {\bibfnamefont {K.}~\bibnamefont {Jansen}}, \ and\ \bibinfo {author}
  {\bibfnamefont {S.}~\bibnamefont {K\"uhn}},\ }\href {\doibase
  10.1103/PhysRevX.7.041046} {\bibfield  {journal} {\bibinfo  {journal} {Phys.
  Rev. X}\ }\textbf {\bibinfo {volume} {7}},\ \bibinfo {pages} {041046}
  (\bibinfo {year} {2017})},\ \Eprint {http://arxiv.org/abs/1707.06434}
  {arXiv:1707.06434 [hep-lat]} \BibitemShut {NoStop}%
\bibitem [{\citenamefont {Paulson}\ \emph
  {et~al.}(2020{\natexlab{b}})\citenamefont {Paulson}, \citenamefont
  {Dellantonio}, \citenamefont {Haase}, \citenamefont {Celi}, \citenamefont
  {Kan}, \citenamefont {Jena}, \citenamefont {Kokail}, \citenamefont {van
  Bijnen}, \citenamefont {Jansen}, \citenamefont {Zoller},\ and\ \citenamefont
  {Muschik}}]{paulson2020simulating}%
  \BibitemOpen
  \bibfield  {author} {\bibinfo {author} {\bibfnamefont {D.}~\bibnamefont
  {Paulson}}, \bibinfo {author} {\bibfnamefont {L.}~\bibnamefont
  {Dellantonio}}, \bibinfo {author} {\bibfnamefont {J.~F.}\ \bibnamefont
  {Haase}}, \bibinfo {author} {\bibfnamefont {A.}~\bibnamefont {Celi}},
  \bibinfo {author} {\bibfnamefont {A.}~\bibnamefont {Kan}}, \bibinfo {author}
  {\bibfnamefont {A.}~\bibnamefont {Jena}}, \bibinfo {author} {\bibfnamefont
  {C.}~\bibnamefont {Kokail}}, \bibinfo {author} {\bibfnamefont
  {R.}~\bibnamefont {van Bijnen}}, \bibinfo {author} {\bibfnamefont
  {K.}~\bibnamefont {Jansen}}, \bibinfo {author} {\bibfnamefont
  {P.}~\bibnamefont {Zoller}}, \ and\ \bibinfo {author} {\bibfnamefont {C.~A.}\
  \bibnamefont {Muschik}},\ }\href@noop {} {\enquote {\bibinfo {title} {Towards
  simulating 2d effects in lattice gauge theories on a quantum computer},}\ }
  (\bibinfo {year} {2020}{\natexlab{b}}),\ \Eprint
  {http://arxiv.org/abs/2008.09252} {arXiv:2008.09252 [quant-ph]} \BibitemShut
  {NoStop}%
\bibitem [{cre()}]{creativecommons4}%
  \BibitemOpen
  \href {https://creativecommons.org/licenses/by/4.0/} {\enquote {\bibinfo
  {title} {Creative commons attribution 4.0 international (cc by 4.0)},}\
  }\BibitemShut {NoStop}%
\bibitem [{\citenamefont {Meyer}(2011)}]{Meyer_2011}%
  \BibitemOpen
  \bibfield  {author} {\bibinfo {author} {\bibfnamefont {H.~B.}\ \bibnamefont
  {Meyer}},\ }\href {\doibase 10.1140/epja/i2011-11086-3} {\bibfield  {journal}
  {\bibinfo  {journal} {The European Physical Journal A}\ }\textbf {\bibinfo
  {volume} {47}} (\bibinfo {year} {2011}),\
  10.1140/epja/i2011-11086-3}\BibitemShut {NoStop}%
\bibitem [{\citenamefont {Roggero}\ and\ \citenamefont
  {Reddy}(2016)}]{Roggero_2016}%
  \BibitemOpen
  \bibfield  {author} {\bibinfo {author} {\bibfnamefont {A.}~\bibnamefont
  {Roggero}}\ and\ \bibinfo {author} {\bibfnamefont {S.}~\bibnamefont
  {Reddy}},\ }\href {\doibase 10.1103/physrevc.94.015803} {\bibfield  {journal}
  {\bibinfo  {journal} {Physical Review C}\ }\textbf {\bibinfo {volume} {94}}
  (\bibinfo {year} {2016}),\ 10.1103/physrevc.94.015803}\BibitemShut {NoStop}%
\bibitem [{\citenamefont {Ciavarella}(2020)}]{PhysRevD.102.094505}%
  \BibitemOpen
  \bibfield  {author} {\bibinfo {author} {\bibfnamefont {A.}~\bibnamefont
  {Ciavarella}},\ }\href {\doibase 10.1103/PhysRevD.102.094505} {\bibfield
  {journal} {\bibinfo  {journal} {Phys. Rev. D}\ }\textbf {\bibinfo {volume}
  {102}},\ \bibinfo {pages} {094505} (\bibinfo {year} {2020})}\BibitemShut
  {NoStop}%
\bibitem [{\citenamefont {Endo}\ \emph {et~al.}(2020)\citenamefont {Endo},
  \citenamefont {Kurata},\ and\ \citenamefont
  {Nakagawa}}]{PhysRevResearch.2.033281}%
  \BibitemOpen
  \bibfield  {author} {\bibinfo {author} {\bibfnamefont {S.}~\bibnamefont
  {Endo}}, \bibinfo {author} {\bibfnamefont {I.}~\bibnamefont {Kurata}}, \ and\
  \bibinfo {author} {\bibfnamefont {Y.~O.}\ \bibnamefont {Nakagawa}},\ }\href
  {\doibase 10.1103/PhysRevResearch.2.033281} {\bibfield  {journal} {\bibinfo
  {journal} {Phys. Rev. Research}\ }\textbf {\bibinfo {volume} {2}},\ \bibinfo
  {pages} {033281} (\bibinfo {year} {2020})}\BibitemShut {NoStop}%
\bibitem [{\citenamefont {Roggero}\ \emph
  {et~al.}(2020{\natexlab{b}})\citenamefont {Roggero}, \citenamefont {Gu},
  \citenamefont {Baroni},\ and\ \citenamefont
  {Papenbrock}}]{PhysRevC.102.064624}%
  \BibitemOpen
  \bibfield  {author} {\bibinfo {author} {\bibfnamefont {A.}~\bibnamefont
  {Roggero}}, \bibinfo {author} {\bibfnamefont {C.}~\bibnamefont {Gu}},
  \bibinfo {author} {\bibfnamefont {A.}~\bibnamefont {Baroni}}, \ and\ \bibinfo
  {author} {\bibfnamefont {T.}~\bibnamefont {Papenbrock}},\ }\href {\doibase
  10.1103/PhysRevC.102.064624} {\bibfield  {journal} {\bibinfo  {journal}
  {Phys. Rev. C}\ }\textbf {\bibinfo {volume} {102}},\ \bibinfo {pages}
  {064624} (\bibinfo {year} {2020}{\natexlab{b}})}\BibitemShut {NoStop}%
\bibitem [{\citenamefont {Wolfenstein}(1978)}]{Wolfenstein:1977ue}%
  \BibitemOpen
  \bibfield  {author} {\bibinfo {author} {\bibfnamefont {L.}~\bibnamefont
  {Wolfenstein}},\ }\href {\doibase 10.1103/PhysRevD.17.2369} {\bibfield
  {journal} {\bibinfo  {journal} {Phys. Rev. D}\ }\textbf {\bibinfo {volume}
  {17}},\ \bibinfo {pages} {2369} (\bibinfo {year} {1978})}\BibitemShut
  {NoStop}%
\bibitem [{\citenamefont {Mikheyev}\ and\ \citenamefont
  {Smirnov}(1985)}]{Mikheev:1986gs}%
  \BibitemOpen
  \bibfield  {author} {\bibinfo {author} {\bibfnamefont {S.}~\bibnamefont
  {Mikheyev}}\ and\ \bibinfo {author} {\bibfnamefont {A.}~\bibnamefont
  {Smirnov}},\ }\href@noop {} {\bibfield  {journal} {\bibinfo  {journal} {Sov.
  J. Nucl. Phys.}\ }\textbf {\bibinfo {volume} {42}},\ \bibinfo {pages} {913}
  (\bibinfo {year} {1985})}\BibitemShut {NoStop}%
\bibitem [{\citenamefont {{Fuller}}\ \emph {et~al.}(1987)\citenamefont
  {{Fuller}}, \citenamefont {{Mayle}}, \citenamefont {{Wilson}},\ and\
  \citenamefont {{Schramm}}}]{1987ApJ...322..795F}%
  \BibitemOpen
  \bibfield  {author} {\bibinfo {author} {\bibfnamefont {G.~M.}\ \bibnamefont
  {{Fuller}}}, \bibinfo {author} {\bibfnamefont {R.~W.}\ \bibnamefont
  {{Mayle}}}, \bibinfo {author} {\bibfnamefont {J.~R.}\ \bibnamefont
  {{Wilson}}}, \ and\ \bibinfo {author} {\bibfnamefont {D.~N.}\ \bibnamefont
  {{Schramm}}},\ }\href {\doibase 10.1086/165772} {\bibfield  {journal}
  {\bibinfo  {journal} {Astrophys. J.}\ }\textbf {\bibinfo {volume} {322}},\
  \bibinfo {pages} {795} (\bibinfo {year} {1987})}\BibitemShut {NoStop}%
\bibitem [{\citenamefont {Savage}\ \emph {et~al.}(1991)\citenamefont {Savage},
  \citenamefont {Malaney},\ and\ \citenamefont {Fuller}}]{Savage:1990by}%
  \BibitemOpen
  \bibfield  {author} {\bibinfo {author} {\bibfnamefont {M.~J.}\ \bibnamefont
  {Savage}}, \bibinfo {author} {\bibfnamefont {R.~A.}\ \bibnamefont {Malaney}},
  \ and\ \bibinfo {author} {\bibfnamefont {G.~M.}\ \bibnamefont {Fuller}},\
  }\href {\doibase 10.1086/169665} {\bibfield  {journal} {\bibinfo  {journal}
  {Astrophys. J.}\ }\textbf {\bibinfo {volume} {368}},\ \bibinfo {pages} {1}
  (\bibinfo {year} {1991})}\BibitemShut {NoStop}%
\bibitem [{\citenamefont {Pehlivan}\ \emph {et~al.}(2011)\citenamefont
  {Pehlivan}, \citenamefont {Balantekin}, \citenamefont {Kajino},\ and\
  \citenamefont {Yoshida}}]{Pehlivan2011}%
  \BibitemOpen
  \bibfield  {author} {\bibinfo {author} {\bibfnamefont {Y.}~\bibnamefont
  {Pehlivan}}, \bibinfo {author} {\bibfnamefont {A.~B.}\ \bibnamefont
  {Balantekin}}, \bibinfo {author} {\bibfnamefont {T.}~\bibnamefont {Kajino}},
  \ and\ \bibinfo {author} {\bibfnamefont {T.}~\bibnamefont {Yoshida}},\ }\href
  {\doibase 10.1103/PhysRevD.84.065008} {\bibfield  {journal} {\bibinfo
  {journal} {Phys. Rev. D}\ }\textbf {\bibinfo {volume} {84}},\ \bibinfo
  {pages} {065008} (\bibinfo {year} {2011})}\BibitemShut {NoStop}%
\bibitem [{\citenamefont {Patwardhan}\ \emph {et~al.}(2019)\citenamefont
  {Patwardhan}, \citenamefont {Cervia},\ and\ \citenamefont
  {Balantekin}}]{PhysRevD.99.123013}%
  \BibitemOpen
  \bibfield  {author} {\bibinfo {author} {\bibfnamefont {A.~V.}\ \bibnamefont
  {Patwardhan}}, \bibinfo {author} {\bibfnamefont {M.~J.}\ \bibnamefont
  {Cervia}}, \ and\ \bibinfo {author} {\bibfnamefont {A.~B.}\ \bibnamefont
  {Balantekin}},\ }\href {\doibase 10.1103/PhysRevD.99.123013} {\bibfield
  {journal} {\bibinfo  {journal} {Phys. Rev. D}\ }\textbf {\bibinfo {volume}
  {99}},\ \bibinfo {pages} {123013} (\bibinfo {year} {2019})}\BibitemShut
  {NoStop}%
\bibitem [{\citenamefont {Duan}\ \emph {et~al.}(2010)\citenamefont {Duan},
  \citenamefont {Fuller},\ and\ \citenamefont {Qian}}]{Duan:2010bg}%
  \BibitemOpen
  \bibfield  {author} {\bibinfo {author} {\bibfnamefont {H.}~\bibnamefont
  {Duan}}, \bibinfo {author} {\bibfnamefont {G.~M.}\ \bibnamefont {Fuller}}, \
  and\ \bibinfo {author} {\bibfnamefont {Y.-Z.}\ \bibnamefont {Qian}},\ }\href
  {\doibase 10.1146/annurev.nucl.012809.104524} {\bibfield  {journal} {\bibinfo
   {journal} {Ann. Rev. Nucl. Part. Sci.}\ }\textbf {\bibinfo {volume} {60}},\
  \bibinfo {pages} {569} (\bibinfo {year} {2010})},\ \Eprint
  {http://arxiv.org/abs/1001.2799} {arXiv:1001.2799 [hep-ph]} \BibitemShut
  {NoStop}%
\bibitem [{\citenamefont {Chakraborty}\ \emph {et~al.}(2016)\citenamefont
  {Chakraborty}, \citenamefont {Hansen}, \citenamefont {Izaguirre},\ and\
  \citenamefont {Raffelt}}]{Chakraborty2016b}%
  \BibitemOpen
  \bibfield  {author} {\bibinfo {author} {\bibfnamefont {S.}~\bibnamefont
  {Chakraborty}}, \bibinfo {author} {\bibfnamefont {R.}~\bibnamefont {Hansen}},
  \bibinfo {author} {\bibfnamefont {I.}~\bibnamefont {Izaguirre}}, \ and\
  \bibinfo {author} {\bibfnamefont {G.}~\bibnamefont {Raffelt}},\ }\href
  {\doibase https://doi.org/10.1016/j.nuclphysb.2016.02.012} {\bibfield
  {journal} {\bibinfo  {journal} {Nuclear Physics B}\ }\textbf {\bibinfo
  {volume} {908}},\ \bibinfo {pages} {366 } (\bibinfo {year} {2016})},\
  \bibinfo {note} {neutrino Oscillations: Celebrating the Nobel Prize in
  Physics 2015}\BibitemShut {NoStop}%
\bibitem [{\citenamefont {Friedland}\ and\ \citenamefont
  {Lunardini}(2003)}]{Friedland2003}%
  \BibitemOpen
  \bibfield  {author} {\bibinfo {author} {\bibfnamefont {A.}~\bibnamefont
  {Friedland}}\ and\ \bibinfo {author} {\bibfnamefont {C.}~\bibnamefont
  {Lunardini}},\ }\href {\doibase 10.1088/1126-6708/2003/10/043} {\bibfield
  {journal} {\bibinfo  {journal} {Journal of High Energy Physics}\ }\textbf
  {\bibinfo {volume} {2003}},\ \bibinfo {pages} {043} (\bibinfo {year}
  {2003})}\BibitemShut {NoStop}%
\bibitem [{\citenamefont {Cervia}\ \emph {et~al.}(2019)\citenamefont {Cervia},
  \citenamefont {Patwardhan}, \citenamefont {Balantekin}, \citenamefont
  {Coppersmith},\ and\ \citenamefont {Johnson}}]{Cervia:2019}%
  \BibitemOpen
  \bibfield  {author} {\bibinfo {author} {\bibfnamefont {M.~J.}\ \bibnamefont
  {Cervia}}, \bibinfo {author} {\bibfnamefont {A.~V.}\ \bibnamefont
  {Patwardhan}}, \bibinfo {author} {\bibfnamefont {A.~B.}\ \bibnamefont
  {Balantekin}}, \bibinfo {author} {\bibfnamefont {S.~N.}\ \bibnamefont
  {Coppersmith}}, \ and\ \bibinfo {author} {\bibfnamefont {C.~W.}\ \bibnamefont
  {Johnson}},\ }\href {\doibase 10.1103/PhysRevD.100.083001} {\bibfield
  {journal} {\bibinfo  {journal} {Phys. Rev. D}\ }\textbf {\bibinfo {volume}
  {100}},\ \bibinfo {pages} {083001} (\bibinfo {year} {2019})}\BibitemShut
  {NoStop}%
\bibitem [{\citenamefont {Rrapaj}(2020)}]{Rrapaj2020}%
  \BibitemOpen
  \bibfield  {author} {\bibinfo {author} {\bibfnamefont {E.}~\bibnamefont
  {Rrapaj}},\ }\href {\doibase 10.1103/PhysRevC.101.065805} {\bibfield
  {journal} {\bibinfo  {journal} {Phys. Rev. C}\ }\textbf {\bibinfo {volume}
  {101}},\ \bibinfo {pages} {065805} (\bibinfo {year} {2020})}\BibitemShut
  {NoStop}%
\bibitem [{\citenamefont {Monroe}\ \emph {et~al.}(2021)\citenamefont {Monroe},
  \citenamefont {Campbell}, \citenamefont {Duan}, \citenamefont {Gong},
  \citenamefont {Gorshkov}, \citenamefont {Hess}, \citenamefont {Islam},
  \citenamefont {Kim}, \citenamefont {Linke}, \citenamefont {Pagano},
  \citenamefont {Richerme}, \citenamefont {Senko},\ and\ \citenamefont
  {Yao}}]{RevModPhys.93.025001}%
  \BibitemOpen
  \bibfield  {author} {\bibinfo {author} {\bibfnamefont {C.}~\bibnamefont
  {Monroe}}, \bibinfo {author} {\bibfnamefont {W.~C.}\ \bibnamefont
  {Campbell}}, \bibinfo {author} {\bibfnamefont {L.-M.}\ \bibnamefont {Duan}},
  \bibinfo {author} {\bibfnamefont {Z.-X.}\ \bibnamefont {Gong}}, \bibinfo
  {author} {\bibfnamefont {A.~V.}\ \bibnamefont {Gorshkov}}, \bibinfo {author}
  {\bibfnamefont {P.~W.}\ \bibnamefont {Hess}}, \bibinfo {author}
  {\bibfnamefont {R.}~\bibnamefont {Islam}}, \bibinfo {author} {\bibfnamefont
  {K.}~\bibnamefont {Kim}}, \bibinfo {author} {\bibfnamefont {N.~M.}\
  \bibnamefont {Linke}}, \bibinfo {author} {\bibfnamefont {G.}~\bibnamefont
  {Pagano}}, \bibinfo {author} {\bibfnamefont {P.}~\bibnamefont {Richerme}},
  \bibinfo {author} {\bibfnamefont {C.}~\bibnamefont {Senko}}, \ and\ \bibinfo
  {author} {\bibfnamefont {N.~Y.}\ \bibnamefont {Yao}},\ }\href {\doibase
  10.1103/RevModPhys.93.025001} {\bibfield  {journal} {\bibinfo  {journal}
  {Rev. Mod. Phys.}\ }\textbf {\bibinfo {volume} {93}},\ \bibinfo {pages}
  {025001} (\bibinfo {year} {2021})}\BibitemShut {NoStop}%
\bibitem [{\citenamefont {Zhang}\ \emph {et~al.}(2017)\citenamefont {Zhang},
  \citenamefont {Pagano}, \citenamefont {Hess}, \citenamefont {Kyprianidis},
  \citenamefont {Becker}, \citenamefont {Kaplan}, \citenamefont {Gorshkov},
  \citenamefont {Gong},\ and\ \citenamefont {Monroe}}]{Zhang2017}%
  \BibitemOpen
  \bibfield  {author} {\bibinfo {author} {\bibfnamefont {J.}~\bibnamefont
  {Zhang}}, \bibinfo {author} {\bibfnamefont {G.}~\bibnamefont {Pagano}},
  \bibinfo {author} {\bibfnamefont {P.~W.}\ \bibnamefont {Hess}}, \bibinfo
  {author} {\bibfnamefont {A.}~\bibnamefont {Kyprianidis}}, \bibinfo {author}
  {\bibfnamefont {P.}~\bibnamefont {Becker}}, \bibinfo {author} {\bibfnamefont
  {H.}~\bibnamefont {Kaplan}}, \bibinfo {author} {\bibfnamefont {A.~V.}\
  \bibnamefont {Gorshkov}}, \bibinfo {author} {\bibfnamefont {Z.-X.}\
  \bibnamefont {Gong}}, \ and\ \bibinfo {author} {\bibfnamefont
  {C.}~\bibnamefont {Monroe}},\ }\href {\doibase 10.1038/nature24654}
  {\bibfield  {journal} {\bibinfo  {journal} {Nature}\ }\textbf {\bibinfo
  {volume} {551}},\ \bibinfo {pages} {601} (\bibinfo {year}
  {2017})}\BibitemShut {NoStop}%
\end{thebibliography}%

\end{document}